\documentclass[longauth,traditabstract]{aa} 
\usepackage[varg]{txfonts}
\usepackage{graphicx}
\usepackage{txfonts}
\usepackage{natbib}
\usepackage{longtable}
\usepackage{enumerate}

\begin{document}

\title{LOFAR: The LOw-Frequency ARray}

\author{
M.~P.~van Haarlem\inst{\ref{astron}} \and 
M.~W.~Wise\thanks{For questions or comments concerning this paper, please contact the corresponding author M. Wise directly at {\tt wise@astron.nl}.}\inst{\ref{astron} \and \ref{uva}} \and 
A.~W.~Gunst\inst{\ref{astron}} \and 
G.~Heald\inst{\ref{astron}} \and 
J.~P.~McKean\inst{\ref{astron}} \and
J.~W.~T.~Hessels\inst{\ref{astron} \and \ref{uva}} \and
A.~G.~de Bruyn\inst{\ref{astron} \and \ref{kapteyn}} \and
R.~Nijboer\inst{\ref{astron}} \and 
J.~Swinbank\inst{\ref{uva}} \and
R.~Fallows\inst{\ref{astron}} \and
M.~Brentjens\inst{\ref{astron}} \and 
A.~Nelles\inst{\ref{nijmegen}} \and 
R.~Beck\inst{\ref{mpifr}} \and 
H.~Falcke\inst{\ref{nijmegen} \and \ref{astron}} \and 
R.~Fender\inst{\ref{soton}} \and
J.~H\"orandel\inst{\ref{nijmegen}} \and
L.~V.~E.~Koopmans\inst{\ref{kapteyn}} \and 
G.~Mann\inst{\ref{aip}} \and 
G.~Miley\inst{\ref{leiden}} \and 
H.~R\"ottgering\inst{\ref{leiden}} \and 
B.~W.~Stappers\inst{\ref{jod}} \and 
R.~A.~M.~J.~Wijers\inst{\ref{uva}} \and
S.~Zaroubi\inst{\ref{kapteyn}} \and 
M.~van den Akker\inst{\ref{nijmegen}} \and
A. ~Alexov\inst{\ref{uva}} \and
J.~Anderson\inst{\ref{mpifr}} \and 
K.~Anderson\inst{\ref{uva}} \and
A.~van Ardenne\inst{\ref{astron} \and \ref{chalmers}} \and 
M.~Arts\inst{\ref{astron}} \and 
A.~Asgekar\inst{\ref{astron}} \and 
I.~M.~Avruch\inst{\ref{astron} \and \ref{kapteyn}} \and
F.~Batejat\inst{\ref{oso}} \and 
L.~B\"ahren\inst{\ref{uva}} \and
M.~E.~Bell\inst{\ref{soton}} \and
M.~R.~Bell\inst{\ref{mpifa}} \and 
I.~van Bemmel\inst{\ref{astron}} \and 
P.~Bennema\inst{\ref{astron}} \and 
M.~J.~Bentum\inst{\ref{astron}} \and
G.~Bernardi\inst{\ref{kapteyn}} \and
P.~Best\inst{\ref{roe}} \and 
L.~B\^{i}rzan\inst{\ref{leiden}} \and
A.~Bonafede\inst{\ref{bremen}} \and
A.-J.~Boonstra\inst{\ref{astron}} \and
R.~Braun\inst{\ref{csiro}} \and
J.~Bregman\inst{\ref{astron}} \and
F.~Breitling\inst{\ref{aip}} \and 
R.~H.~van de Brink\inst{\ref{astron}} \and 
J.~Broderick\inst{\ref{soton}} \and
P.~C.~Broekema\inst{\ref{astron}} \and 
W.~N.~Brouw\inst{\ref{astron} \and \ref{kapteyn}} \and
M.~Br\"uggen\inst{\ref{hamburg}} \and
H.~R.~Butcher\inst{\ref{astron} \and \ref{anu}} \and 
W.~van Cappellen\inst{\ref{astron}} \and 
B.~Ciardi\inst{\ref{mpifa}} \and 
T.~Coenen\inst{\ref{uva}} \and
J.~Conway\inst{\ref{oso}} \and 
A.~Coolen\inst{\ref{astron}} \and 
A.~Corstanje\inst{\ref{nijmegen}} \and
S.~Damstra\inst{\ref{astron}} \and 
O.~Davies\inst{\ref{stfc}} \and
A.~T.~Deller\inst{\ref{astron}} \and
R.-J.~Dettmar\inst{\ref{raiub}} \and
G.~van Diepen\inst{\ref{astron}} \and
K.~Dijkstra\inst{\ref{groningen}} \and  
P.~Donker\inst{\ref{astron}} \and 
A.~Doorduin\inst{\ref{astron}} \and 
J.~Dromer\inst{\ref{astron}} \and 
M.~Drost\inst{\ref{astron}} \and
A.~van Duin\inst{\ref{astron}} \and
J.~Eisl\"offel\inst{\ref{tls}} \and 
J.~van Enst\inst{\ref{astron}} \and
C.~Ferrari\inst{\ref{cotedazur}} \and
W.~Frieswijk\inst{\ref{astron}} \and 
H.~Gankema\inst{\ref{kapteyn}} \and 
M.~A.~Garrett\inst{\ref{astron} \and \ref{leiden}} \and 
F.~de Gasperin\inst{\ref{mpifa}} \and
M.~Gerbers\inst{\ref{astron}} \and 
E.~de Geus\inst{\ref{astron}} \and
J.-M.~Grie\ss{}meier\inst{\ref{cnrs} \and \ref{astron}} \and 
T.~Grit\inst{\ref{astron}} \and 
P.~Gruppen\inst{\ref{astron}} \and 
J.~P.~Hamaker\inst{\ref{astron}} \and
T.~Hassall\inst{\ref{jod}} \and
M.~Hoeft\inst{\ref{tls}} \and 
H.~Holties\inst{\ref{astron}} \and 
A.~Horneffer\inst{\ref{mpifr} \and \ref{nijmegen}} \and
A.~van der Horst\inst{\ref{uva}} \and
A.~van Houwelingen\inst{\ref{astron}} \and 
A.~Huijgen\inst{\ref{astron}} \and
M.~Iacobelli\inst{\ref{leiden}} \and 
H.~Intema\inst{\ref{leiden} \and \ref{nrao}} \and
N.~Jackson\inst{\ref{jod}} \and 
V.~Jelic\inst{\ref{astron}} \and 
A.~de Jong\inst{\ref{astron}} \and
E.~Juette\inst{\ref{raiub}} \and  
D.~Kant\inst{\ref{astron}} \and
A.~Karastergiou\inst{\ref{ox}} \and
A.~Koers\inst{\ref{astron}} \and
H.~Kollen\inst{\ref{astron}} \and
V.~I.~Kondratiev\inst{\ref{astron}} \and
E.~Kooistra\inst{\ref{astron}} \and
Y.~Koopman\inst{\ref{astron}} \and
A.~Koster\inst{\ref{astron}} \and
M.~Kuniyoshi\inst{\ref{mpifr}} \and
M.~Kramer\inst{\ref{mpifr} \and \ref{jod}} \and
G.~Kuper\inst{\ref{astron}} \and
P.~Lambropoulos\inst{\ref{astron}} \and 
C.~Law\inst{\ref{berkley} \and \ref{uva}} \and  
J.~van Leeuwen\inst{\ref{astron} \and \ref{uva}} \and
J.~Lemaitre\inst{\ref{astron}} \and 
M.~Loose\inst{\ref{astron}} \and 
P.~Maat\inst{\ref{astron}} \and
G.~Macario\inst{\ref{cotedazur}} \and
S.~Markoff\inst{\ref{uva}} \and
J.~Masters\inst{\ref{nrao} \and \ref{uva}} \and
D.~McKay-Bukowski\inst{\ref{stfc}} \and 
H.~Meijering\inst{\ref{astron}} \and 
H.~Meulman\inst{\ref{astron}} \and 
M.~Mevius\inst{\ref{kapteyn}} \and
E.~Middelberg\inst{\ref{raiub}} \and   
R.~Millenaar\inst{\ref{astron}} \and 
J.~C.~A.~Miller-Jones\inst{\ref{curtin} \and \ref{uva}} \and 
R.~N.~Mohan\inst{\ref{leiden}} \and 
J.~D.~Mol\inst{\ref{astron}} \and
J.~Morawietz\inst{\ref{astron}} \and
R.~Morganti\inst{\ref{astron} \and \ref{kapteyn}} \and 
D.~D.~Mulcahy\inst{\ref{mpifr}} \and
E.~Mulder\inst{\ref{astron}} \and
H.~Munk\inst{\ref{astron}} \and
L.~Nieuwenhuis\inst{\ref{astron}} \and 
R.~van Nieuwpoort\inst{\ref{astron} \and \ref{escience}} \and 
J.~E.~Noordam\inst{\ref{astron}} \and 
M.~Norden\inst{\ref{astron}} \and 
A.~Noutsos\inst{\ref{mpifr}} \and
A.~R.~Offringa\inst{\ref{kapteyn}} \and 
H.~Olofsson\inst{\ref{oso}} \and 
A.~Omar\inst{\ref{astron}} \and
E.~Orr\'{u}\inst{\ref{nijmegen} \and \ref{astron}} \and 
R.~Overeem\inst{\ref{astron}} \and
H.~Paas\inst{\ref{groningen}} \and  
M.~Pandey-Pommier\inst{\ref{leiden} \and \ref{lyon}} \and 
V.~N.~Pandey\inst{\ref{kapteyn}} \and 
R.~Pizzo\inst{\ref{astron}} \and 
A.~Polatidis\inst{\ref{astron}} \and 
D.~Rafferty\inst{\ref{leiden}} \and
S.~Rawlings\inst{\ref{ox}} \and
W.~Reich\inst{\ref{mpifr}} \and 
J.-P.~de Reijer\inst{\ref{astron}} \and 
J.~Reitsma\inst{\ref{astron}} \and 
A.~Renting\inst{\ref{astron}} \and 
P.~Riemers\inst{\ref{astron}} \and 
E.~Rol\inst{\ref{uva}} \and
J.~W.~Romein\inst{\ref{astron}} \and
J.~Roosjen\inst{\ref{astron}} \and 
M.~Ruiter\inst{\ref{astron}} \and 
A.~Scaife\inst{\ref{soton}} \and
K.~van der Schaaf\inst{\ref{astron}} \and 
B.~Scheers\inst{\ref{uva} \and \ref{cwi}} \and 
P.~Schellart\inst{\ref{nijmegen}} \and
A.~Schoenmakers\inst{\ref{astron}} \and 
G.~Schoonderbeek\inst{\ref{astron}} \and
M.~Serylak\inst{\ref{nancay} \and \ref{cnrs}} \and 
A.~Shulevski\inst{\ref{kapteyn}} \and 
J.~Sluman\inst{\ref{astron}} \and 
O.~Smirnov\inst{\ref{astron}} \and
C.~Sobey\inst{\ref{mpifr}} \and 
H.~Spreeuw\inst{\ref{uva}} \and 
M.~Steinmetz\inst{\ref{aip}} \and 
C.~G.~M.~Sterks\inst{\ref{groningen}} \and  
H.-J.~Stiepel\inst{\ref{astron}} \and 
K.~Stuurwold\inst{\ref{astron}} \and 
M.~Tagger\inst{\ref{cnrs}} \and 
Y.~Tang\inst{\ref{astron}} \and 
C.~Tasse\inst{\ref{meudon}} \and
I.~Thomas\inst{\ref{astron}} \and
S.~Thoudam\inst{\ref{nijmegen}} \and 
M.~C.~Toribio\inst{\ref{astron}} \and 
B.~van der Tol\inst{\ref{leiden}} \and 
O.~Usov\inst{\ref{leiden}} \and 
M.~van Veelen\inst{\ref{astron}} \and 
A.-J.~ van der Veen\inst{\ref{astron}} \and 
S.~ter Veen\inst{\ref{nijmegen}} \and
J.~P.~W.~Verbiest\inst{\ref{mpifr}} \and 
R.~Vermeulen\inst{\ref{astron}} \and 
N.~Vermaas\inst{\ref{astron}} \and 
C.~Vocks\inst{\ref{aip}} \and 
C.~Vogt\inst{\ref{astron}} \and 
M.~de Vos\inst{\ref{astron}} \and 
E.~van der Wal\inst{\ref{astron}} \and 
R.~van Weeren\inst{\ref{leiden} \and \ref{astron}} \and 
H.~Weggemans\inst{\ref{astron}} \and 
P.~Weltevrede\inst{\ref{jod}} \and
S.~White\inst{\ref{mpifa}} \and 
S.~J.~Wijnholds\inst{\ref{astron}} \and 
T.~Wilhelmsson\inst{\ref{mpifa}} \and 
O.~Wucknitz\inst{\ref{ubonn}} \and 
S.~Yatawatta\inst{\ref{kapteyn}} \and 
P.~Zarka\inst{\ref{meudon}} \and
A.~Zensus\inst{\ref{mpifr}} \and
J.~van Zwieten\inst{\ref{astron}}
}

\institute{
Netherlands Institute for Radio Astronomy (ASTRON), Postbus 2, 7990 AA Dwingeloo, The Netherlands\label{astron}
\and 
Astronomical Institute 'Anton Pannekoek', University of Amsterdam, Postbus 94249, 1090 GE Amsterdam, The Netherlands\label{uva}
\and 
Kapteyn Astronomical Institute, P.O. Box 800, 9700 AV Groningen, The Netherlands\label{kapteyn}
\and
Leiden Observatory, Leiden University, P.O. Box 9513, 2300 RA Leiden, The Netherlands\label{leiden}
\and
Department of Astrophysics/IMAPP, Radboud University Nijmegen, P.O. Box 9010, 6500 GL Nijmegen, The Netherlands\label{nijmegen}
\and
Jodrell Bank Center for Astrophysics, School of Physics and Astronomy, The University of Manchester, Manchester M13 9PL,UK\label{jod}
Astrophysics, University of Oxford, Denys Wilkinson Building, Keble Road, Oxford OX1 3RH\label{ox}
\and
Max-Planck-Institut f\"ur Radioastronomie, Auf dem H\"ugel 69, 53121 Bonn, Germany\label{mpifr}
\and 
School of Physics and Astronomy, University of Southampton, Southampton, SO17 1BJ, UK\label{soton}
\and 
Max Planck Institute for Astrophysics, Karl Schwarzschild Str. 1, 85741 Garching, Germany\label{mpifa}
\and 
Department of Physics \& Astronomy, Hicks Building, Hounsfield Road, Sheffield S3 7RH, United Kingdom\label{sheff}
\and
Onsala Space Observatory, Dept. of Earth and Space Sciences, Chalmers University of Technology, SE-43992 Onsala, Sweden\label{oso}
\and
International Centre for Radio Astronomy Research - Curtin University, GPO Box U1987, Perth, WA 6845, Australia\label{curtin}
\and
STFC Rutherford Appleton Laboratory,  Harwell Science and Innovation Campus,  Didcot  OX11 0QX, UK\label{stfc}
\and
Institute for Astronomy, University of Edinburgh, Royal Observatory of Edinburgh, Blackford Hill, Edinburgh EH9 
3HJ, UK\label{roe}
\and
LESIA, Observatoire de Paris, CNRS, UPMC, Universit\'{e} Paris Diderot, 92190 Meudon, France\label{meudon}
\and
Argelander-Institut f\"ur Astronomie, University of Bonn, Auf dem H\"ugel 71, 53121, Bonn, Germany\label{ubonn}
\and
Leibniz-Institut für Astrophysik Potsdam (AIP), An der Sternwarte 16, 14482 Potsdam, Germany\label{aip}
\and
Th\"uringer Landessternwarte, Sternwarte 5, D-07778 Tautenburg, Germany\label{tls}
\and
Astronomisches Institut der Ruhr-Universit\"at Bochum, Universit\"atsstrasse 150, 44780 Bochum, Germany\label{raiub}
\and
Universit\"at Hamburg, Hamburger Sternwarte, Gojenbergsweg 112, 21029 Hamburg, Germany\label{hamburg}
\and
Jacobs University Bremen, Campus Ring 1, 28759 Bremen, Germany\label{bremen}
\and
Laboratoire de Physique et Chimie de l'Environnement et de l'Espace, CNRS/Universit\'e d'Orl\'eans, LPC2E UMR 7328 CNRS, 45071 Orl\'eans Cedex 02, France\label{cnrs}
\and
Center for Information Technology (CIT), University of Groningen, The Netherlands\label{groningen}
\and
Radio Astronomy Lab, UC Berkeley, CA, USA\label{berkley}
\and
Centre de Recherche Astrophysique de Lyon, Observatoire de Lyon, 9 av Charles Andr\'e, 69561 Saint Genis Laval Cedex, France\label{lyon}
\and
Mt. Stromlo Obs., Research School of Astronomy and Astrophysics, Australian National University, Weston, A.C.T. 2611, Australia\label{anu}
\and
CSIRO Australia Telescope National Facility, P.O. Box 76, Epping NSW 1710, Australia\label{csiro}
\and
National Radio Astronomy Observatory, 520 Edgemont Road, Charlottesville, VA 22903-2475, USA\label{nrao}
\and
Chalmers University of Technology, SE-412 96 Gothenburg, Sweden\label{chalmers}
\and
Observatoire de la C\^{o}te d'Azur, D\'{e}partement Lagrange, Boulevard de l'Observatoire, B.P. 4229, F-06304 NICE Cedex 4, France\label{cotedazur}
\and
Station de Radioastronomie de Nan\c{c}ay, Observatoire de Paris, CNRS/INSU, 18330 Nan\c{c}ay, France\label{nancay}
\and
Netherlands eScience Center, Science Park 140, 1098 XG Amsterdam, The Netherlands\label{escience}
\and
Centrum Wiskunde \& Informatica, P.O. Box 94079, 1090 GB Amsterdam, The Netherlands\label{cwi}
}

\date{Received December 7, 2012; accepted May 9, 2013}

\titlerunning{LOFAR: The Low-Frequency Array}
\authorrunning{van Haarlem et al. ~~}

\abstract{
\vspace{-0.35cm}
LOFAR, the LOw-Frequency ARray, is a new-generation radio interferometer constructed in the north of the Netherlands and across europe. Utilizing a novel phased-array design, LOFAR covers the largely unexplored low-frequency range from 10--240\,MHz and provides a number of unique observing capabilities. Spreading out from a core located near the village of Exloo in the northeast of the Netherlands, a total of 40 LOFAR stations are nearing completion. A further five stations have been deployed throughout Germany, and one station has been built in each of France, Sweden, and the UK. Digital beam-forming techniques make the LOFAR system agile and allow for rapid repointing of the telescope as well as the potential for multiple simultaneous observations. With its dense core array and long interferometric baselines, LOFAR achieves unparalleled sensitivity and angular resolution in the low-frequency radio regime. The LOFAR facilities are jointly operated by the International LOFAR Telescope (ILT) foundation, as an observatory open to the global astronomical community. LOFAR is one of the first radio observatories to feature automated processing pipelines to deliver fully calibrated science products to its user community. LOFAR's new capabilities, techniques and modus operandi make it an important pathfinder for the Square Kilometre Array (SKA). We give an overview of the LOFAR instrument, its major hardware and software components, and the core science objectives that have driven its design. In addition, we present a selection of new results from the commissioning phase of this new radio observatory.
}


\maketitle

\begin{figure*}[ht]
\centering
\includegraphics[width=\textwidth]{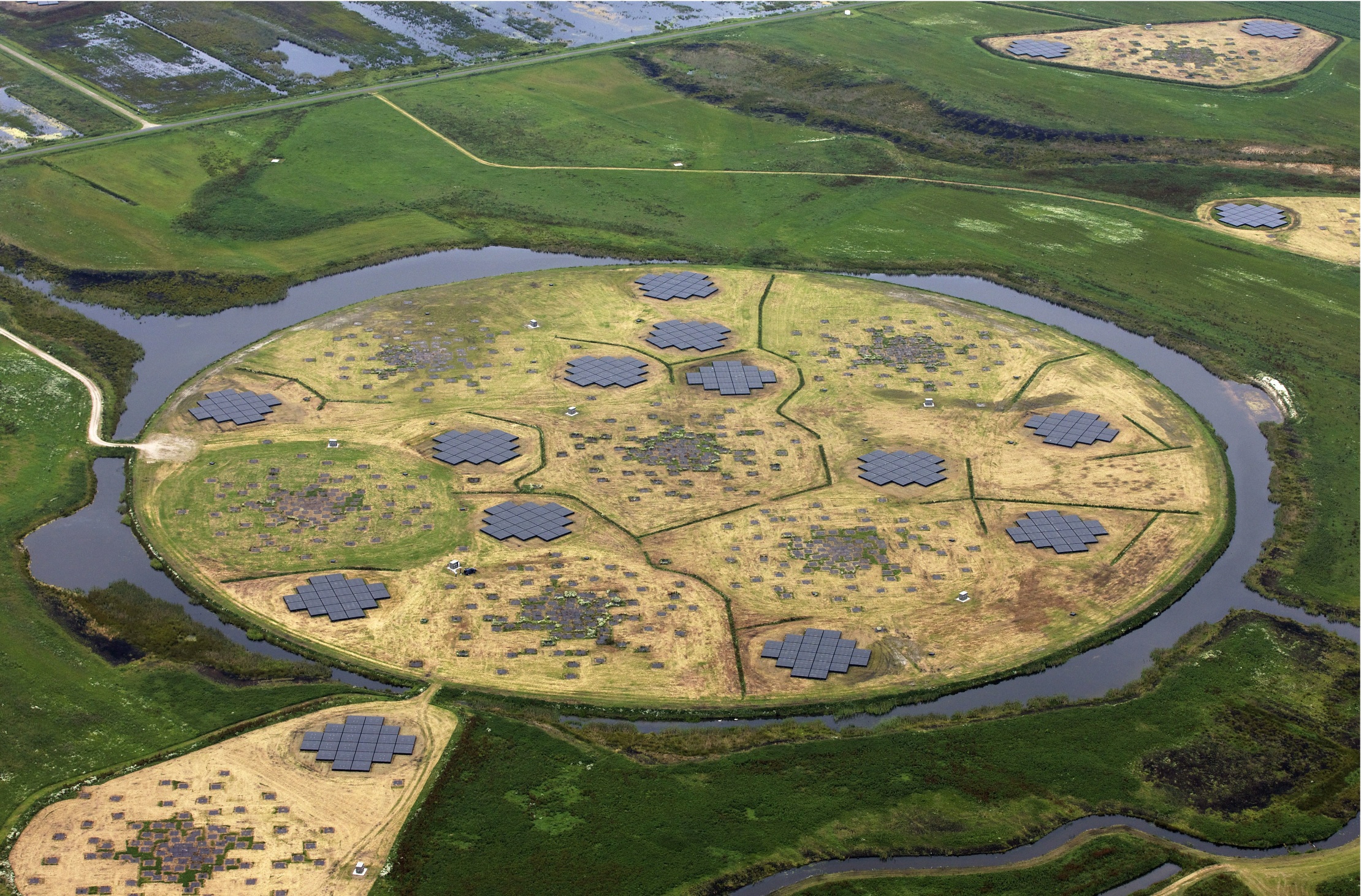}
\caption{\small
Aerial photograph of the Superterp, the heart of the LOFAR core, from August 2011. The large circular island encompasses the six core stations that make up the Superterp. Three additional LOFAR core stations are visible in the upper right and lower left of the image. Each of these core stations includes a field of 96 low-band antennas and two sub-stations of 24 high-band antenna tiles each.}
\label{fig:superterp}
\end{figure*}

\section{Introduction}
\label{sec:intro}

During the last half century, our knowledge of the Universe has been revolutionized by the opening of observable windows outside the narrow visible region of the electromagnetic spectrum. Radio waves, infrared, ultraviolet, X-rays, and most recently $\gamma$-rays have all provided new, exciting, and completely unexpected information about the nature and history of the Universe, as well as revealing a cosmic zoo of strange and exotic objects.   One spectral window that as yet remains relatively unexplored is the low-frequency radio domain below a few hundred MHz, representing the lowest frequency extreme of the accessible spectrum.

Since the discovery of radio emission from the Milky Way \citep{Jansky33}, now 80 years ago, radio astronomy has made a continuous stream of fundamental contributions to astronomy. Following the first large-sky surveys in Cambridge, yielding the 3C and 4C catalogs \citep{Edge59,Bennett62,Pilkington65,Gower67} containing hundreds to thousands of radio sources, radio  astronomy has blossomed.  Crucial events in those early years were the identifications of the newly discovered radio sources in the optical waveband. Radio astrometric techniques, made possible through both interferometric and lunar occultation techniques, led to the systematic classification of many types of radio sources:  Galactic supernova remnants (such as the Crab Nebula and Cassiopeia A), normal galaxies (M31), powerful radio galaxies (Cygnus A), and quasars (3C48 and 3C273).  

During this same time period, our understanding of the physical processes responsible for the radio emission also progressed rapidly. The discovery of powerful very low-frequency coherent cyclotron radio emission from Jupiter \citep{Burke55} and the nature of radio galaxies and quasars in the late 1950s was rapidly followed by such fundamental discoveries as the Cosmic Microwave Background \citep{Penzias65}, pulsars \citep{Bell67}, 
and apparent superluminal motion in compact extragalactic radio sources by the 1970s \citep{Whitney71}.   

Although the first two decades of radio astronomy were dominated by observations below a few hundred MHz, the prediction and subsequent detection of the 21cm line of hydrogen at 1420 MHz \citep{vandeHulst45,Ewen51}, as well as the quest for higher angular resolution, shifted attention to higher frequencies. This shift toward higher frequencies was also driven in part by developments in receiver technology, interferometry, aperture synthesis, continental and intercontinental very long baseline interferometry (VLBI). Between 1970 and 2000, discoveries in radio astronomy were indeed dominated by the higher frequencies using aperture synthesis arrays in Cambridge, Westerbork, the VLA, MERLIN, ATCA and the GMRT in India as well as large monolithic dishes at Parkes, Effelsberg, Arecibo, Green Bank, Jodrell Bank, and Nan\c{c}ay. 

By the mid 1980s to early 1990s, however, several factors combined to cause a renewed interest in low-frequency radio astronomy. Scientifically, the realization that many sources have inverted radio spectra due to synchrotron self-absorption or free-free absorption as well as the detection of (ultra-) steep spectra in pulsars and high redshift radio galaxies highlighted the need for data at lower frequencies. Further impetus for low-frequency radio data came from early results from Clark Lake \citep{Erickson74, Kassim88}, the Cambridge sky surveys at 151 MHz, and the 74 MHz receiver system at the VLA \citep{Kassim93, Kassim07}. In this same period, a number of arrays were constructed around the world to explore the sky at frequencies well below 1 GHz \citep[see Table 2 in][and references therein]{sha+11}.

Amidst all this progress, radio astronomers nonetheless began to look toward the future and one ambition that emerged was the proposed construction of an instrument capable of detecting neutral hydrogen at cosmological distances. A first order analysis, suggested that a telescope with a collecting area of about one square kilometer was required \citep{Wilkinson91}. The project, later to be known as the Square Kilometre Array \citep[SKA;][]{Ekers12}, was adopted by the community globally, and around the world various institutes began to consider potential technologies that might furnish such a huge collecting area at an affordable cost. 

At ASTRON in the Netherlands, the concept of Phased or Aperture Arrays was proposed as a possible solution to this problem \citep{vanArdenne00}, and in the slip-stream of those early developments, the idea of constructing a large low frequency dipole array also emerged \citep{Bregman00,Miley10}. The concept of a large, low frequency array had arisen previously \citep{Perley84}, and been revisited several times over the years \citep[e.g., see][]{Kassim98}. These plans were greatly aided by the revolution then taking place in other fields, in particular major advances in digital electronics, fibre-based data networks, high performance computing and storage capacity, made it possible to consider the construction of a transformational radio telescope design that would operate between 10--200\,MHz with unprecedented sensitivity and angular resolution. This telescope would be a major scientific instrument in its own right, bridging the gap to the even more ambitious SKA \citep{Miley10}. This international initiative became known as the LOw Frequency ARray or LOFAR \citep{Bregman00, Kassim03, Butcher04}.

As originally envisioned, LOFAR was intended to surpass the power of previous interferometers in its frequency range by 2-3 orders of magnitude providing a square kilometer of collecting area at 15 MHz, millijansky sensitivity, and arcsecond resolution \citep{Kassim03}. Due to funding constraints, the original collaboration split in 2004 resulting in three currently ongoing low-frequency array development projects: the European LOFAR project described here; the US-led, Long Wavelength Array \citep[LWA;][]{LWA09, LWA13}; and the international Murchison Widefield Array (MWA) collaboration \citep{MWA09,Tingay13}.

The scientific motivation for the construction of these arrays has become very broad.  Among the most interesting application of the low-frequency arrays is the detection of highly redshifted 21cm line emission from the epoch of reionization (HI redshifts z=6 to 20) and a phase called Cosmic Dawn \citep[HI redshifts from z=50 to z=20; see][]{zaroubi12}. However, the science case for LOFAR has continued to broaden since 2000 to include the detection of nanosecond radio flashes from ultra-high energy cosmic rays \citep[CRs;][]{Falcke05}, deep surveys of the sky in search for high redshift radio sources \citep{Rottgering11}, surveys of pulsars and cosmic radio transients \citep{sha+11}, or the radio detection of exoplanets \citep{Zarka11}. The great sensitivity and broad low-frequency bandwidth may also prove crucial for studies of cosmic magnetic fields (see Sect.\,\ref{sec:magnetism}).

In this paper, we present an overview and reference description of the LOFAR telescope. We aim to give the potential LOFAR user a general working knowledge of the main components and capabilities of the system. More detailed descriptions of individual components or subsystems will be published elsewhere. The paper continues in Sect.\,\ref{sec:overview} with a general overview of the system and descriptions of the overall layout of the array and the antenna fields themselves in Sect.\,\ref{sec:array} and Sect.\,\ref{sec:stations}. The LOFAR processing hardware and data-flow through the system are summarized in Sect.\,\ref{sec:wan} and Sect.\,\ref{sec:cep}. An overview of the software infrastructure including a description of LOFAR's primary observational modes and science pipelines is given in Sect.\,\ref{sec:control}, Sect.\,\ref{sec:modes}, and Sect.\,\ref{sec:pipelines}, respectively. In Sect.\,\ref{sec:performance}, an initial set of performance metrics are presented. LOFAR's key science drivers are reviewed in Sect.\,\ref{sec:ksps} along with examples of recent results that demonstrate the potential of this new facility. A discussion of ongoing construction plans and possible future enhancements to the system are given in Sect.\,\ref{sec:future}.  Lastly, Sect.\,\ref{sec:conclude} offers some brief conclusions.

\begin{figure*}[ht]
\centering
\includegraphics[height=3.2in]{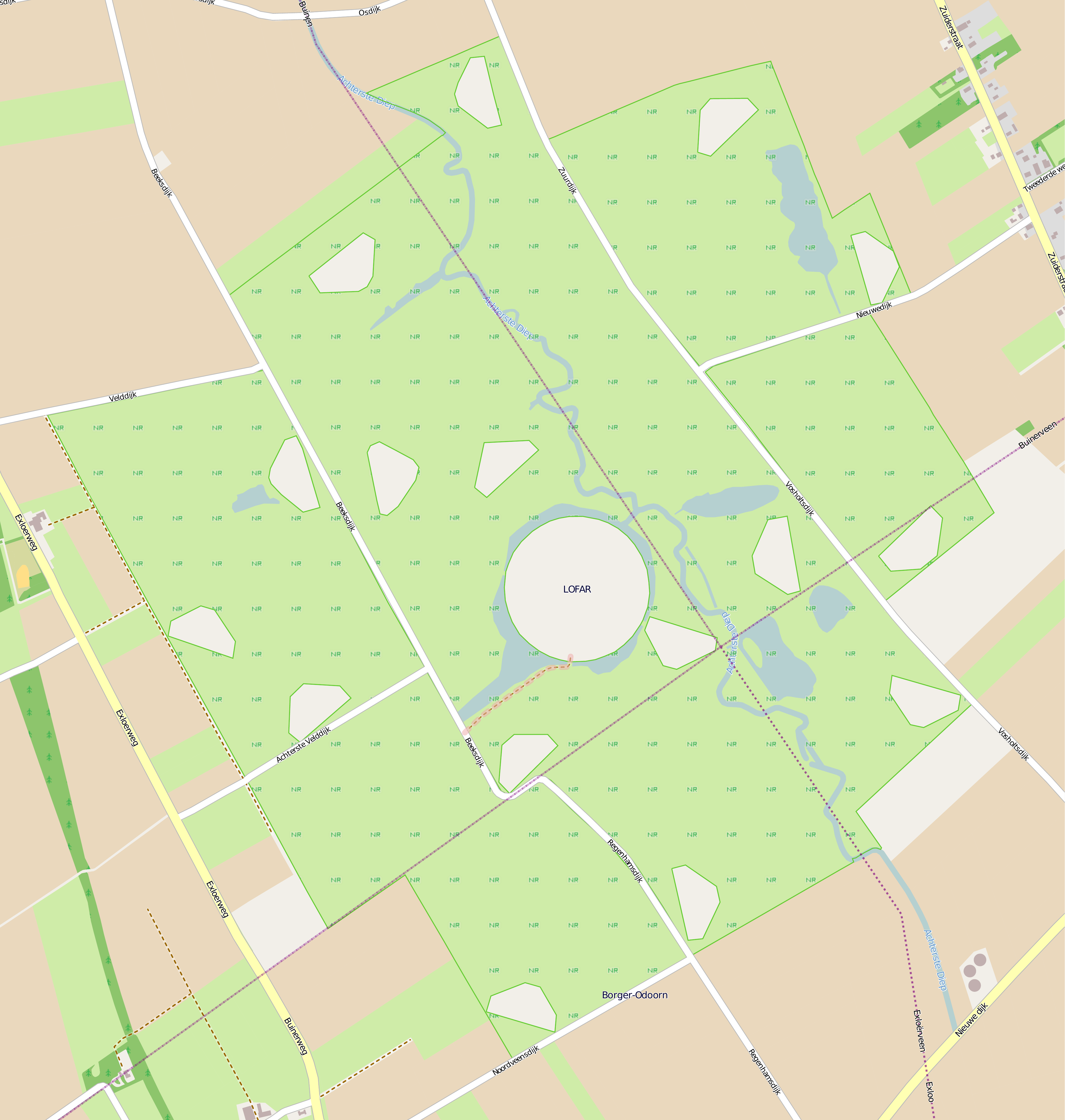}
\hspace{0.1in}
\includegraphics[height=3.2in]{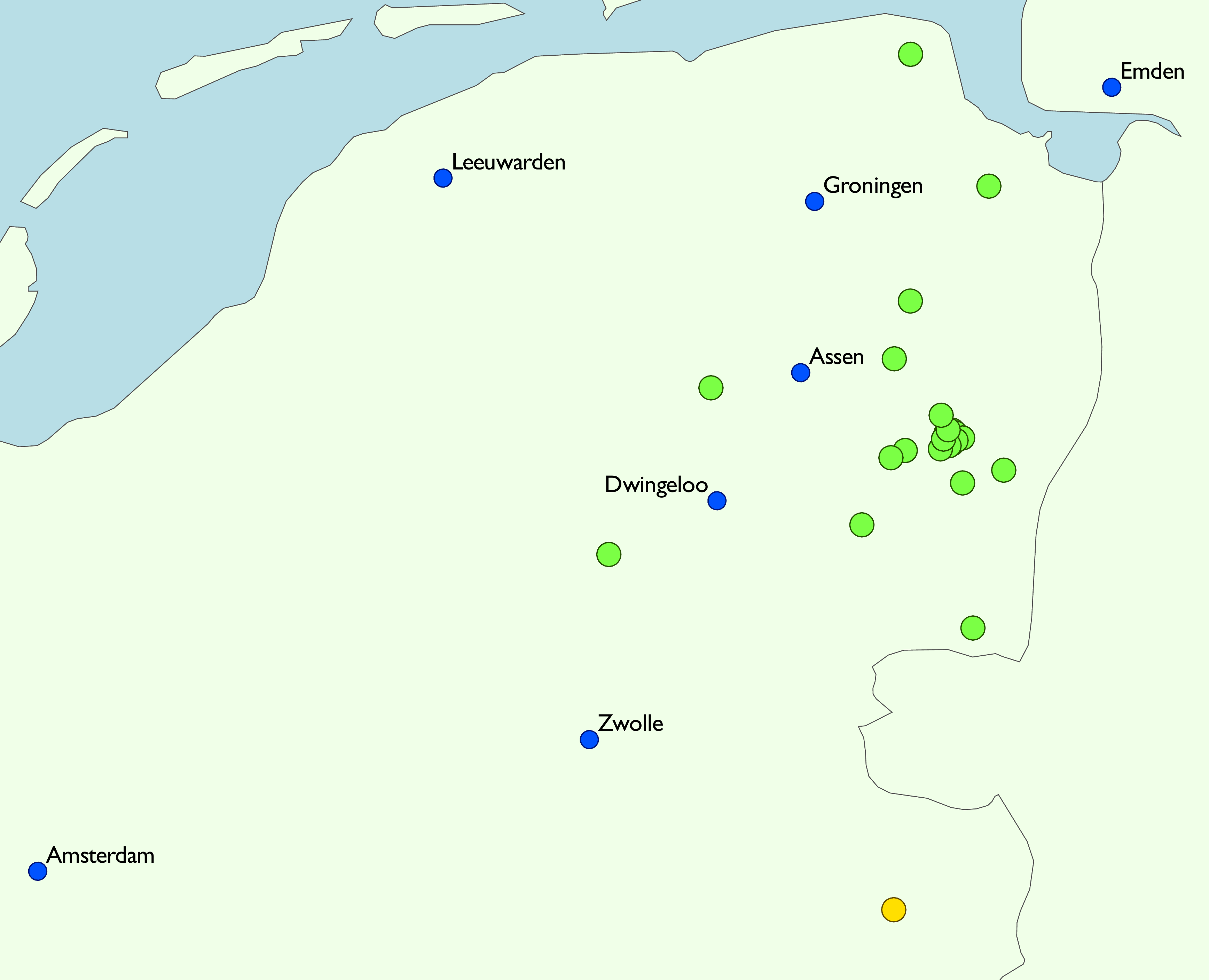}
\caption{\small Geographic distribution of LOFAR stations within the Netherlands. {\bf Left:} This panel shows the distribution for the majority of the stations within the LOFAR core. The central, circular area contains the six Superterp stations described in the text. The white, polygonal areas mark the location of LOFAR core stations. In addition to the Superterp stations, 16 of the remaining 18 core stations are shown. {\bf Right:} This panel shows the distribution of remote stations within the Netherlands located at distances of up to 90 km from the center of the array. Stations shown in green are complete and operational while yellow depicts stations that are under construction as of March 2013 (see Sect.\,\ref{sec:rollout}).}
\label{fig:nl}
\end{figure*}

\section{System overview}
\label{sec:overview}

LOFAR, the LOw-Frequency ARray, is a new and innovative radio telescope designed and constructed by ASTRON to open the lowest frequency radio regime to a broad range of astrophysical studies.  Capable of operating in the frequency range from 10--240\,MHz (corresponding to wavelengths of 30--1.2\,m), LOFAR consists of an interferometric array of dipole antenna stations distributed throughout the Netherlands and Europe. These stations have no moving parts and, due to the effectively all-sky coverage of the component dipoles, give LOFAR a large field-of-view (FoV). At station level, the signals from individual dipoles are combined digitally into a phased array. Electronic beam-forming techniques make the system agile and allow for rapid repointing of the telescope as well as the simultaneous observation of multiple, independent areas of the sky. Brief descriptions of the LOFAR system have been presented previously in \cite{Bregman00, Falcke06, Falcke07, deVos09}.

In the Netherlands, a total of 40 LOFAR stations are being deployed with an additional 8 international stations currently built throughout Europe. The densely sampled, 2-km-wide, core hosts 24 stations and is located $\sim$30 km from ASTRON's headquarters in Dwingeloo. The datastreams from all LOFAR stations are sent via a high-speed fiber network infrastructure to a central processing (CEP) facility located in Groningen in the north of the Netherlands. At the computing center of the University of Groningen, data from all stations are aligned, combined, and further processed using a flexible IBM Blue Gene/P supercomputer offering about 28 Tflop/s of processing power. In the Blue Gene/P, a variety of processing operations are possible including correlation for standard interferometric imaging, tied-array beam-forming for high time resolution observations, and even real-time triggering on incoming station data-streams. Combinations of these operations can also be run in parallel.

After processing in the Blue Gene/P, raw data products are written to a storage cluster for additional post-processing. This cluster currently hosts 2 Pbyte of working storage. Once on the storage cluster, a variety of reduction pipelines are then used to further process the data into the relevant scientific data products depending on the specific type of observation. In the case of the standard imaging pipeline, subsequent processing includes flagging of the data for the presence of radio frequency interference, averaging, calibration, and creation of the final images. This and other science-specific pipelines run on a dedicated compute cluster with a total processing power of approximately 10 Tflop/s. After processing, the final scientific data products are transferred to the LOFAR long-term archive (LTA) for cataloging and distribution to the community.

In order to fully exploit this new wavelength regime with unprecedented resolution and sensitivity, LOFAR must meet several non-trivial technical challenges. For example, the meter-wave wavelength regime is prone to high levels of man-made interference.  Excising this interference requires high spectral and time resolution, and high dynamic range analog to digital (A/D) converters. Furthermore, for the typical sampling rate of 200 MHz, the raw data-rate generated by the entire LOFAR array is 13 Tbit/s, far too much to transport in total. Even utilizing beam-forming at the station level, the long range data transport rates over the array are of order 150 Gbit/s requiring dedicated fibre networks. Such large data transport rates naturally also imply data storage challenges. For example, typical interferometric imaging observations can easily produce 35 Tbyte/h of raw, correlated visibilities. LOFAR is one of the first of a number of new astronomical facilities coming online that must deal with the transport, processing, and storage of these large amounts of data. In this sense, LOFAR represents an important technological pathfinder for the SKA and data intensive astronomy in the coming decade.

In addition to hardware and data transport challenges, LOFAR faces many technical challenges that are conceptual or algorithmic in nature.  Low-frequency radio signals acquire phase-shifts due to variations in the total electron content of the ionosphere. For baselines longer than a few kilometers, the dynamic and non-isoplanatic nature of the ionosphere has a dramatic impact on the quality of the resulting scientific data. Correcting for these effects in LOFAR data has required improving existing calibration techniques that can simultaneously determine multi-directional station gain solutions to operate in the near, real-time regime. Likewise, LOFAR's huge FoV means the traditional interferometric assumption of a coplanar array is no longer valid. Consequently, highly optimized versions of imaging algorithms that recognize that the interferometric response and the sky brightness are no longer related by a simple 2-D Fourier transform were required. These and similar issues have driven much of the design for LOFAR's processing software and computational architecture.

Scientifically, this new technology makes LOFAR a powerful and versatile instrument. With the longer European baselines in place, LOFAR can achieve sub-arcsecond angular resolution over most of its 30--240\,MHz nominal operating bandpass, limited primarily by atmospheric effects and scattering due to interplanetary scintillation (IPS). This resolution, when combined with the large FoV, makes LOFAR an excellent instrument for all-sky surveys. Exploiting this potential has been one of LOFAR's key science drivers from its inception. The large effective area of LOFAR's densely populated core, support for multi-beaming, and inherent high time resolution also make LOFAR a breakthrough instrument for the detection and all-sky monitoring of transient radio sources. Finally, the ability to buffer large amounts of data at the dipole level provides a unique capability to perform retrospective imaging of the entire sky on short timescales. Among other applications, these buffers are used to detect radio emission from CR air showers. As discussed later, this versatility is apparent in the wide array of key science projects (KSPs) that have driven the initial design and commissioning phase.

\begin{figure*}[ht]
\centering
\includegraphics[width=\textwidth]{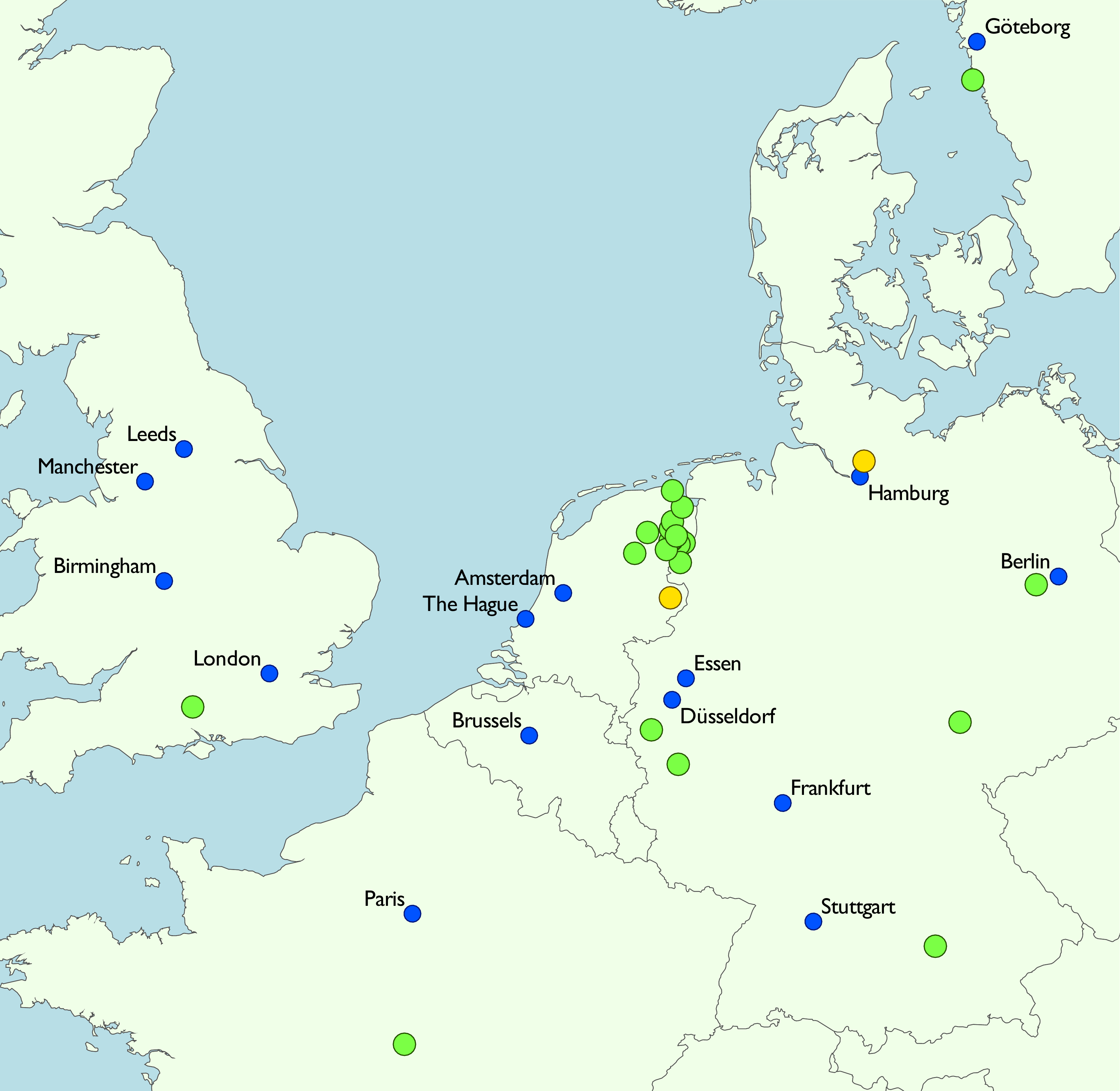}
\caption{\small Current distribution of the European LOFAR stations that have been built in Germany (5), France (1), Sweden (1) and the UK (1). The color scheme for the stations is the same as in Fig.\,\ref{fig:nl}. A sixth German station located near Hamburg (shown in yellow) has recently begun construction and is expected to be online by the end of 2013. Data from all international stations is routed through Amsterdam before transfer to CEP in Groningen, NL. For the German stations, data are first routed through J\"{u}lich before transfer on to Amsterdam (see Sect.\,\ref{sec:wan}).}
\label{fig:europe}
\end{figure*}

\section{Array configuration}
\label{sec:array}

The fundamental receiving elements of LOFAR are two types of small, relatively low-cost antennas that together cover the 30--240\,MHz operating bandpass. These antennas are grouped together into 48 separate stations distributed over the northeastern part of the Netherlands as well as in Germany, France, the UK, and Sweden. The majority of these stations, 40 in total, are distributed over an area roughly 180 km in diameter centered near the town of Exloo in the northeastern Dutch province of Drenthe. This area was chosen because of its low population density and relatively low levels of radio frequency interference (RFI). The feasibility of obtaining the land required to build the stations ($\sim$20000 m$^2$ per station) also played an important part in the final decision to site the array here. 

For the majority of the array located in the Netherlands, the geographic distribution of stations shows a strong central concentration with 24 stations located within a radius of 2\,km referred to as the ``core". Within the core, the land was purchased to allow maximum freedom in choosing station locations. This freedom allowed the core station distribution to be optimized to achieve the good instantaneous {\it uv} coverage required by many of the KSPs including the epoch of reionization (EoR) experiment and radio transients searches (see Sect.\,\ref{sec:ksps}). At the heart of the core, six stations reside on a 320 m diameter island referred to as the ``Superterp"; ``terp'' is a local name for an elevated site used for buildings as protection against rising water. These Superterp stations, shown in Fig.\,\ref{fig:superterp}, provide the shortest baselines in the array and can also be combined to effectively form a single, large station as discussed in Sect.\,\ref{sec:bfmodes}.

Beyond the core, the 16 remaining LOFAR stations in the Netherlands are arranged in an approximation to a logarithmic spiral distribution. Deviations from this optimal pattern were necessary due to the availability of land for the stations as well as the locations of existing fiber infrastructure. These outer stations extend out to a radius of 90\,km and are generally classified as ``remote" stations. As discussed below, these remote stations also exhibit a different configuration to their antenna distributions than core stations. The full distribution of core and remote stations within the Netherlands is shown in Fig.\,\ref{fig:nl}.

For the 8 international LOFAR stations, the locations were provided by the host countries and institutions that own them. Consequently, selection of their locations was driven primarily by the sites of existing facilities and infrastructure. As such, the longest baseline distribution has not been designed to achieve optimal {\it uv} coverage, although obvious duplication of baselines has been avoided. Fig.\,\ref{fig:europe} shows the location of the current set of international LOFAR stations. Examples of the resulting {\it uv} coverage for the array can be found in Sect.\,\ref{sec:performance}.

\section{Stations}
\label{sec:stations}

LOFAR antenna stations perform the same basic functions as the dishes of a conventional interferometric radio telescope. Like traditional radio dishes, these stations provide collecting area and raw sensitivity as well as pointing and tracking capabilities.  However, unlike previous generation, high-frequency radio telescopes, the antennas within a LOFAR station do not physically move. Instead, pointing and tracking are achieved by combining signals from the individual antenna elements to form a phased array using a combination of analog and digital beam-forming techniques \citep[see][]{Thompson2007}. Consequently, all LOFAR stations contain not only antennas and digital electronics, but significant local computing resources as well. 

This fundamental difference makes the LOFAR system both flexible and agile. Station-level beam-forming allows for rapid repointing of the telescope as well as the potential for multiple, simultaneous observations from a given station. The resulting digitized, beam-formed data from the stations can then be streamed to the CEP facility (see Sect.\,\ref{sec:cep}) and correlated to produce visibilities for imaging applications, or further combined into array beams (i.e. the sum of multiple stations) to produce high resolution time-series (e.g. for pulsar, CR, and solar studies). In effect, each individual LOFAR station is a fully functional radio telescope in its own right and a number of the main science drivers exploit this flexibility \citep[e.g., see Sect.\,5.3 of][]{sha+11}. In the following section, we review the major hardware and processing components of a LOFAR station.

\begin{figure*}[ht]
\centering
\includegraphics[width=\textwidth]{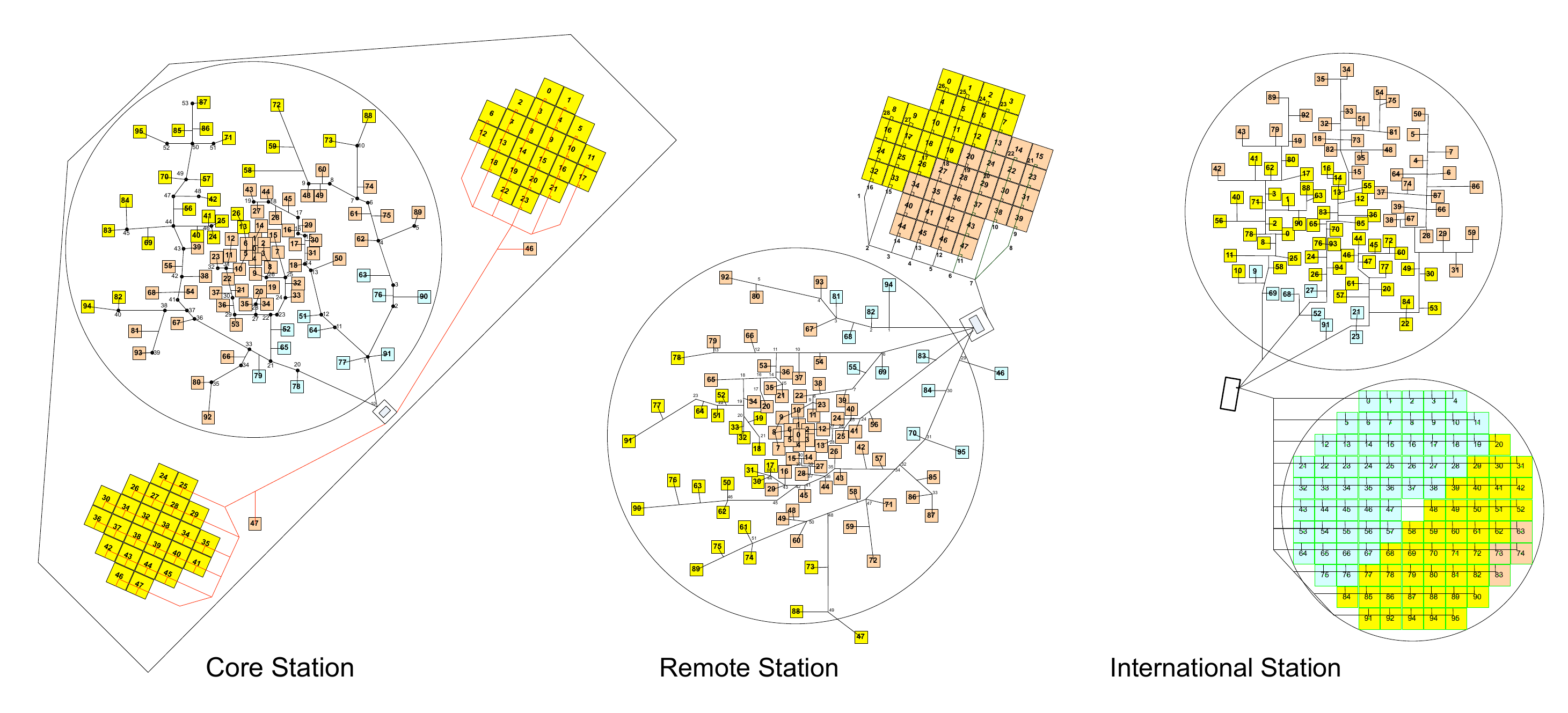}
\caption{\small Station layout diagrams showing core, remote and international stations. The large circles denote the LBA antennas while the arrays of small squares indicate the HBA tiles. Note that the station layouts are not shown on the same spatial scale.}
\label{fig:stations}
\end{figure*}

\begin{table*}[t]                                                                                                
\begin{minipage}[t]{\textwidth}
\caption{Overview of stations and antennas}
\label{tab:stations}                                                                                           
\centering
\renewcommand{\footnoterule}{}  
\begin{tabular}{lccccccc}                                                                                       
\hline \hline                                                                                                 
Station Configurations		&	~~~~~Number of Stations & 	LBA dipoles  & HBA tiles & Signal Paths
                            & Min. baseline (m) & Max. baseline (km) \\
\hline \hline
Superterp            	&	6	&			2x48	&		2x24	&   96  & 68  & 0.24    \\
NL Core Stations    	&	24	&			2x48	&		2x24	&   96  & 68  & 3.5     \\
NL Remote Stations	    &	16	&			2x48	&		48      &   96  & 68  & 121.0   \\
International Stations  &	8	&			96	    &		96	    &  192  & 68  & 1158.0  \\
\hline \hline
\end{tabular}
\tablefoot{The 6 stations comprising the central Superterp are a subset of the total 24 core stations. Please note that the tabulated baseline lengths represent unprojected values. Both the LBA dipoles and the HBA tiles are dual polarization.}
\end{minipage}
\end{table*}

\subsection{Station configurations}
\label{sec:layouts}

As discussed in Sect.\,\ref{sec:array}, LOFAR stations are classified as either core, remote, or international, nominally corresponding to their distance from the center of the array. More fundamentally, each of these three types of stations have different antenna field configurations. In its original design, all LOFAR stations were envisioned to be identical to simplify both construction and deployment as well as subsequent calibration. Due to funding considerations, this design was altered in 2006 to reduce costs while preserving the maximum number of stations possible and the corresponding quality of the {\it uv} coverage. This decision led to different choices for the antenna configurations and underlying electronics in the core, remote, and international stations. Consequently all LOFAR stations in the Netherlands have 96 signal paths that can be used to simultaneously process signals from either 48 dual-polarized or 96 single-polarized antennas. To provide sufficient sensitivity on the longest baselines, international LOFAR stations are equipped with 192 signal paths. These three station types are summarized in Table\,\ref{tab:stations}.

The geometric distribution of low-band antennas (LBAs) and high-band antennas (HBAs) for each of the three LOFAR station configurations is shown in Fig.\,\ref{fig:stations}. All stations in the Netherlands have 96 LBAs, 48 HBAs, and a total of 48 digital receiver units (RCUs). These RCUs represent the beginning of the digital signal path and feature three distinct inputs per board (see Sect.\,\ref{sec:rcu} below). For core and remote stations in the Netherlands, two of these inputs are assigned to the 96 LBA dipoles while the remaining input is used for the 48 HBA tiles. Only one of these three RCU inputs, however, can be active at any one time. As a result, whereas all 48 HBA tiles can be used at once, only half the 96 signals coming from the LBA dipoles can be processed at any given time. Operationally, the LBA dipoles are grouped into an inner circle and an outer annulus each consisting of 48 dipoles and identified as the ``LBA Inner" and ``LBA Outer" configurations, respectively. These two configurations result in different FoVs, and potentially sensitivity (due to mutual coupling of closely spaced antennas), and can be selected by the observer during the observation specification process.

As Fig.\,\ref{fig:stations} illustrates, a further distinction is apparent in the layout of HBA tiles within the core and remote stations in the Netherlands. In contrast to remote stations, where the HBAs are contained within a single field, the HBA dipoles in LOFAR core stations are distributed over two sub-stations of 24 tiles each. These core HBA sub-stations can be used in concert as a single station or separately as independent LOFAR stations. The latter option has the advantage of providing many more short baselines within the core and by extension a significantly more uniform {\it uv} coverage. In addition, many of the short baselines that result from the dual HBA sub-stations are redundant and therefore yield additional diagnostics for identifying bad phase and gain solutions during the calibration process. These advantages are especially important for science cases that depend critically on the use of the LOFAR core such as the EoR experiment or the search for radio transients.

Since the stations are constructed with a finite number of individual elements, the digitally formed station beams have non-negligible sidelobe structure. The sidelobe pattern is particularly strong for the HBA stations, because the tiles are laid out on a uniform grid. In order to reduce the effect of bright off-axis sources contributing strongly to the visibility function when located in a sidelobe, the layout of each individual station is rotated by a particular angle. This rotation in turn causes the sidelobe pattern of each station to be projected differently on the sky from the others, so that the sensitivity to off-axis sources is reduced on any particular baseline. Note that only the station {\it layout} is rotated. Each of the individual dipole pairs are oriented at the same angle with respect to a commonly defined polarimetric axis.

Unlike stations in the Netherlands, international LOFAR stations are uniform and most closely follow the original station design. These stations consist of a full complement of 96 HBAs, 96 LBAs, and 96 RCUs. The additional RCUs in these stations provide a total of 192 digital signal paths such that the full set of HBA tiles or LBA dipoles are available during any given observation. In these stations, the third RCU input is currently not used and therefore available for possible future expansion. Several proposals are already under consideration that would take advantage of this unused capacity in order to expand the capabilities of the international LOFAR stations (see Sect.\,\ref{sec:enhance}).

\begin{figure*}[ht]
\centering
\includegraphics[width=\textwidth]{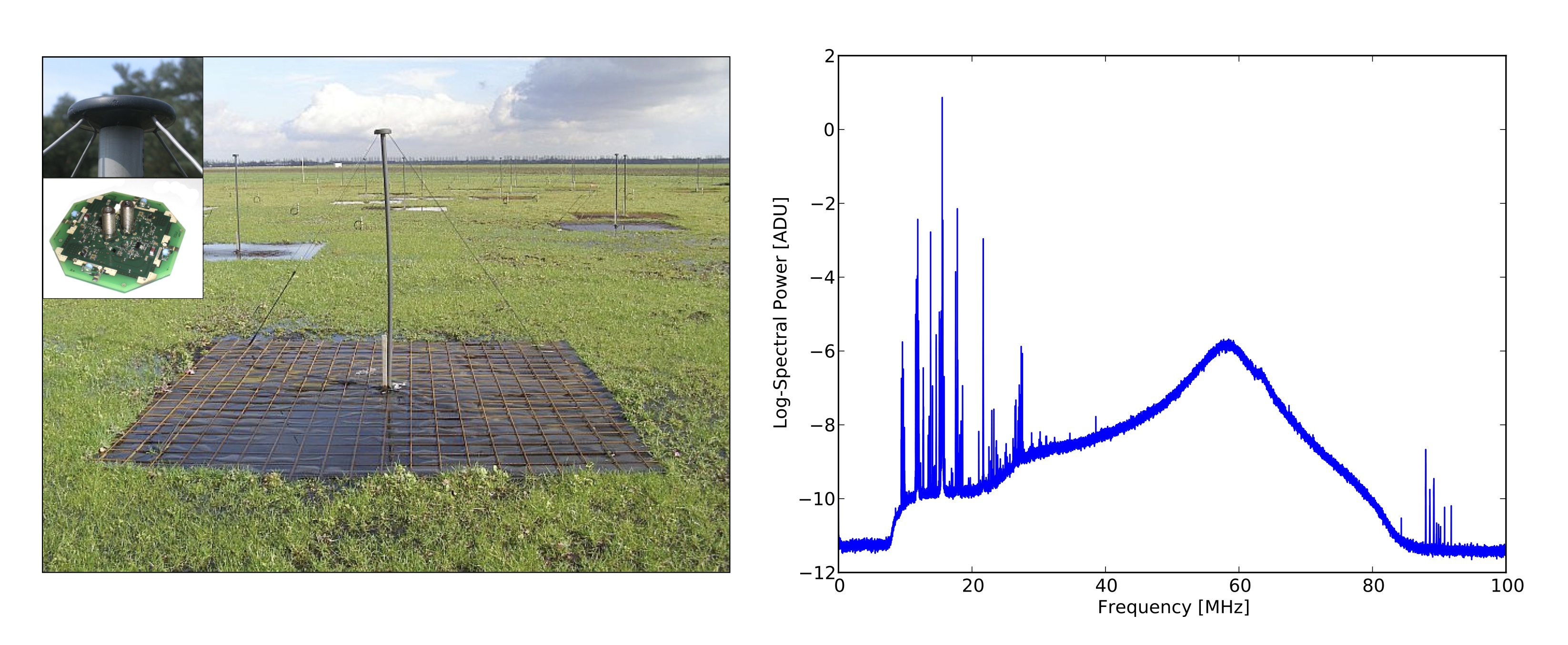}
\caption{\small {\bf Left:} Image of a single LOFAR LBA dipole including the ground plane. The inset images show the molded cap containing the LNA electronics as well as the wire attachment points. {\bf Right:} Median averaged spectrum for all LBA dipoles in station CS003. The peak of the curve near 58\,MHz is clearly visible as well as strong RFI below 30\,MHz, partly because of ionospheric reflection of sub-horizon RFI back toward the ground, and above 80\,MHz, due to the FM band.}
\label{fig:lba}
\end{figure*}

\begin{figure*}[ht]
\centering
\includegraphics[width=3.4in]{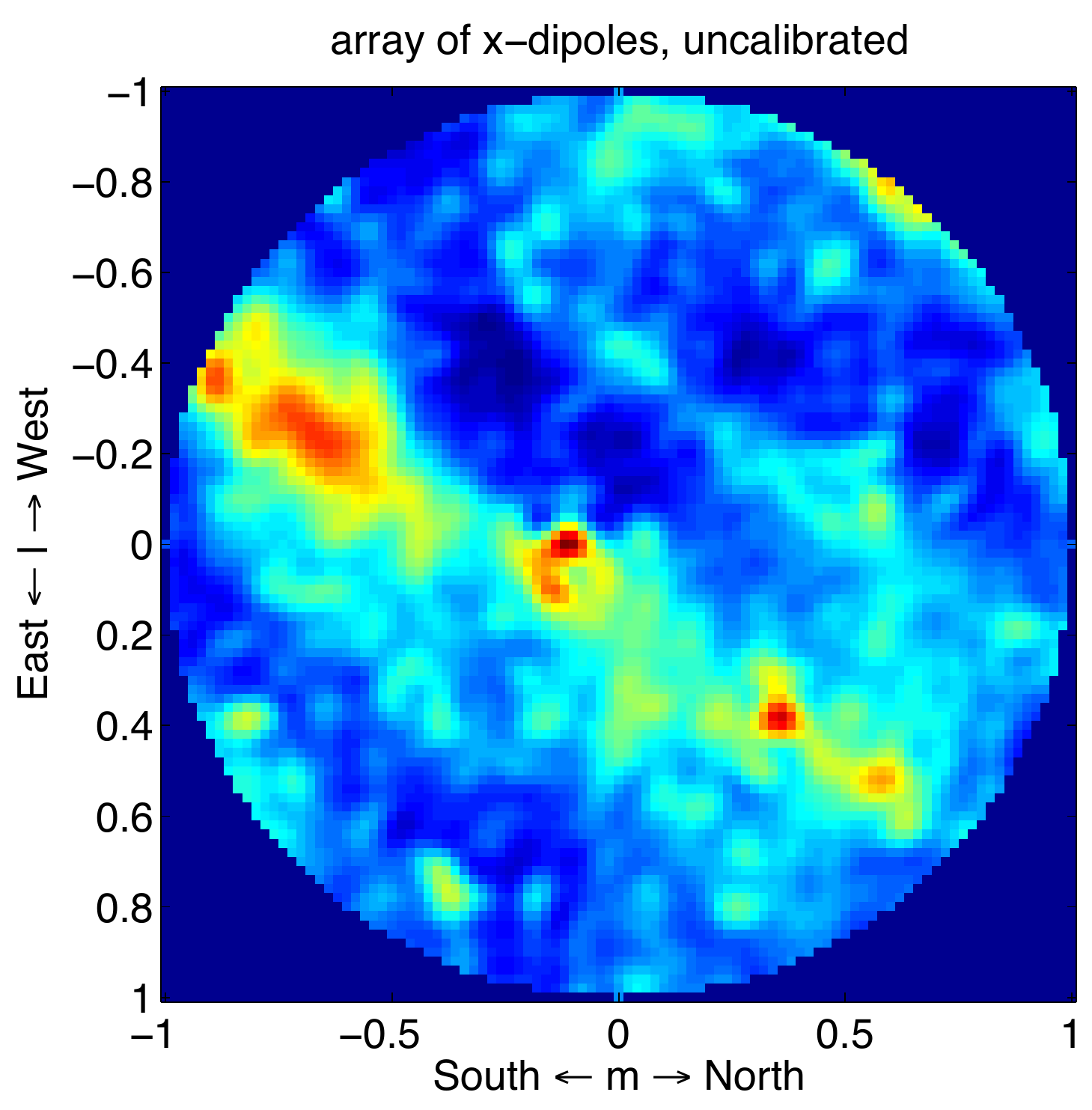}
\includegraphics[width=3.4in]{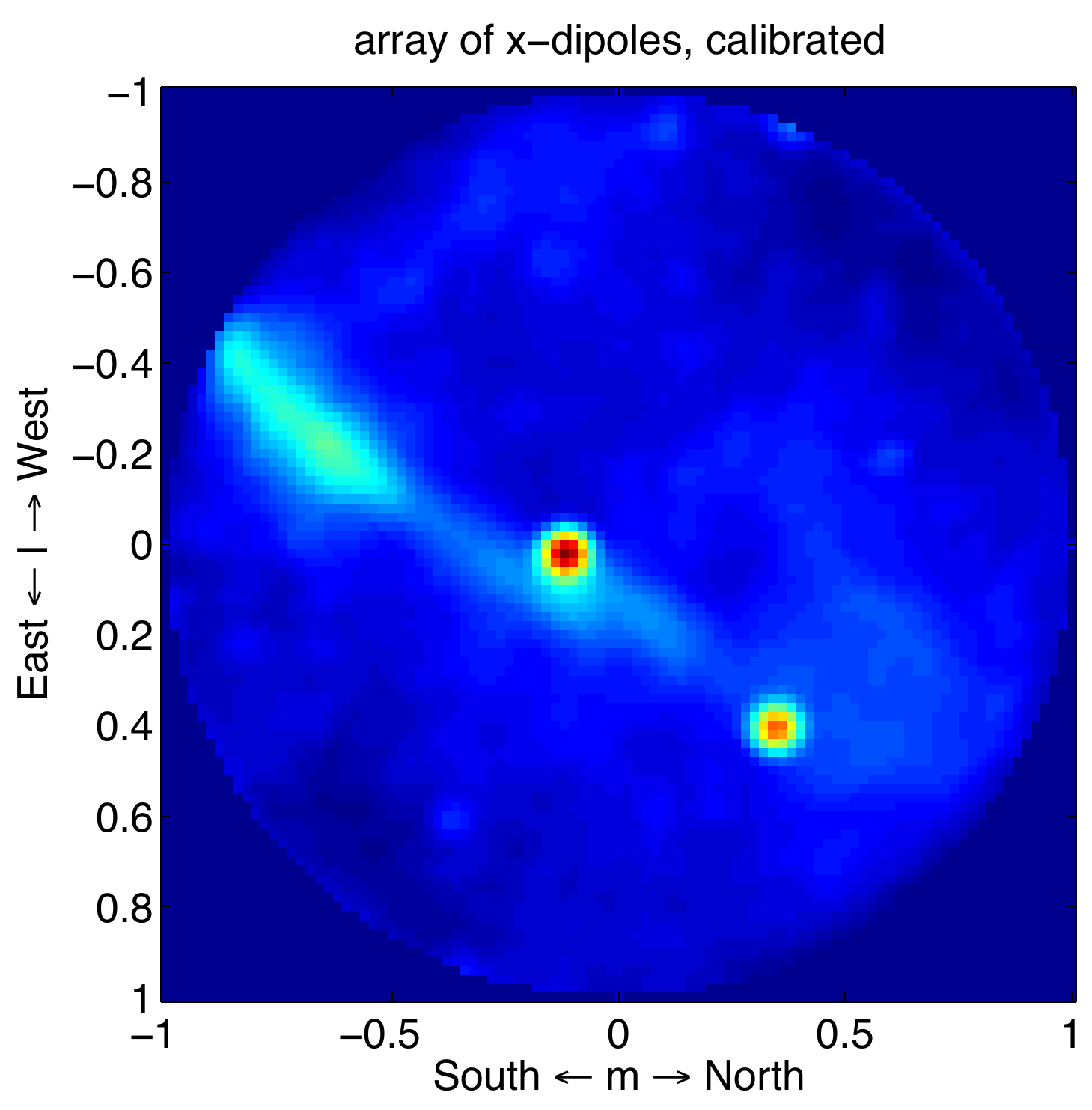}
\includegraphics[width=3.4in]{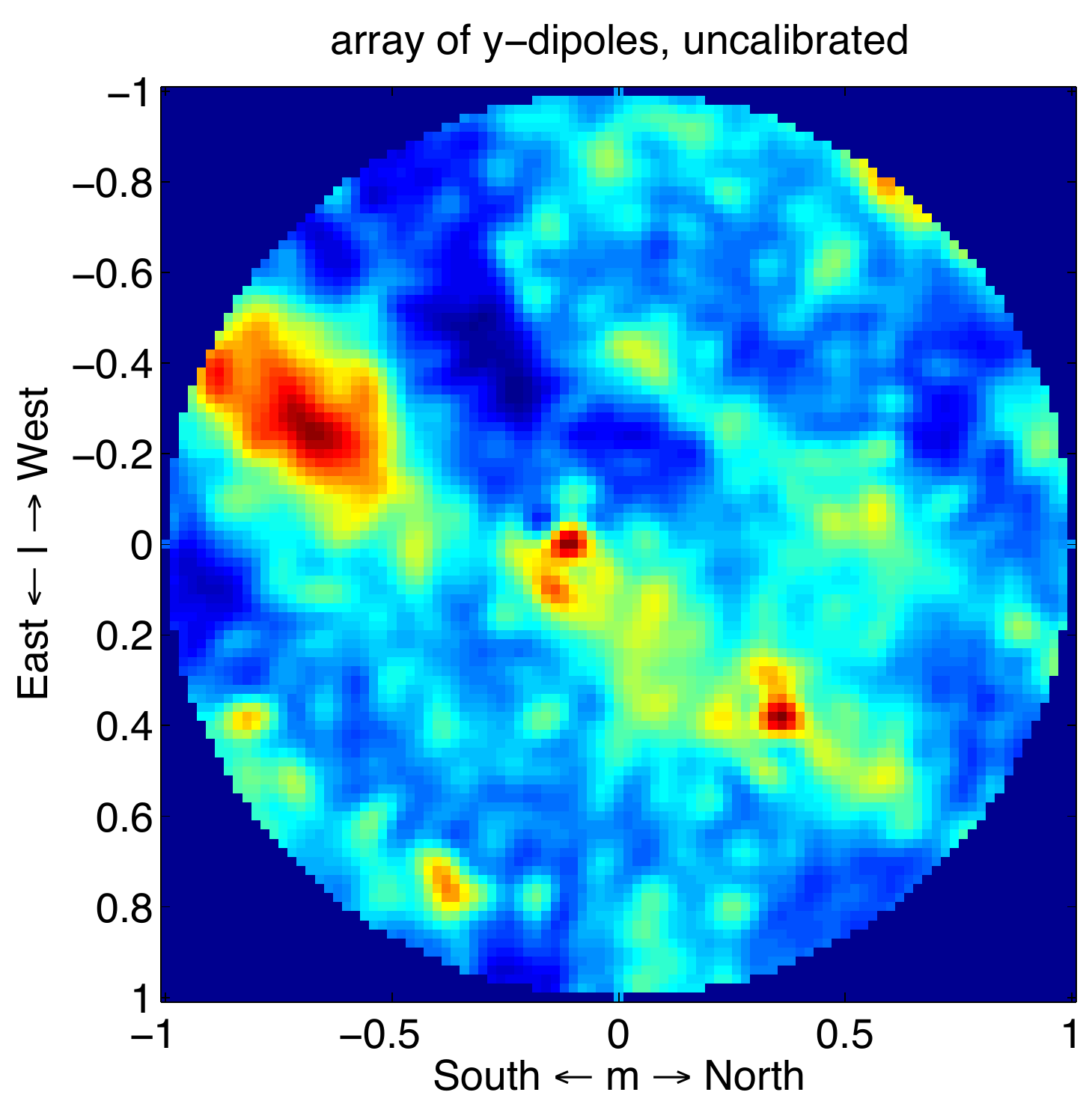}
\includegraphics[width=3.4in]{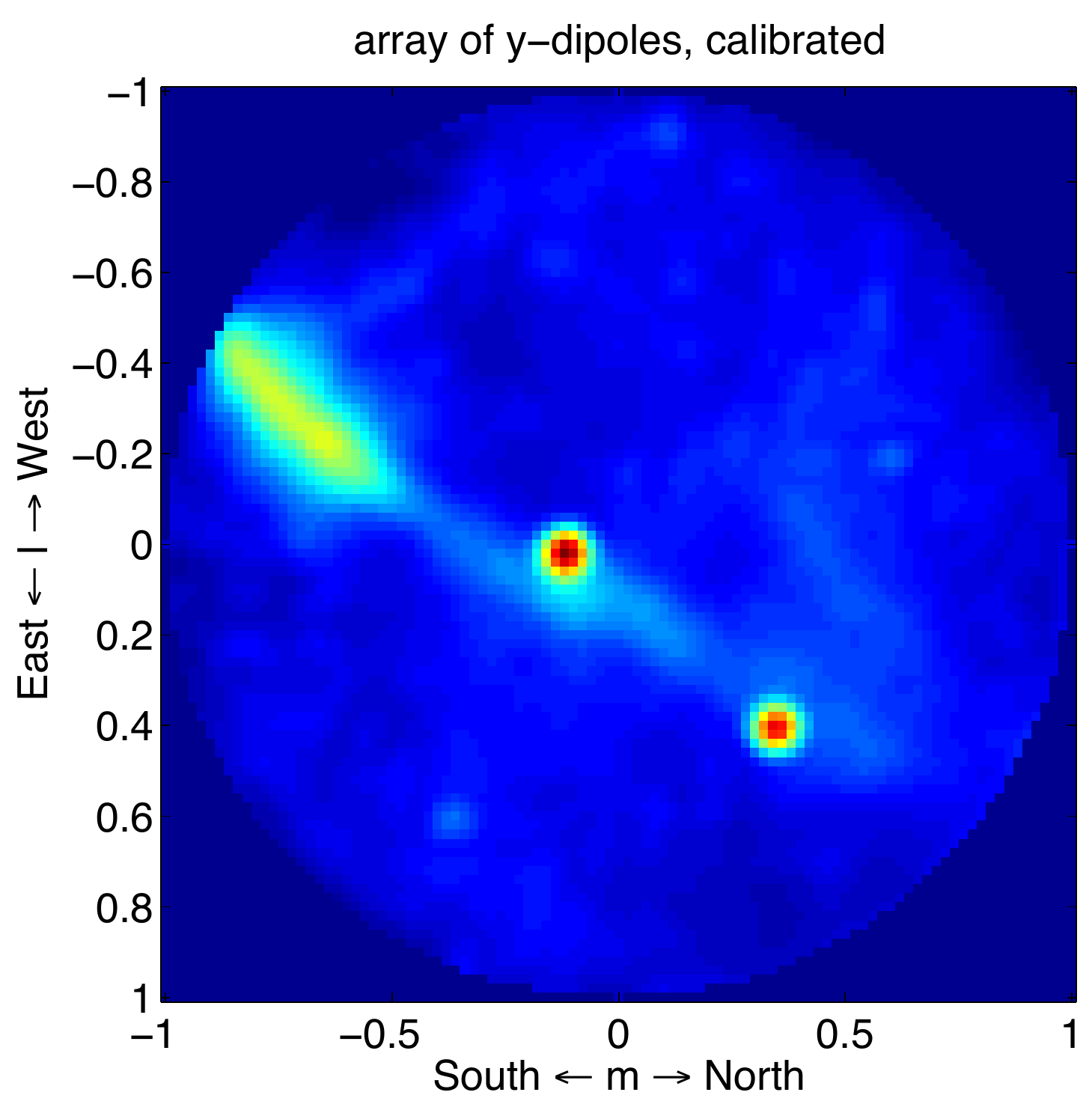}
\caption{\small All-sky observation produced by a single LOFAR station (station FR606 in Nan\c{c}ay, France) and created offline by correlating the signals from each of the individual dipoles in the station. The station level data collection and processing is described in Sect.\,\ref{sec:rcu}$-$\ref{sec:lcu}. The observation was taken at a frequency of 60\,MHz, with a bandwidth of only 195.3125\,kHz (1 subband). The integration time was 20 seconds. Even with this limited dataset, Cassiopeia A, Cygnus A, and the Galactic plane are all clearly visible. The left panels show images made from uncalibrated station data while the calibrated images are shown on the right. The upper and lower panels give images for the X and Y polarizations, respectively.}
\label{fig:allsky}
\end{figure*}

\subsection{Low-band antenna}
\label{sec:lba}

At the lowest frequencies, LOFAR utilizes the LBAs, which are designed to operate from the ionospheric cutoff of the ``radio window" near 10\,MHz up to the onset of the commercial FM radio band at about 90\,MHz. Due to the presence of strong RFI at the lowest frequencies and the proximity of the FM band at the upper end, this range is operationally limited to 30--80\,MHz by default. An analog filter is used to suppress the response below 30\,MHz, although observers wishing to work at the lowest frequencies have the option of deselecting this filter \citep[see][]{vanWeeren12}. In designing the LOFAR LBAs, the goal was to produce a sky-noise dominated receiver with all-sky sensitivity and that goal has largely been achieved over $\sim 70$\% of the bandpass (see Sect.\,\ref{sec:sensitivity}). At the same time, the resulting antenna needed to be sturdy enough to operate at least 15 years in sometimes harsh environmental conditions as well as be of sufficiently low cost that it could be mass produced. The resulting LBA is shown in Fig.\,\ref{fig:lba}.

\begin{figure*}[ht]
\centering
\includegraphics[width=\textwidth]{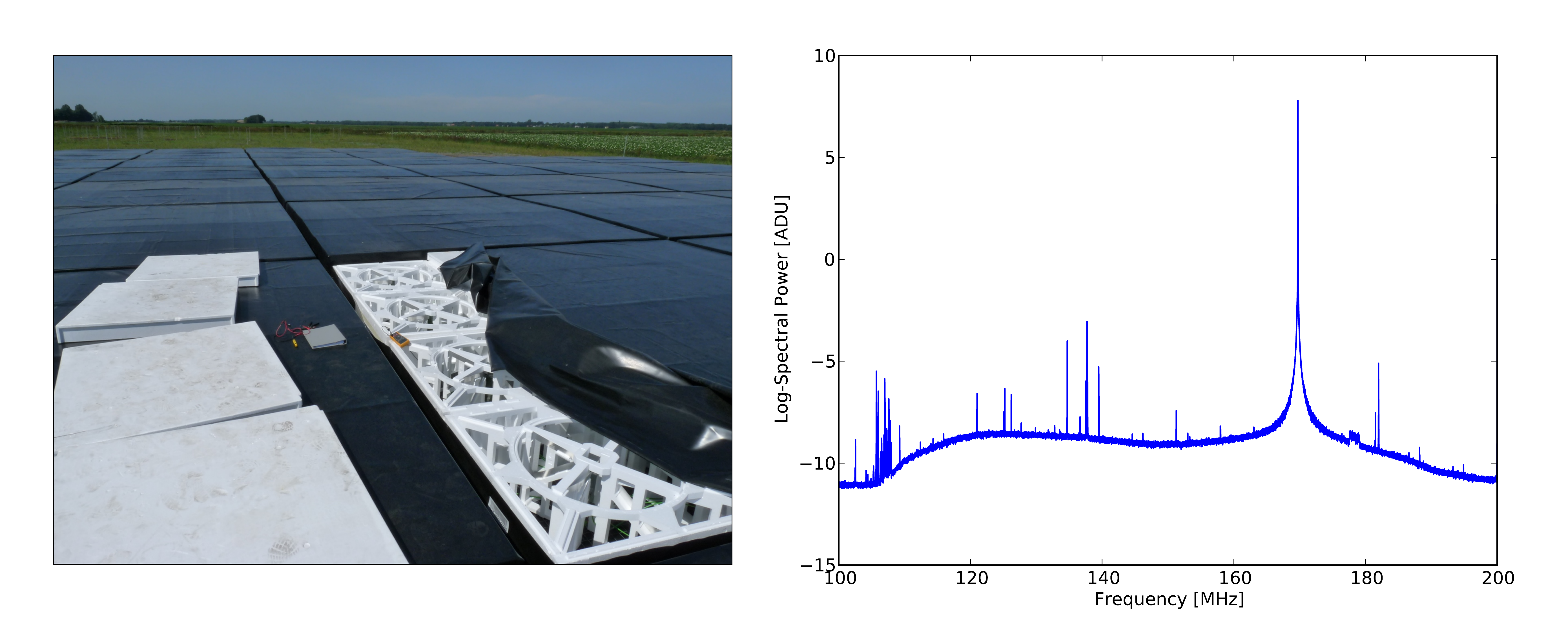}
\caption{\small {\bf Left:} Closeup image of a single LOFAR HBA tile. The protective covering has been partially removed to expose the actual dipole assembly. The circular dipole rotation mechanism is visible. {\bf Right:} Median averaged spectrum for all HBA tiles in station CS003. Various prominent RFI sources are clearly visible distributed across the band including the strong peak near 170 MHz corresponding to an emergency pager signal.}
\label{fig:hba}
\end{figure*}

The LBA element, or dipole, is sensitive to two orthogonal linear polarizations. Each polarization is detected using two copper wires that are connected at the top of the antenna to a molded head containing a low-noise amplifier (LNA). At the other end, these copper wires terminate in either a synthetic, rubber spring or a polyester rope and are held in place by a ground anchor. The molded head of the LBA rests on a vertical shaft of PVC pipe. The tension of the springs and the ground anchor hold the antenna upright and also minimize vibrations in the wires due to wind loading. The dipole itself rests on a ground plane consisting of a metal mesh constructed from steel concrete reinforcement rods. A foil sheet is used to minimize vegetation growth underneath the antenna. Each polarization has its own output and hence two coaxial cables per LBA element run through the vertical PVC pipe. Power is supplied to the LNA over these same coaxial cables. The dipole arms have a length of 1.38 meter corresponding to a resonance frequency of 52 MHz. The additional impedance of the amplifier shifts the peak of the response curve to 58\,MHz, however, as shown in the right panel of Fig.\,\ref{fig:lba}.

Despite the deceptively simple design, when coupled with digital beam-forming techniques, the LOFAR LBA dipole provides a powerful detection system at low frequencies. In particular, the omnidirectional response of the LBA antennas allows for the simultaneous monitoring of the entire visible sky. The LBA dipoles in a given LOFAR station can easily be correlated to provide all-sky maps on timescales of seconds (see Fig.\,\ref{fig:allsky}). This novel capability is useful for a number of scientific objectives including studies of the large scale Galactic emission from the Milky Way and all-sky monitoring for radio transients.

\subsection{High-band antenna}
\label{sec:hba}

To cover the higher end of the LOFAR spectral response, an entirely different mechanical design has been utilized. The LOFAR HBA has been optimized to operate in the 110--250\,MHz range. In practice, the frequency range above 240\,MHz is heavily contaminated by RFI so operationally the band is limited to 110--240\,MHz. At these frequencies, sky noise no longer dominates the total system noise as is the case for the LBAs. Consequently, another design topology for the HBA antennas was required in order to minimize contributions to the system noise due to the electronics.  Nonetheless, the HBA design was of course subject to the similar constraints on environmental durability and low manufacturing cost as the LBA design. An image of the final HBA tile is shown in Fig.\,\ref{fig:hba}.

In order to minimize cost while maintaining adequate collecting area, the HBA design clusters 16 antenna elements together into ``tiles" that include initial analog amplification and a first stage of analog beam-forming. A single ``tile beam'' is formed by combining the signals from these 16 antenna elements in phase for a given direction on the sky. Hence, while the LBAs are effectively passive (requiring power but no active control and synchronization), the HBAs contain tile-level beam forming and are subject to control by the Monitoring and Control system MAC (see Sect.\,\ref{sec:mac}).

A single HBA tile consists of a square, 4x4 element (dual polarized) phased array with built-in amplifiers and an analog beam-former consisting of delay units and summators. The 5\,bit time delay can be up to 15 ns long and is set by a signal received from the MAC system. Each 16 element tile measures 5x5 meter and is made of an expanded polystyrene structure which supports the aluminum antenna elements. The distance between tile centers is 5.15 m resulting in a spacing between tiles of 15 cm. The contents of the tile are protected from weather by two overlapping flexible polypropylene foil layers. A light-weight ground plane consisting of a 5x5 cm wire mesh is integrated into the structure. As with the LBAs, the resulting signals are transported over coaxial cables to the receiver unit in the electronics cabinet.

\subsection{Receiver unit}
\label{sec:rcu}

At the receiver unit (RCU), the input signals are filtered, amplified, converted to base-band frequencies and digitized. A sub-sampling architecture for the receiver is used. This choice implies a larger required analog bandwidth and multiple band-pass filters to select the frequency band of interest. The receiver is designed to be sky noise limited so a 12 bit A/D converter is used with 3 bits reserved to cover the anticipated range of sky noise and the rest available for RFI headroom. This number of bits is sufficient to observe signals, including strong RFI sources, with strengths up to 40 dB over and above the integrated sky noise in a bandwidth of at least 48\,MHz.

Because observing in the FM band is not feasible, a sampling frequency of 200\,MHz has been chosen for most of the receiver modes. This sampling results in a Nyquist edge almost at the center of the FM band. To cover the region around 200 MHz in the HBA band, which will suffer from aliasing due to the flanks of the analog filter, an alternative sampling frequency of 160 MHz is also supported. These choices result in several possible observing bands to cover the total HBA frequency range. The available frequency bands are summarized in Table\,\ref{tab:system}.

As discussed above in Sect.\,\ref{sec:layouts}, three main signal paths can be distinguished in the RCU. For stations in the Netherlands, two of these are allocated to the two sets of LBAs, although only one can be used at any given time. One of these signal paths was originally intended for a (not currently planned) low-band antenna optimized for the 10--30\,MHz frequency range. For the present LBA, either a 10-MHz or 30-MHz high-pass filter can be inserted to suppress the strong RFI often encountered below 20\,MHz. The remaining signal path is used for the HBA. It is first filtered to select the 110--250\,MHz band and then again by one of three filters that select the appropriate Nyquist zones listed in Table\,\ref{tab:system}.

\begin{table*}                                                                                                
\begin{minipage}[t]{\textwidth}
\caption{Overview of LOFAR system parameters}
\label{tab:system}                                                                                           
\centering
\renewcommand{\footnoterule}{}  
\begin{tabular}{llrl}                                                                                       
\hline \hline 
System characteristic & Options & Values & Comments \\
\hline \hline
Frequency range & Low-band Antenna	 & 10-90 MHz & \\
\hspace{5.0cm}  & \hspace{3.0cm}     & 30-90 MHz & With analog filter \\
                & High-band Antenna  & 110-190 MHz & 200 MHz sampling (2nd Nyquist zone) \\
                &                    & 170-230 MHz & 160 MHz sampling (3rd Nyquist zone) \\
                &                    & 210-250 MHz & 200 MHz sampling (3rd Nyquist zone) \\
\hline
Number of polarizations & &			2 & \\
\hline
Bandwidth				& Default &		48 MHz & 16-bit mode \\
                        & Maximum &     96 MHz & 8-bit mode \\
\hline
Number of simultaneous beams & Minimum &   1 & \\
                             & Maximum & 244 &  16 bit mode, one per sub-band \\ 
                             & Maximum & 488 &  8 bit mode, one per sub-band \\ 
\hline
Sample bit depth			& &	12 & \\
\hline
Sample rate					& Mode 1 & 	160 MHz & \\
							& Mode 2 & 	200 MHz & \\
\hline
Beamformer spectral resolution	& Mode 1 & 156 kHz & \\
								& Mode 2 & 195 kHz & \\
\hline
Channel width					& Mode 1 & 610 Hz & \\
(raw correlator resolution)		& Mode 2 & 763 Hz & \\
\hline \hline
\end{tabular}
\end{minipage}
\end{table*}

\subsection{Digital signal processing}
\label{sec:dsp}

\begin{figure*}[ht]
\centering
\includegraphics[width=\textwidth]{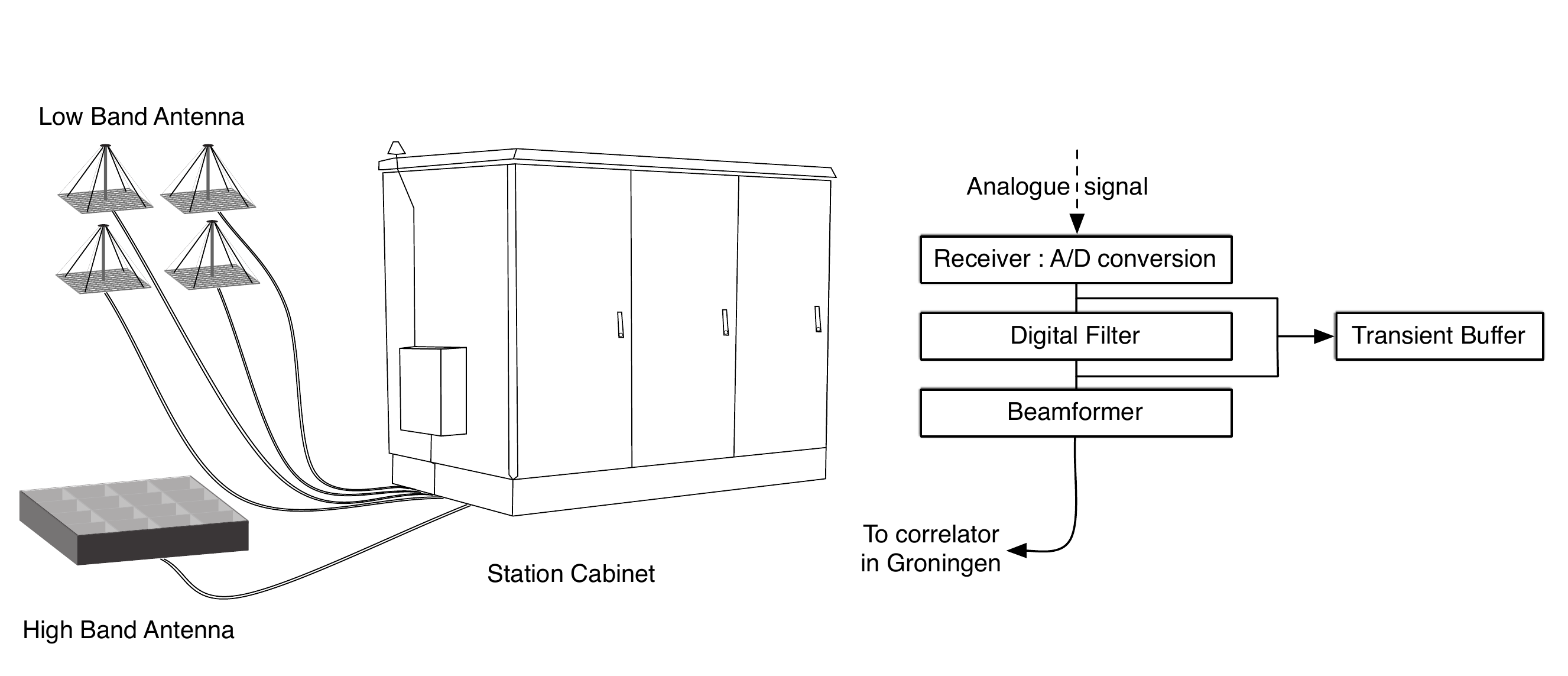}
\caption{\small Schematic illustrating the signal connections at station level as well as the digital processing chain. After the beam-forming step, the signals are transferred to the correlator at the CEP facility in Groningen.}
\label{fig:dsp}
\end{figure*}

Both the LBA and HBA antennas are connected via coaxial cables to the electronics housed in a cabinet located on the edge of each LOFAR station. This cabinet is heavily shielded and contains the RCUs, digital signal processing (DSP) hardware, local control unit (LCU), and other equipment used to perform the first data processing stage. After digitization by the RCUs, the datastreams enter the digital electronics section. This section is mainly responsible for beam-forming although either raw or filtered signals can also be stored in a circular buffer in order to trap specific events (see Sect.\,\ref{sec:tbb} below). Further processing is done by the remote station processing (RSP) boards utilizing low-cost, field programmable gate arrays (FPGAs). These FPGAs provide sufficient computing power to keep up with the datastream and can also be updated remotely allowing for easy patches and enhancements to be applied. Following the beam-forming step, the data packets are streamed over the wide-area network (WAN) to the CEP facility in Groningen. A schematic of this data flow is given in Fig.\,\ref{fig:dsp}.

Once digitized, the RSP boards first separate the input signals from the RCUs into 512 sub-bands via a polyphase filter (PPF). Further processing is done per sub-band. The sub-bands have widths of 156\,kHz or 195\,kHz depending on whether the 160\,MHz or 200\,MHz sampling clock is selected, respectively. By default sample values are stored using 16 bit floating point representations allowing up to 244 of these sub-bands to be arbitrarily distributed over the bandpass for a total bandwidth of 48\,MHz per polarization. Alternatively, the station firmware may be configured to utilize an 8 bit representation for the sample values yielding up to 488 sub-bands for a total bandwidth of 96\,MHz per polarization. Although providing increased bandwidth, this 8 bit mode is potentially more vulnerable to periods of strong RFI. The frequency selection can vary for each station and is configured by the user during the initial observation specification.

After formation of the sub-bands, the primary processing step is the digital, phase rotation-based beam-former. This beam-former sums the signals from all selected RCUs after first multiplying them by a set of complex weights that reflect the phase rotation produced by the geometrical and other delays toward a certain direction. The weights are calculated in the local control unit (see Sect.\,\ref{sec:lcu}) and sent to the RSP boards during the observation. The update rate of the beam-former is set to 1 second by default resulting in about 0.3\% gain variation for a station beam of 3$\degr$ in diameter. The beam-forming is done independently per sub-band and the resulting beam for each sub-band is referred to as a ``beamlet". Multiple beamlets with the same pointing position can be combined to produce beams with larger bandwidth.

The number of simultaneous beams that may be constructed can in principle be as high as the number of beamlets since all operate independently of each other. Operationally, the number of independent beams per station is currently limited to 8, although this limit will ultimately increase. Successful experiments utilizing the maximum 244 beams available in 16 bit mode have already been conducted. Similarly, for 8 bit observations, a maximum of 488 beams are can be formed. The 48\,MHz (16 bit mode) or 96\,MHz (8 bit mode) total bandwidth can be distributed flexibly over the number of station beams by exchanging beams for bandwidth. In the case of the LBA, simultaneous beams can be formed in any combination of directions on the sky. While strictly true for the HBA as well, HBA station beams can only usefully be formed within pointing directions covered by the single HBA tile beam, corresponding to a FWHM of $\sim$20$\degr$ at 140 MHz.

\subsection{Transient buffer boards}
\label{sec:tbb}

In addition to the default beam-forming operations, the LOFAR digital processing also provides the unique option of a RAM (Random Access Memory) buffer at station level. These RAM buffers provide access to a snapshot of the running data-streams from the HBA or LBA antennas. As depicted in Fig.\,\ref{fig:dsp}, a dedicated transient buffer board (TBB) is used that operates in parallel with the normal streaming data processing. Each TBB can store 1 Gbyte of data for up to 8 dual-polarized antennas either before or after conversion to sub-bands. This amount is sufficient to store 1.3\,s of raw data allowing samples to be recorded at LOFAR's full time resolution of 5 ns (assuming the 200-MHz sampling clock). Following successful tests for various science cases (see Sect.\, \ref{sec:crpipe}), an upgrade of the RAM memory to store up to 5\,s of raw-data has been approved and is currently being installed. The temporal window captured by the TBBs can be further extended by up to a factor of 512 by storing data from fewer antennas or by storing sub-band data. We note that while the TBBs may operate in either raw timeseries or sub-band mode, they can not operate in both at the same time.

Upon receiving a dump command, the TBB RAM buffer is frozen and read out over the WAN network directly to the storage section of the CEP post-processing cluster (see Sect.\,\ref{sec:offline}). These commands can originate locally at the station level, from the system level, or even as a result of triggers received from other telescopes or satellites. At the station level, each TBB is constantly running a monitoring algorithm on the incoming data-stream. This algorithm generates a continuous stream of event data that is received and processing by routines running on the local control unit (LCU). If the incoming event stream matches the pre-defined criteria, a trigger is generated and the TBBs are read out. As discussed in Sect.\,\ref{sec:crpipe}, this local trigger mechanism gives LOFAR the unique ability to respond to ns-scale events associated with strong CRs. The Transients KSP also intends to utilize this functionality to study fast radio transients (see Sect.\,\ref{sec:transients}).

\subsection{Local control unit}
\label{sec:lcu}

Each LOFAR station, regardless of configuration, contains computing resources co-located adjacent to the HBA and LBA antenna fields. This local control unit (LCU) is housed inside the RF-shielded cabinet containing the other digital electronics and consists of a commodity PC with dual Intel Xeon 2.33 GHz quad-core CPUs, 8 Gbyte of RAM, and 250 Gbyte of local disk storage. The station LCUs run a version of Linux and are administered remotely over the network from the LOFAR operations center in Dwingeloo. Processes running on the LCU can include control drivers for the TBBs, RCUs, and other hardware components as well as additional computational tasks. All processes running on the LCUs are initialized, monitored, and terminated by the MAC/SAS control system discussed below in Sect.\,\ref{sec:control}. 

Computationally the LCU provides several crucial computing tasks at the station level. Chief among these are the beam-former computations mentioned previously in Sect.\,\ref{sec:dsp}. The number of independent beams that may be supported is limited by the processing power of the LCU since it must calculate the appropriate weights for each direction on the sky every second. 

Equally important, the LCU runs a station-level calibration algorithm to correct for gain and phase differences in all the individual analog signal paths. The correlation matrix of all dipoles in the station is calculated for one sub-band each second as input to this calibration and the procedure runs in real-time during an observation \citep{2009ITSP...57.3512W, stationcal10a, stationcal10b}. The algorithm cycles through the selected sub-bands, with a new sub-band calibrated each second, resulting in an updated calibration for the complete band every 512 seconds. This active calibration is necessary to compensate for environmental temperature variations that cause gain and phase drifts in the signal paths (see the discussion in Sect.\,\ref{sec:stability}). The array correlation matrix can also be used for RFI detection and mitigation \citep{statrfifilter05}.

Additional computational tasks can also be run on the LCU subject to the constraint that they do not impact the performance of the core calibration and beam-forming capabilities. Current examples of these station-level applications include the TBB trigger algorithms discussed previously in Sect.\,\ref{sec:tbb}. We note that adding additional compute capacity to the LCU is a fairly straightforward way to expand the capabilities of the LOFAR array (see Sect.\,\ref{sec:enhance} for some currently planned enhancements).

\section{Wide-area network}
\label{sec:wan}

The function of the LOFAR Wide-Area Network (WAN) is to transport data between the LOFAR stations and the central processor in Groningen. The main component is the streaming of observational data from the stations. A smaller part of the LOFAR datastream consists of Monitoring And Control (MAC) related data and management information of the active network equipment. Connections of the LOFAR stations in the Netherlands to Groningen run over light-paths (also referred to as managed dark fibers) that are either owned by LOFAR or leased. This ensures the required performance and security of the entire network and the equipment connected to it. Signals from all stations in the core and an area around it are first sent to a concentrator node and subsequently patched through to Groningen.

The LOFAR stations outside the Netherlands are connected via international links that often involve the local NRENs (National Research and Education Networks). In some cases, commercial providers also play a role for part of the way. 

For the communication over the light-paths 10 Gigabit Ethernet (GbE) technology has been adopted. The high bandwidth connection between the concentrator node in the core and Groningen uses Course Wavelength Division Multiplexing (CWDM) techniques to transfer multiple signals on a single fiber, thereby saving on costs. Since the availability requirement for LOFAR is relatively low (95\%), when compared with commercial data communication networks, redundant routing has not been implemented.

\section{Central processing (CEP)}
\label{sec:cep}

LOFAR's CEP facility is located at the University of Groningen's Centre for Information Technology (CIT). The CIT houses the hardware for the CEP system but also part of the distributed long term archive (LTA) discussed in Sect.\,\ref{sec:lta}. With the exception of standalone operation where a given LOFAR station can be used locally independent from the rest of the array, data from all LOFAR stations, including the international stations, is received at CEP in a streaming mode. At CEP these raw datastreams are subsequently processed into a wide variety of data products as discussed in Sect.\,\ref{sec:pipelines} below.

The CEP facility can be broadly divided into two essentially autonomous sections. The ``online" section collects and processes the incoming station datastreams in real-time and all operations on the data are completed before it is written to disk. Once the initially processed data-streams are stored, additional, less time-critical processing is done on the ``offline" section to produce the final set of LOFAR data products. A large storage cluster connects these two distinct processing phases. The same Monitoring and Control system discussed in Sect.\,\ref{sec:control} and used to operate the stations themselves also manages the allocation of processing and storage resources at CEP. Multiple observations and processing streams on both the online and offline sections can be performed in parallel. In the following, we briefly review the major features of these two components.

\begin{figure*}
\centering
\includegraphics[width=6.5in]{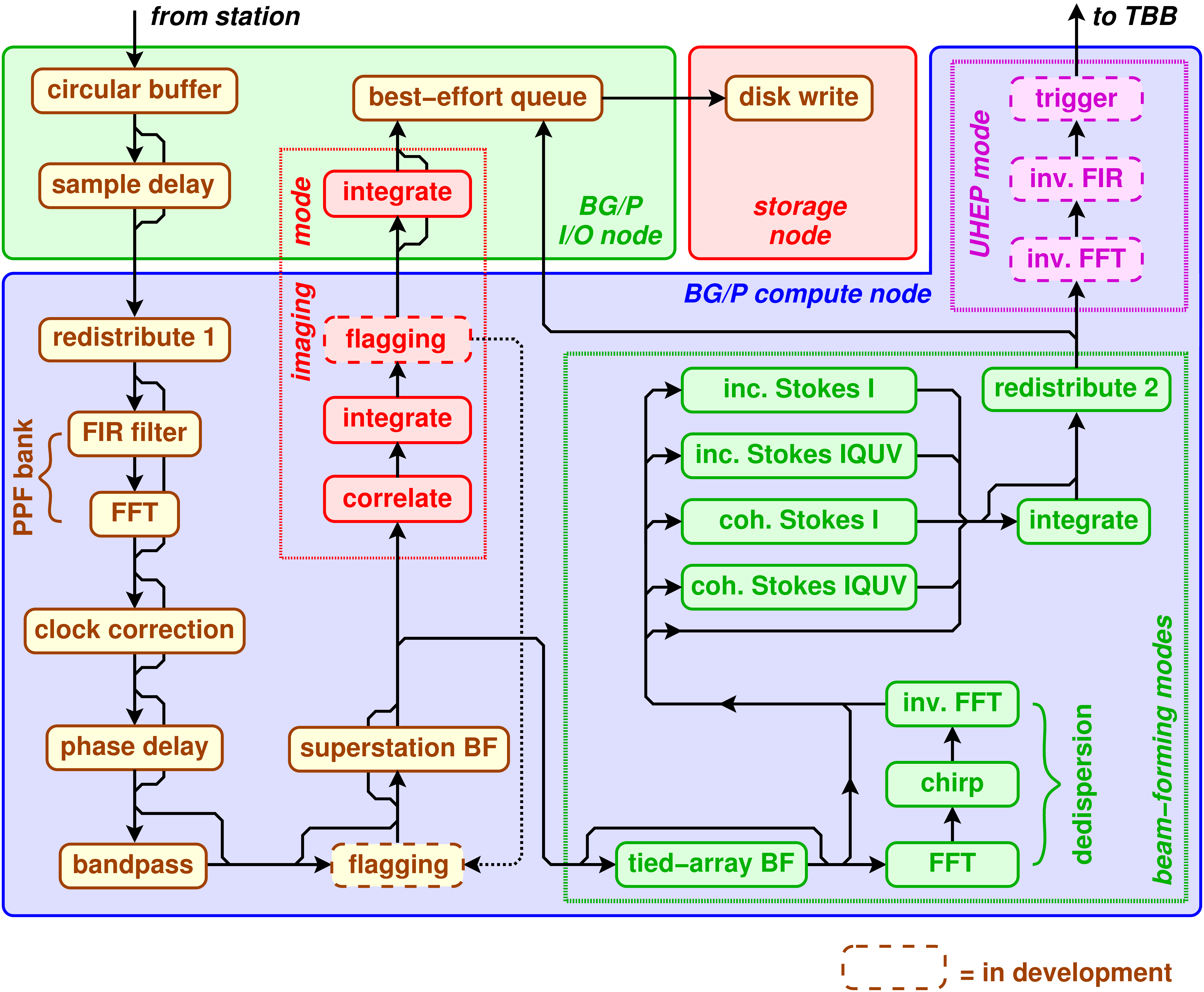}
\caption{\small Schematic showing the possible online data processing paths currently available or under development. These pipelines run in real-time on the IBM Blue Gene/P supercomputer that comprises the core of LOFAR's online processing system (see Sect.\,\ref{sec:online}). This schematic illustrates that many processing steps can be selected or deselected as necessary. Pipelines can also be run in parallel with, for example, the incoming station datastream being split off to form both correlated and beam-formed data simultaneously. The imaging and beam-formed data pipelines are indicated separately. The online triggering component of the CR UHEP experiment currently under development is also shown (see Sect.\,\ref{sec:cr}).}
\label{fig:online}
\end{figure*}

\subsection{Online central processing}
\label{sec:online}

The online processing section handles all real-time aspects of LOFAR and is built around a three-rack IBM Blue Gene/P (BG/P) supercomputer. Current LOFAR operations are limited to one rack of the three available. Each rack of the BG/P is equipped with 64 individual 10 GbE interfaces (I/O nodes). A single LOFAR station can be mapped to one I/O node. The peak performance of each rack is 14 Tflop/s. The processing power and I/O bandwidth of one rack is sufficient to correlate 2048 baselines at full-polarization for the maximum bandwidth of 48 MHz with an integration time of one second. 

Each BG/P I/O node receives data from a station and runs a data-handling application that buffers the input data and synchronizes its output stream with the other input nodes based on the timestamps contained in the data. For imaging observations, the BG/P performs its main function as the correlator of the array.  As Fig.\,\ref{fig:online} shows, it can also support a variety of other processing streams including the formation of tied-array beams and real-time triggering. Combinations of these processing streams can be run simultaneously subject to resource constraints.

The current set of supported online processing streams is depicted in Fig.\,\ref{fig:online}. Most of these represent the initial processing stages in the observing modes discussed in Sect.\,\ref{sec:modes}. Several common transformations are applied to all incoming station datastreams regardless of subsequent processing. For example, time offsets are applied to each incoming datastream to account for geometric delays caused by differing station distances from the array phase center. These offsets must be calculated on-the-fly since the rotation of the Earth alters the orientation of the stations continuously with respect to the sky. For observations with multiple beams, unique delays must be calculated for each beam.

Once the geometric delays are applied, a transpose operation is performed to reorder the now aligned station data packets. Incoming data packets from the stations are grouped as a set of sub-bands per station. After the transpose, the data are rearranged such that all station data for a given sub-band is grouped. At this point, a second polyphase filter is applied to resample the data to the kHz level. The filter-bank implemented on the BG/P splits a 195 kHz (or 156 kHz) sub-band datastream into, typically, 256 frequency channels of 763 Hz (or 610 Hz) each. Splitting the data into narrow frequency channels allows the offline processing to flag narrow-band RFI, so that unaffected channels remain usable.

In classical radio telescopes an XF correlator was generally used, meaning that first the correlation and integration of the signals was done in the time domain (X) and afterwards the Fourier transform (F) was accomplished to get a cross power spectrum out of the correlator \citep{Romney99}. This option is still an economically attractive technique for radio telescopes with a limited number of antennas (input signals to the correlator). However, for LOFAR an FX correlator (first Fourier transform and then correlating the resulting channels) is favorable in terms of processing at the expense of data transport (the signals must be regrouped per channel instead of per antenna, resulting in a transpose operation). Using only a Fourier transform in the FX correlator leads to a significant amount of leakage between the channels. Therefore it was chosen to use filter banks before the correlator. This architecture is also known as an HFX (Hybrid FX correlator) architecture \citep{Romney99}.

The correlator calculates the auto and cross correlations between all pairs of stations, for each channel and for each polarization (XX, XY, YX, and YY). A correlation is the complex product of a sample from one station and the complex conjugate of a sample from the other station. By default, the results are integrated (accumulated) over one second of data; however, smaller integration times are possible for applications such as full-field imaging with the international stations or fast solar imaging. Since the correlation of station S1 and S2 is the conjugate of the correlation of station S2 and S1, we only compute the correlations for S1 $\leq$ S2. The output data rate of the correlator is significantly lower than the input data rate. To achieve optimal performance, the correlator consists of a mix of both C++ and assembler code, with the critical inner loops written entirely in assembly language \citep{Romein06, Romein10}.

\subsection{Offline central processing}
\label{sec:offline}

The offline central processing cluster provides disk space for the collection of datastreams and storage of complete observation datasets for offline processing. This storage is intended for temporary usage (typically a week) until the final data products are generated and archived or the raw data themselves are exported or archived. In addition to the storage part the offline cluster offers general-purpose compute power and high bandwidth interconnections for the offline processing applications.

The offline cluster is a Linux cluster that is optimized for cost per flop and cost per byte. The cluster consists of 100 hybrid storage / compute nodes. Each node has 12 disks of 2 Tbyte each providing 20\,Tbyte of usable disk space per node. Furthermore, each node contains 64\,Gbyte of memory and 24, 2.1\,GHz cores. Thus the cluster has 2\,Pbyte of storage capacity total and 20.6 Tflop/s peak performance. The offline tasks differ depending on the application at hand. For example for the imaging application the offline tasks are typically flagging of bad data, self-calibration and image creation.

In addition to the offline cluster extra processing power will be available in GRID networks in Groningen or at remote sites. GRID networks also provide the basic infrastructure for the LOFAR archive enabling data access and data export to users.

\section{LOFAR long-term archive}
\label{sec:lta}

The LOFAR Long-Term Archive (LTA) is a distributed information system created to store and process the large data volumes generated by the LOFAR radio telescope. When in full operation, LOFAR can produce observational data at rates up to 80 Gbit/s. Once analyzed and processed, the volume of data that are to be kept for a longer period (longer than the CEP storage is able to support) will be reduced significantly. These data will be stored in the LTA and the archive of LOFAR science data products is expected to grow by up to 5 Pbyte per year. The LTA currently involves sites in the Netherlands and Germany. 

For astronomers, the LOFAR LTA provides the principal interface not only to LOFAR data retrieval and data mining but also to processing facilities for this data. Each site involved in the LTA provides storage capacity and optionally processing capabilities. To allow collaboration with a variety of institutes and projects, the LOFAR LTA merges different technologies (EGI, global file systems, Astro-WISE dataservers). Well-defined interfaces ensure that to both the astronomer and the LOFAR observatory the LTA behaves as a coherent information system. Access and utilization policies are managed via the central LOFAR identity management system that is designed to allow federation with organizational user administrations. The network connecting LOFAR to the LTA sites in Groningen, Amsterdam and J\"{u}lich, Germany consists of light-path connections, currently utilizing 10GbE technology, that are shared with LOFAR station connections and with the European eVLBI network \citep[e-EVN;][]{Szomoru08}.

The 10 Gbit/s bandwidth between the sites is sufficient for regular one-time LTA data transports but to allow transparent processing within the LTA it may grow to 60--80 Gbit/s bandwidth in the future. Such bandwidths will enable two major new processing modes: 1) Streaming of realtime or buffered observation data to a remote HPC system; 2) Streaming of stored data from one LTA site to a compute cluster located at another site. With these modes an optimal utilization of storage and processing facilities can be realized. If additional processing capacity is required for a given observing mode or for large-scale data processing, existing resources at partner institutes can be brought in without having to store (multiple copies of) datasets before processing can commence. For LOFAR datasets, which can grow up to hundreds of Tbyte, this capability will be essential.

\section{Operations and management}
\label{sec:operations}

Everyday LOFAR operations are coordinated and controlled from ASTRON's headquarters in Dwingeloo. Operators perform the detailed scheduling and configuration of the instrument, which includes setting up the appropriate online processing chain and destination of the data. The proper functioning of the stations, WAN and CEP system can be verified remotely. The monitoring and control system also collects and analyses the meta data gathered throughout the system in order to trace (impending) problems. Maintenance and repair of systems in the field is carried out under supervision of ASTRON personnel or the staff of an international station owner. Central systems maintenance is performed by staff of the University of Groningen's Centre for Information Technology. Advice and support is also given to the staff of the international partners who retain overall responsibility for their stations.

The International LOFAR Telescope (ILT) is a foundation established in Dwingeloo, the Netherlands, to coordinate the exploitation of the LOFAR resources under a common scientific policy. ASTRON provides the central operational entity for the ILT and the foundation is governed by a board consisting of delegates from each of the national consortia as well as a separate delegate from ASTRON itself. In relative proportion to their number of stations, the national owners put together the central exploitation budget. All observing proposals utilizing ILT facilities are reviewed on scientific merit by an independent ILT programme committee (PC). In the first and second years of operation, 10\% and 20\%, respectively, of the LOFAR observing and processing capacity will be distributed directly under Open Skies conditions and available to the general astronomical community. For the remainder, the national consortia each play a role in distributing reserved access shares, partly following national priorities, and partly taking into account the PC rankings. The fractions of time for open and reserved access in later years will be set by the ILT board.

\section{Software control infrastructure}
\label{sec:control}

\subsection{Monitoring and control system}
\label{sec:mac}

In the data processing pipeline of LOFAR, real time control is required to set the instrument in a certain state at a defined time. Furthermore, the instrument needs to be able to quickly switch between observing modes and be able to track sources. Hence, a distinction is made between real-time control during data taking and processing on the one hand, and control prior to this phase (mainly specification) and after that phase (mainly inspection) on the other hand. This separation is motivated by the different types of database technology and software design issues related to real-time operation requirements. 

The Specification, Administration and Scheduling (SAS) subsystem takes care of the specification and configuration of all observations and instrument settings. In contrast, the Monitoring and Control (MAC) subsystem is responsible for the operation of the instrument and the execution of observations, while collecting meta-data about those operations and observations. All user interaction is through the SAS and MAC systems. MAC is used to interface to running observations or processing pipelines. SAS is used for all other interaction prior to execution and afterwards. There is no “direct” interaction with applications. Interfaces to specific processing applications are implemented through the MAC layer and via a set of SAS GUIs. Finally, the SAS subsystem is used to provide an interface to the users for the collected meta-data and possible snapshots to inspect the observation performance and quality. 

At system level the choice has been made to control LOFAR centrally so that information is collected (and accessible) in a single place as much as possible. However, one of the design requirements is that the stations should be able to function for at least one hour autonomously. Hence, in each station a Local Control Unit (LCU) is present which controls the complete station (see Sect.\,\ref{sec:lcu}). In practice, the stations can operate autonomously indefinitely. All LCU functions are controlled remotely from the LOFAR operations center via the MAC system.

\subsection{System health monitoring (SHM)}

The percentage of time during which the LOFAR system is effectively operational, i.e. the system uptime, is an important issue that warrants considerable attention.  Due to the complexity of the LOFAR system and the harsh operating environment, it is almost certain that at any moment in time several of LOFAR's components (antennas, amplifiers, network links, computing nodes, etc.) will be non-functional. In the Netherlands, for example, the moisture levels and high humidity can lead to higher rates of component failure. Within reasonable bounds, this fact should not impact the usability of LOFAR for performing useful scientific measurements; rather, the system performance should gracefully degrade with each failing component until repairs can be effected.

Any faulty component may affect the quality of the measurements in a negative way, and may also jeopardize the operational capabilities of the LOFAR network. The objective of the SHM module of the LOFAR system is to support the efforts to maximize the system uptime. The main functions of the module will be the early detection of system failure, the accurate identification of failing components, and the support for remedial actions.

Daily or weekly on-site inspections cannot be performed in an economically viable way (at least for the remote stations). Hence the SHM subsystem will primarily be guided by the data that is generated by the LOFAR system in an automated fashion. This information consists of both the scientific data (generated by the antennas) and the ``housekeeping'' data of the equipment that controls the sensor network. Deviations in system health will be reflected in sensor data that deviate from the normative measurements. These deviations are called symptoms and are used by the SHM module to detect system failure and identify the responsible system component.

\subsection{Event triggers}
\label{sec:trigger}

LOFAR's digital nature makes it an inherently responsive telescope. With few moving parts, the ability to observe multiple targets simultaneously, and a software-driven control system, it is possible to make the telescope react intelligently to events (such as the detection of a fast transient, Sect.\,\ref{sec:tkp}, or CR, Sect.\,\ref{sec:cr}) as they happen, enabling the full capabilities of the telescope (long baselines, TBBs) to be rapidly brought to bear and ultimately maximizing scientific output.

The LOFAR pipeline system will make it possible to generate triggers as part of regular data processing, or in response to notifications from other facilities. The scheduler and control systems will then be able to insert appropriate follow-up actions into the schedule on the fly. Such actions will include, for example, reconfiguring the array, performing a new observation, or re-running a data processing pipeline with modified parameters.

For exchanging information about transient events with other facilities, LOFAR has standardized on the International Virtual Observatory Alliance (IVOA) VOEvent system \citep{VOEvent08,VOEvent11}. VOEvent provides a convenient and flexible way of representing and publishing information about events in a structured form that is well suited for machine processing. Although the full LOFAR VOEvent system is still under development, a VOEvent-based trigger has already been used to initiate LOFAR follow-up observations of gravitational wave event candidates detected by LIGO during September and October 2010 \citep{LIGO12}.

\begin{table*}[t]                                                                                                
\begin{minipage}[t]{\textwidth}
\caption{Current LOFAR observing modes}
\label{tab:modes}                                                                                           
\centering
\renewcommand{\footnoterule}{}  
\begin{tabular}{llll}                                                                                       
\hline \hline                                                                                                 
Type & Mode & Outputs & Description \\
\hline \hline
Interferometric & Correlated & Visibilities~~~~~~~ & Arbitrary number of stations, 8 beams per station, full Stokes \\
Beam-formed     & Incoherent stokes~~~ & BF data file & Incoherent summation, arbitrary stations, 8 station beams, full Stokes \\ 
   & Coherent stokes & BF data file & Coherent summation, Superterp only, ~20 full-resolution beams, full Stokes \\
   & Complex voltage & BF data file & Coherent summation, Superterp only, bypasses 2nd PPF, raw voltage output \\ 
   &  Station level & BF data file & Arbitrary stations, individual pointing and frequency settings per station \\
   &                &              & 8 station beams, Stokes I  \\ 
Direct storage  & Raw voltage &  TBB data file & Station level triggering of TBB dumps, direct storage to CEP cluster \\
\hline \hline
\end{tabular}
\end{minipage}
\end{table*}

\section{Observing modes}
\label{sec:modes}

\subsection{Interferometric imaging}
\label{sec:imaging}

The interferometric imaging mode provides correlated visibility data, just like traditional aperture synthesis radio telescope arrays consisting of antenna elements. The goal of the LOFAR imaging mode is to achieve high fidelity, low noise images of a range of astronomical objects, using customizable observing parameters. In this operating mode, station beams are transferred to the CEP facility where they are correlated to produce raw visibility data. The raw $uv$ data are stored on the temporary storage cluster. Further processing, which consists of calibration and imaging (see Sect.\,\ref{sec:sip}), is handled off-line. Calibration is an iterative process of obtaining the best estimates of instrumental and environmental effects such as electronic station gains and ionospheric delays. 

The final data products for this mode include the calibrated $uv$ data, optionally averaged in time and frequency, and corresponding image cubes. The visibility averaging is performed to a level which reduces the data volume to a manageable level, while minimizing the effects of time and bandwidth smearing. It will be possible to routinely export datasets to investigators for reduction and analysis at their Science Centre or through the use of suitable resources on the GRID.

For imaging observations, a wide range of user interaction will be supported. Experienced users will require control over the calibration and imaging stages of data reduction, while more typical users will not wish to recalibrate the visibility data, but may need to control imaging parameters. Many users may require only a fully processed image. The MAC system will provide personalized control over key aspects of the calibration and imaging pipelines. For expert users, interactive control of this processing will be available using the SAS and MAC GUIs over the network.

This mode requires medium to long-term storage of un-calibrated or partially calibrated data at the CEP facility, to allow reprocessing of data following detailed inspection of results by the user. The resulting storage and processing requirements may impose limits on the amount of such customized reprocessing which may be conducted in the early years of LOFAR operation.

\begin{figure*}
\centering
\includegraphics[width=\textwidth]{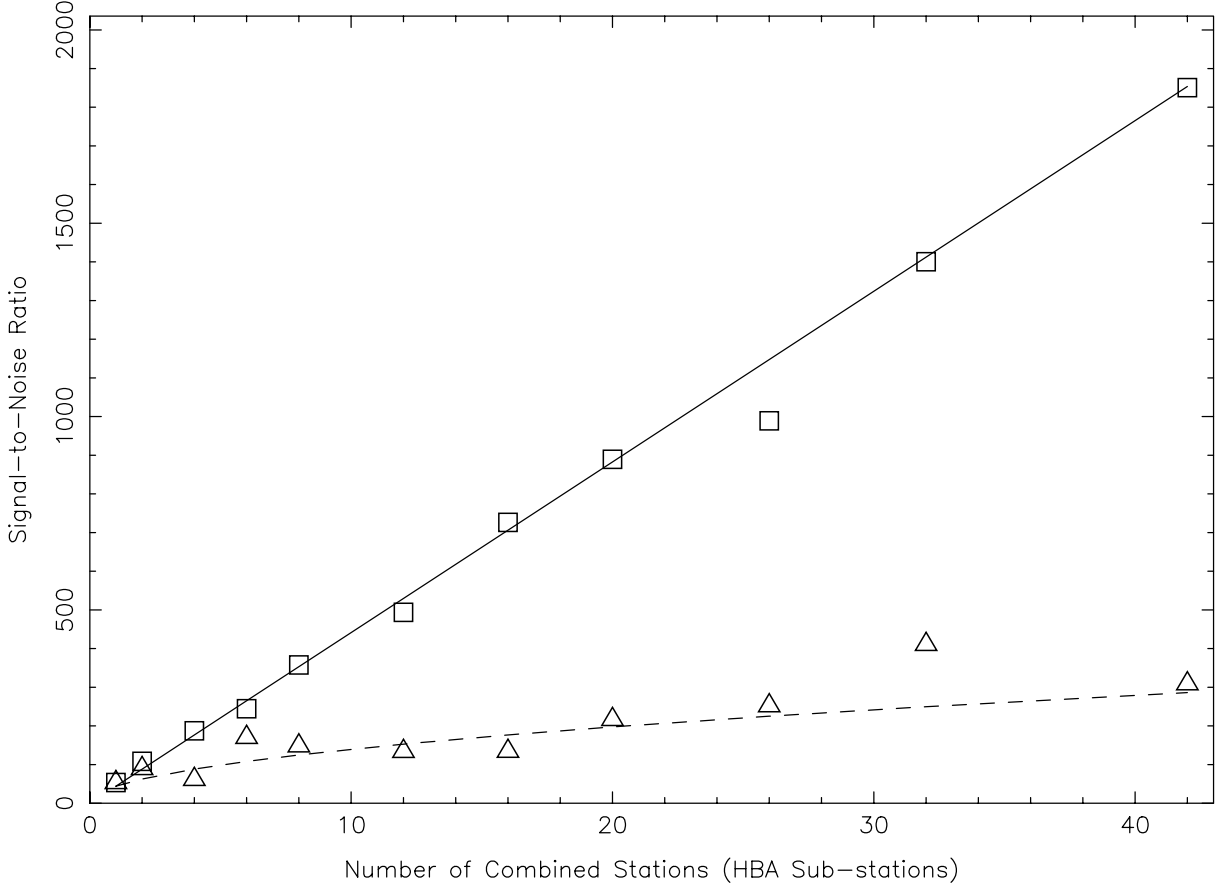}
\caption{\small Increase in the signal-to-noise ratio (S/N) of the pulsar PSR B1530+27 as a function of the number of coherently (squares) and incoherently (triangles) added HBA sub-stations in the LOFAR core.  The S/N is seen to increase linearly (solid line) in the case of coherent addition and as the square-root (dashed line) of the number of stations in the case of incoherent addition - as expected for sources that do not contribute significantly to the system temperature.  Coherently and incoherently summed data were acquired simultaneously in 11 separate observations that summed between 1 to 42 HBA sub-stations. Note that the typical error on the S/N ratio measurements is $\sim 10$\% and these measurements are also systematically affected by the intrinsic brightness of the source (pulse-to-pulse brightness variations) as well as RFI.}
\label{fig:coh_inc_sum}
\end{figure*}

\subsection{Beam-formed modes}
\label{sec:beamformed}

Instead of producing interferometric visibilities, LOFAR's beam-formed modes can either combine the LOFAR collecting area into ``array beams''- i.e. the coherent or incoherent sum of multiple station beams - or return the un-correlated station beams from one or more stations \citep[see also][]{sha+11,ml11}.  These data are used to produce time-series and dynamic spectra for high-time-resolution studies of, e.g., pulsars, (exo)planets, the Sun, flare stars, and CRs. These modes are also useful for system characterization and commissioning (e.g. beam-shape characterization, offline correlation at high time resolution, etc.).  In the current implementation, there are several beam-formed sub-modes: i) Coherent Stokes, ii) Incoherent Stokes, and iii) Fly's Eye.  These can all be run in parallel in order to produce multiple types of data products simultaneously.

The Coherent Stokes sub-mode produces a coherent sum of multiple stations (also known as a ''tied-array" beam) by correcting for the geometric and instrumental time and phase delays. This produces a beam with restricted FoV, but with the full, cumulative sensitivity of the combined stations. This sub-mode can currently be used with all 24 LOFAR core stations which all receive the same clock signal and hence do not require a real-time clock calibration loop for proper phase alignment.  This coherent summation results in a huge increase in sensivity, but with a limited FoV of only $\sim 5^{\prime}$ (see Fig.\,\ref{fig:coh_inc_sum}). The Superterp and in fact the entire 2-km LOFAR core are compact enough that ionospheric calibration is also not likely to be a major limitation to coherently combining these stations, at least not for the high band. In practice, experience has shown that the calibration tables used to correct for the delays are stable over timescales of many months and need only be updated occasionally.

\begin{figure*}
\centering
\includegraphics[width=6.5in]{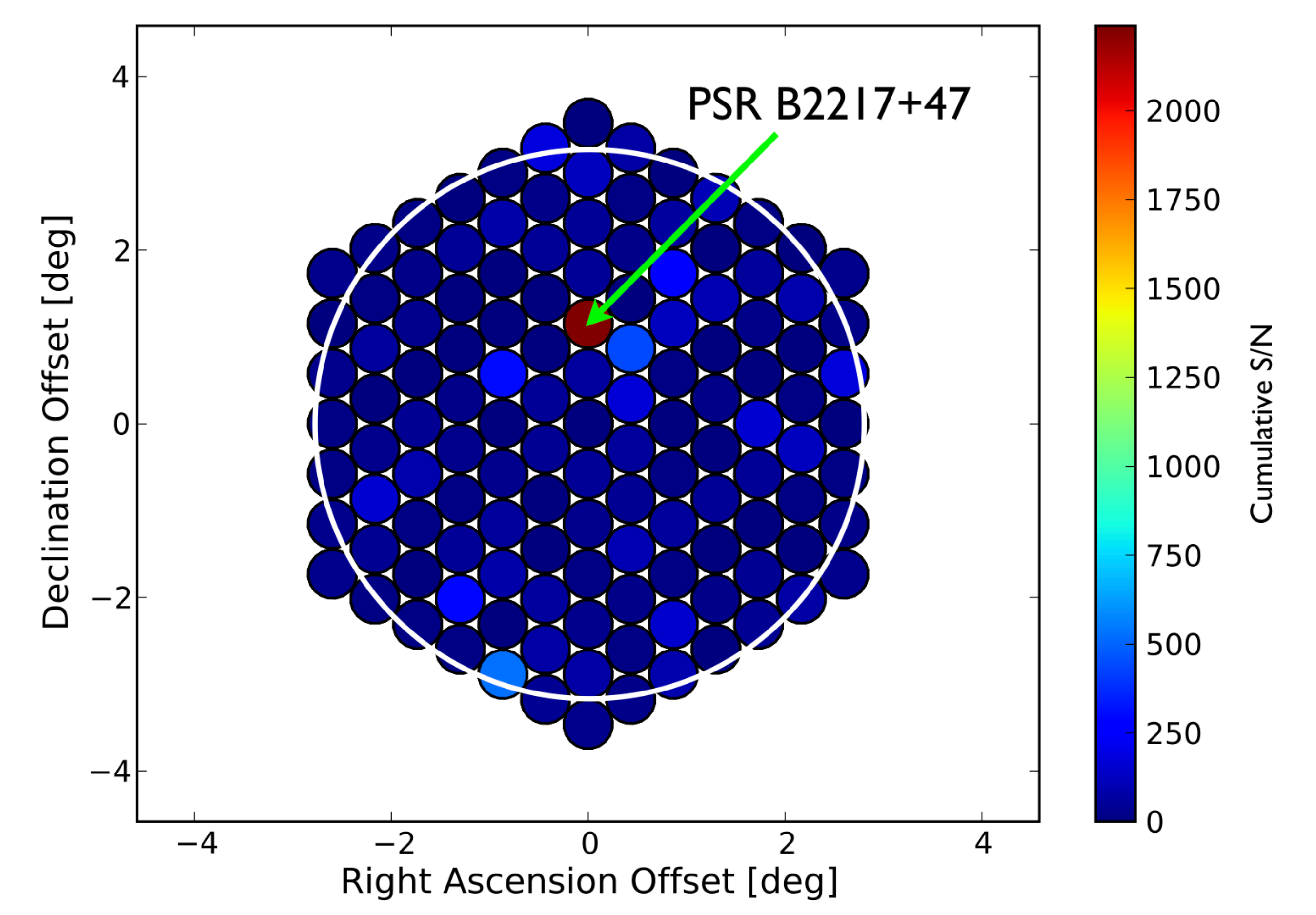}
\caption{\small Cumulative S/N map from a 127-beam tied-array observation of pulsar B2217+47, using the 6 HBA Superterp stations.  These beams have been arranged into a ``honeycomb" pattern in order to completely cover the 5.5$\degr$ station beam FoV.  The circle sizes represent the tied-array beam full-width half maxima (roughly 0.5$\degr$), and the color scale reflects the S/N of the pulsar in each beam.}
\label{fig:tabs}
\end{figure*}

In this mode, one can write up to $\sim 300$ simultaneous, full-bandwidth tied-array beams as long as the time and frequency resolution are modest \citep[for limitations and system benchmarking results, see][]{ml11}.  Note that the Superterp tied-array beams have a FWHM of $\sim 0.5^{\circ}$ and roughly 127 are required to cover the full single station FoV (see Fig.\,\ref{fig:tabs}). Depending on the scientific goal of the observations, either Stokes I or Stokes I,Q,U,V can be recorded, with a range of possible frequency (0.8--195\,kHz) and time ($\gid 5.12$\,$\mu$s) resolutions.  It is also possible to record the two Nyquist-sampled linear polarizations separately, which is referred to as `Complex Voltage' mode.  This mode is necessary for applications such as offline coherent dedispersion, fast imaging, or inverting the initial, station-level poly-phase filter to achieve the maximum possible time resolution.

The Incoherent Stokes sub-mode produces an incoherent combination of the various station beams by summing the powers after correcting for the geometric delay.  This produces beams with the same FoV as a station beam, but results in a decrease in sensitivity compared with a coherently added tied-array beam - i.e. the gain in sensitivity scales with the square-root of the number of stations as opposed to linearly (see Fig.\,\ref{fig:coh_inc_sum}).  One incoherent array beam can be formed for each of the beams created at station level - e.g., if all the stations being summed split their recorded bandwidth across 8 pointing directions, then 8 incoherent array beams can also be formed from these.  All LOFAR stations, including the international stations, can be summed in this sub-mode, which can be run in parallel with the Coherent Stokes sub-mode.  As in Coherent Stokes, one can record either Stokes I or Stokes I,Q,U,V with the same range of frequency and time resolutions.

The Fly's Eye sub-mode records the individual station beams (one or multiple per station) {\it without} summing.  As with the Coherent and Incoherent Stokes modes, the normal online BG/P processing steps (e.g. channelization and bandpass correction) are still applied.  This mode is useful for diagnostic comparisons of the stations, e.g. comparing station sensitivities, but can also be used for extremely wide-field surveys if one points each station in a different direction.  In combination with the Complex Voltage sub-mode, Fly's Eye can also be used to record the separate station voltages as input for offline fast-imaging experiments.  It is also possible to simultaneously record a coherent and incoherent sum of all the stations used in this mode.

These modes, and in some cases even a combination of these modes, can be run in parallel with the standard imaging mode described above. This allows one to simultaneously image a field while recording high-time-resolution dynamic spectra to probe sub-second variations of any source in the field \citep[see for example Fig.\,11 in][]{sha+11}.

\subsection{Direct storage modes}
\label{sec:direct}

Direct storage modes refer to observing modes that bypass the BG/P and deliver station data directly to the storage nodes of the offline cluster. These modes typically correspond either to triggered, short-term observations run in parallel with other observing modes, such as dumping the TBB boards following a CR or transient detection, or data taken by a single station in standalone mode. Types of data that can currently be stored in this manner include TBB data dumps using either full resolution or sub-band mode, station level beamformed data, and station level metadata. A variety of metadata are produced on the stations such as event triggers from the TBB boards (see Sect.\,\ref{sec:tbb}) as well as diagnostic output from the calibration algorithm running on the LCU. Any or all of these data and metadata may be streamed directly from the stations to the storage nodes where they are incorporated into LOFAR standard data products. Once on the offline cluster, these data products can then be archived or further processed depending upon the scientific objective as with all LOFAR outputs. Example astronomical applications that utilize data from direct storage modes include all-sky imaging using intra-station baselines, single station observations of bright pulsars, dynamic spectral monitoring of the Sun or planets, and the detection of CR air showers.

\begin{table*}[t]                                                                                                
\begin{minipage}[t]{\textwidth}
\caption{Current LOFAR processing pipelines}
\label{tab:pipelines}                                                                                           
\centering
\renewcommand{\footnoterule}{}  
\begin{tabular}{lllll}                                                                                       
\hline \hline                                                                                                 
Pipeline & Mode & Inputs & Outputs & Description \\
\hline \hline
Standard & Interferometric & Visibilities 
         & Image cubes, source lists  & Limited angular resolution, full FOV \\
     & & & sky models, quality metrics \\
Long-baseline & Interferometric & Visibilities 
         & Image cubes, source lists  & Highest angular resolution, limited FOV \\  
     & & & sky models, quality metrics \\     
Known pulsar & Beam-formed & BF data file 
             & Folded pulse profiles & Arbitrary number of stations, \\
         & & & de-dispersed time series & 8 beams per station, full Stokes \\
         
CR event & Direct storage & TBB data file 
             & CR characteristics& Single or multiple station event triggering \\
         & & & event database \\
Transient detection & Interferometric & Image cubes 
                    & Source lists, light curves & Can run in dedicated mode or commensal \\
                & & & classifications, triggers & with other imaging observations \\
\hline \hline
\end{tabular}
\end{minipage}
\end{table*}

\section{Processing pipelines}
\label{sec:pipelines}

\begin{figure*}
\centering
\includegraphics[width=\textwidth]{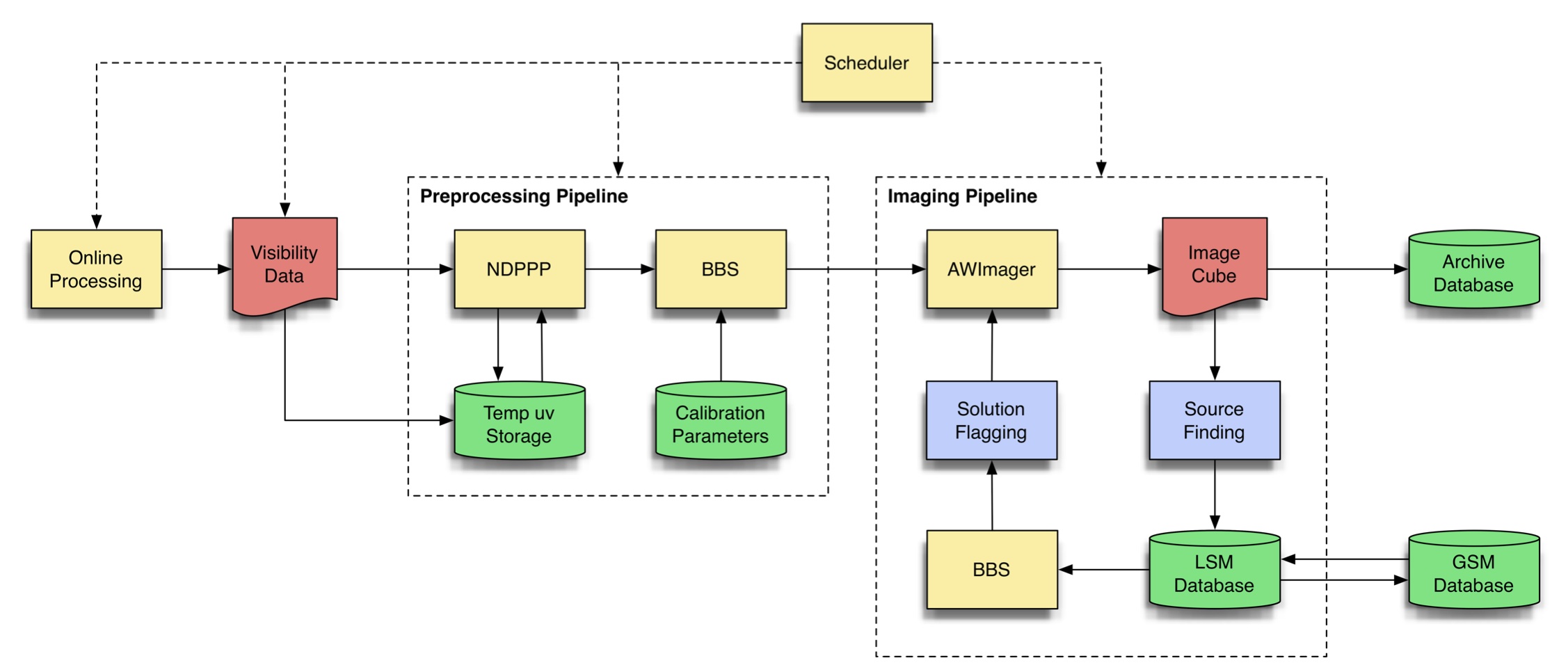}
\caption{\small The LOFAR imaging pipeline presented in schematic form \citep{Heald10}. See the text for a description of the various software components and the data path.}
\label{fig:sip}
\end{figure*}

\subsection{Standard imaging}
\label{sec:sip}

The standard imaging pipeline (SIP) is shown schematically in Fig.\,\ref{fig:sip}. A short overview of the pipeline is given by \citet{Heald11}, and a more in-depth description of the pipeline, its components, and the intermediate data products is in preparation (Heald et al., in prep.). Here, we outline the main pipeline features.

Following the data path from the left, visibility data are produced in the form of measurement sets at CEP, and recorded to multiple nodes in the LOFAR offline CEP cluster. The first standard data processing steps are encapsulated within a sub-pipeline called the pre-processing pipeline. Its role is to flag the data in time and frequency, and optionally to average the data in time, frequency, or both. The software that performs this step is labelled new default pre-processing pipeline, or NDPPP, and includes flagging using the {\tt AOFlagger} routine (see Sect.\,\ref{sec:rfi}). 

This first stage of the processing also includes a subtraction of the contributions of the brightest sources in the sky (Cygnus A, Cassiopeia A, etc.) from the visibilities, using the demixing technique described by \citet{vdTol2007} and implemented in NDPPP. Next, an initial set of calibration parameters is applied. In the current system, the initial calibration comes from an observation of a standard flux reference source \citep[as characterized by][]{scaife_heald_2012} which may have been performed in parallel with, or immediately preceding, the main observation. An initial phase calibration is achieved using the BlackBoard Selfcal (BBS) package developed for LOFAR. 

The local sky model (LSM) used for the phase calibration is generated from the LOFAR Global Sky Model (GSM) that is stored in a database. The LOFAR GSM contains entries from the VLA Low-frequency Sky Survey \citep[VLSS and VLSSr;][]{VLSS2007, VLSS2012}, the Westerbork Northern Sky Survey \citep[WENSS;][]{WENSS1997}, and the NRAO VLA Sky Survey \citep[NVSS;][]{NVSS1998} catalogs, and is being supplemented with entries from the Multifrequency Snapshot Sky Survey (MSSS, see Sect.\,\ref{sec:calib}). Finally, additional flagging and filtering operations (not shown in the figure) are performed in order to remove any remaining RFI or bad data.

\begin{figure*}
\centering
\includegraphics[width=\textwidth]{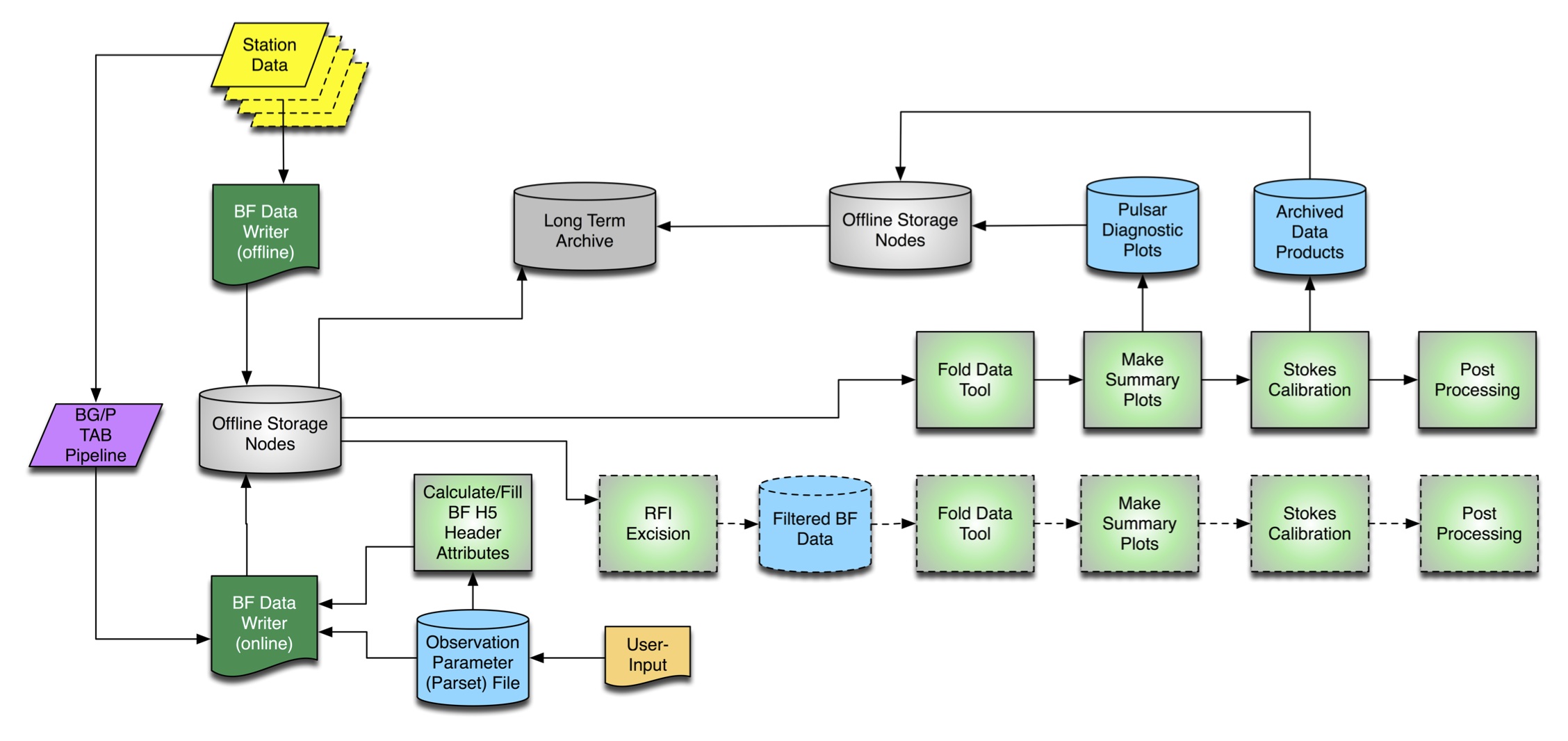}
\caption{\small Schematic diagram of the overall Pulsar Pipeline, as it runs online on the BG/P (see also Fig.\,\ref{fig:online}) followed by offline scientific processing on the offline cluster. Offline pipeline processing can be run on data directly out of the BG/P or on RFI-filtered data.}
\label{fig:pulp}
\end{figure*}

Following the pre-processing stage, the calibrated data are further processed in the Imaging Pipeline, which begins with an imaging step that uses a modified version of the CASA imager \citep{Tasse13}. This imager applies the w-projection algorithm \citep{Cornwell08} to remove the effects of non-coplanar baselines when imaging large fields and the new A-projection algorithm \citep{Bhatnagar08} to take into account the varying primary beam during synthesis observations. Source finding software is used to identify the sources detected in the image, and generate an updated LSM. One or more `major cycle' loops of calibration (with BBS), flagging, imaging, and LSM updates are performed. At the end of the process, the final LSM will be used to update the GSM, and final image products will be made available via the LTA.

The Scheduler oversees the entire end-to-end process, from performing the observation through obtaining the final images. In addition to scheduling the observing blocks at the telescope level, it keeps an overview of the storage resources in order to decide where to store the raw visibilities. It also keeps an overview of the computational resources on the cluster, so that runs of the Pre-processing Pipeline and Imaging Pipeline can be scheduled and distributed over cluster nodes with available processing power.

\subsection{Pulsar processing}
\label{sec:pulsarpipe}

The raw beam-formed data written by BG/P (see Sect.\,\ref{sec:beamformed}) are stored on the LOFAR offline processing cluster and Long-Term Archive in the HDF5 format (Hierarchical Data Format).  The exact structure of these files as well as the metadata are fully described in the appropriate LOFAR Interface Control Document (ICD) available from the LOFAR website.

Since the beam-formed data serve a variety of different science cases, several pipelines exist, e.g., to create dynamic spectra, search in real-time for fast transients, and for performing standard pulsar processing.  The most advanced of these pipelines is the standard pulsar pipeline, `Pulp', which is shown schematically in Fig.\,\ref{fig:pulp}, and is described in more detail by \citet{sha+11}.  Pulp is currently implemented within a python-based framework that executes the various processing steps.  The framework is sufficiently flexible that it can be extended to include other processing steps in the future.

Several conversion tools have been developed to convert these data into other formats, e.g. {\tt PSRFITS} \citep{PSRCHIVE04}, suitable for direct input into standard pulsar data reduction packages, such as {\tt PSRCHIVE} \citep{PSRCHIVE04}, {\tt PRESTO} \citep{PRESTO11}, and {\tt SIGPROC} \citep{SIGPROC11}.  The long-term goal is to adapt these packages to all natively read HDF5, using the LOFAR Data Access Layer (DAL) for interpreting the HDF5 files.  We have already successfully done this adaptation with the well-known program {\tt DSPSR} \citep{DSPSR11}, which now natively reads LOFAR HDF5.  Among other things, these reduction packages allow for RFI masking, dedispersion, and searching of the data for single pulses and periodic signals.  Already, a test-mode exists to perform coherent dedispersion online, also for multiple beams/dispersion measures.  Likewise, online RFI excision is also being implemented in order to excise corrupted data from individual stations before it is added in to form an array beam.

\begin{figure*}
\centering
\includegraphics[width=\textwidth]{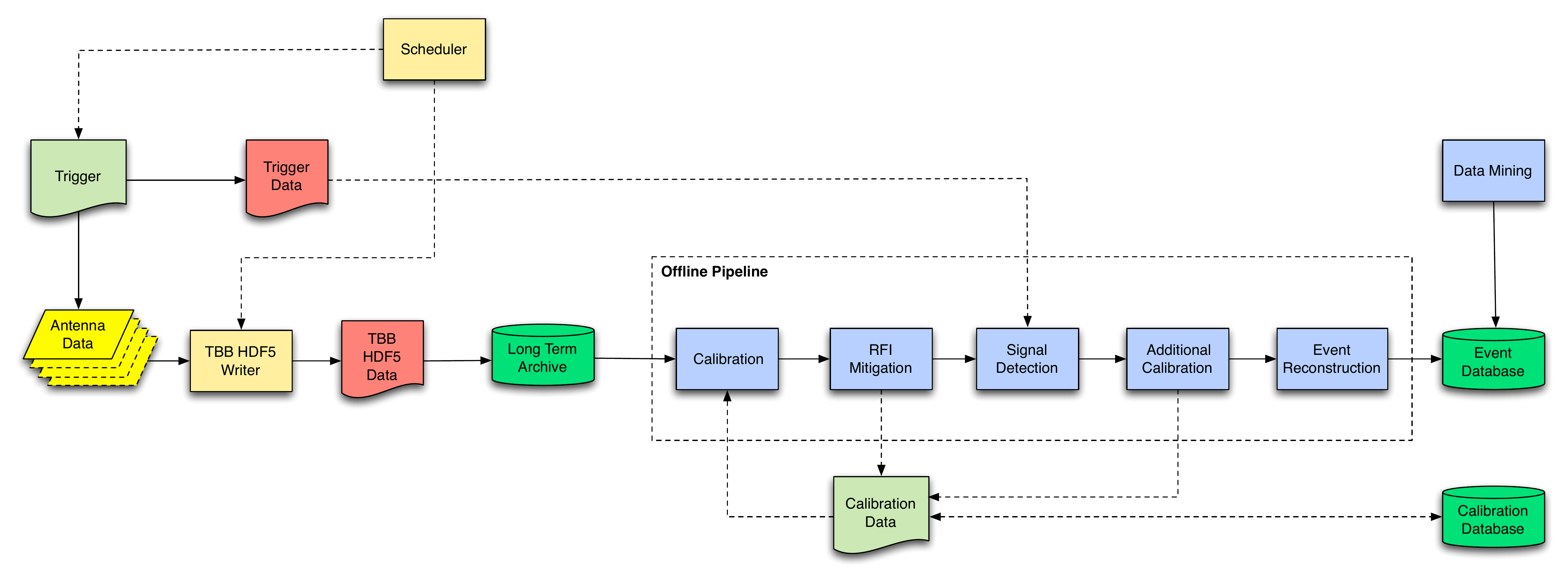}
\caption{\small Schematic view of the CR pipeline. The HDF5 data are the standardized output. The offline pipeline can be adapted to the purpose and type of the observation.}
\label{fig:vhecr}
\end{figure*}

\begin{figure*}[ht]
\centering
\includegraphics[width=0.485\textwidth]{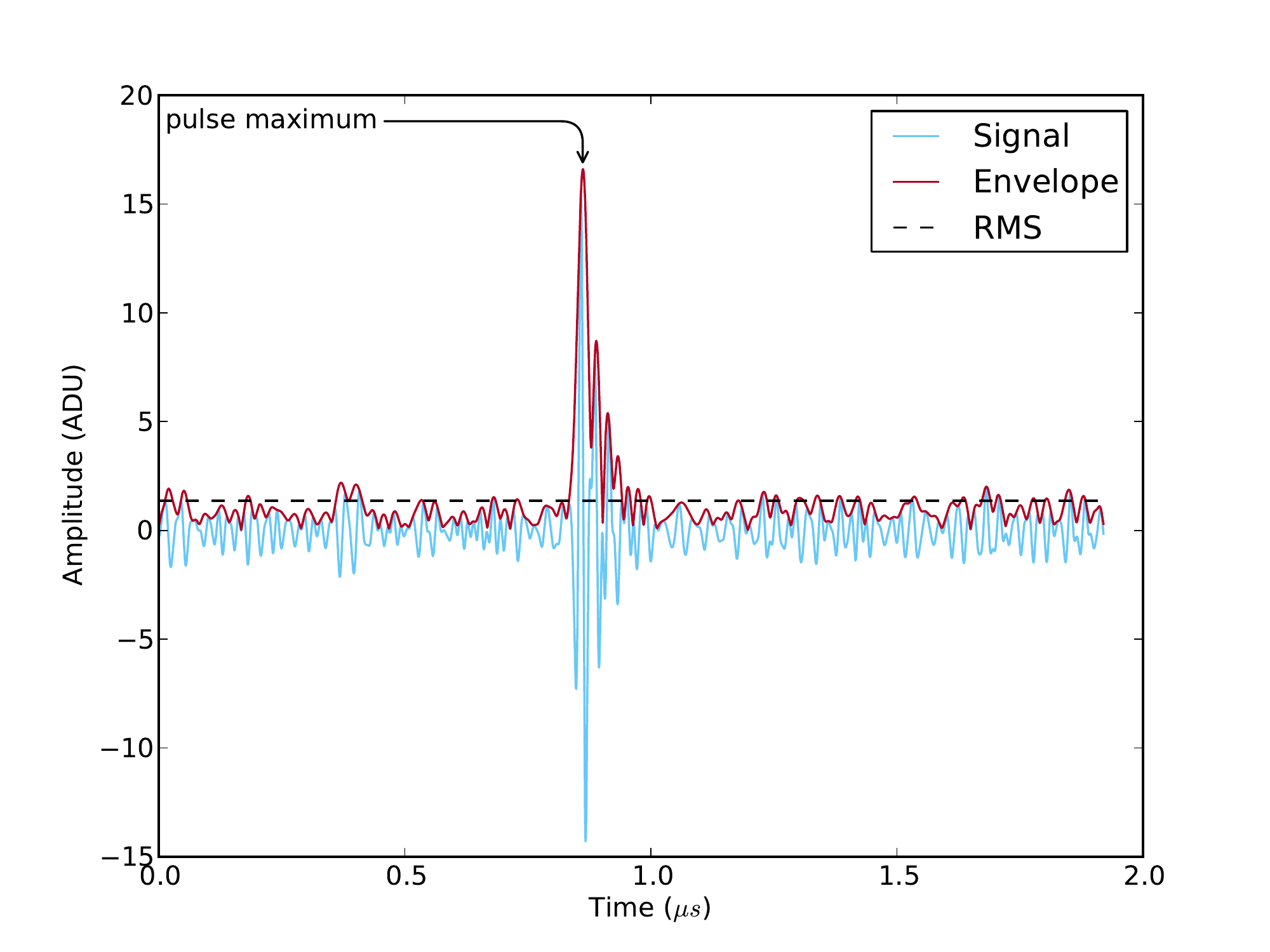}
\hspace{0.1in}
\includegraphics[width=0.485\textwidth]{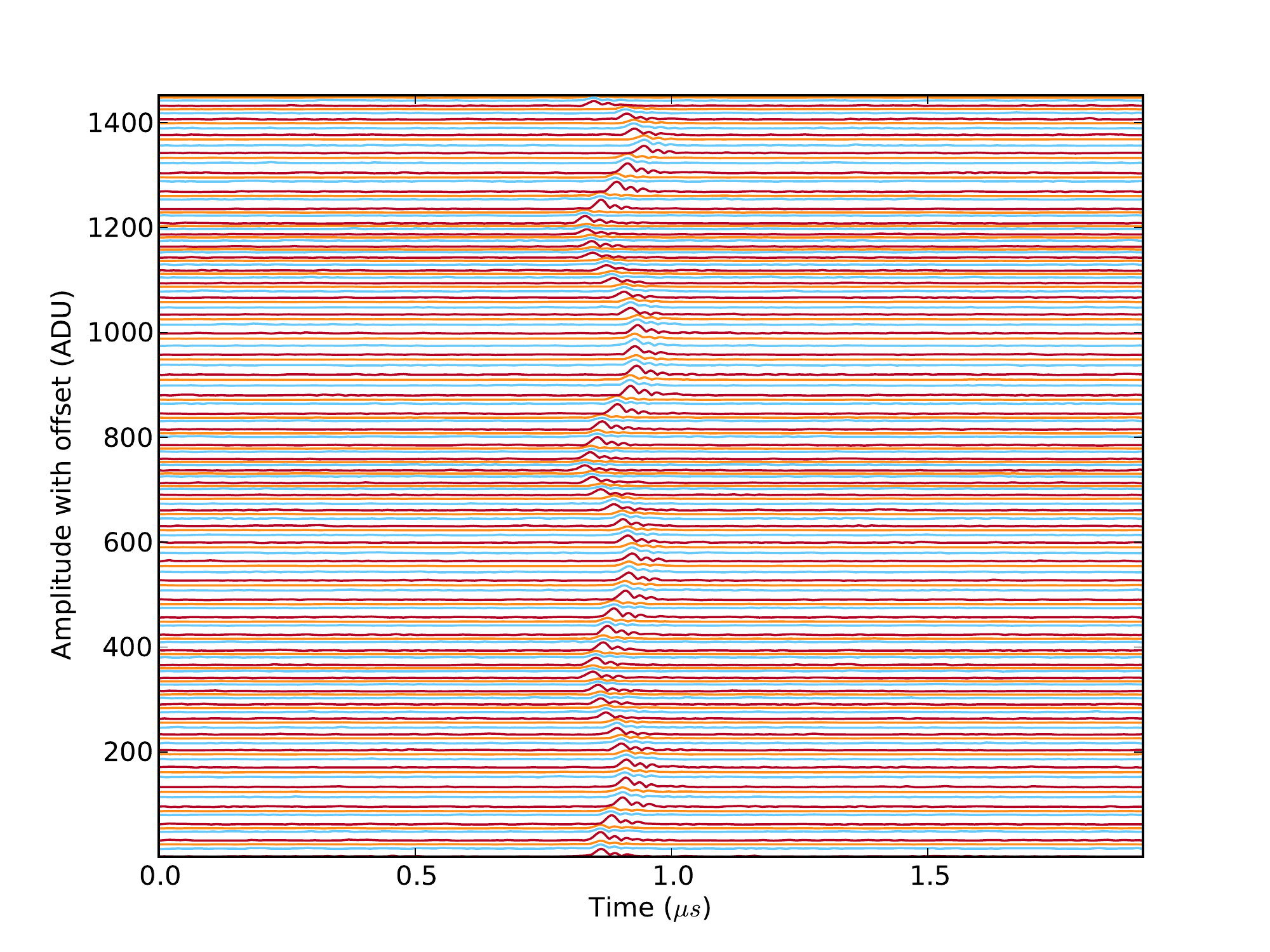}
\caption{\small Illustrative results of the CR pipeline. {\bf Left:} CR pulse as recorded by one LBA antenna along with the reconstructed Hilbert envelope. The square of the Hilbert envelope corresponds to the sum of the squares of the original signal and the squares of the Hilbert transform. The Hilbert envelope is the amplitude of the analytic signal and essentially captures the amplitude of the pulse. {\bf Right:} Hilbert envelopes for all antennas of one station ordered by their RCU number. One can clearly see the time delay of the air shower signal between different antennas as the antennas are numbered in a circular layout \citep{Nelles2013}.}
\label{fig:cr_pulses}
\end{figure*}

\subsection{CR event processing}
\label{sec:crpipe}

The high digital sampling rate of LOFAR (5 ns or 6.25 ns for the 200 MHz or 160 MHz clock, respectively) combined with the wide-field nature of its receivers make it a uniquely powerful instrument for the detection and study of CRs. Air showers of charged particles produced by CRs striking the Earth's upper atmosphere can generate bright, extremely short duration radio pulses \citep{FalckeGorham2003}. Depending on the energy and direction of the incident CR, these pulses can be detected by the antennas in one or more LOFAR stations, as shown in Fig.\,\ref{fig:cr_pulses}. 

Due to their short duration, in order to measure radio pulses from CRs, LOFAR must be be triggered. When triggered the TBB RAM buffers in the station are frozen and the data are transferred directly to the CEP post-processing cluster (see Sect.\,\ref{sec:tbb} and Sect.\,\ref{sec:direct}). Such a trigger can be initiated in several ways. Either a pulse-finding algorithm is run on the FPGA and if a pulse is recorded by multiple dipoles simultaneously within a specified time window a dump is initiated. Alternatively, a dump of the TBB RAM buffers in a given station (or stations) can be triggered from the system level by triggers external to the station itself. These external triggers may come from outside LOFAR, as in the case of VOEvents from other observatories (see Sect.\,\ref{sec:trigger}), or from within the LOFAR system.

\begin{figure*}[ht]
\centering
\includegraphics[width=\textwidth]{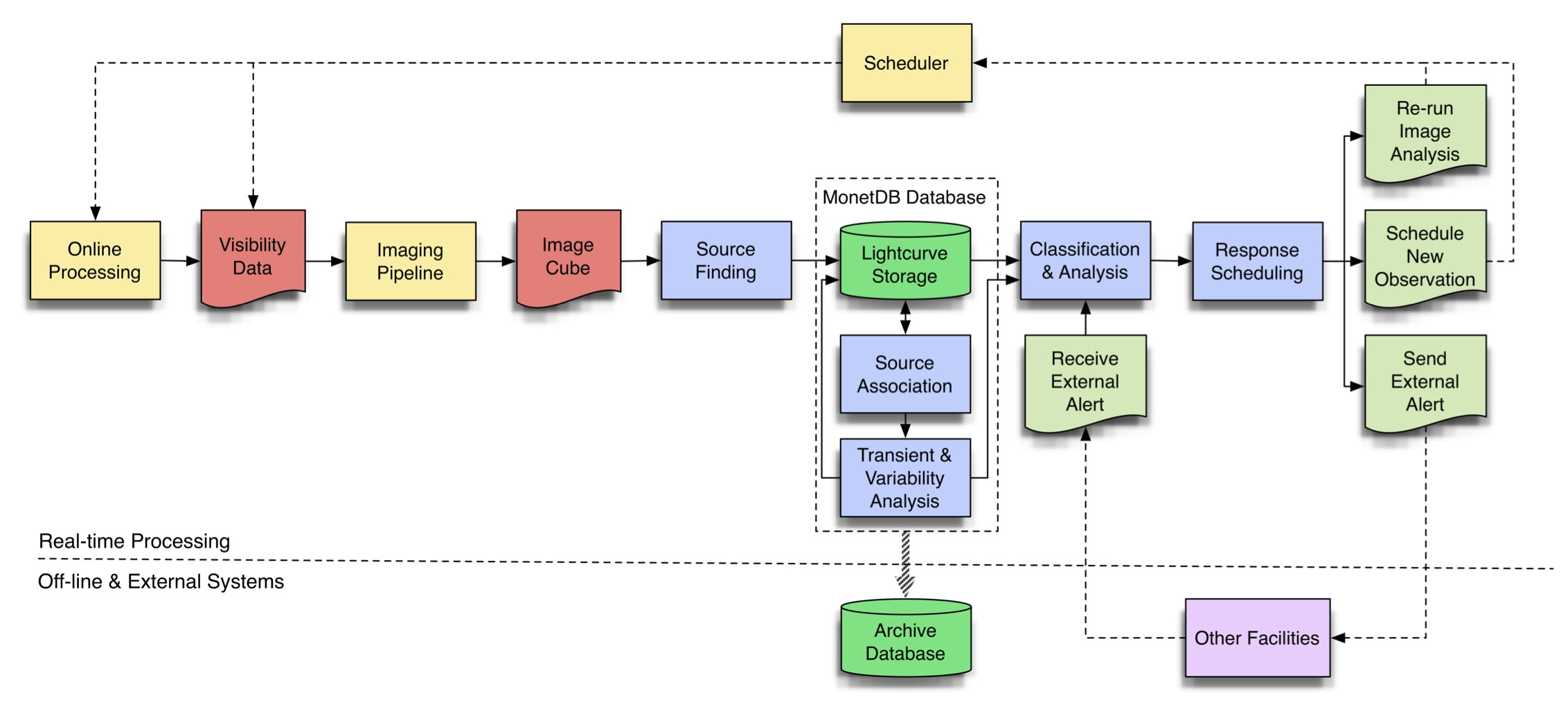}
\caption{\small Schematic outline of the main components in the LOFAR transients detection pipeline. Data are ingested from a modified version of the standard imaging pipeline (Sect.\,\ref{sec:sip}), while transients analysis is performed using a combination of custom source-finding and analysis routines and a high-performance MonetDB database.}
\label{fig:tkp}
\end{figure*}

As a CR produces its signal in the atmosphere a single CR pulse in an individual antenna does not largely differ from a RFI pulse. A large training set of detected CRs is needed in order to program the pulse-finding algorithm to only send a minimal amount of false triggers. In order to achieve this, one of the internal triggers is sent to LOFAR by an array of particle detectors, which is set up at the Superterp \citep{Thoudam11}. These detectors trigger LOFAR only on CRs and also allow a cross-calibration of the measured characteristics of the air shower. Other types of internal triggers from for example other processing pipelines are foreseen. All CR detection modes place a strong constraint on the response time of the LOFAR system.  The system must be able to process a detected pulse and freeze the contents of the TBB RAM buffer within a time interval which is smaller than the length of the buffer itself, otherwise the data from the event will be lost.

Following a trigger, raw voltage time series data from the TBB RAM buffers are stored in the HDF5 format using a data structure similar to the beam-formed data files mentioned in Sect.\,\ref{sec:pulsarpipe}. The relevenat ICD describing this TBB format is available form the LOFAR website. From the CEP post-processing cluster, these TBB data files are sent to the Long-term Archive (see Sect.\,\ref{sec:lta}) where they can then be accessed for offline processing as described in Fig.\,\ref{fig:vhecr}. The processing pipeline applies filtering and calibration corrections and characterizes the original CR event itself (in terms of direction and signal distribution). Finally the event information is stored in an SQL database in order to provide fast access for further study. The pipeline is implemented as a mixture of C++ libraries with Python bindings and Python scripts and accounts for the fact that the source is in the near-field, not at infinity as is assumed in other LOFAR processing pipelines. The pipeline can be run as a post-processing step performed automatically following any CR observations that result in data dumps from the TBBs, as well as interactively. The pipeline will be described in detail in a forthcoming publication.

Currently, the LOFAR system supports two types of CR observing modes differing only in whether the detection trigger is generated at the station level or by an external event at the system level. A number of additional modes are, however, envisioned for future development. These include modifications to the station level detection algorithm to tune the trigger mechanism to characteristic pulse profiles from different phenomena such as lightning for example. Furthermore, this type of pulse search can be tuned to detect single dispersed pulses originating from fast radio transients like pulsars or other astronomical objects \citep{2008ASPC..395..271F}. This method has already been successfully tested at LOFAR by detecting a giant pulse from the Crab Nebula \citep{2012IAUS..285..411T}.

Similarly for pulses too faint to be detected by individual antennas, a trigger mechanism employing an anti-coincidence check between multiple on/off-source tied-array beams is envisioned. Such a trigger mechanism could in principle be used to detect faint radio flashes due to neutrinos interacting with the lunar regolith (see Sect.\,\ref{sec:cr}). Although much more sensitive, tied-array beam-based trigger mechanisms will also necessarily have more limited fields of view as opposed to dipole-based triggering that is essentially omnidirectional.

\subsection{Transient detection}
\label{sec:transients}

Beyond the pipelines already deployed as part of the operational LOFAR system, an additional science pipeline is currently under development tailored to detect transient and variable radio sources. The digital nature of the LOFAR system makes it inherently agile and an ideal instrument for detecting and, perhaps more crucially, responding to transient sources. Unlike the modes discussed previously, the transient detection pipeline will consist of a near real-time imaging pipeline that monitors the incoming stream of correlated data for both known variable sources and previously unknown transients. When a new source is detected, or a known source undergoes a rapid change in state, this mode will make it possible to respond on short timescales. Through a mixture of processing performance improvements and data buffering, the ultimate goal is to deploy a system capable of detecting radio transients down to timescales $\sim$1\,s with a response time latency of order $\sim$10\,s.

Possible responses include triggering actions within the LOFAR system such as switching to a different, targeted observational mode, adjusting the sub-band selection for an optimal frequency coverage, or dumping the data from the transient buffer; or, alternatively sending notifications to other observatories to initiate coordinated observations. LOFAR will also be capable of receiving and responding to triggers from external facilities in much the same way. This section presents a brief overview of the main components of the pipeline. A more comprehensive description is available in \cite{Swinbank11} and Swinbank et al. (in prep.).

An overview of the design of the transients detection pipeline is shown in Fig.\,\ref{fig:tkp}. Image cubes produced by a variant of the standard imaging pipeline (Sect.\,\ref{sec:sip}) are ingested into the system, which identifies transients both by image plane analysis and by comparing the list of sources found in the images with those in previous LOFAR observations and other catalogs. Measurements from individual images are automatically associated across time and frequency to form light-curves, which are then analyzed for variability. Cross-catalog comparison, light-curve construction and variability analysis take place within a high-performance MonetDB database \citep{monetdb06, monetdb07}. A classification system, based initially on simple, astronomer-defined decision trees, but later to be augmented by machine learning-based approaches, is then used to identify events worthy of response. The primary data products of this mode will be rapid notifications to the community of transient events and a database of light-curves of all point sources observed by LOFAR (around 50--100\,Tbyte\,year$^{-1}$). In addition, snapshot images integrated over different time-scales as well as {\it uv} datasets suitably averaged in frequency will be archived.

High-speed response to transients is essential for the best scientific outcome so low-latency operation is therefore crucial in this mode. Achieving these low system latencies and response times may ultimately require adaptation of new, non-imaging algorithms for transient detection that rely on phase closure quantities \citep{Law2012a, Law2012b}. Experiments employing these new algorithms are already underway as part of the commissioning process. If sufficiently low latency can be achieved, the TBBs (Sect.\,\ref{sec:tbb}) can be dumped in response to a new transient, providing a look-back capability at the highest possible time, frequency and angular resolution. These requirements preclude human intervention, so all processing is fully automated. Efforts are also underway to minimize the time taken to transport and process data within the LOFAR system.

\begin{figure*}[p]
\centering
\vspace{0.2in}
\begin{center}
\hspace{0.01in}
\hbox{
\hspace{0.125in}
\includegraphics[height=2.77778in]{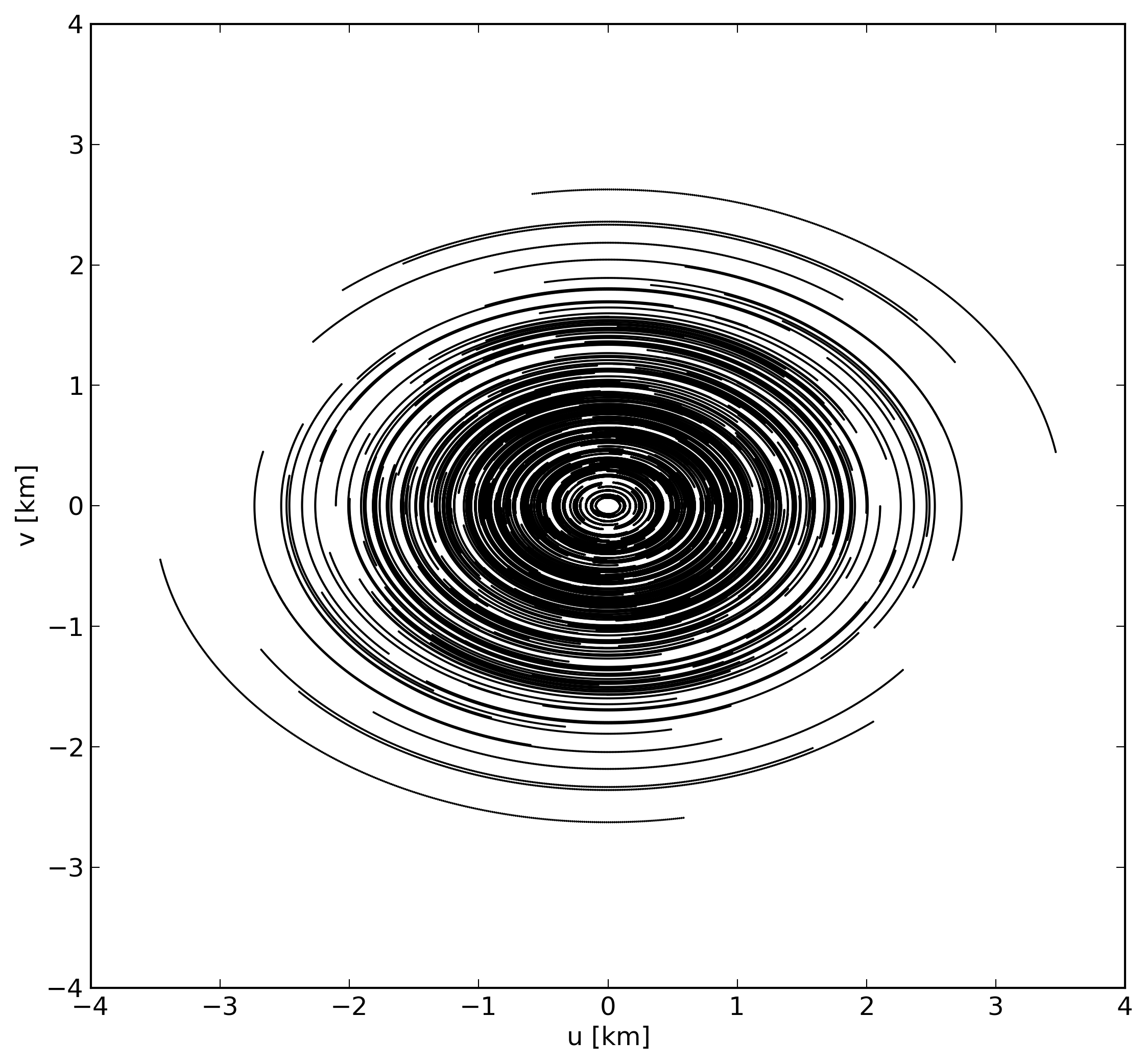}
\hspace{0.2250in}
\includegraphics[height=2.75in]{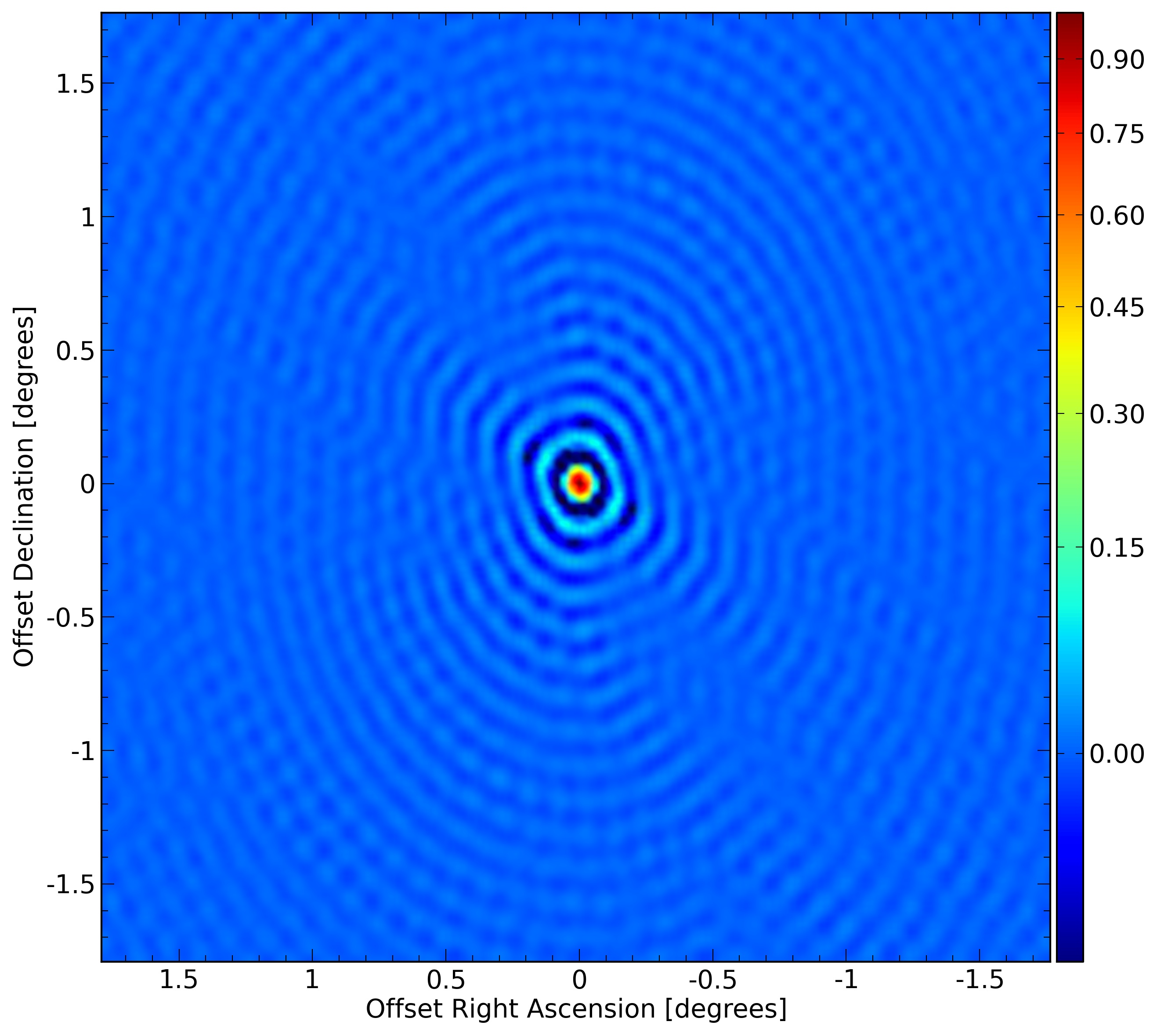}
}
\end{center}

\begin{center}
\hspace{0.01in}
\hbox{
\hspace{0.025in}
\includegraphics[height=2.77778in]{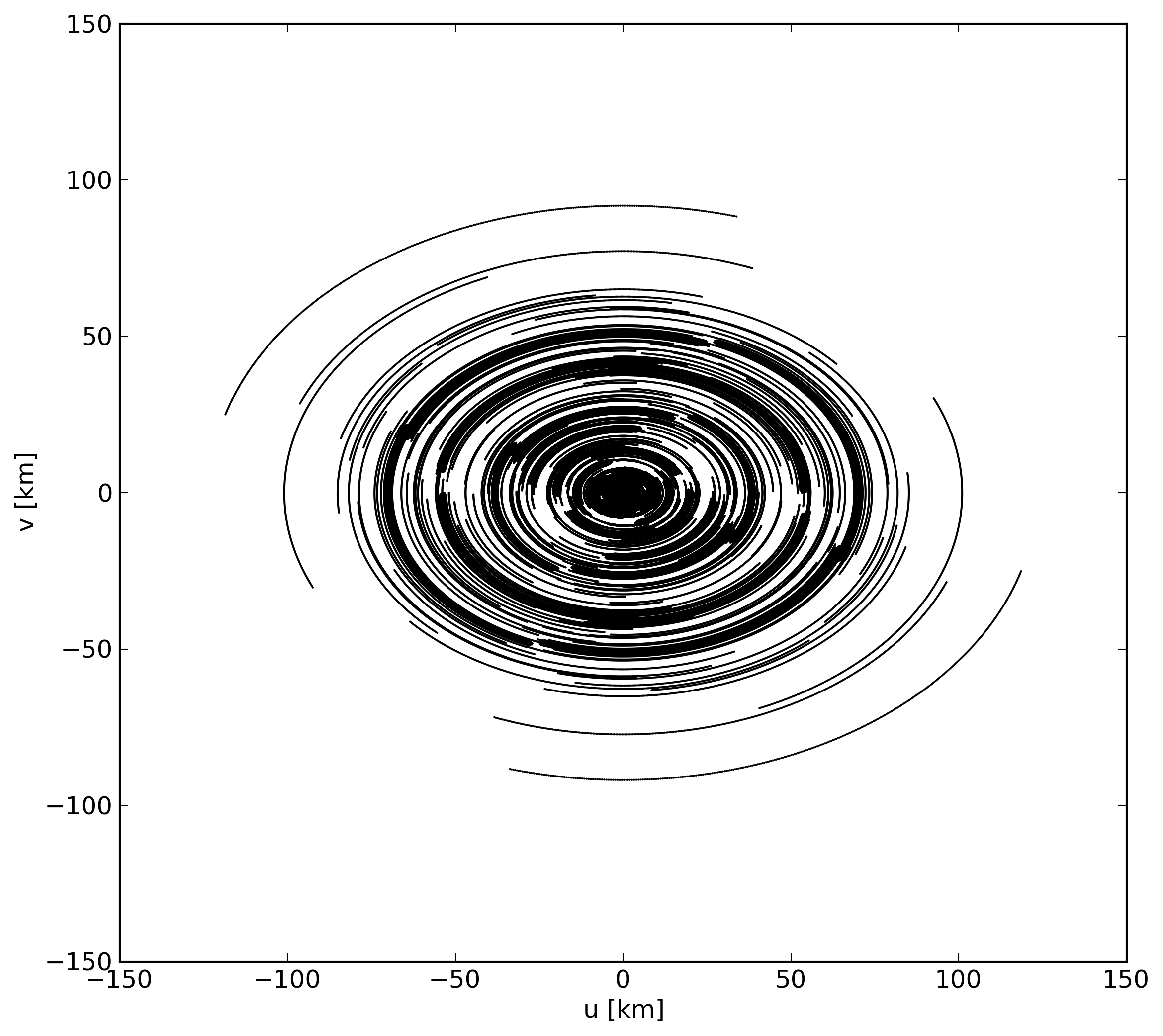}
\hspace{0.275in}
\includegraphics[height=2.75in]{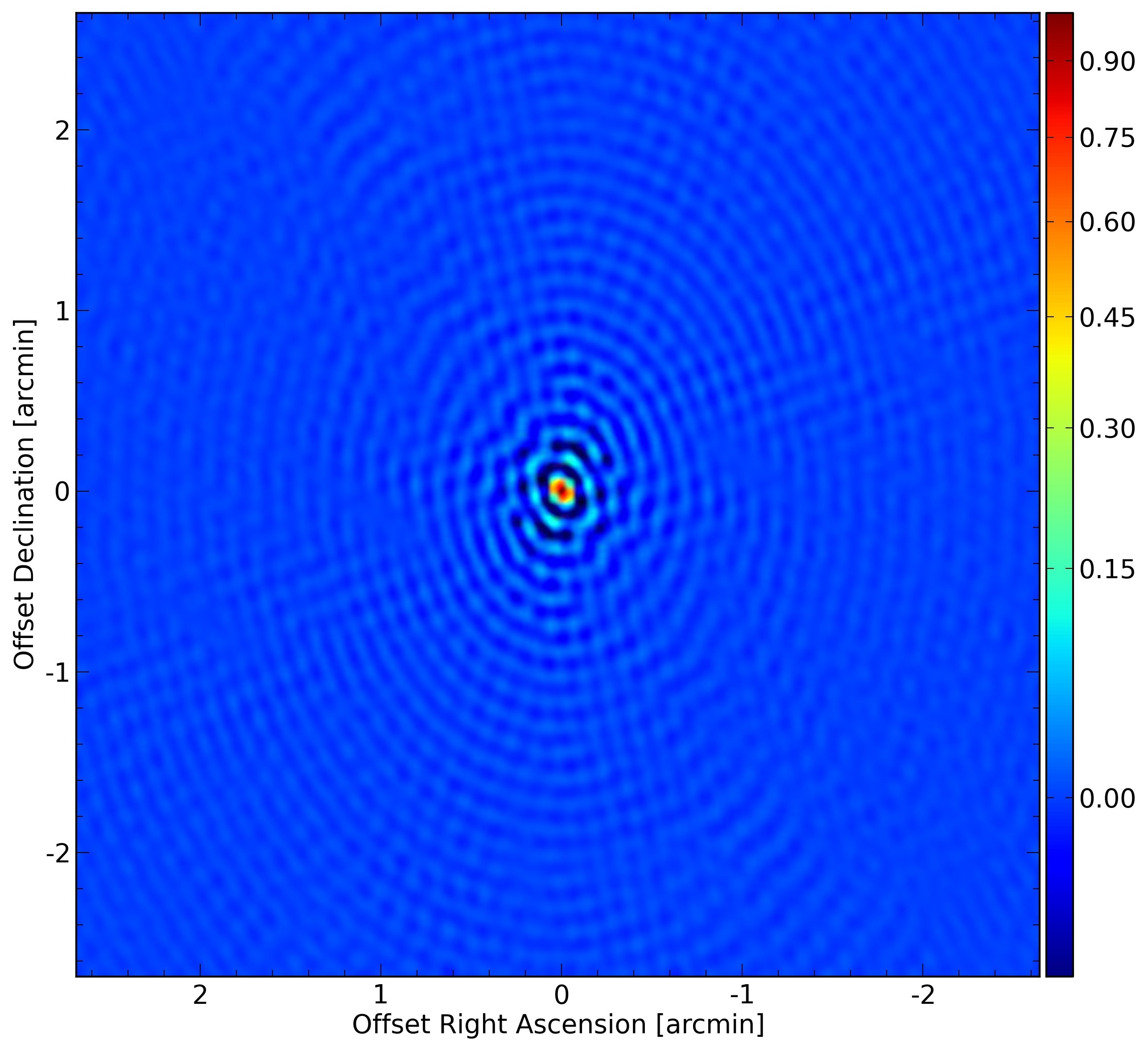}
}
\end{center}

\begin{center}
\hspace{0.01in}
\hbox{
\includegraphics[height=2.77778in]{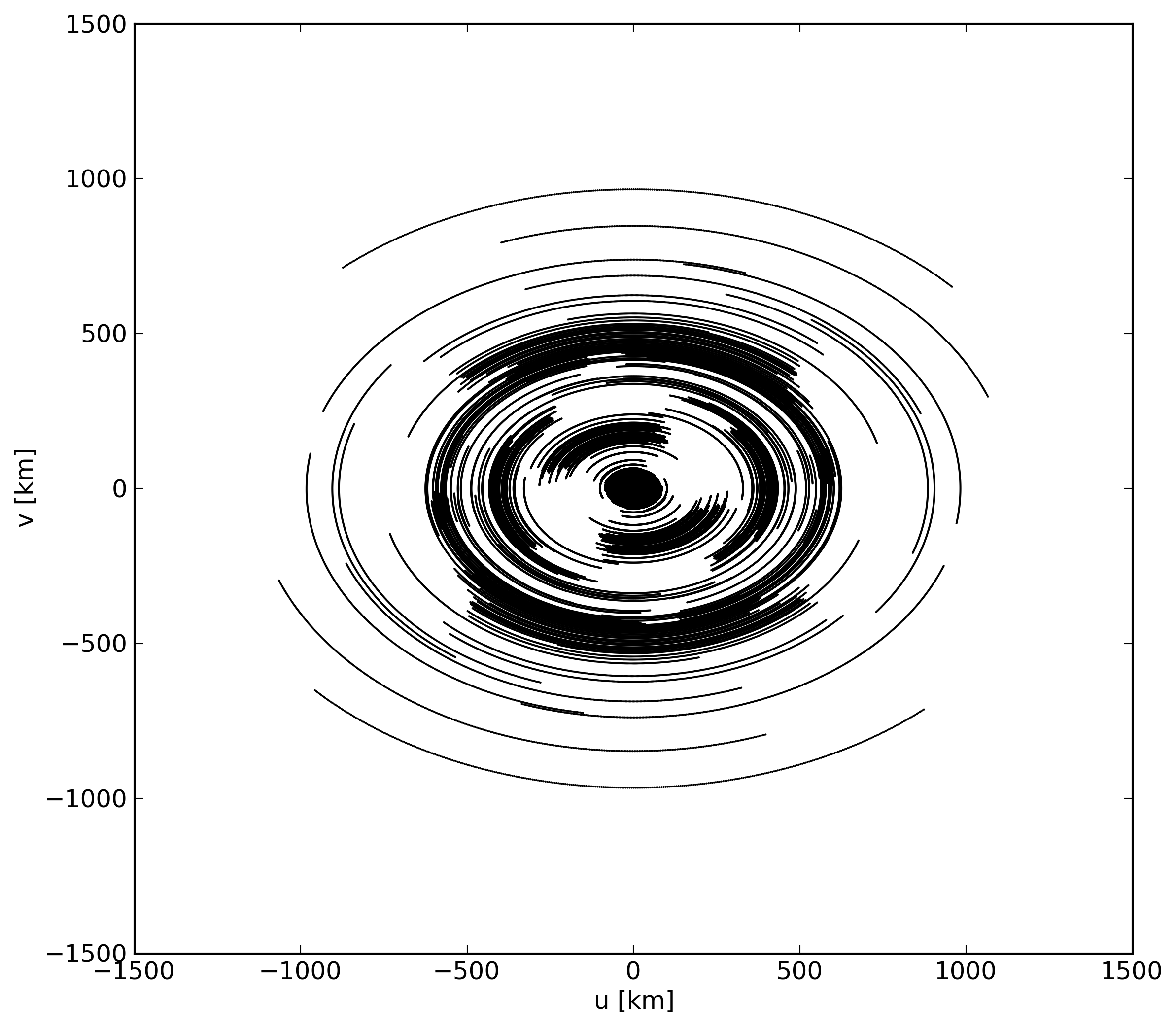}
\hspace{0.200in}
\includegraphics[height=2.75in]{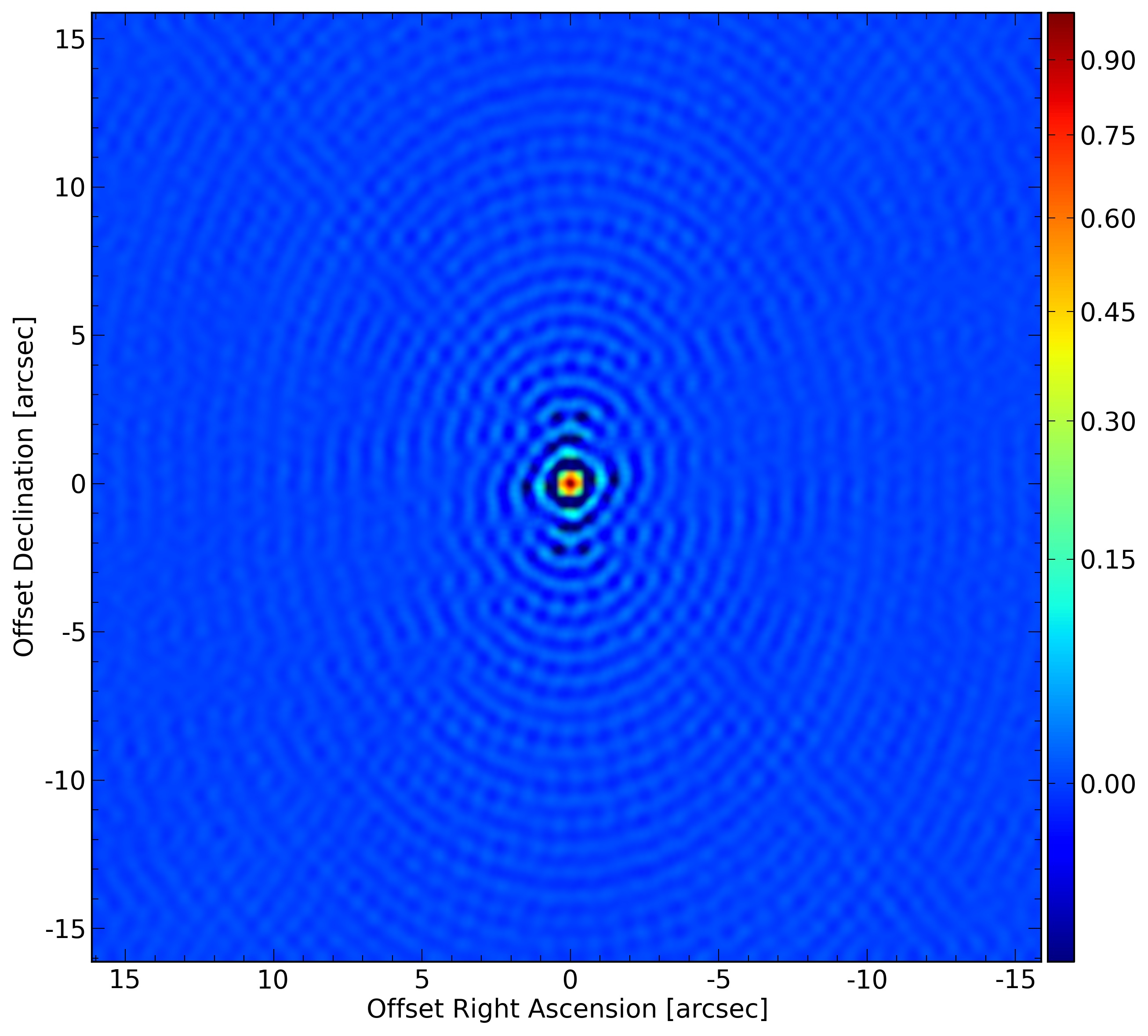}
}
\end{center}
\caption{\small Sample uv coverage plots ({\it left}) and synthesized beams ({\it right}) including all present and planned LOFAR stations. The {\it uv} coverage is calculated for a source at declination 48$\degr$, and covers a 6\,h track between hour angles of approximately -3 to +3 hours. One point is plotted every minute. Synthesized beams are calculated using uniform weighting, and using multi-frequency synthesis over the full LBA frequency range from 30--78 MHz. The top frames are for the 24 core stations only, middle frames include all 40 core and remote stations, and the bottom frames include all 48 core, remote, and international stations.}
\label{fig:uv}
\end{figure*}

\begin{figure*}[p]
\centering
\vspace{0.2in}
\begin{center}
\hspace{0.01in}
\hbox{
\hspace{0.125in}
\includegraphics[height=2.77778in]{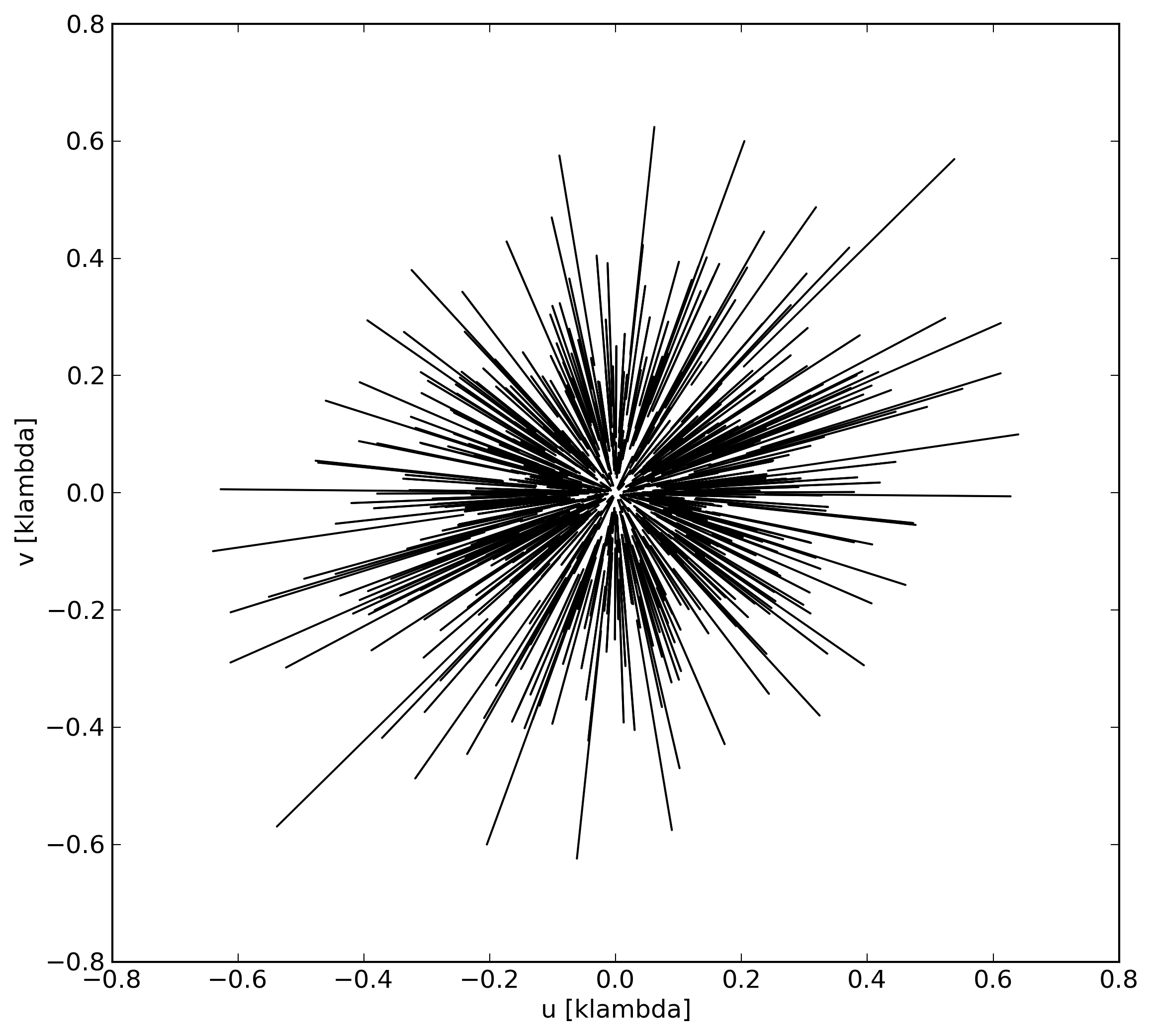}
\hspace{0.2250in}
\includegraphics[height=2.75in]{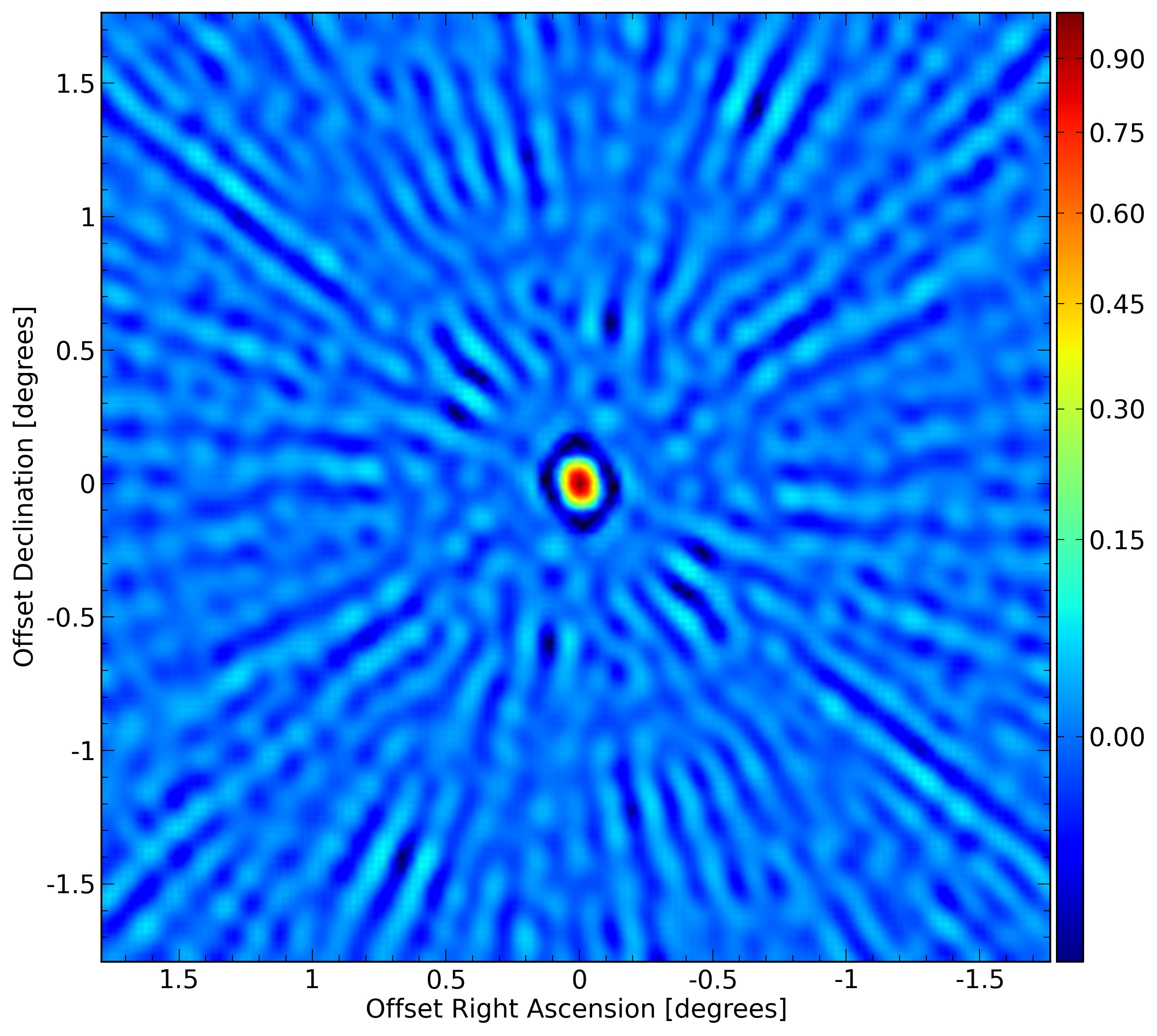}
}
\end{center}

\begin{center}
\hspace{0.1in}
\hbox{
\hspace{0.025in}
\includegraphics[height=2.77778in]{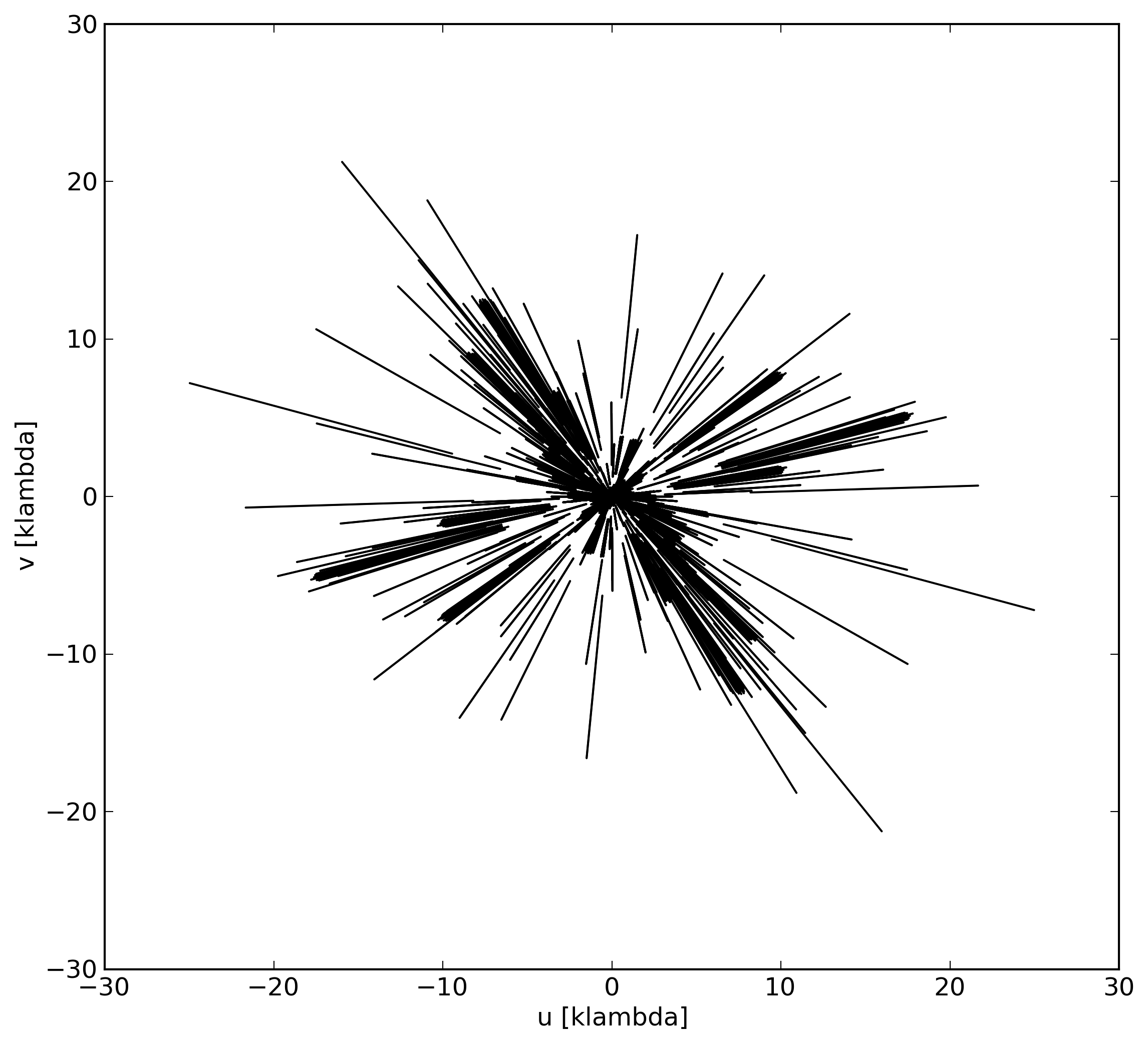}
\hspace{0.275in}
\includegraphics[height=2.75in]{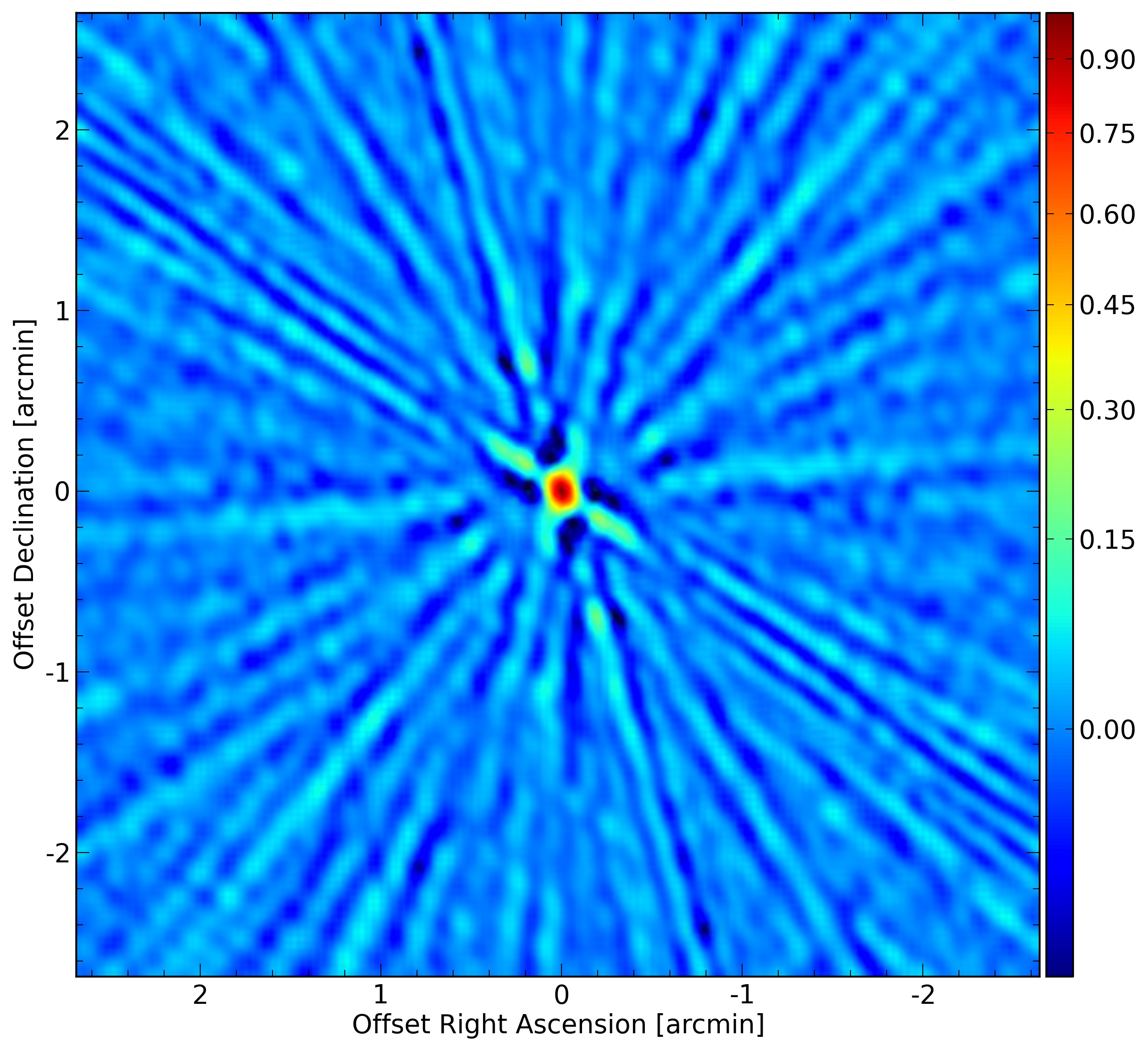}
}
\end{center}

\begin{center}
\hspace{0.1in}
\hbox{
\includegraphics[height=2.77778in]{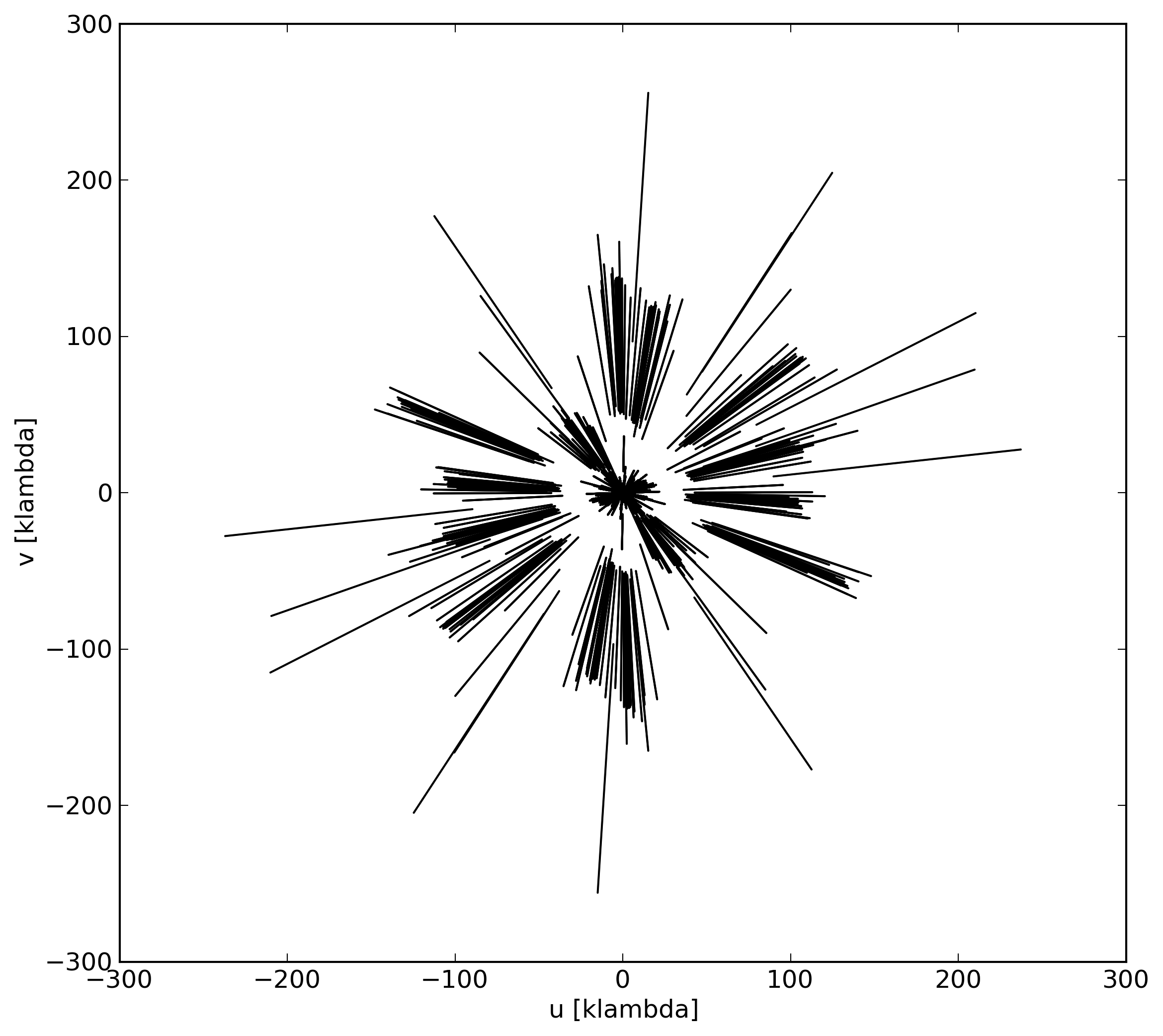}
\hspace{0.200in}
\includegraphics[height=2.75in]{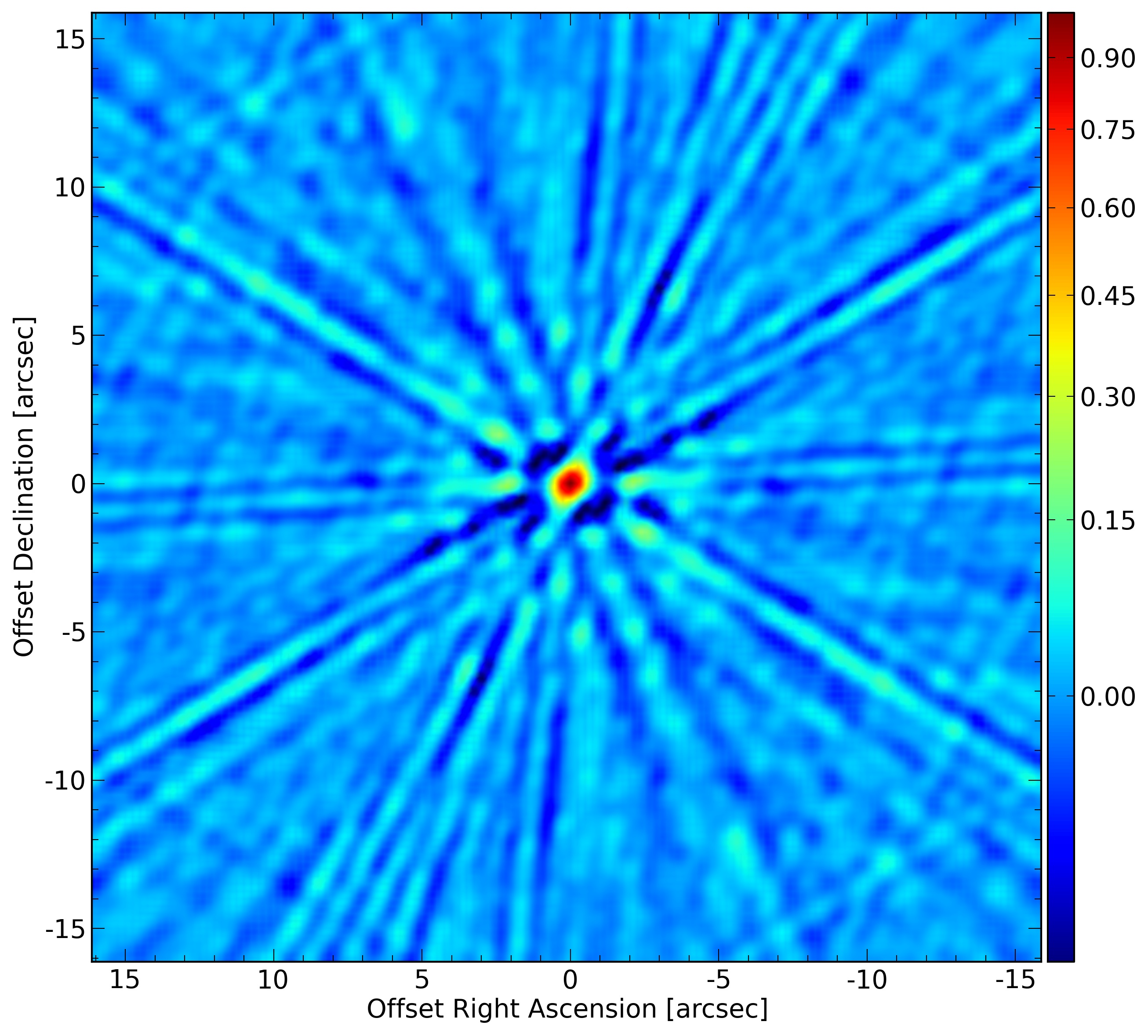}
}
\end{center}
\caption{\small As in Fig.\,\ref{fig:uv}, but illustrating the instantaneous {\it uv} coverage (near transit at the same declination) for the same complement of stations, and showing the effect of the full 48 MHz bandwidth from 30-78 MHz on the {\it uv} coverage. In the left frames, one point is plotted every 0.2 MHz.}
\label{fig:uv_snap}
\end{figure*}

\section{System performance}
\label{sec:performance}

\subsection{System stability}
\label{sec:stability}

There are several effects that can lead to the deterioration of the phase and amplitude stability of LOFAR stations. For example, the ionospheric phase above the Netherlands often changes by one radian per 15 seconds in the 110--190\,MHz band, and one radian per 5 seconds around 50 MHz on LOFAR NL baselines. Because typical ionospheric disturbances have scales of order $\sim 100$\,km, the ionospheric phase on EU baselines fluctuates similarly. The GPS-corrected rubidium clocks at Dutch remote stations and most international stations can typically drift up to 20 ns per 20 minutes, which corresponds to about a radian per minute at 150 MHz. This drift is much less than the ionospheric changes under solar maximum conditions, but comparable at solar minimum.

The HBA amplitudes are generally very stable. Although early experiments with prototype tiles showed up to $10--30$\% reduction in gain when the tiles were covered with a few cm of water and held in place by improvised edges, in practice these circumstances never occur in reality with the production tiles. The LBA antennas on the other hand, \emph{are} sensitive to water under fairly normal operating conditions. If wet, and covered in water droplets, the resonance frequency can shift by several MHz, increasing the gain on one side of the peak by of order $\sim 10$\%, and decreasing the gain on the other side by a similar amount.

Fortunately, these effects are all station-based, hence easily corrected by self calibration given sufficient flux in the FoV and enough equations per unknown. LOFAR's tremendous sensitivity and large number of stations are therefore key. The MSSS surveys in both LBA and HBA clearly show that there is more than enough flux to calibrate within an ionospheric coherence time in the vast majority of fields, even during solar maximum (see Sect.\,\ref{sec:calib} and Heald et al., in prep.). The EoR group has furthermore demonstrated image sensitivities better than 100 $\mu$Jy in the 110--190\,MHz range, thereby demonstrating that system stability issues are not the limiting factor in achieving quality images (see Sect.\,\ref{sec:calib}).

\subsection{{\it uv}-coverage}
\label{sec:uv}

The image fidelity of an aperture synthesis array is dependent on how well the {\it uv}-plane is sampled. A poorly sampled {\it uv}-plane can result in strong side-lobes in the synthesized beam that will limit the overall dynamic range of an image. Also, since an interferometer samples discrete points in the {\it uv}-plane, incomplete {\it uv}-coverage can result in a loss of information on particular angular scales in the sky brightness distribution, which is important for imaging extended radio sources.  For LOFAR, the {\it uv}-coverage has been optimized by choosing suitable locations for the stations throughout the Netherlands and by taking advantage of the large fractional bandwidth that is available. The positions of the international stations have not been chosen to maximize the filling of the {\it uv}-coverage, but care has been taken to avoid duplicate baseline lengths.

In Fig.\,\ref{fig:uv}, the {\it uv}-coverage for the completed LOFAR has been simulated using the known and expected positions of the 40 core and remote stations in the Netherlands and the 8 currently existing international stations. This simulation is based on a hypothetical 6 hour observation of a radio source at declination $48^{\circ}$ between 30 and 78 MHz and uses a single beam with a total contiguous bandwidth of 48 MHz. The {\it uv}-coverage for an array comprised of only the core stations, only the core and remote stations, and all of the LOFAR stations are shown. For clarity, the {\it uv}-distances are given in meters, since for {\it uv}-distances shown in $\lambda$ the {\it uv}-coverage is densely sampled due to the large fractional bandwidth ($\sim0.88$ between 30 and 78 MHz; $\sim0.33$ between 120 and 168 MHz). Also shown in Fig.\,\ref{fig:uv} are the synthesized beams for each of the different array configurations between 30 and 78 MHz using uniform weighting, which show the side-lobe response pattern. The excellent {\it uv}-coverage results in first side-lobes that are $\sim$5\%, $\sim$5\% and $\sim$7\% of the synthesized beam peak for the core, core and remote, and the full array, respectively. Similar values are obtained for a simulation that is carried out with the HBA frequencies between 120 and 168 MHz, and between 210 and 250 MHz. As normal, sidelobe levels can be reduced at the expense of angular resolution through the use of other visibility weighting schemes.

The simulations above are for a standard long-track observation. The sensitivity and the large FoV of LOFAR will also allow surveys of the radio sky to be carried out efficiently using snapshot observations. The point source response of LOFAR in snapshot mode has also been simulated by calculating the instantaneous {\it uv}-coverage for the hypothetical observation described above, for a radio source at 0 hour angle (transit). The resulting instantaneous {\it uv}-coverages for the core, core and remote, and full LOFAR arrays are also shown in Fig.\,\ref{fig:uv_snap}. Note that for these simulations the data for the full 48 MHz bandwidth are presented, which highlights the excellent large fractional bandwidth of LOFAR. For the core, the {\it uv}-plane is well sampled, but for the core and remote, and for the full array, multiple snapshot observations over several hour-angles are needed to fill the gaps in the {\it uv}-coverage for the $>5$~km baselines.

\begin{figure*}[ht]
\centering
\includegraphics[height=3.2in]{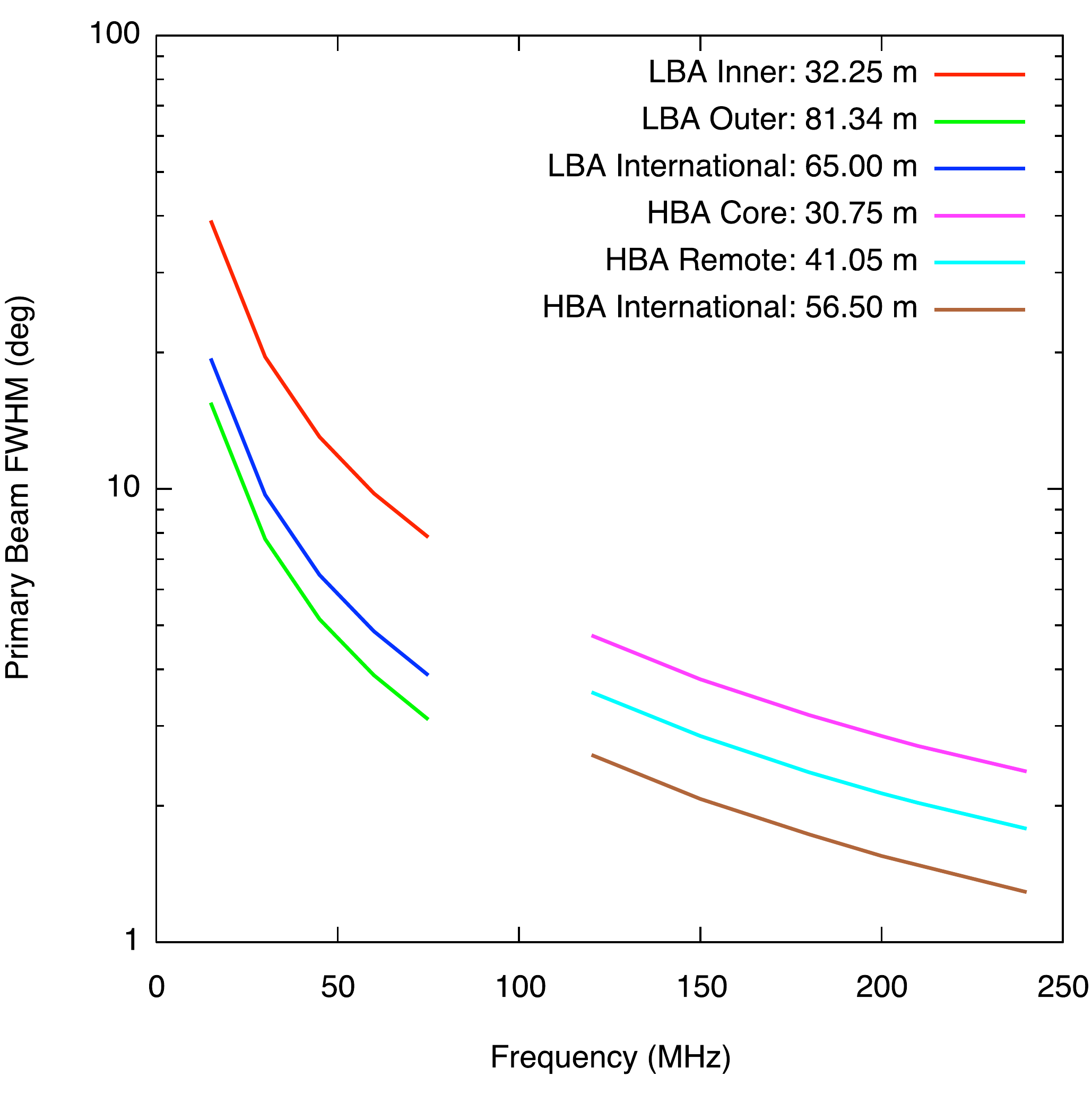}
\hspace{0.1in}
\includegraphics[height=3.2in]{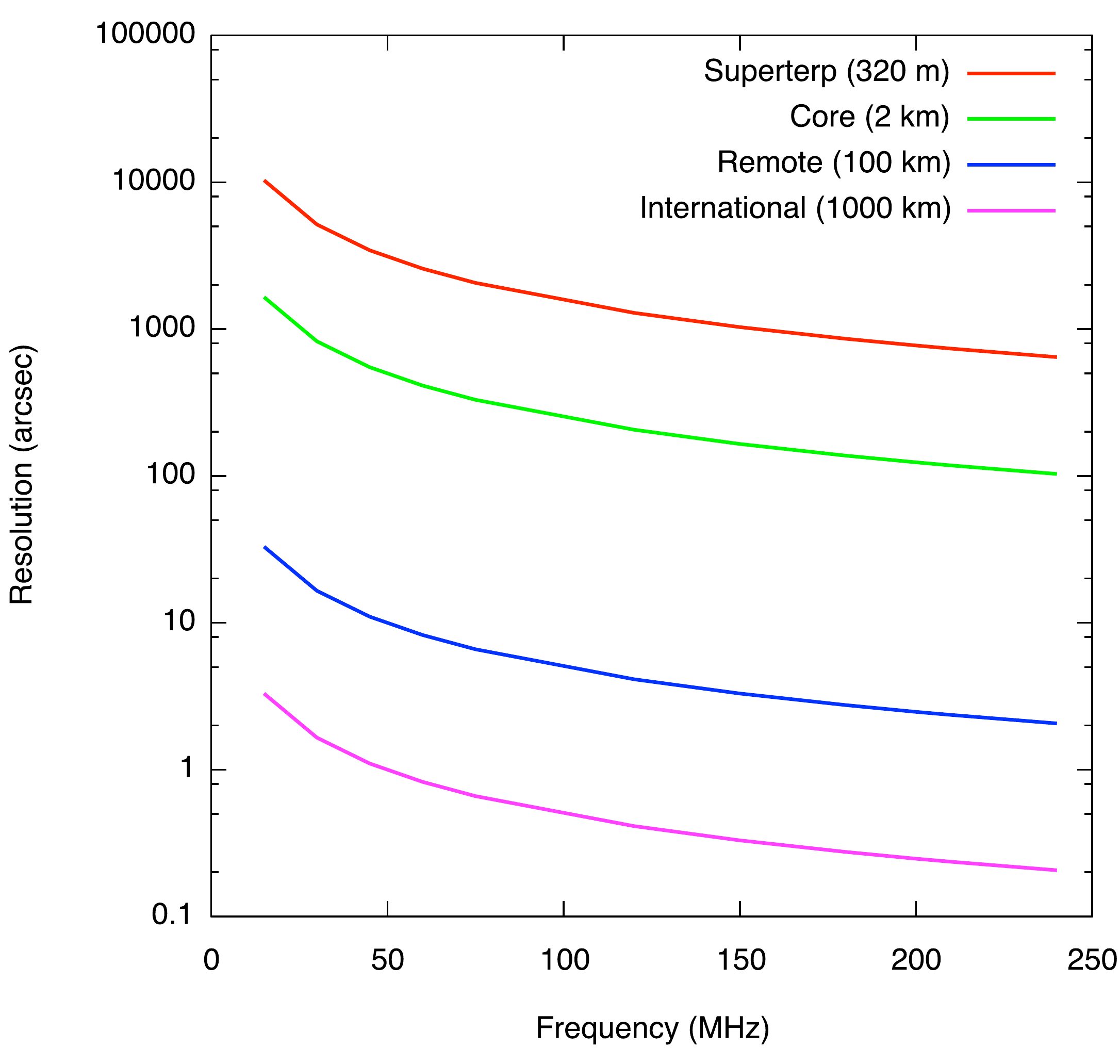}
\caption{\small {\bf Left:} 
Full-width half-maximum (FWHM) of a LOFAR Station beam as a function of frequency for the different station configurations. The curves are labeled by the type of LOFAR antenna field, whether LBA or HBA, as well as by either core, remote, or international. The effective size of each field is also indicated in meters. {\bf Right:} Effective angular resolution as a function of frequency for different subsets of the LOFAR array.}
\label{fig:resolution}
\end{figure*}

\subsection{Angular resolution}

The ability to identify and characterize structures of different angular sizes depends on the angular resolution of an interferometric array. One of the transformational aspects of LOFAR is the unprecedented range of angular scales that are achievable at low observing frequencies. In Table\,\ref{tab:resolution}, and also shown in Fig.\,\ref{fig:resolution}, the angular resolution for various baseline lengths as a function of frequency are given. These angular resolutions have been calculated using,
\begin{equation}
\theta_{\rm res} = \alpha\,\frac{\lambda}{D}, ~~~~[{\rm rad}] 
\label{eqn:beam}
\end{equation}
where $\theta_{\rm res}$ is the full width at half maximum (FWHM) of the synthesized beam in radians, $\lambda$ is the observing wavelength, $D$ is the maximum baseline length and $\alpha$ depends on the array configuration and the imaging weighting scheme (natural, uniform, robust, etc). The angular-resolutions given in Table\,\ref{tab:resolution} are based on a value of $\alpha = 0.8$, corresponding to a uniform weighting scheme for the Dutch array. Note that this is for the ideal case of a source that has a maximum projected baseline length. 

In reality, the angular resolution of an interferometric observation will be dependent on the declination of the source, the composition of the array, the observing frequency and the visibility weighting that is used. With baseline lengths ranging from a few tens of meters to over one thousand kilometers, the angular resolution of LOFAR extends from 0.5$\degr$ to sub-arcsecond scales.

\subsection{Bandpass}
\label{sec:bandpass}

\begin{figure*}
\centering
\includegraphics[width=3.395in]{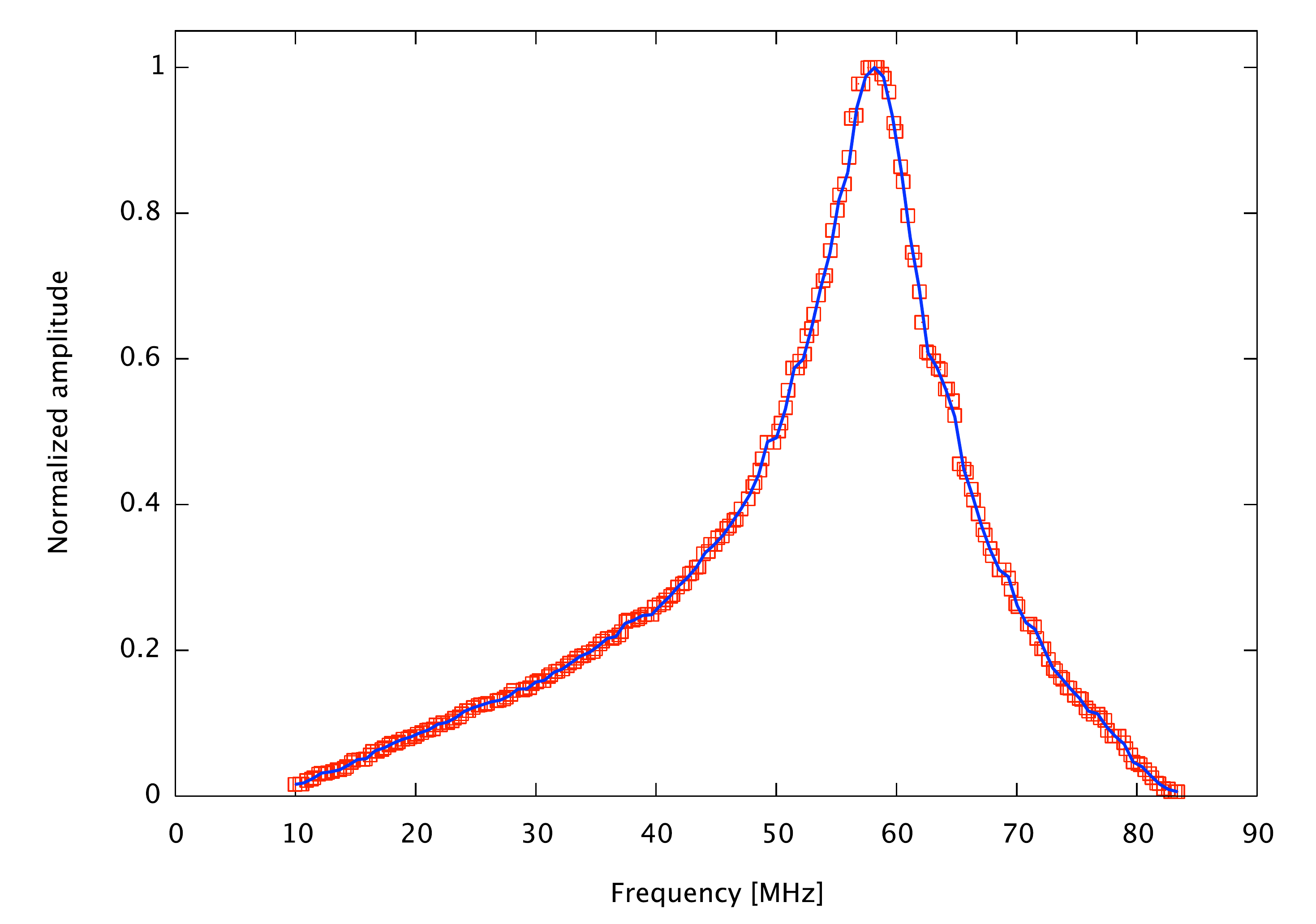}
\includegraphics[width=3.395in]{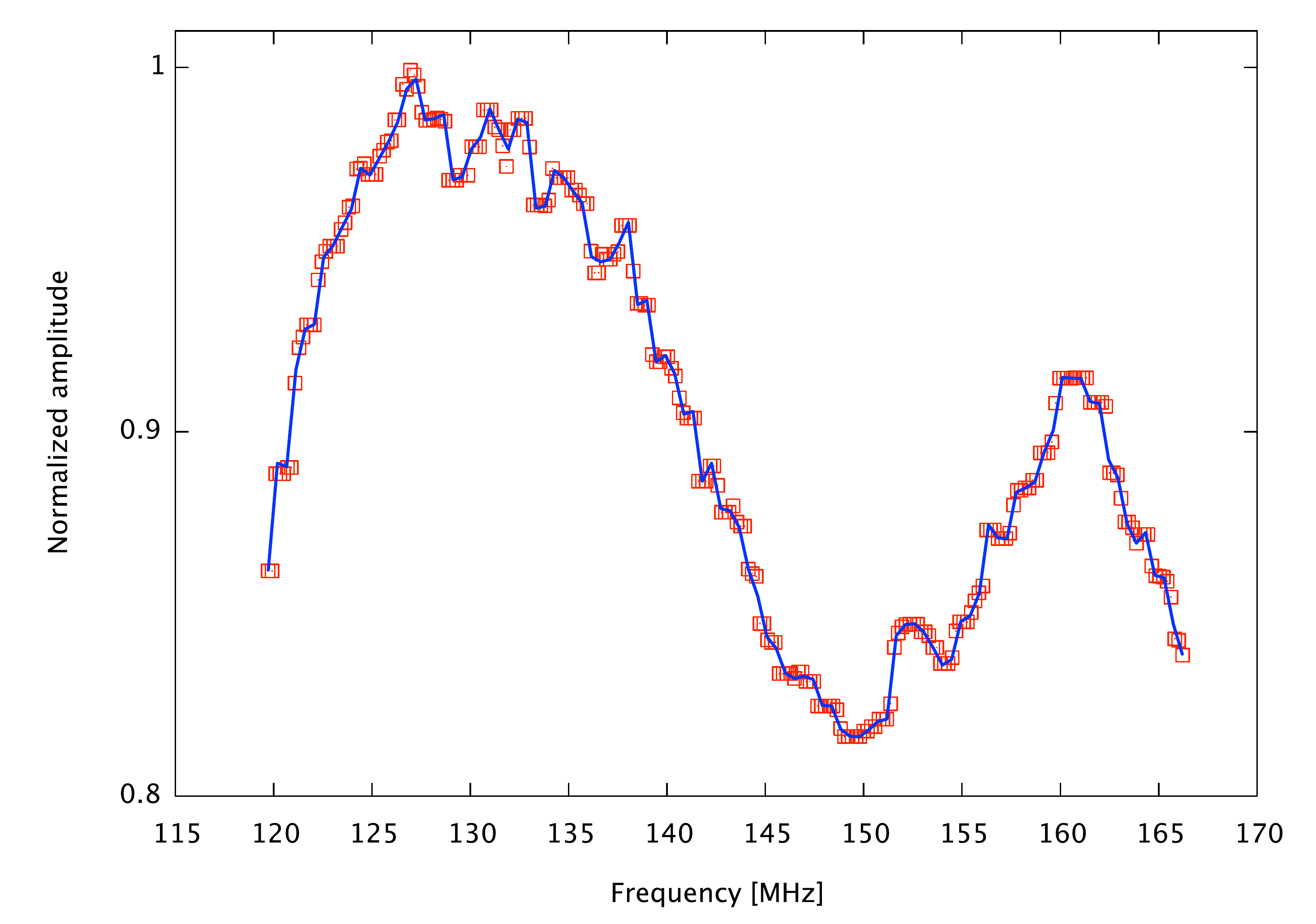}
\includegraphics[width=3.395in]{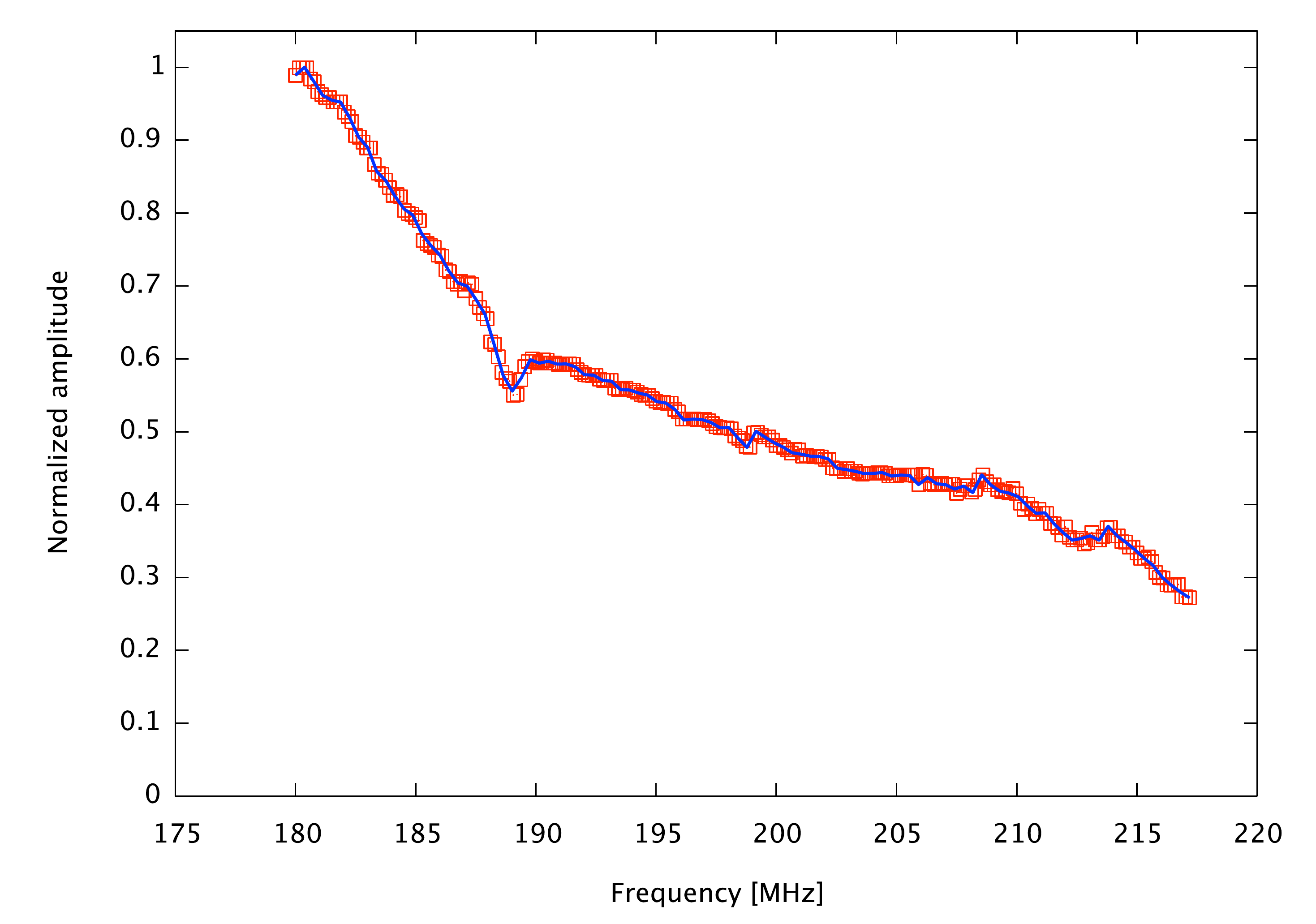}
\includegraphics[width=3.395in]{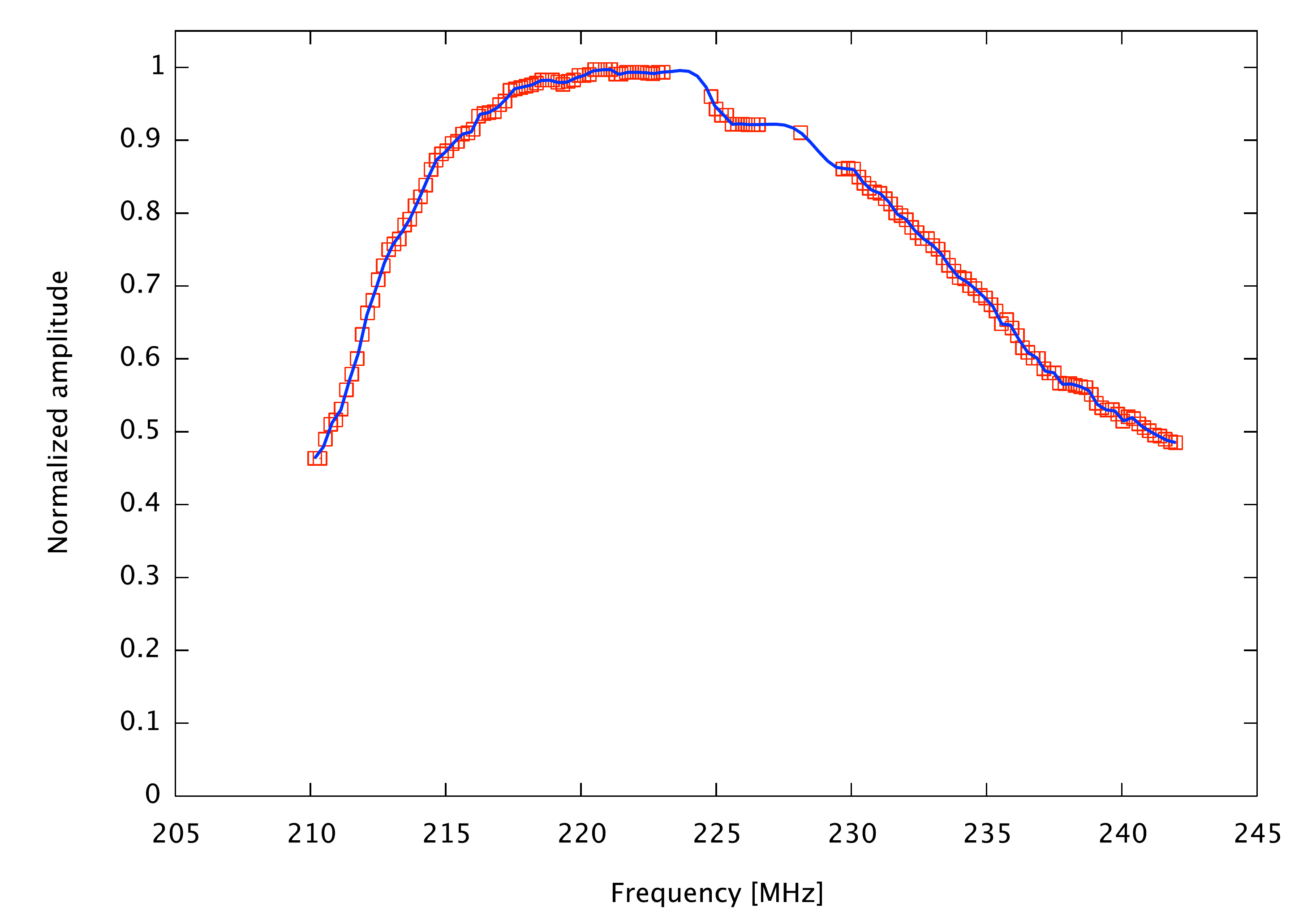}
\caption{\small Normalized global bandpass for several of the LOFAR bands. The global bandpass is defined as the total system gain converting measured voltage units to flux on the sky. These bandpass measurements were determined using short observations of the bright source 3C196 and calculating the mean gain for all stations in each sub-band. The values have been corrected for the intrinsic spectrum of 3C196 assuming a spectral index of -0.70 \citep{scaife_heald_2012}. Curves are shown for ({\bf upper left}) LBA from 10-90 MHz with 200 MHz clock sampling, ({\bf upper right}) HBA low from 110-190 MHz with 200 MHz clock sampling, ({\bf lower left}) HBA mid from 170-230 MHz with 160 MHz clock sampling, and ({\bf lower right}) HBA high from 210-250 MHz with 200 MHz clock sampling.}
\label{fig:bandpass}
\end{figure*}

There are several contributions to the frequency dependent sensitivity of LOFAR to incoming radiation (the bandpass). At the correlator, a digital correction is applied within each 0.2 MHz subband to remove the frequency-dependent effects of the conversion to the frequency domain. The station beam is also strongly frequency dependent, except at the beam pointing center. Finally, the physical structure of the individual receiving elements (described in Sect.\,\ref{sec:lba} and Sect.\,\ref{sec:hba}) causes a strongly peaked contribution to the bandpass near the resonance frequency of the dipole. In the case of the LBA dipoles, the nominal resonance frequency is at 52~MHz. However, as can already be seen in Fig.\,\ref{fig:lba} and Fig.\,\ref{fig:bandpass}, the actual peak of the dipole response is closer to 58~MHz in dry conditions (see also Sect.\,\ref{sec:lba}). This shift in the peak is caused by the interaction between the low-noise amplifiers (LNAs) and the antenna.

Determining the combined or "global" bandpass can be achieved during the calibration step post-correlation.
This global bandpass combines all frequency dependent effects in the system that have not already been corrected following correlation. To illustrate this point, the bright quasar 3C196 has been observed using the core and remote stations in the LBA and all three HBA bands. 3C196 is unresolved on the angular scales sampled by those baselines at LBA frequencies. 3C196 has a known spectral energy distribution down to the lowest LOFAR observing frequencies \citep{scaife_heald_2012}. Moreover, 3C196 is the dominant source in its field. Observations of 3C196 between 15--78~MHz were obtained in two observing sessions (15--30~MHz in one session, and 30--78~MHz in the other). BBS was used to calibrate the data. 

For each subband, a system gain was determined, and the gain amplitude was taken as the value of the global bandpass at the frequency of the particular subband. The median of all stations is shown in Fig.\,\ref{fig:bandpass} for a typical 10~minutes of data. Curves are shown for all four LOFAR observing bands. While most stations exhibit individual bandpasses which are similar to the median value, some stations deviate significantly due to RFI or incomplete calibration information. In the future, station-level calibration information will be updated in near-realtime during observations.

\subsection{Beam characterization}
\label{sec:beam}

The response pattern, or beam, of a LOFAR observation is determined by the combination of several effects. The first is the sensitivity pattern of the individual dipoles themselves. This pattern changes relatively slowly across the sky. Electromagnetic simulations of the LBA and HBA dipoles have been performed, and parameterized descriptions of the results of these simulations form the basis of the dipole beam model used in the calibration of LOFAR data.

For imaging observations, the dominant effect in determining the sensitivity within the FoV (analogous to the `primary beam' of traditional radio telescopes) is the electronically formed station beam. The size of the LOFAR primary beam determines the effective FoV for a given observation. The pointing of the station beam is determined by digital delays applied to the elements that make up an individual station. In the LBA, the elements are the dipoles themselves. In the HBA, groups of 16 dipole pairs are combined into HBA tiles. Each tile contains an analog beam former, which adds physical delay lines to each dipole and thus ``points'' the tile in a particular direction (as described in Sect.\,\ref{sec:hba}). The HBA station beam is formed from the combined signal from the tiles rather than directly from the dipoles. A description of the full HBA beam thus includes a term for the tile response pattern, which is of intermediate angular scale when compared to the dipole beam and the station beam.

Delays within the station (i.e. delays between individual dipoles or tiles) are calibrated by observing a bright source and determining the complex gain for each element that maximizes the response toward that bright source. These complex gains are stored as a calibration table at the station level and applied when forming the station beam. In future, it will be possible to control the shape of the station beam by applying a tapering function to the individual elements that are combined.

The nominal FWHM of a LOFAR station beam is determined using Equation~\ref{eqn:beam}, where $D$ is now the diameter of the station and the value of $\alpha$ will depend on the tapering intrinsic to the layout of the station, and any additional tapering which may be used to form the station beam. No electronic tapering is presently applied to LOFAR station beamforming. For a uniformly illuminated circular aperture, $\alpha$ takes the value of 1.02, and the value increases with tapering \citep{napier1999}. The FoV of a LOFAR station can then be approximated by
\begin{equation}
{\rm FoV} = \pi \left(\frac{FWHM}{2}\right)^2. 
\label{eqn:fov}
\end{equation}

An overview of the expected beam sizes for the various LOFAR station configurations is presented in Table\,\ref{tab:beams} and is also shown in Fig.\,\ref{fig:resolution}. In the Dutch LBA stations, 48 dipoles must be selected out of the total 96. Selecting the innermost dipoles results in a large-FoV configuration with a diameter of 32.25 meters. Selecting the outermost dipoles results in a small-FoV configuration with a diameter of 81.34 meters. The European stations always use all 96 dipoles in the low-band, which corresponds to a station diameter of 65 meters. In the high-band, the core stations are split into two sub-stations, each with 24 tiles and a diameter of 30.75 meters. The Dutch remote stations have 48 tiles and a diameter of 41.05 meters. The European stations consist of 96 tiles and have a diameter of 56.5 meters. In addition, individual stations are rotated relative to one another in order to suppress sensitivity in the sidelobes (see Sect.\,\ref{sec:layouts}). As a result of this wide variation in station configurations and orientations, the beam modeling software is required to treat each station independently. The FoV of LOFAR imaging observations can range from $\sim$2--1200 deg$^2$, depending on the observing frequency and the station configuration.

We have observationally verified the station beam shapes and diameters using a strategy that takes advantage of LOFAR's multi-beaming capabilities. A grid of $15\times15$ pointings, centered on Cygnus A, was observed simultaneously in interferometric imaging mode for 2 minutes at each of a sequence of frequencies in LBA and HBA. Calibration solutions were determined independently in each of the 225 directions to Cygnus A. Since Cygnus A is so bright, it dominates the visibility function in all grid points, and allows a good calibration solution. The influence of the distant bright source Cassiopeia A was overcome by using a long solution interval (in both frequency and time). All baselines were used to determine the gain solutions, since removal of the long and/or short baselines was found to have no affect on the quality of the output. 

\begin{figure*}
\centering
\includegraphics[width=3.375in]{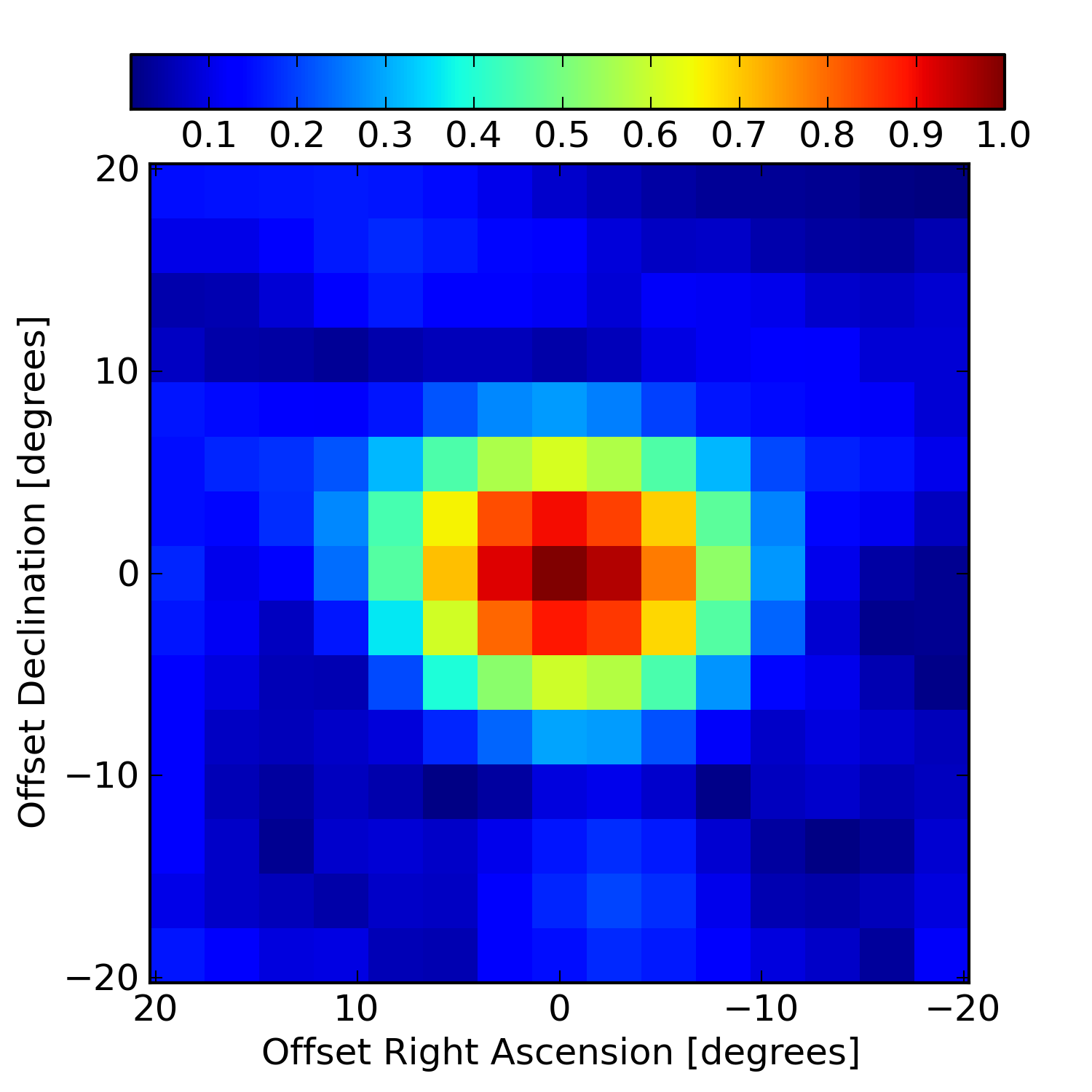}
\includegraphics[width=3.375in]{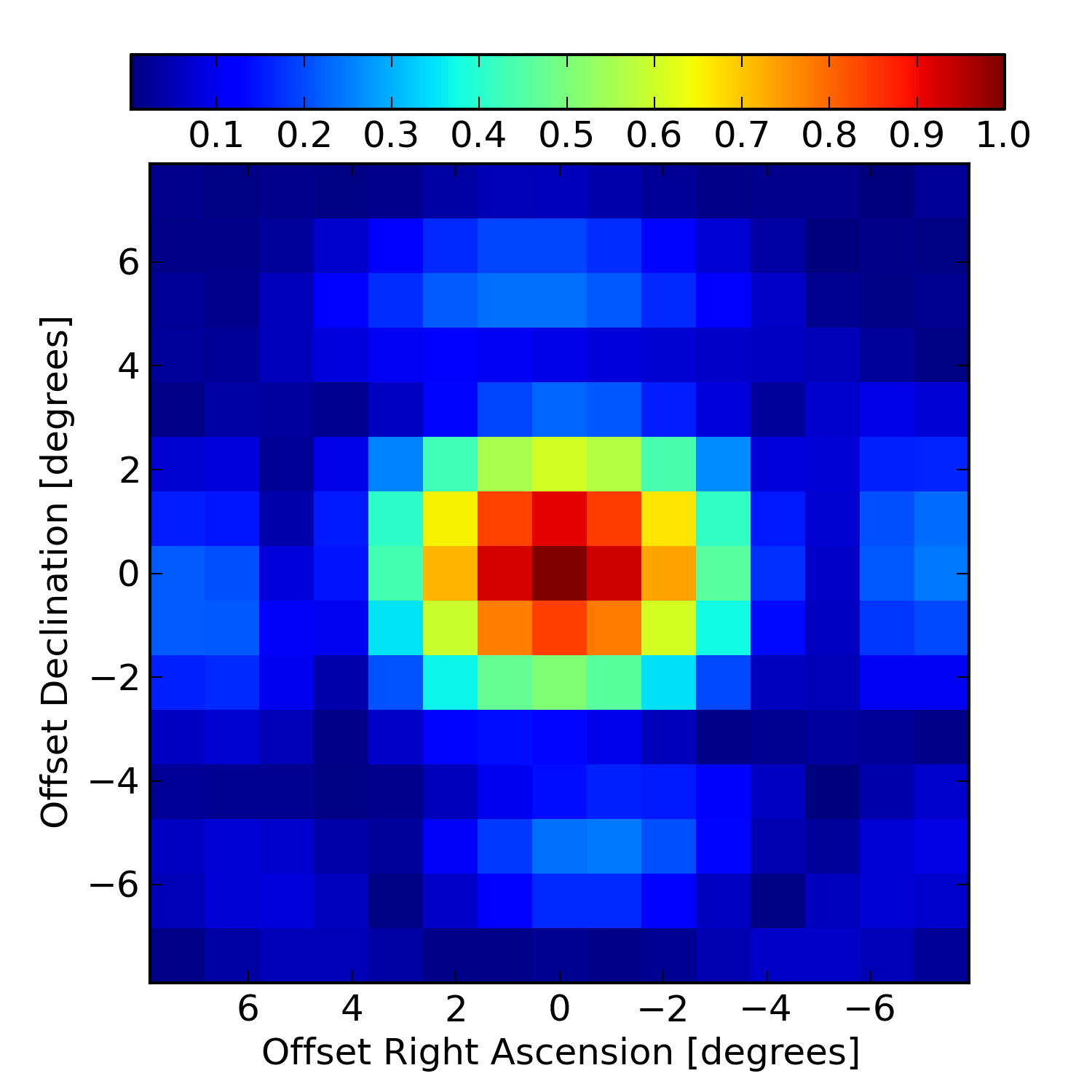}
\includegraphics[width=3.375in]{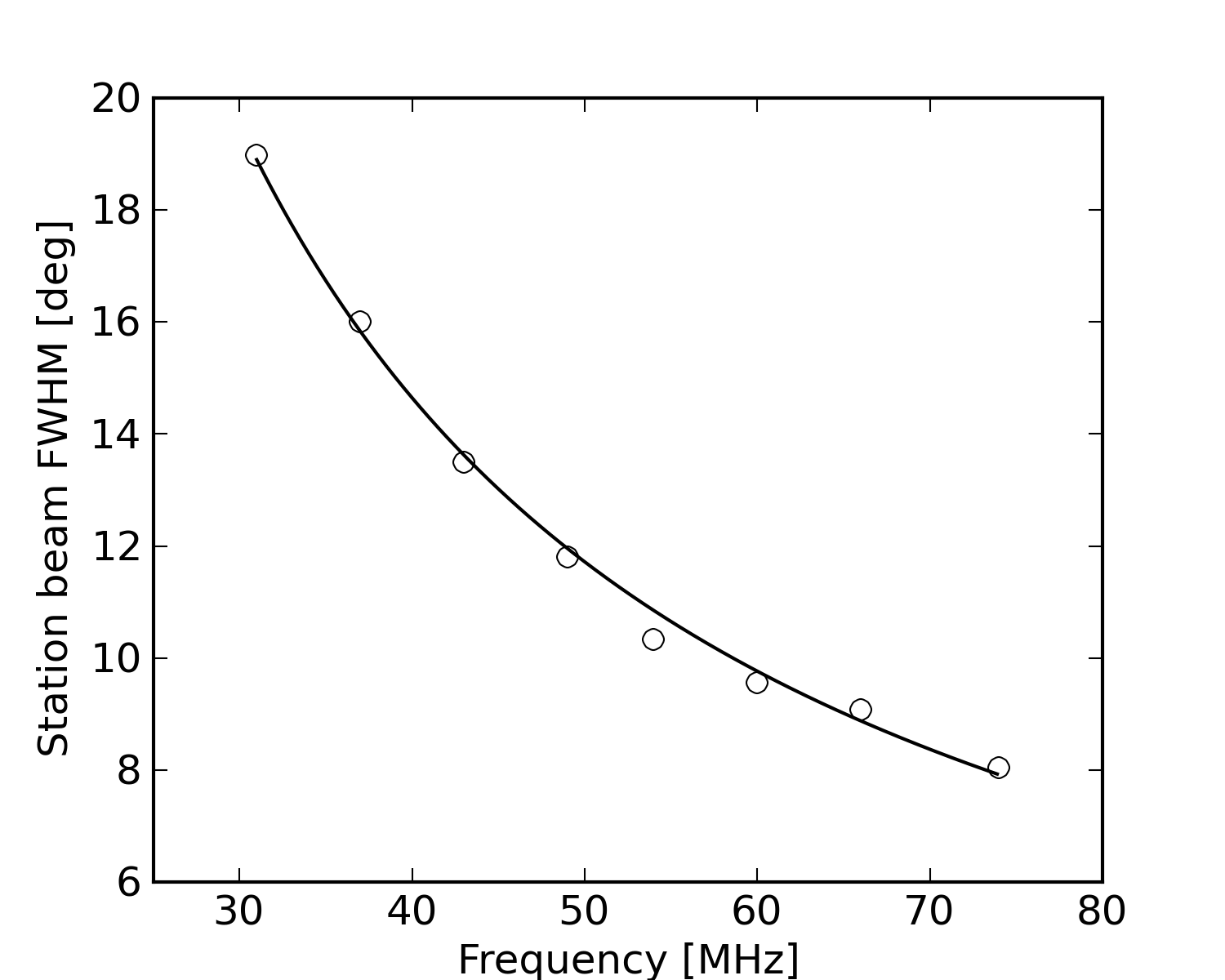}
\includegraphics[width=3.375in]{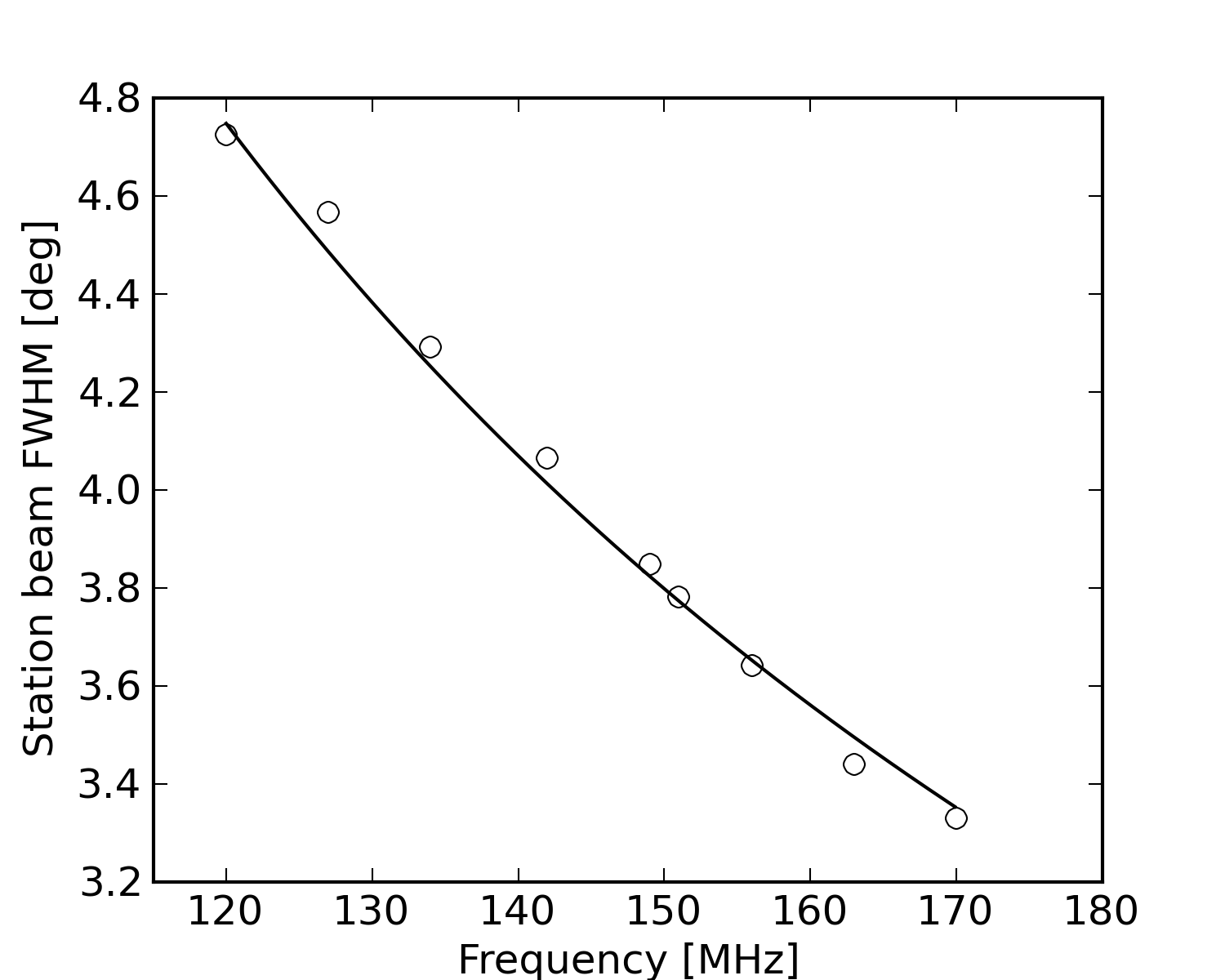}
\caption{\small The upper row of figures show the observationally determined station beam patterns as described in the text. On the left, the beam is shown for core station CS004 in its LBA\_INNER configuration. On the right, the beam is shown for a single HBA ear of the same core station. The peak response is normalized to unity in both plots. Note the low-level sidelobe structure apparent in the figures. The bottom row of plots give the FWHM for a Gaussian fit to the main station beam lobe, plotted as a function of frequency. Solid lines indicate the fitted $\alpha\lambda/D$ relations explained in the text.}
\label{fig:beamsize}
\end{figure*}

The resulting gain amplitudes were mapped onto a complex beam pattern that was in turn used to derive a ``power beam'' (the square of the complex beam pattern) for each station. Examples of the power beams observed at 60 MHz and 163 MHz, for the core station CS004, are shown in the top panels of Fig.\,\ref{fig:beamsize}. Gaussians were fitted to vertical cuts through the center of the power beams, and resulted in FWHM values shown in the bottom panels of Fig.\,\ref{fig:beamsize}. By fitting Equation~\ref{eqn:beam} for $\alpha$, we determined that the actual values are $1.02\pm0.01$ for HBA (core stations) and $1.10\pm0.02$ for LBA (in the LBA\_INNER configuration). Since the LBA stations are less uniformly distributed than the HBA stations, the value of $\alpha$ is expected to be larger, as the observations confirm. We note that these values are only indicative since the stations are not circular apertures. The LOFAR processing software includes a beam model that directly computes the instantaneous station beam pattern for each station using the appropriate pointing direction and observing frequency.

Although the beam mapping observations discussed here were not specifically designed to carefully study the sidelobe pattern, they were sufficient to quantify the strength of the innermost sidelobes. Typical sidelobes levels of $20-25\%$ were found for both the LBA and HBA with structure consistent with a Bessel sinc function
\citep[see][]{napier1999}. A more detailed discussion of the beam structure can be found in Heald et al. (in prep.).

\subsection{Sensitivity}
\label{sec:sensitivity}

Given estimates for the system equivalent flux density (SEFD) of a LOFAR station, one can calculate the expected sensitivity for different configurations of the array. The SEFD of a LOFAR station, in turn, depends on the ratio of the system noise temperature ($T_\mathrm{sys}$) and the total effective area ($A_\mathrm{eff}$) \citep{tcp99}. Since LOFAR consists of stations with different numbers of receiving elements, $A_\mathrm{eff}$ differs for the various types of stations and hence their SEFD also varies. The adopted values of the effective area, $A_\mathrm{eff}$, were obtained from numerical simulations that account for the overlap of dipoles in the different station layouts and are provided in Table\,\ref{tab:beams}. 

\begin{figure*}[]
\centering
\includegraphics[width=3.375in]{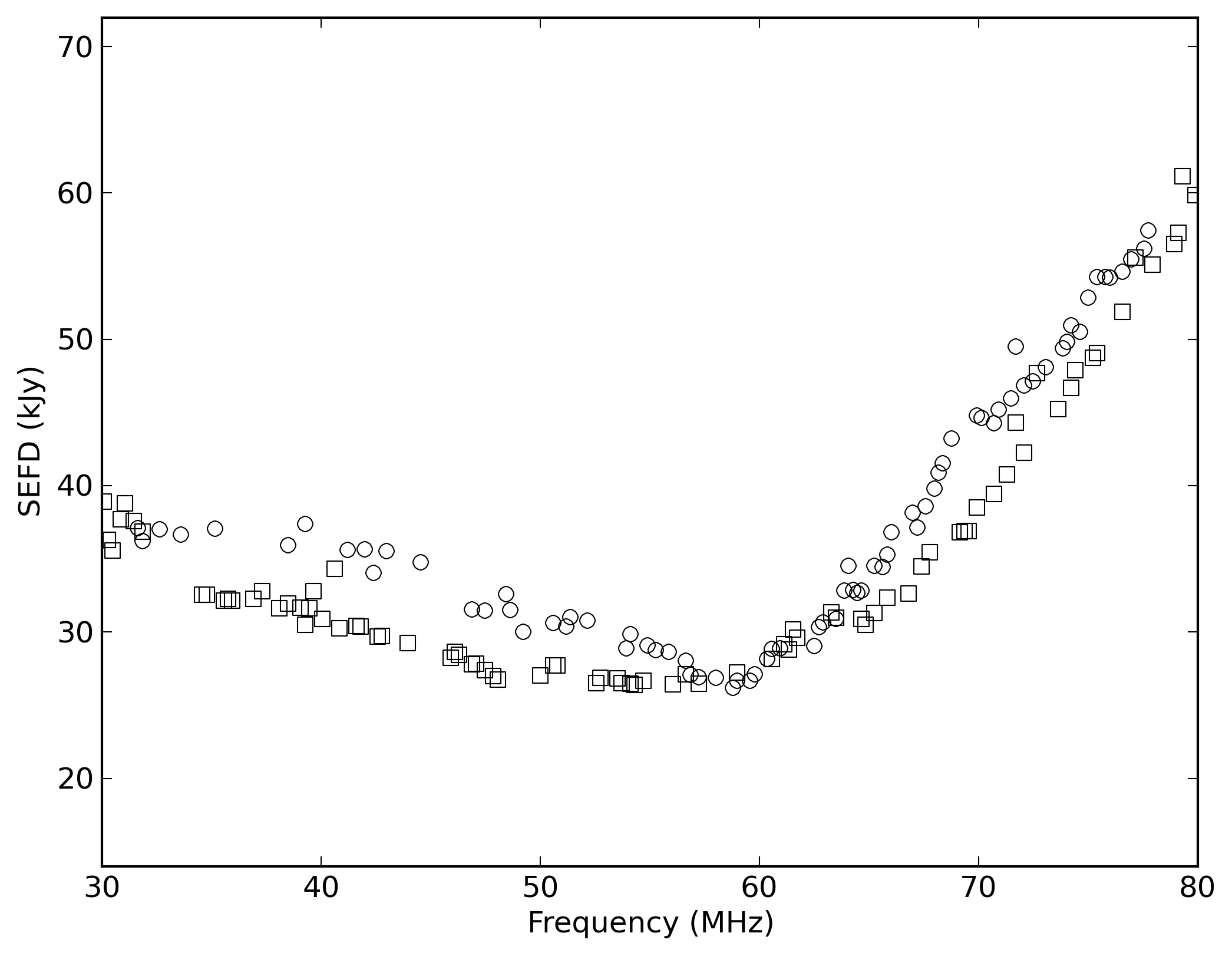} \hspace{0.15in}
\includegraphics[width=3.375in]{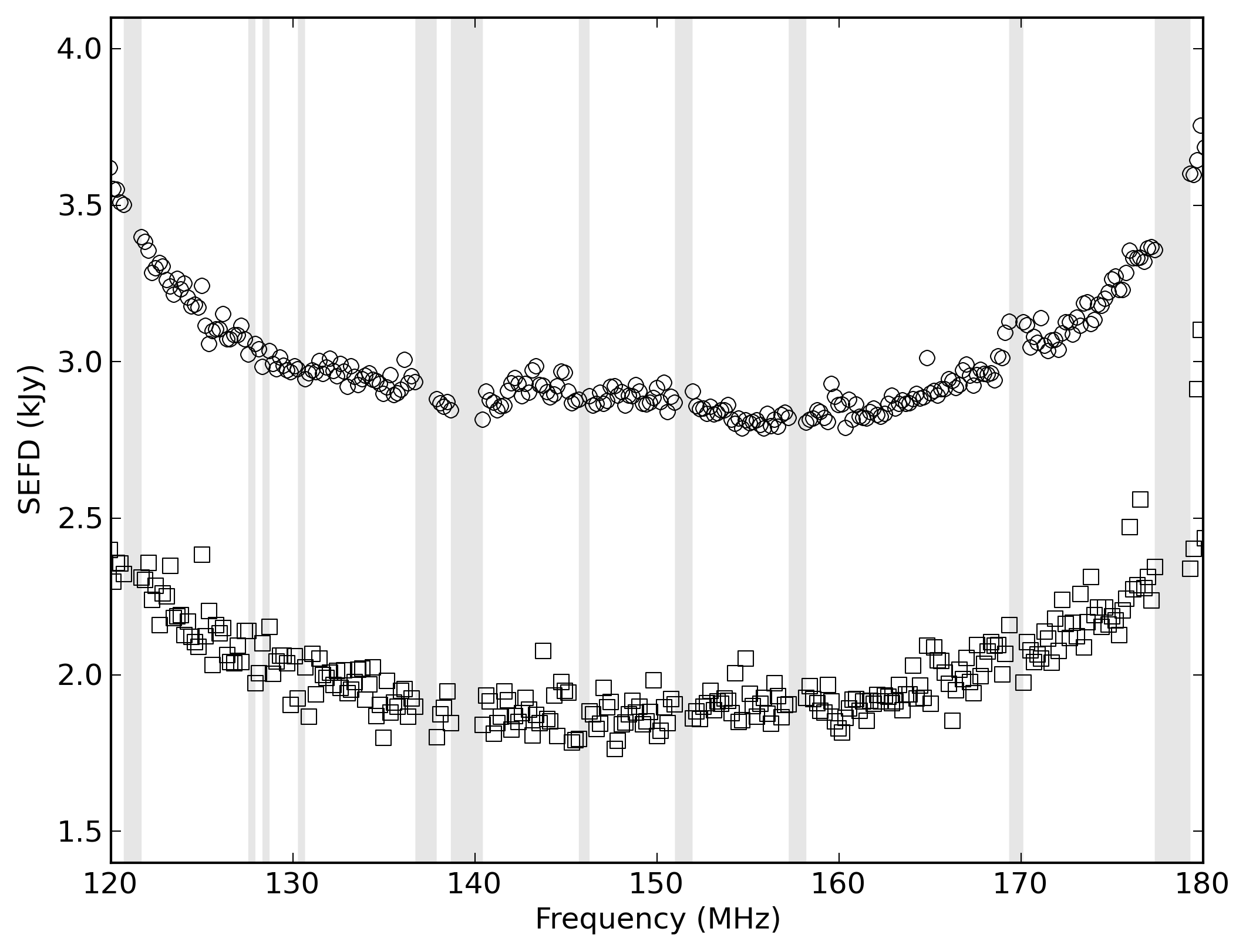}
\includegraphics[width=3.375in]{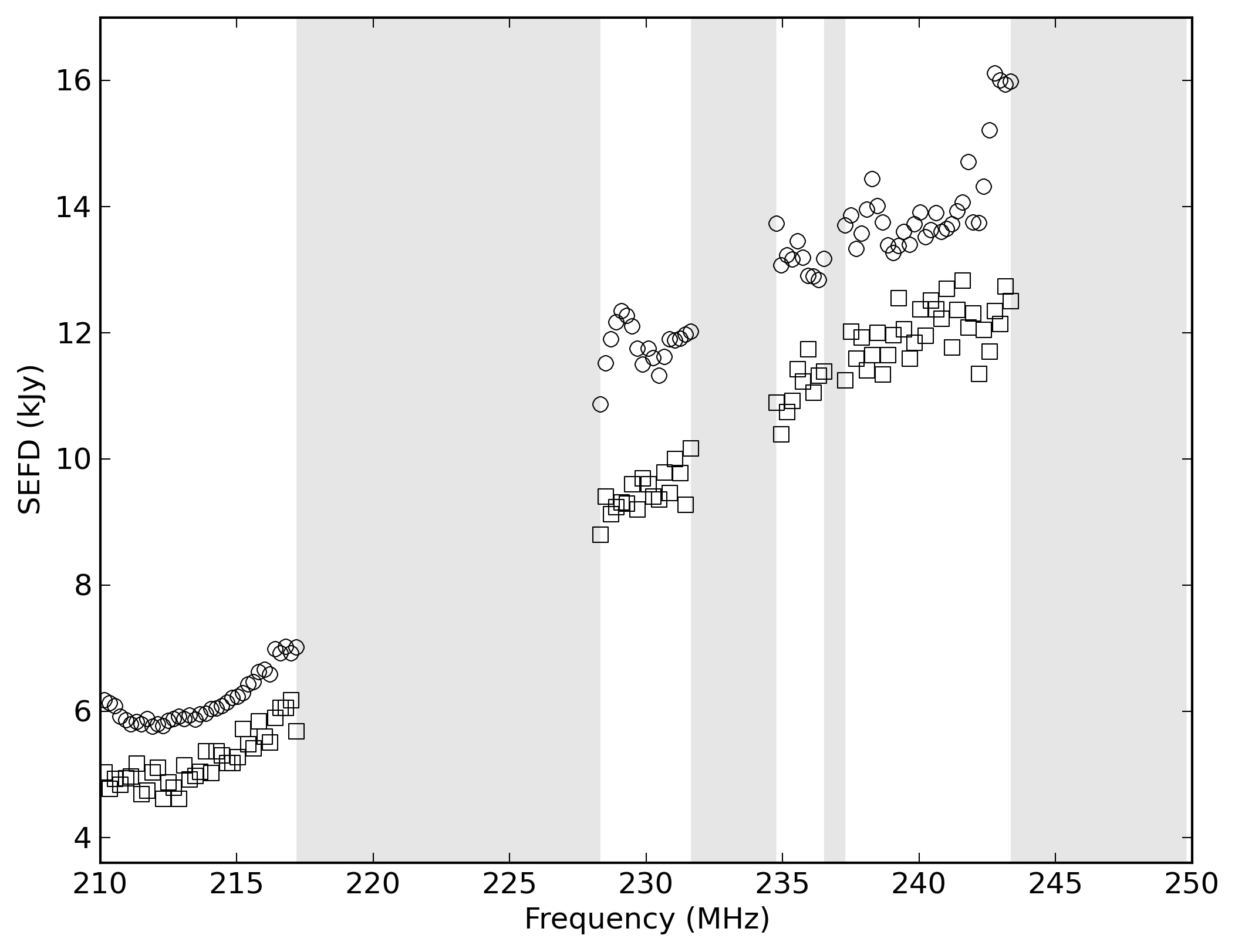} \hspace{0.15in}
\includegraphics[width=3.375in]{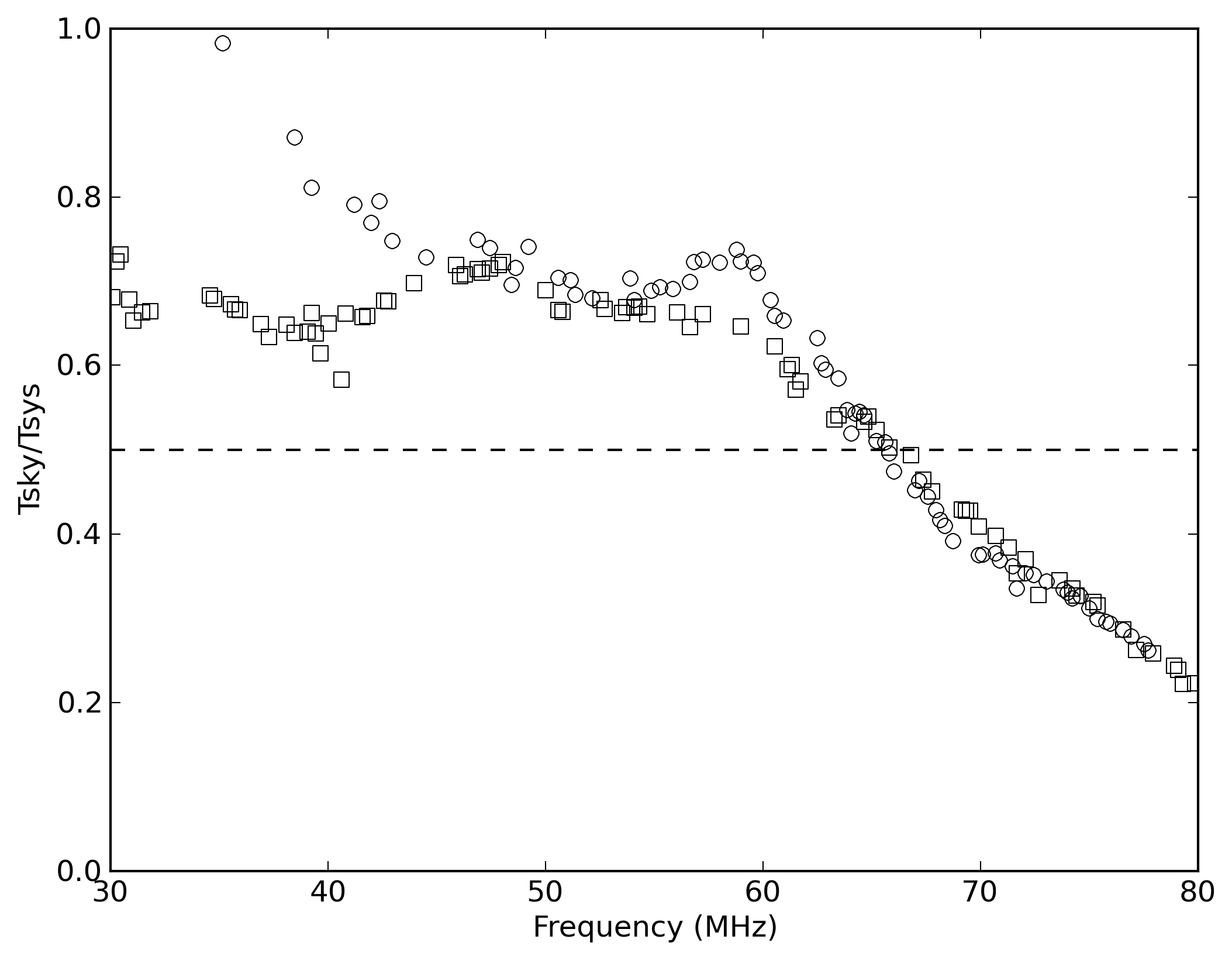}
\caption{\small 
{\it Top and bottom-left panels}: Plots of the SEFD as a function of frequency for the various LOFAR operating bands and station configurations. These curves are derived using the procedure described in Sect.\,\ref{sec:sensitivity}. The grayed regions are excluded from plotting due to strong post-flagging RFI contamination. In the case of HBA, the circles are for core stations and squares are for remote stations. In the LBA, the circles are LBA\_INNER core stations and the squares are LBA\_OUTER core stations. {\it Bottom-right panel}: Contribution of sky temperature to the total system temperature of the LOFAR LBA system. Values are plotted separately for the LBA\_INNER ({\it circles}) and LBA\_OUTER ({\it squares}) station configurations. Empirical system temperature values and sky temperatures were determined as described in the text. The LBA system is clearly sky dominated at frequencies below $\approx65\,\mathrm{MHz}$. Above that frequency, the instrumental noise term dominates.}
\label{fig:sensitivity}
\end{figure*}

To obtain empirical SEFD values for the Dutch stations, we have utilized 2-minute imaging-mode observations of 3C295, taken near transit. The visibilities were flagged to remove RFI, and the contributions of Cygnus~A and Cassiopeia~A were modeled and removed (in the case of the LBA). From these pre-processed data, we determined the S/N ratio of the visibilities for each baseline between similar stations (i.e., core-core and remote-remote baselines for the HBA). The S/N was defined as the mean of the parallel-hand (XX,YY) visibilities, divided by the standard deviation of the cross-hand (XY,YX) visibilities. These S/N values were then combined with the spectral model of 3C295 from \cite{scaife_heald_2012}, and taking the bandwidth and integration time of the individual visibilities into account, we directly obtain an estimate of the SEFD for the type of station comprising this baseline selection. The median contribution of all stations is plotted in Fig.\,\ref{fig:sensitivity}. The most distant remote stations are excluded from this analysis as 3C295 is resolved on all baselines to those stations, thereby invalidating our S/N ratio proxy. For the same reason, we have not attempted to determine empirical SEFDs for international stations using this procedure.

Starting with the empirical LBA SEFDs shown in the top-left panel of Fig.\,\ref{fig:sensitivity}, we have derived the corresponding $T_\mathrm{sys}$ values, using $\mathrm{SEFD}=2760\,T_\mathrm{sys}/A_\mathrm{eff}$
and the adopted values for the effective area, $A_\mathrm{eff}$, given in Table\,\ref{tab:beams}. To determine what fraction of the $T_\mathrm{sys}$ of the LBA system can be attributed to sky flux, we have compared our measured system temperatures with the standard equation from \citet{Thompson2007},
\begin{equation}
T_\mathrm{sky}=60\,\lambda^{2.55},
\end{equation}
where $T_\mathrm{sky}$ is in K, and $\lambda$ in meters. This expression corresponds to the average sky contribution. In the Galactic plane, the value of $T_\mathrm{sky}$ will be higher, and lower at the Galactic pole. The result is shown in the bottom-right panel of Fig.\,\ref{fig:sensitivity} and clearly illustrates that the LOFAR LBA system is sky-noise dominated below $65\,\mathrm{MHz}$, in parts of the sky where the adopted sky spectrum is appropriate or an overestimate.

With these derived SEFD values, one can now compute the expected sensitivity of the array during a typical observation \citep{tcp99}. In Table\,\ref{tab:sensitivity}, sensitivities are quoted for an 8 hour integration time and an effective bandwidth of 3.66 MHz (20 subbands) for the cases of a 6-station Superterp, a 24-station core array, a 40-station Dutch array, and a 48-station full array. The quoted sensitivities are for image noise and assume a factor of 1.3 loss in sensitivity due to time-variable station projection losses for a declination of 30 degrees, as well as a factor 1.5 to take into account losses for ``robust'' weighting of the visibilities, as compared to natural weighting. Note that this robustness factor is very strongly dependent on how the various stations, which all have different sensitivity, are weighted during the imaging process. 

The values quoted for the HBA in Table\,\ref{tab:sensitivity} agree with empirical values derived from recent observations on 3C196 and the North Celestial Pole (NCP) where all NL remote stations were tapered to match 24-tile core stations. With improved station calibration, these estimates can likely be improved in the future by a factor of about 1.2. For the more compact LOFAR configurations, confusion noise will exceed the quoted values (see Sect.\,\ref{sec:confusion}). The quoted sensitivities for the lower LBA frequencies have not yet been achieved in practice. At the lowest frequencies below 30 MHz, values have not yet been determined awaiting a final station calibration. Similarly, the quoted values at 200, 210 and 240 MHz should be viewed as preliminary and are expected to improve with revised station calibration as well. For more recent values of the estimated array sensitivies and updates on the status of the station calibration, the reader is referred to the online documentation\footnote{See http://www.astron.nl/radio-observatory/astronomers for current updates on the calibration status of the LOFAR array, including up-to-date estimates for the achievable sensitivities.}.

\subsection{Confusion noise}
\label{sec:confusion}

The presence of faint, unresolved extragalactic sources in the synthesized beam produces ``confusion'' fluctuations in deep radio maps and represents a fundamental limit to the achievable sensitivity of a radio telescope. Confusion is normally said to occur when more than one source falls within the telescope beam and the classical confusion limit, $\sigma_c$, is defined as the flux density level where this condition is met taking into account the underlying population of faint sources. Formally, this condition can be written
\begin{equation}
M ~\Omega_b ~N(\sigma_c) = 1
\label{eqn:confusion}
\end{equation}
where $N(S)$ specifies the number of sources per steradian with a flux density greater than $S$ and $\Omega_b$ represents the solid angle of the synthesized beam. The parameter $M$ represents the number of beam solid angles per source and depends on the assumed form for the underlying distribution of sources. 

In order to estimate $\sigma_c$, we first adopt a parameterization for $N(S)$ determined from the VLSS sky survey at 74 MHz \citep{VLSS2007, VLSS2012} and given by
\begin{equation}
N(>S) = A \,S^{\beta} \,\biggl( \frac{\lambda}{\lambda_0} \biggr)^{\alpha \beta} = 1.14 ~S^{-1.30} \,\biggl( \frac{\lambda}{\rm{4 \,m}} \biggr)^{0.91}
\label{eqn:numdist}
\end{equation}
where $\beta$ represents the intrinsic slope of the underlying source distrbution as a function of flux density $S$ and $\alpha$ is the mean spectral index of a source at these wavelengths \citep{Cohen04, Cohen06}. Based on the VLSS catalog, \cite{Cohen06} estimates values of $-0.7$ and $-1.30$ for $\alpha$ and $\beta$, respectively. 
The normalization constant is $A = 1.14 ~\rm{Jy} ~\rm{beam}^{-1}$ where the beam size is given in degrees.

Following \citet{Condon74}, the solid angle for a Gaussian beam with FWHM, $\theta$, is given by $\Omega_b = \pi \theta^2 / [4 ~\rm{ln}(2)] \sim 1.133 ~\theta^2$. The final term, $M$, corresponds to the number of synthesized beams per source assuming a given flux density limit cutoff of $S = q \,\sigma_c$ and is given by $M = q^2 /(2+\beta)$ ~\citep{Condon74}. In the following, we have selected a cutoff of $q=3$ yielding a value for $M = 12.8571$.

Combining these expressions with Equation~\ref{eqn:confusion}, we can derive an expression for the expected confusion limit in the LOFAR band for different array configurations. Substituting these values, we obtain
\begin{equation}
\sigma_c = 30 ~~\biggl( \frac{\theta}{1\arcsec} \biggr)^{1.54} ~\biggl( \frac{\nu}{\rm{74 ~MHz}} \biggr)^{-0.7}
 ~~~~[\,\mu\rm{Jy} ~\rm{beam}^{-1}]
\label{eqn:limit}
\end{equation}
for the classical confusion limit. To put this expression in the context of LOFAR, for a frequency of 60 MHz near the peak of the LBA band, we would estimate values for $\sigma_c$ of 150 mJy, 0.7 mJy, and 20 $\mu$Jy for observations using the NL core, full NL array, and full European array, respectively. Similarly for 150 MHz in the HBA band, we would estimate confusion limits of 20 mJy, 80 $\mu$Jy, and 3 $\mu$Jy for the core, NL, and international baseline configurations, respectively.

It is worth noting that these estimates rely on the VLSS source counts which are 100\% complete down to flux density levels of only 1 Jy \citep[see Fig.\,15 in][]{VLSS2007}. The source distribution at much lower flux densities and lower frequencies may be significantly different than seen by the VLSS. Ultimately the source catalog produced by LOFAR's first all-sky, calibration survey, the Multifrequency Snapshot Sky Survey (MSSS) (see Sect.\,\ref{sec:calib} below and Heald et al., in prep.), will provide better constraints on the actual degree of source confusion in LOFAR images.

\begin{figure*}[ht]
\centering
\includegraphics[width=6.750in]{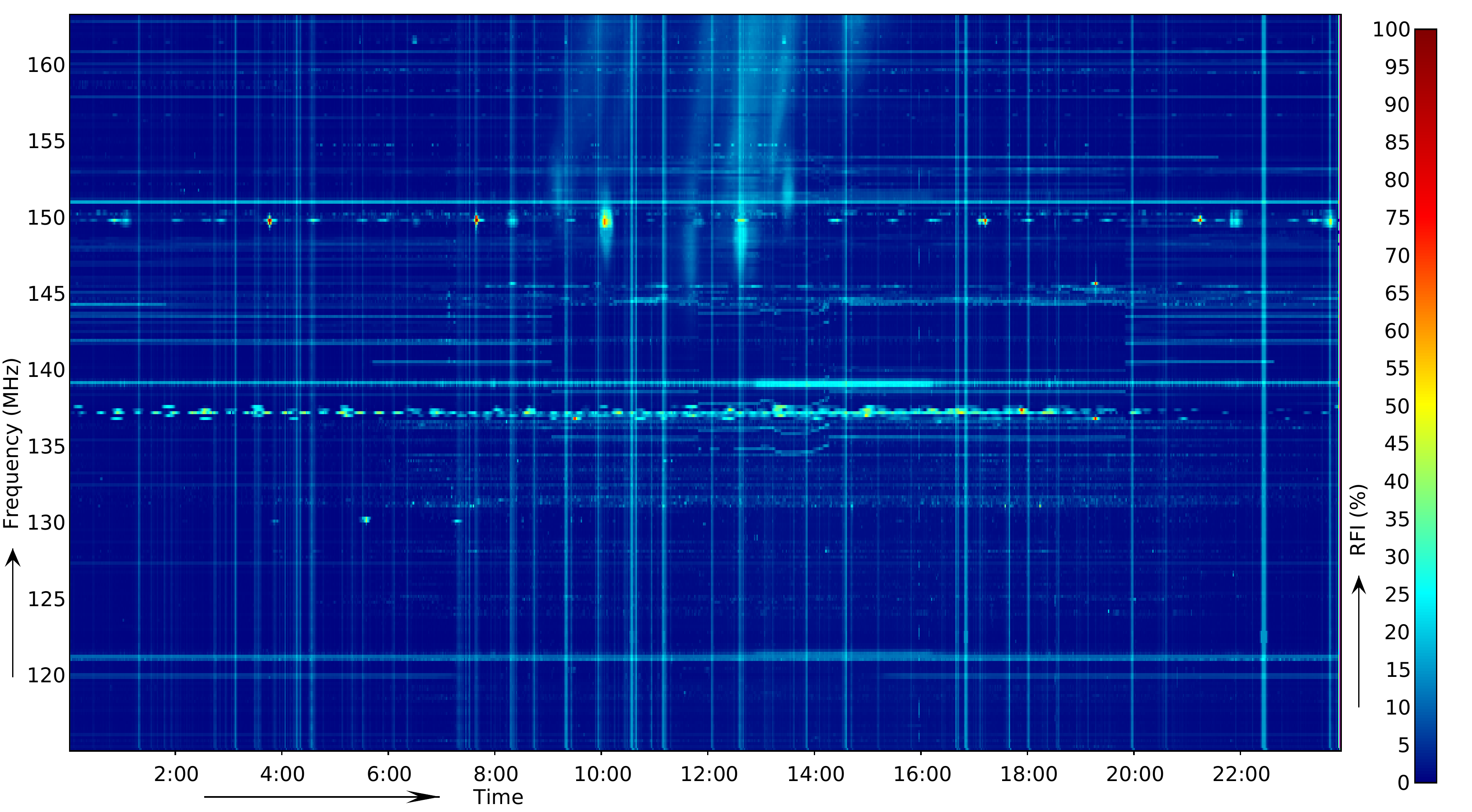}
\includegraphics[width=6.750in]{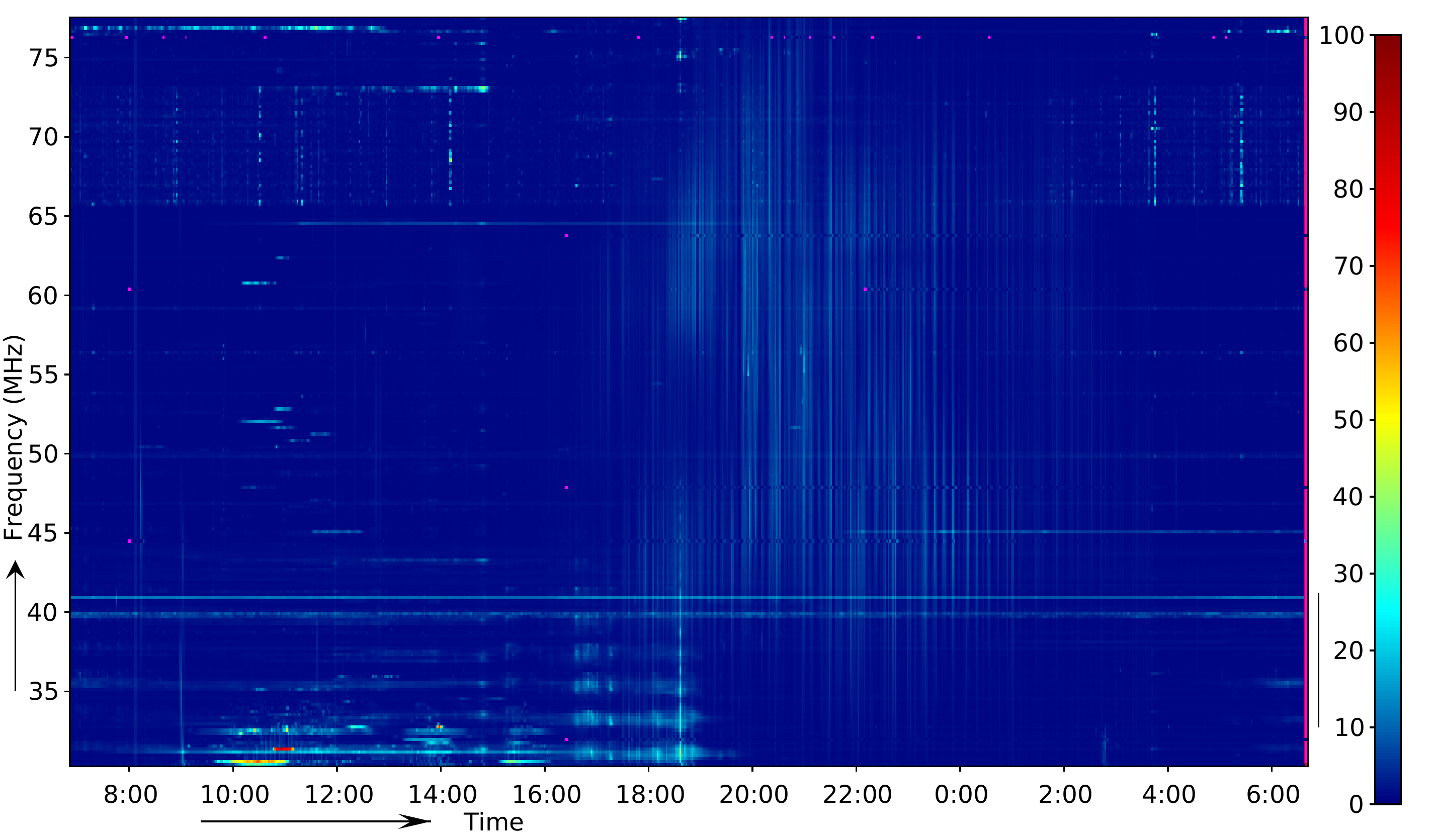}
\caption{\small 
Dynamic spectrum of RFI occupancy during the LBA and HBA survey. The median baseline RFI is around 1 to 2\% across the bands, although there are regions of the spectrum with significant narrow and broad band RFI features \citep{Offringa13}.}
\label{fig:rfi}
\end{figure*}

\begin{figure*}[ht]
\centering
\includegraphics[width=6.5in]{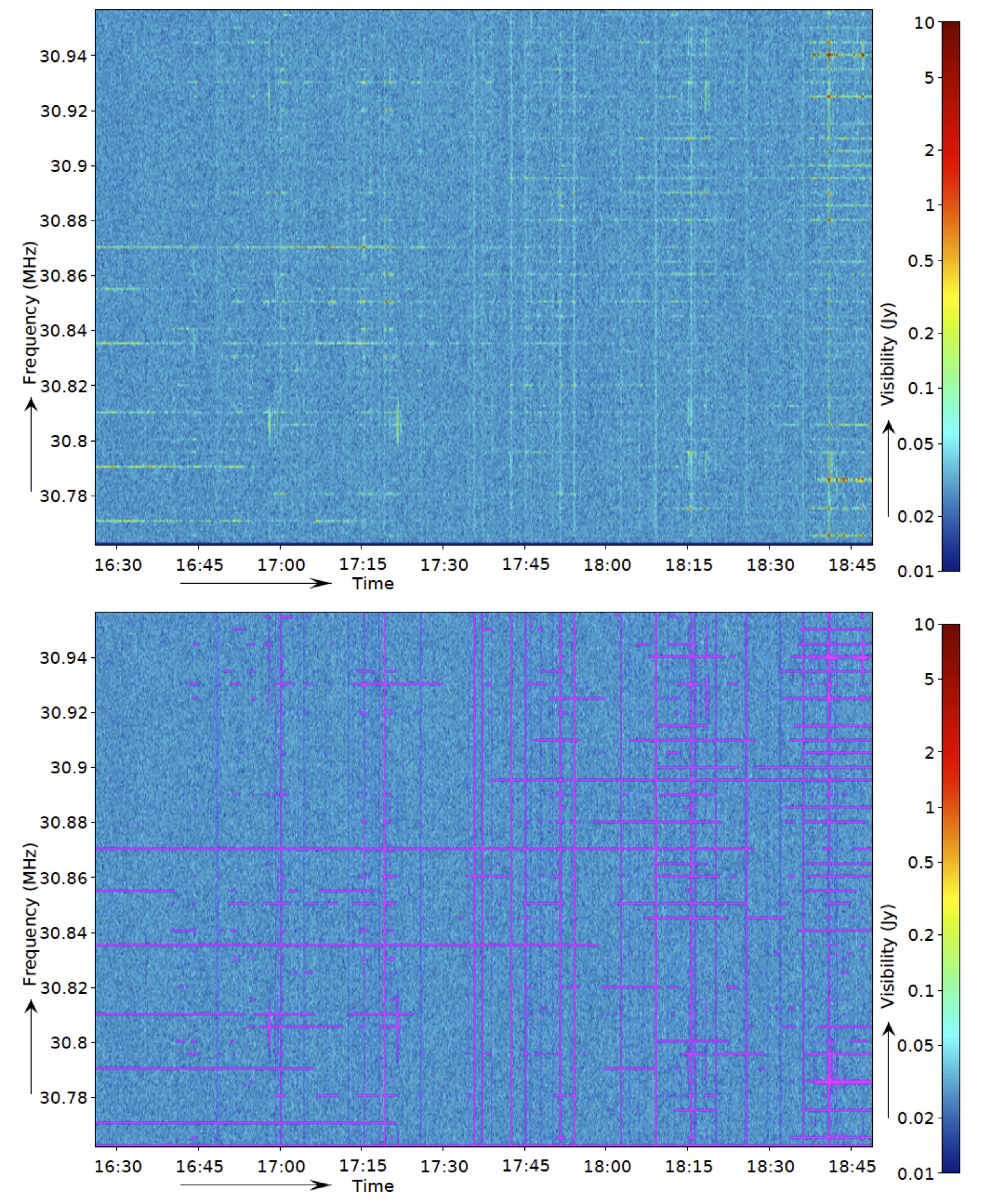}
\caption{Dynamic spectrum of data from one sub-band of the LBA survey, formed by the correlation coefficients of baseline CS001--CS002 at the original frequency resolution of 0.76 kHz. The displayed sub-band is one of the worst sub-bands in terms of the detected level of RFI. The top image shows the original spectrum, while the bottom image shows with purple what has been detected as interference \citep{Offringa13}.}
\label{fig:flagging}
\end{figure*}

\subsection{RFI environment}
\label{sec:rfi}

A possible concern with the construction of LOFAR in the high population density environment of the Netherlands and surrounding countries is terrestrial RFI in the local low-frequency radio spectrum. To overcome this, LOFAR has been designed to provide extremely high frequency- and time-resolution data during normal interferometric operations. The default frequency resolution is 610 or 763 Hz (each subband is subsequently divided into 256 channels), depending on the clock setting, and the typical visibility integration times are either 1 second in the low-band (10--80~MHz) or 3 seconds in the high-band (120--240~MHz). Even though 256 channels are available, in practice typical observations are performed using only 64 channels per subband. This choice lowers the resulting data volume by a factor of 4 without additional loss of data due to RFI flagging.

The flagging of the full resolution data in both time and frequency is carried out using the {\tt AOFlagger}, a post-correlation RFI mitigation pipeline developed by \cite{Offringa10,Offringa12a,Offringa12b}. This routine uses an iterative method to determine the true sky brightness by applying a high-pass filter to the visibility amplitudes in the time–frequency plane. Subsequently, it flags line-shaped features with the SumThreshold method \citep{Offringa10}. Finally, the scale-invariant rank operator, a morphological technique to search for contaminated samples, is applied on the two-dimensional flag mask \citep{Offringa12b}. Additional developments, for example, pre-correlation RFI mitigation, will be incorporated into the LOFAR analysis routines in the future.

As an example of the RFI environment of LOFAR, the percentages of RFI that have been identified and removed from 24 hour datasets taken with the LBA and HBA-low systems are shown in Fig.\,\ref{fig:rfi}. For the low-band system, the median level of RFI is estimated to be around 2\% of the data, although this increases to around 10\% at the lowest frequencies. These values represent the maxima per sub-band since within a sub-band any given channel can be 100\% contaminated. For some sub-bands in the 30--80 MHz range, the median level of RFI can spike to as high as 7 to 20\% of the data. For the HBA-low case, the median level of RFI that is identified over a 24 hour period is around 1\% of the data. However, there are several frequencies that show RFI spikes between 5 and 17\% of the data across the band. For the HBA system, the median baseline RFI is again around 1 to 2\% of the data, but there are also significant broadband RFI signals that range from 5 to above 50\% of the data. In general, the level of RFI that is identified and removed from LOFAR datasets during the commissioning phase is not severe over the standard 30--80\,MHz and 110--240\,MHz observing frequencies in which LOFAR operates. Additional results of the automated flagging algorithm are shown for a more extreme case of RFI contamination in Fig.\,\ref{fig:flagging}.

\begin{figure*}
\centering
\includegraphics[width=6.5in]{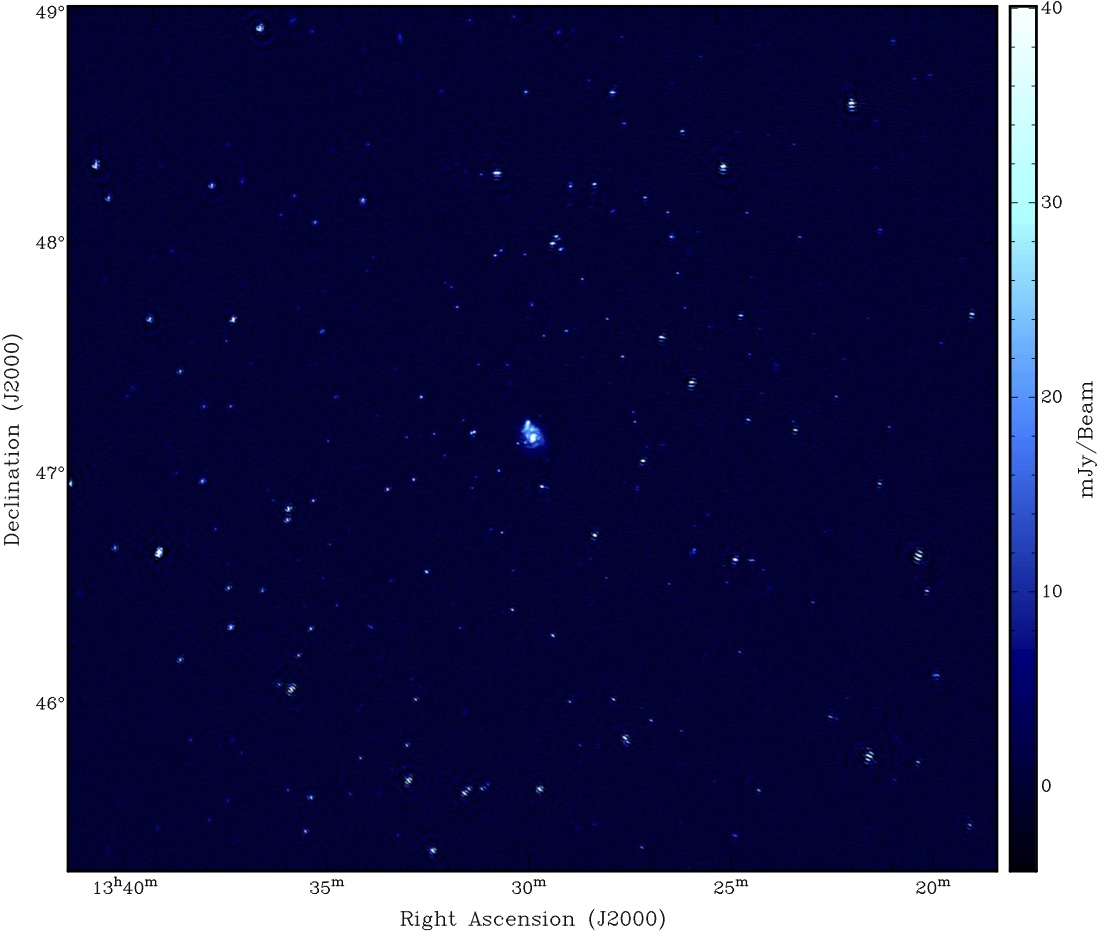}
\caption{\small LOFAR wide-field image of the area around M51. This image represents a 6\,h integration using 34 MHz (204 sub-bands) of bandwidth centered on 151 MHz taken with the HBA. The noise in the image is $\sigma \sim 1$ mJy beam$^{-1}$ in the vicinity of M51 and $\sigma \sim 0.6$ mJy beam$^{-1}$ away from bright sources with an effective beamsize of $20\arcsec$ (Mulcahy et al., in prep.).}
\label{fig:m51}
\end{figure*}

\subsection{Image quality}
\label{sec:calib}

The calibration step is performed using BlackBoard Selfcal (BBS). This calibration package is based on the Hamaker-Bregman-Sault measurement equation \citep[ME; see][]{1996A&AS..117..137H,1996A&AS..117..149S, 1996A&AS..117..161H, 2000A&AS..143..515H, 2006A&A..456..395H, Smirnov11} which expresses the instrumental response to incoming electromagnetic radiation within the framework of a matrix formalism. Here, the various instrumental effects are identified, their effect on the signal is characterized in full polarization, and are quantified and parameterized as separate Jones matrices. Each of these terms may depend on different dimensions: frequency (e.g. the bandpass); time (e.g. the station gains); or direction (e.g. the station beam). Because it is based on the general form of the ME, BBS can natively handle difficult problems such as direction dependent effects and full polarization calibration, using parameterized models based on the physics of the signal path.

\begin{figure*}
\centering
\includegraphics[width=\textwidth]{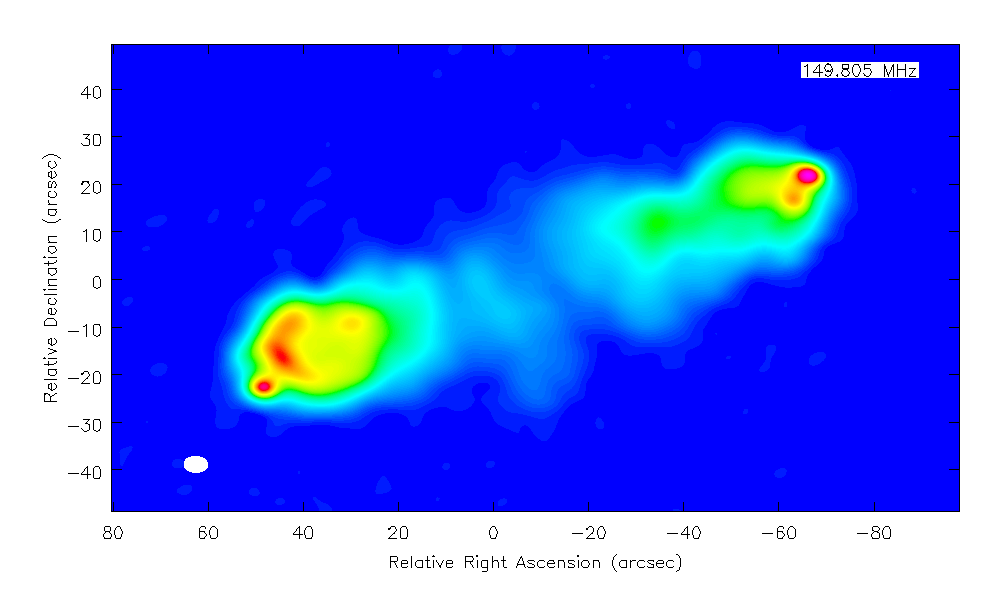}
\includegraphics[width=\textwidth]{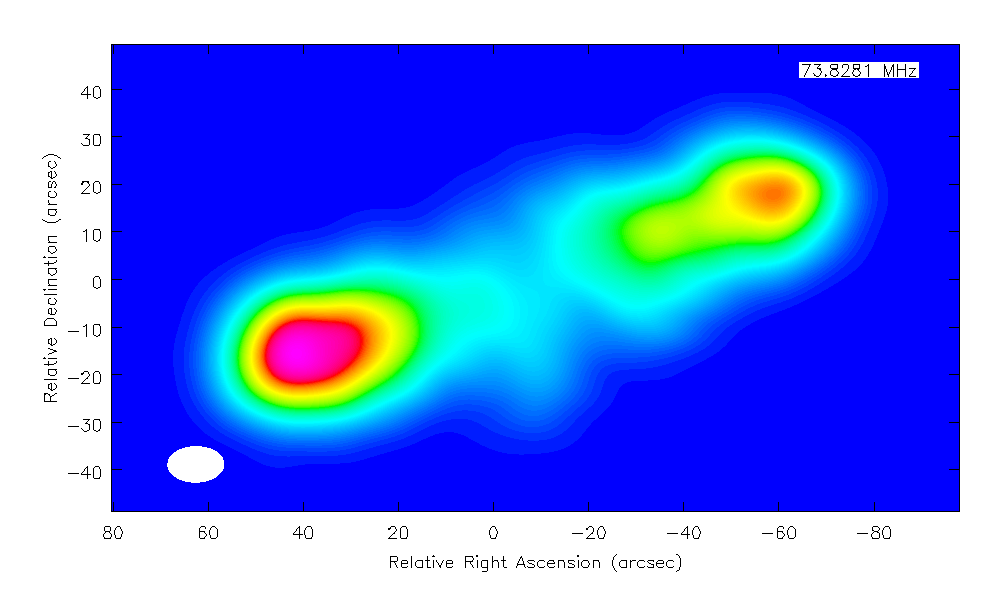}
\caption{\small Example of LOFAR single sub-band (0.2 MHz bandwidth) imaging of the radio galaxy Cygnus A with the HBA system at 150 MHz (top) and the LBA system at 74 MHz (bottom), made using 24 core stations and 9 remote stations during the commissioning phase (McKean et al., in prep.). Both datasets consist of 12\,h synthesis observations. These images show the expected combination of compact and extended structure that has previously been seen in this source at these frequencies using the VLA and MERLIN, c.f. with the images of \cite{lazio06} and \cite{leahy89}, respectively. The beam-size of the images are $5.7\times3.5$~arcsecond and $11.7\times7.4$~arcsecond, respectively, and are shown as the white ellipses in the bottom left corner of each image. The dynamic ranges are $\sim 3500$ and $\sim 2000$ for the 74 MHz and 150 MHz maps, respectively.}
\label{fig:cyga}
\end{figure*}

A critical input to BBS is the sky model that is used to predict the visibilities. Early in the commissioning process, this input to the BBS stage of the pipeline was a hand-crafted listing of the brightest sources in the field of interest. The current SIP (see Sect.\,\ref{sec:sip}) automatically constructs an initial sky model based on cataloged values from the VLSS, WENSS, and NVSS. Note that this should be considered the ``Mark-0'' LOFAR GSM; the ``Mark-1'' LOFAR GSM is being generated by the Multifrequency Snapshot Sky Survey (MSSS). MSSS is a broadband survey of the northern ($\delta>0^{\circ}$) sky, using multiple simultaneous station beams to increase the survey speed. MSSS provides a higher areal density of sources than the VLSS catalog, and more importantly includes well-sampled spectral information in 16 bands spanning 30\,MHz to 160\,MHz. The primary goal of the survey is to provide a broadband catalog of the brightest population of sources in the LOFAR sky, creating a low-frequency calibration database for future imaging observations. MSSS observations began in autumn 2011 and were nearly half completed during 2012. A detailed description of the survey setup, data processing, and results is in preparation (Heald et al., in prep.).

\begin{figure*}[ht]
\centering
\includegraphics[width=3.5in]{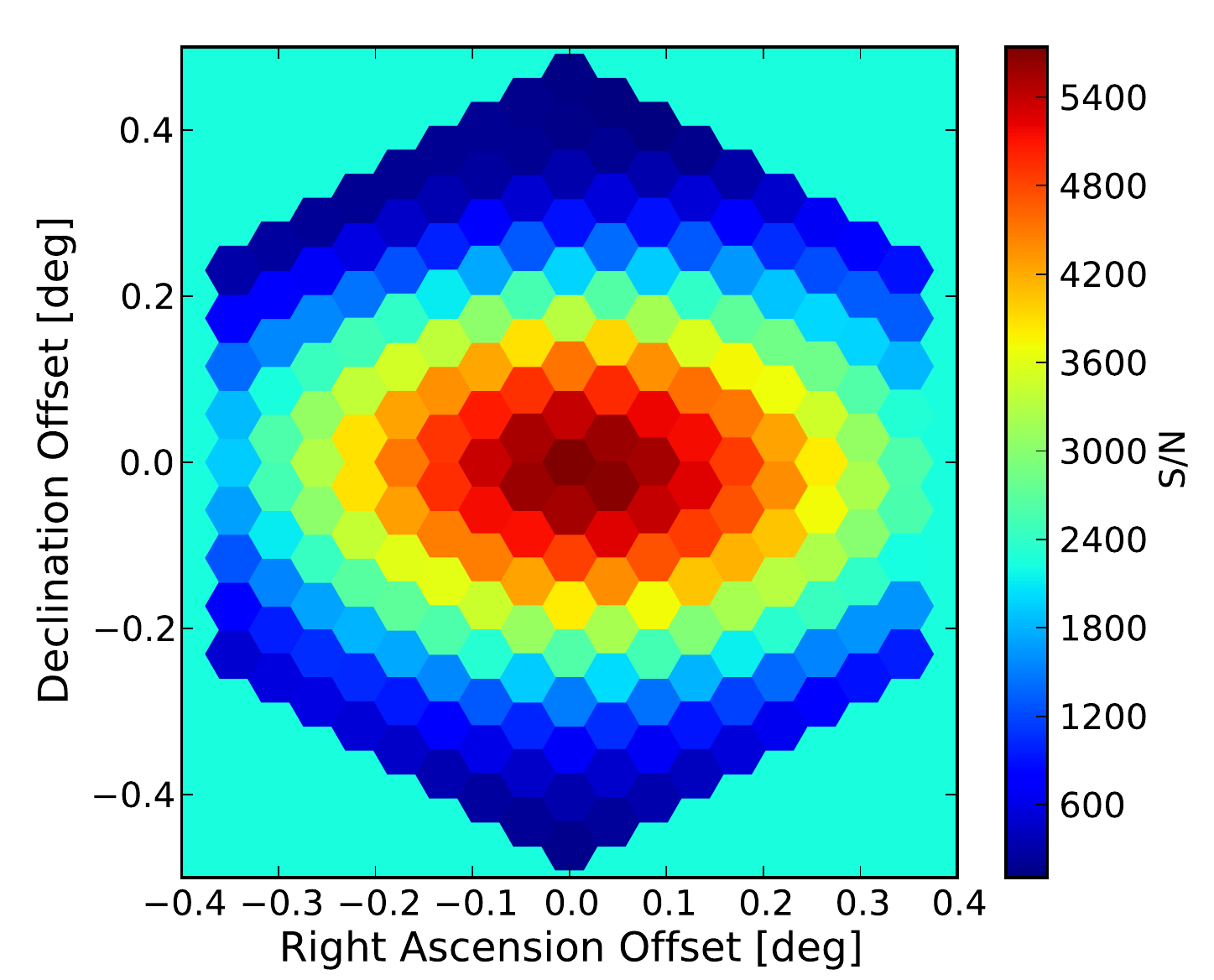}
\includegraphics[width=3.5in]{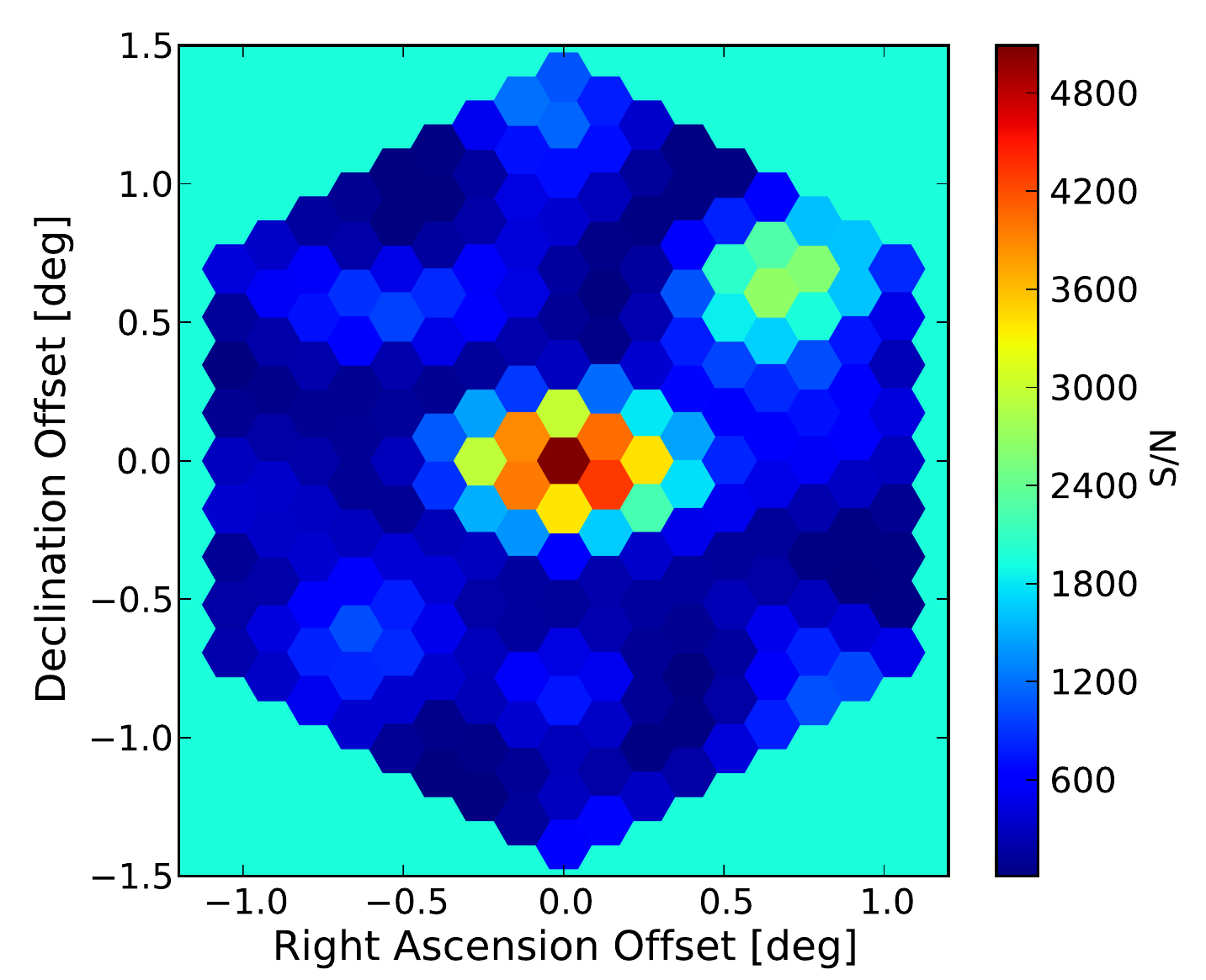}
\caption{\small Maps of the coherent Superterp beam, also known as a tied-array beam. These were made by simultaneously recording 217 coherent and 1 incoherent beam for all 12 Superterp HBA sub-stations from
120--168\,MHz.  The bright pulsar B0329+54 was observed twice for 5 minutes, near zenith and the color maps reflect the S/N ratio of the dedispersed and folded signal in various directions.  The background color reflects the S/N ratio of the simultaneously acquired incoherent beam.  {\bf Left:} Observation L57554 in which the tied-array beams were arranged in a hexogonal grid with spacing $0.05^{\circ}$.  This densely samples the main lobe of the Superterp beam.  {\bf Right:} Observation L57553 in which the tied-array beams were more widely spaced ($0.15^{\circ}$ apart) in order to probe the sidelobes. Asymmetry in the sidelobe pattern is due to imperfect phasing of the coherent beam.}
\label{fig:ta-beam}
\end{figure*}

The quality of images produced by LOFAR's interferometric imaging mode is dependent on many factors. Novel techniques must be brought to bear in order to achieve imaging at high dynamic range and fidelity over a large FoV. There are many factors that may limit the achievable dynamic range (DR) in LOFAR images. In fields where there are no unusually bright sources (see below) the main limiting factors are direction-dependent effects, namely issues related to variable beam response as a function of time, and the ionosphere. The former is being handled by the inclusion of a comprehensive beam modeling library in both the calibration and imaging software, while techniques to address the latter are based on the method used by \citet{2009A&A...501.1185I}. Polarization calibration will include the prediction (to within $\sim0.1\,\mathrm{rad\,m^{-2}}$) and application of ionospheric rotation measure values as described by \citet{Sotomayor-Beltran13}. Fig.\,\ref{fig:m51} shows a relatively wide-field ($\sim 4 \times 4$ square degrees) HBA image of the field surrounding the bright galaxy M51 (Mulcahy et al., in prep.).

A large number of commissioning observations have now been obtained by the LOFAR EoR project team on fields containing the bright compact sources 3C196 and 3C295, which have a flux density of 100 Jy at 115 and 144\,MHz, respectively.  The dynamic range achieved in these fields exceeds more than 500,000:1 at distances at least several arcmin from the sources.  In the neighbourhood of these sources, the dynamic range is currently still restricted to 10,000-100,000:1, depending on the PSF used, and appears to be limited only by our imperfect knowledge of the (sub-)arcsecond structure of the sources themselves. We note that this knowledge will likely improve in the very near future with the inclusion of structural information obtained using LOFAR's international baselines. The correlator itself, therefore, does not appear to introduce any errors that limit the LOFAR's achievable dynamic range.

Although the dynamic ranges already achieved in images of select LOFAR fields are impressive, the more relevant number characterizing the quality of LOFAR images is the achievable noise level as given in Table\,\ref{tab:sensitivity}, with the caveats listed in Sect.\,\ref{sec:sensitivity}. The many factors influencing the actual noise in real observations were already listed above. In practice, deep observations (3 nights, of 6\,h each) of the NCP field have reached noise levels of 100 $\mu$Jy or better corresponding to a factor of only 1.4 above the thermal limit set by the noise from our Galaxy and the receivers \citep{Yatawatta13}. For more complicated fields of course, the noise levels can be higher.

The $uv$-coverage of the LOFAR array is also designed to provide excellent imaging of extended sources. When combined with its high sensitivity, LOFAR can deliver high quality images of faint, diffuse objects \citep{vanWeeren12, deGasperin12}. For example, Fig.\,\ref{fig:cyga} shows an HBA image of the diffuse emission associated with the bright AGN Cygnus A obtained during the commissioning phase (McKean et al., in prep.). During the commissioning period, the available calibration and imaging software has been shown to deliver on-axis image rms levels near the expected thermal noise. Predicted values for the achievable sensitivities are given in Table\ref{tab:sensitivity}. Significantly deeper images are achievable by utilizing the full bandwidth \citep{Yatawatta13}. A discussion of the tradeoff between sensitivity and resolution is given by \citet{Heald11}.

\begin{figure*}
\centering
\includegraphics[width=7.25in]{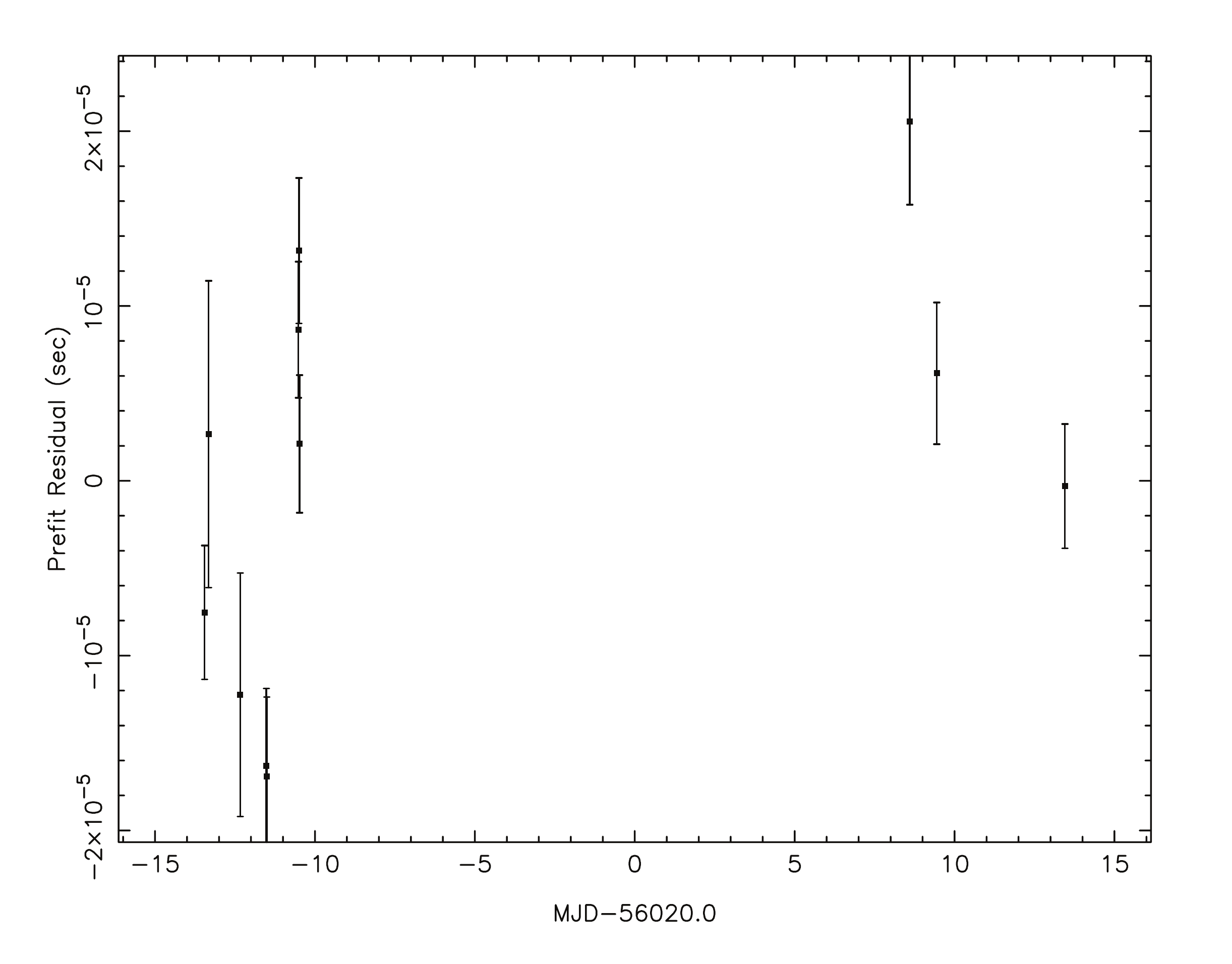}
\caption{\small Pre-fit timing residuals from a 1-month LOFAR campaign on PSR J0034-0534.  The individual time-of-arrival measurements reflect the deviations from a Westerbork-derived timing model for this pulsar (which includes astrometric, spin, and orbital parameters), with no additional re-fitting of these parameters.  The measurements have an rms of 11\,$\mu$s, and it is possible that the deviations of individual points reflect small changes in the dispersion and/or scattering measure of the pulsar.  In any case, these data clearly demonstrate LOFAR's capability to do precision pulsar timing.}
\label{fig:timing}
\end{figure*}

\subsection{Beam-formed modes}
\label{sec:bfmodes}

LOFAR's beam-formed modes share many of the same system requirements imposed by the interferometric imaging mode, but they also have some unique requirements of their own.  As such, beam-formed observations provide a complementary, and sometimes orthogonal, means with which to test both generic and mode-specific system performance \citep[see][for several examples]{hsa+10, sha+11}.

Wide-band observations of continuum sources like pulsars are well-suited to measuring the instrumental bandpass and overall sensitivity (e.g., Fig.\,\ref{fig:coh_inc_sum}). They also serve as important polarimetric calibration sources, both using interferometric imaging and beam-formed data. For example, Fig.\,\ref{fig:rmsf} shows the rotation measure spread functions (RMSF) for two pulsars in the HBA and LBA bands.

The microsecond to millisecond time resolution typically used in the beam-formed modes also probes a different RFI regime than that apparent from the $>1$\,s time resolution used to record visibilities. RFI occupancy histograms for LOFAR beam-formed data can be found in \citet{sha+11}.

The various possible beam shapes (e.g. element beam, station beam, and tied-array beam) can also be mapped by the beam-formed mode by scanning across relatively bright point sources \citep[e.g. see][]{sha+11}. This knowledge can then be used as an input for imaging calibration.  Alternatively, using LOFAR's multi-beam tied-array survey mode, one can map the instantaneous beam in two dimensions.  Fig.\,\ref{fig:ta-beam} shows the result of an observation in which 217 simultaneous tied-array beams where pointed in a honey-comb pattern around the phase center in order to map the shape of the coherent Superterp beam.  The observed beam shape agrees well with that predicted by a simple model that takes into account station sizes and positions.

Clock calibration is vital to interferometric imaging, but these corrections can largely be made in the post-processing.  For tied-array observations, however, the instrumental, geometrical, and environmental (e.g. ionospheric) delays must be applied in real time in order to form a properly coherent sum of the station beams \citep[e.g., see][and Fig.\,\ref{fig:tabs} and \ref{fig:coh_inc_sum} for demonstrations of tied-array beams using the LOFAR Superterp]{sha+11}. The implementation of a single clock signal shared between all 24 Core stations has greatly simplified this process.

Furthermore, for applications like long-term, phase-coherent pulsar timing, the clock reference standard needs to be maintained between observations and systematic offsets and changes to the observatory's time standard need to be well-documented.  LOFAR's station-level Rubidium clocks are guided by local GPS receivers, and thus the long-term LOFAR time standard is automatically in sync with this system.  Comparison of the time stamps between observations is greatly simplified by the use of a consistent geographical reference.  For LOFAR the phase center of all observations is the geographical center of the LBA field of station CS002, regardless of whether this particular station is being used or not (see Table ~\ref{tab:fields}). This reference position is used, e.g., for barycentering, pulsar timing, and phasing-up the array.  Fig.\,\ref{fig:timing} shows an example of phase-coherent timing of a millisecond pulsar using LOFAR.

\section{Key science drivers}
\label{sec:ksps}

\subsection{Epoch of reionization}
\label{sec:eor}

The formation of the first stars and galaxies marks a major transition in the evolution of structure in the Universe.  These galaxies with their zero-metallicity Population-III and second-generation Population-II stars and black-hole driven sources (e.g., mini-quasars, x-ray binaries, etc.) first heated and subsequently transformed the intergalactic medium from neutral to ionized. This period is known as the Cosmic Dawn and epoch of reionization (EoR). Observing and quantifying this poorly observed and little understood process is the main aim of the LOFAR EoR Key-Science Project (KSP).

The last thirty years have witnessed the emergence of an overarching paradigm, the $\Lambda$CDM model, that describes the formation and evolution of the Universe and its structure.  The $\Lambda$CDM model accounts very successfully for most of the available observational evidence on large-scales. According to this paradigm about 400,000 years after the Big Bang ($z\approx1100$), the temperature and density decreased enough to allow ions and electrons to recombine and the Universe to become neutral. As a result, the Universe became almost transparent leaving a relic radiation, known as the cosmic microwave background (CMB) radiation \citep[for recent results see the WMAP papers, e.g.,][]{spergel07,page07,komatsu10}.  The matter-radiation decoupling has ushered the Universe into a period of \textit{darkness} as its temperature dropped below 3000 K and steadily decreased with the Universe's expansion. These \textit{Dark Ages} ended about 400 million years later, when the first radiation emitting objects (stars, black-holes, etc.)  were formed and assembled into protogalaxies during the Cosmic Dawn (CD).

The most accepted picture on how the cosmic dawn and reionization unfolded is simple. The first radiation-emitting objects heated and subsequently ionized their immediate surroundings, forming ionized bubbles that expanded until the neutral intergalactic medium consumed all ionizing photons. As the number of objects increased, so did the number and size of their ionization bubbles. These bubble gradually perculated until eventually they filled the whole Universe.

Most of the details of this scenario, however, are yet to be clarified.  For example: what controled the formation of the first objects and how much ionizing radiation did they produce? How did the bubbles expand into the intergalactic medium and what did they ionize first, high-density or low density regions?  The answers to these questions and many others that arise in the context of studying the CD and EoR touch upon many fundamental questions in cosmology, galaxy formation, quasars and very metal-poor stars; all are foremost research topics in modern astrophysics. A substantial theoretical effort is currently dedicated to understanding the physical processes that triggered this epoch, governed its evolution, and the ramifications it had on subsequent structure formation \citep[c.f.,][]{barkana01, bromm04, ciardi05, choudhury06, furlanetto06, zaroubi13}. Observationally however, this epoch is poorly studied. Still the current constraints strongly suggest that the EoR roughly straddled the redshift range of $z \sim 6$--12 \citep{komatsu10,fan03,fan06,bolton07, theuns02, bolton10, Oesch10, bunker10}.

It is generally acknowledged that the 21-cm emission line from neutral hydrogen at high redshifts is the most promising probe for studying the \textit{Cosmic Dawn} and the EoR in detail \citep{field58, madau97, ciardi03}. HI fills the IGM except in regions surrounding the ionizing radiation of the first objects to condense out of the cosmic flow.  Computer simulations suggest that we may expect an evolving complex patch work of neutral (HI) and ionized hydrogen (HII) regions \citep{gnedin01,ciardi03b,whalen06,mellema06,zahn07,mesinger07,thomas09,thomas11,ciardi12}.

LOFAR with its highly sensitive HBA band is the best available instrument to-date to probe this process from $z=11.4$ (115~MHz) down to $z=6$ (203~MHz). At lower frequencies, both the sensitivity of LOFAR drastically decrease and the sky noise dramatically increase making it very hard to use the telescope for such a measurement. Given the very low brightness temperature and the angular scale of the expected EoR signal, the only part of LOFAR that has the sensitivity to detect the EoR redshifted 21-cm signal is the LOFAR core. The LOFAR core stations give a resolution of about 3 arcminutes over a FoV, given by the HBA LOFAR station size, of about 5$^\circ$ corresponding, at $z=9$, to $\approx 8$ and $800$ comoving Mpc, respectively.  A number of fields ($\sim 5$) with minimal Galactic foreground emission and polarization will be observed for a total of several thousands of hours, reaching brightness temperatures of $50-100$ mK per resolution element per MHz bandwidth, close to that of the redshifted 21-cm emission from the EoR. Ultimately, the EoR KSP hopes to achieve a noise level of approximately 60 mK per resolution element per MHz after 600 hours.

Studying the power-spectra as a function of redshift (or frequency) allows us to probe the EoR as it unfolded over cosmic time.  The EoR power spectrum can be observed over about two orders of magnitude in wave numbers and other higher-order statistical measures can be obtained as well \citep{jelic08, harker09,harker10, panos09}. It might even be possible to image the EoR as it unfolds on very large scales after several thousand hours of integration time on a single field \citep{zaroubi12}. Finally, using total-power measurements, LOFAR might also be able to probe the total (i.e.\ global) intensity signal from neutral hydrogen to even higher redshifts with the LBA, complementing interferometric measurements at lower redshifts with the HBA. The LOFAR EoR KSP is currently investigating, using the LOFAR LBA system in different beam-forming modes, whether the system is suitable to detect, or place stringent upper limits, on the global redshifted 21-cm signal from the Cosmic Dawn around $z\sim20$.

Summarizing, LOFAR observations will allow detection and quantification of the Cosmic Dawn and EoR over wide range in angular scales and redshifts. Such measurement will help answer the main questions surrounding the earliest phases of the formation of the Universe: What is the nature of the first objects that ended the Dark Ages, ushering in the Cosmic Dawn and the reionization of the high-redshift IGM?  What is the relative role of galaxies and AGN, of UV-radiation and X-rays?  When did the EoR start and how did it percolate through the IGM? Which regions re-ionized first: low or high density regions (inside-out versus outside-in scenario)? What are the detectable imprints that the re-ionization process left on the 21-cm signal? What can we learn from 21-cm measurements about the matter density fluctuations on the conditions prior to the EoR? What can we learn about the formation of (supermassive) black holes and the duration of their active phases?

The LOFAR EoR KSP plans to address all these questions over the next years, playing an important role as well in paving part of the way for future more sensitive observations of both the Cosmic Dawn and EoR with the SKA.

\begin{figure*}
\centering
\includegraphics[width=\textwidth]{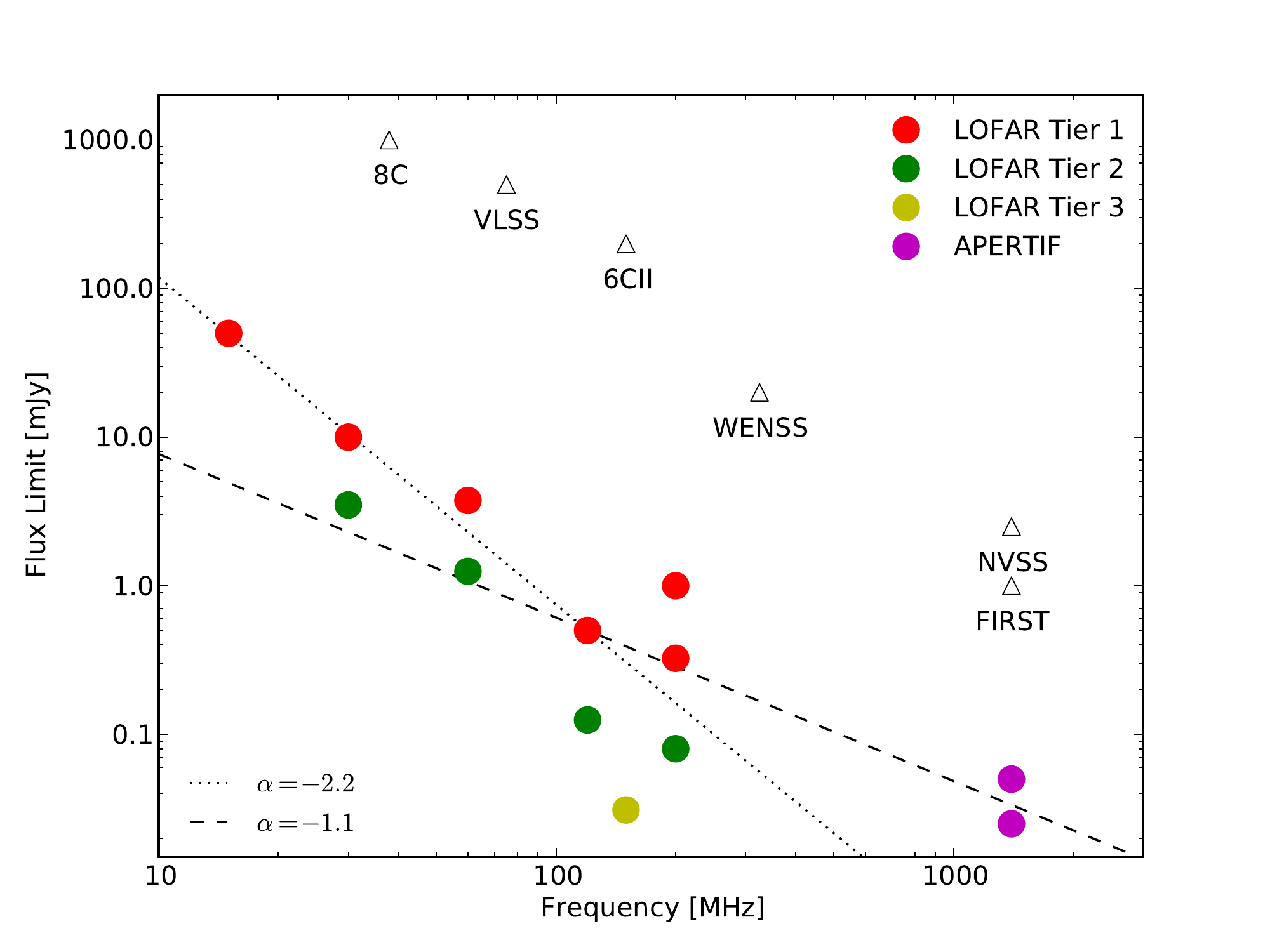}
\caption{\small Flux limits (5$\sigma$) of the proposed LOFAR surveys compared to other existing radio surveys. The triangles represent existing large area radio surveys. The lines represent different power-laws ($S\sim \nu^{\alpha}$, with $\alpha=-1.6$ and $-0.8$) to llustrate how, depending on the spectral indices of the sources, the LOFAR surveys will compare to other surveys.}
\label{fig:surveys}
\end{figure*}

\subsection{Surveying the low-frequency sky}
\label{sec:surveys}

An important goal that has driven the development of LOFAR since its inception is to explore the low-frequency radio sky through several dedicated surveys. The main science driving the design of these surveys will use the unique aspects of LOFAR to advance our understanding of the formation and evolution of galaxies, AGNs and galaxy clusters over cosmic time. Since LOFAR will open a new observational spectral window and is a radio ``synoptic'' telescope, the surveys will explore new parameter space and are well-suited for serendipitous discovery. Furthermore, a carefully designed and easily accessible LOFAR data archive will provide the maximum scientific benefit to the broader astronomical community.

Due to LOFAR's low operating frequencies, and the resultant large beam size on the sky, this radio telescope is an ideal survey facility. For example, at 50 MHz, each beam typically has a FoV of 7--8 deg. With theoretical LOFAR sensitivities and feasible observing times, such a field will typically contain 1 radio galaxy at $z>6$, 5 Abell clusters, 5 NGC galaxies, 5 lensed radio sources and several giant ($>1$ Mpc) radio galaxies. The aimed legacy value of the LOFAR surveys will be comparable to previous high-impact surveys (e.g. Palomar, IRAS, SDSS, GALEX, {\it Spitzer}, NVSS) and will also complement currently planned surveys in other wavebands (e.g. JEDAM, Euclid, Pan-STARRS, {\it Herschel}, {\it Planck}, VISTA, VST, JVLA, ASKAP, MeerKAT, ATA). The surveys described here will provide meter-wave data on up to $10^8$ galaxies and $10^4$ clusters out to $z \sim$ 8 and will address a wide range of topics from current astrophysics. The LOFAR survey key Project has, from the outset, been driven by four key topics. The first three are directly related to the formation of massive black holes, galaxies and clusters. The fourth is the exploration of parameter space for serendipitous discovery. The four key topics are:

(1) {\it High-redshift radio galaxies} (HzRGs, $z > 2$) are unique laboratories for studying the formation and evolution of massive galaxies, rich clusters and massive black holes \citep[see review by][]{miley08}. Presently, the most distant HzRG has a redshift of $z = 5.1$ \citep{vanBreugel99}. However, due to its low operating frequency, LOFAR will detect about 100 radio galaxies at $z>6$, enabling robust studies of massive galaxies and proto-clusters at formative epochs, and provide sufficient numbers of radio sources to probe structure in the neutral IGM near or even within the EoR through HI absorption studies. 

(2) {\it Clusters of galaxies} are the most massive gravitationally bound structures in the Universe, and drive galaxy evolution through mergers and interactions. However, approximately 40 clusters are known to also contain Mpc-sized, steep spectrum synchrotron radio sources that are not clearly associated with individual cluster galaxies. These are classified either as radio halos or radio relics, depending on their location, morphology and polarization properties \citep{Ferrari08}. The LOFAR surveys will allow detailed studies of about 15 local clusters in unprecedented detail, detect about 100 clusters at $z \ga 0.6$, and will contain thousands of diffuse cluster radio sources out to $z\sim1$. These surveys will enable the characteristics of the magnetic fields (strength, topology) in clusters to be determined and test models for the origin and amplification of these fields. Also, the origin and properties of the CR acceleration and evolution within clusters will be studied in detail.

(3) Determining the {\it cosmic star-formation history} of the Universe is a key goal of the LOFAR surveys, the deepest of which will detect radio emission from millions of regular star-forming galaxies at the epoch when the bulk of galaxy formation occurred. The combination of LOFAR and infra-red surveys will yield radio-IR photometric redshifts, enabling studies of the volume-averaged star formation rate as a function of epoch, galaxy type and environment. These studies will cover a sky area large enough to sample diverse environments (from voids to rich proto-clusters) and over a wide range of cosmic epochs.

(4) One of the most exciting aspects of LOFAR is the potential of {\it exploring new parameter space for serendipitous discovery}. The uncharted parameter space with the highest probability of serendipitous discovery is at frequencies $< 30$ MHz, where the radiation mechanisms being probed are not observable at higher radio frequencies, such as coherent plasma emission.

These four key topics drive the areas, depths and frequency coverage of the LOFAR surveys. In addition to the key topics (1--4), the LOFAR surveys will provide a wealth of unique data for the following additional science topics; (5) detailed studies of AGN and AGN physics,  (6) AGN evolution and black hole accretion history studies, (7) observations of nearby radio galaxies, (8) strong gravitational lensing, (9) studies of the cosmological parameters and large-scale structure, and (10) observations of Galactic radio sources. Furthermore,  to maximise the usefulness of the survey data for the Magnetism key project, the LOFAR survey data will be taken with sufficient bandwidth so that the technique of rotation measure synthesis can be applied. In collaboration with members of the Transient key project, the survey observations will be  taken in several passes, to facilitate searches for variable sources on various timescales.

To achieve these science goals, a three-tier approach to the LOFAR surveys has been adopted, using five different frequency setups. For each part of the surveys a minimum total bandwidth of 24~MHz will be used to improve the {\it uv}-coverage through multi-frequency-synthesis, as well as offering a significant benefit for polarization studies.

\begin{enumerate}

\item{\bf Tier 1: The ``large area'' ${\bf {2\pi}}$ steradian surveys}: These shallow wide area surveys will be carried out at 15--40, 40--65 and 120--180 MHz, and will reach an rms of 2, 1 and 0.07 mJy, respectively. It is expected that up to $3\times10^{7}$ radio sources will be detected by these three surveys, including, $\sim 100$ cluster halos at $z>0.6$ and $\sim$ 100 radio galaxies at $z>6$ \citep[cf.][]{Ensslin02,Cassano10a, Cassano10b}. The sensitivity and multi-frequency nature of the wide-area surveys will allow the low-frequency spectral shape of distant galaxy candidate sources with at least $\alpha = -2.0$ to be measured.

\item{\bf Tier 2: The ``deep'' surveys}: These surveys will be carried out at the same frequencies as the shallow all-sky surveys, but will be substantially deeper and over a smaller sky area. The HBA part will cover around 550 square degrees to provide a representative volume of the Universe. To maximise the additional science, the pointings will be centered on 25 well-studied extragalactic-fields that already have excellent degree-scale multi-wavelength data, 15 fields centered on clusters or super-clusters and 15 fields centered on nearby galaxies. The HBA survey will reach an rms of 15 $\mu$Jy at 150 MHz, which will be sensitive enough to detect galaxies with a star formation rate SFR $>10$ and $> 100$~M$_\odot$yr$^{-1}$  (5$\sigma$) out to $z = 0.5$ and 2.5, respectively \citep{Carilli99}. The LBA part of the deep survey will be over 1000--1500 square degrees and will reach a depth of 0.3--1.0 mJy. The pointings will be centered on 6 of the best-studied extragalactic fields and 9 of the most important nearby galaxies and clusters.

\item{\bf Tier 3: The ``ultra-deep'' survey}: Finally, there will be 5 fields which will be observed with the HBA covering a sky-area of 83 square degrees. This part of the survey will be ultra-deep, reaching an rms of $7~\mu$Jy at 150~MHz. This sensitivity will be sufficient to detect 50 proto-clusters at $z>2$, and detect galaxies with a SFR of 10 and 100 M$_\odot$yr$^{-1}$ at $z\sim 1.5$ and $\sim5$, respectively. These sensitivities are  similar to that needed for the EoR Project.

\end{enumerate}

The LOFAR surveys will not only be unique due to their low frequencies, but will also reach 2--3 orders of magnitude deeper in sensitivity than existing large-sky radio surveys, as is illustrated in Fig.\,\ref{fig:surveys}. They will permit a wide range of science goals to be attained and provide a legacy value data set for studies of the low frequency radio sky.

\subsection{The transient radio sky}
\label{sec:tkp}

LOFAR's ability to image very wide fields with good sensitivity, and to eventually do so in nearly real time (see Sect.\,\ref{sec:transients}), opens up a very large discovery space in time-domain astronomy. The known and suspected transients already span a very large range of properties. On the shorter timescales, coherent emitters are the only ones reaching detectable fluxes: Jupiter's radio outbursts reach fluxes of thousands of kJy, with substructure down to below milliseconds, rich polarization variations and narrow structures in frequency \citep{Zarka04}. Giant pulses of regular radio pulsars can reach upwards of 100\,Jy over microseconds, and millisecond single pulses of RRATS are at a wide range of flux levels up to 1\,Jy. Stellar radio flares have similarly rich structures in polarization, time and frequency as Jupiter, but at lower fluxes and with durations from minutes to hours. Elusive Jupiter-like signals from exoplanets, still to be discovered, are expected to be much weaker \citep[][and references therein]{Zarka11}.

At longer timescales, a wide variety of jet sources produce incoherent synchrotron emission with a large range of variability timescales: Galactic microquasars have outbursts that may last from days to months, but with rise times and substructures that can be very much shorter, at flux levels from hundreds of Jy down to the mJy level. AGN flares typically have very much longer time scales due to the scaling of black-hole phenomena with mass. On the patient end of the range, the variability of radio supernovae and gamma-ray bursts (GRBs) at LOFAR frequencies is measured in years to decades, with peak fluxes below a mJy in many cases; here the challenge changes from rapid response and high-volume fast data processing to careful analysis of deep images, and good use of supporting data from other instruments to tell the different types of slow radio transient apart.

Besides its potential as a discovery tool, the fully electronic operation of LOFAR makes it an excellent followup response machine for rapidly variable phenomena (see Sect.\,\ref{sec:trigger}). LOFAR's electronic repointing capability enables it to start a completely new observation (new settings of observing mode and pointing direction) in well under a minute, and to do so fully automatically upon receipt of external triggers from any telescope using, e.g., VOEvent protocols \citep{VOEvent11}. It can thus play a prominent role in the emerging network of wide-field sky monitors at many wavelengths, as well as in the multi-messenger world of gravity-wave and particle telescopes.

That there is still much to discover in the transient radio sky may perhaps be best illustrated by some recent examples of radio transient discoveries that have already greatly expanded the range of known
phenomena. With some of these discoveries has come the realization that many previously known extreme objects also emit radio flares --such as those associated with giant flares of Soft Gamma Repeaters. In other cases, radio transient searches have produced serendipitous finds of completely new types of object. For example, one of these is the discovery of so-called RRATs \citep{mll+06}, apparently pulsars that only emit a pulse of radio emission once per very many rotation periods. Another example is the discovery of an enigmatic radio source close to a supernova remnant near the Galactic Center \citep{2005Natur.434...50H}. This source emits 10 minute bursts in the radio with a very precisely constant 77 minute period (Spreeuw et
al.\ 2009). It is only detected in about 10\% of the previous attempts to observe it; however, emphasizing the importance of long-term monitoring of such objects. Most strange perhaps was the discovery of a millisecond dispersed radio burst from near the direction of the Small Magellanic Cloud \citep{lbm+07}, which --if astrophysical in origin at all-- certainly implies most strange and extreme compact-object astrophysics. Strange transients discovered at other wavelengths \citep[e.g., the puzzling Swift J1955 --][]{2008Natur.455..506C} will often require radio observations to help elucidate their nature.

One of the dominant problems in making progress in understanding what these strange objects are, and by what mechanism they radiate, is the great sparsity of observations we have of each one, and the fact that
only one or a few sources of each class are known. LOFAR's Transient capability will do much to remedy both of these: its very wide FoV combined with good sensitivity will make it likely that many representatives of these classes of object will be found in systematic transient surveys. The trigger and response capability will reliably provide fast (within seconds to minutes) radio data on sources newly discovered by other telescopes.

Due to the effects of dispersion, the signal from a fast transient in the range of 0.1--10\,s will be spread over a large bandwidth in the LOFAR frequency range making it more difficult to detect. In such cases, fast-imaging modes will naturally give way to beam-formed modes as the method of choice for exploring the transient radio sky due to the increased sensivity albiet at the expense of angular resolution. Consequently our standard transient imaging pipelines will target 1\,s as the shortest timescale to detect and characterize. Since crude dedispersion is also possible on sets of narrow-band images at least when dispersion is still modest, the optimal boundary between the two techniques will have to be explored once a specific set of algorithms and computing platforms is in place.

\begin{figure*}[ht]
\centering
\includegraphics[width=\textwidth]{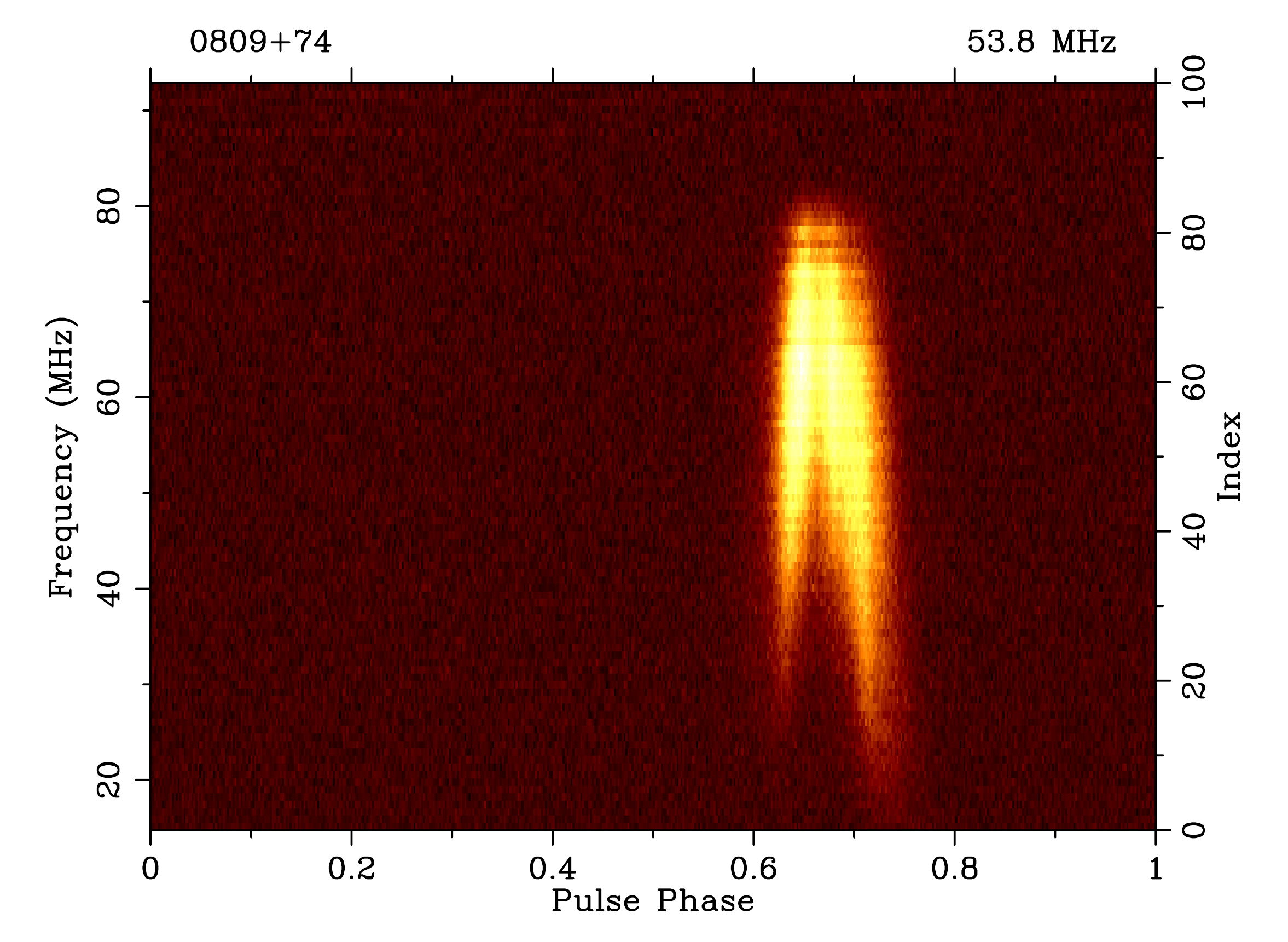}
\caption{\small 
A 1\,h LBA observation (L77295) of pulsar B0809+74 using a coherent addition of all 24 LOFAR core stations from 15--93\,MHz.  The data has been dedispersed and folded using a rotational ephemeris to produce a cumulative pulse profile as a function of frequency.  Given that the central observing frequency is 53.8\,MHz, the fractional bandwidth is 145\%.  This wide bandwidth is key to following the drastic evolution of the cumulative profile with frequency.  At the bottom of the LBA band there are two distinct pulse components that almost completely merge toward the top of the band.  Such data are being used to constrain properties of the emitting region in the pulsar magnetosphere \citep{hsh+12}.}
\label{fig:pulsar}
\end{figure*}

\subsection{Pulsar studies and surveys}
\label{sec:pulsars}

Although pulsars were discovered at 82\,MHz \citep{hbp+68}, the majority of pulsar studies have been at frequencies $> 300$\,MHz, and often at $\sim 1.4$\,GHz, because effects in the interstellar medium (ISM, e.g. dispersion and scattering), coupled with the Galactic synchrotron background and the steep power-law spectra of most pulsars, combine to make these frequencies well suited for the study of typical radio pulsars. Nonetheless, pulsar observations in the LOFAR frequency range of 10--240\,MHz are also very interesting for addressing some long-standing issues about the pulsar emission mechanism, and for studying the ISM. Furthermore, LOFAR's high sensitivity, flexible beam-formed observing modes, multi-beaming, and large FoV are well-suited for pulsar and fast transient searches (e.g. Fig.\,\ref{fig:tabs}).  Here we give a very brief overview of the expected studies of known pulsars and searches for new pulsars and fast transients.  More details and early LOFAR pulsar results can be found in \citet{sha+11}, \citet{hsh+12}, \citet{hsa+10}, \citet{ls10}, and \citet{Hermsen13}.

LOFAR will study the pulsar radio emission mechanism by providing wide-bandwidth, low-frequency spectra at high time resolution.  It is believed that the power-law spectra of most pulsars turn over somewhere in the 10--240\,MHz frequency range, making LOFAR an ideal instrument to study this important aspect of the pulsar emission mechanism.  The roughly 4 octaves of frequency coverage provided by LOFAR allow very detailed studies of profile morphology as a function of observing frequency (see Fig.\,\ref{fig:pulsar}), e.g. the so-called ``radius-to-frequency mapping" phenomenon \citep{cor78,hsh+12}.  Most of the known radio pulsars in the northern hemisphere will be detectable with LOFAR (by summing many hundreds of individual pulses), and we expect to detect single pulses from one-third of the visible pulsars in the high-band and ten percent of visible pulsars in the low-band.  Though low radio frequencies are poorly suited to precision timing tests, LOFAR will allow the frequent monitoring of many pulsars to look for timing anomalies such as glitches \citep[e.g.][]{elsk11} and sudden profile and spin-down changes \citep{lhk+10,klo+06}. There is also the possibility that the radio emission from some neutron stars may {\it  only} be detectable at the lowest radio frequencies \citep[e.g. PSR B0943+10][]{dr94}.

Targeted surveys of, e.g., nearby galaxies, globular clusters, supernova remnants, and the large population of $\gamma$-ray sources recently found with {\it Fermi} are likely to find interesting new radio pulsars.  These sources have a small extent on the sky and can be observed using either one or a few tied-array beams simultaneously (for reference a tied-array beam made from the Superterp/entire LOFAR core has a FWHM of 30/5 arcminutes).  Efficient all-sky surveys have already begun and can be done either using hundreds of tied-array beams (which provides high sensitivity and excellent source location, but produces a large data rate) or with the incoherent sum of the station beams (lower raw sensitivity and poorer source localization, but the single beam FoV is 5.5\,deg across and the data rate is low). These surveys will also search for generic fast transients \citep[e.g.,][]{bbj+11}, and aim to eventually trigger on transient
bursts in real time in order to dump the TBBs for better source localization.

\begin{figure*}[ht]
\centering
\includegraphics[width=\textwidth]{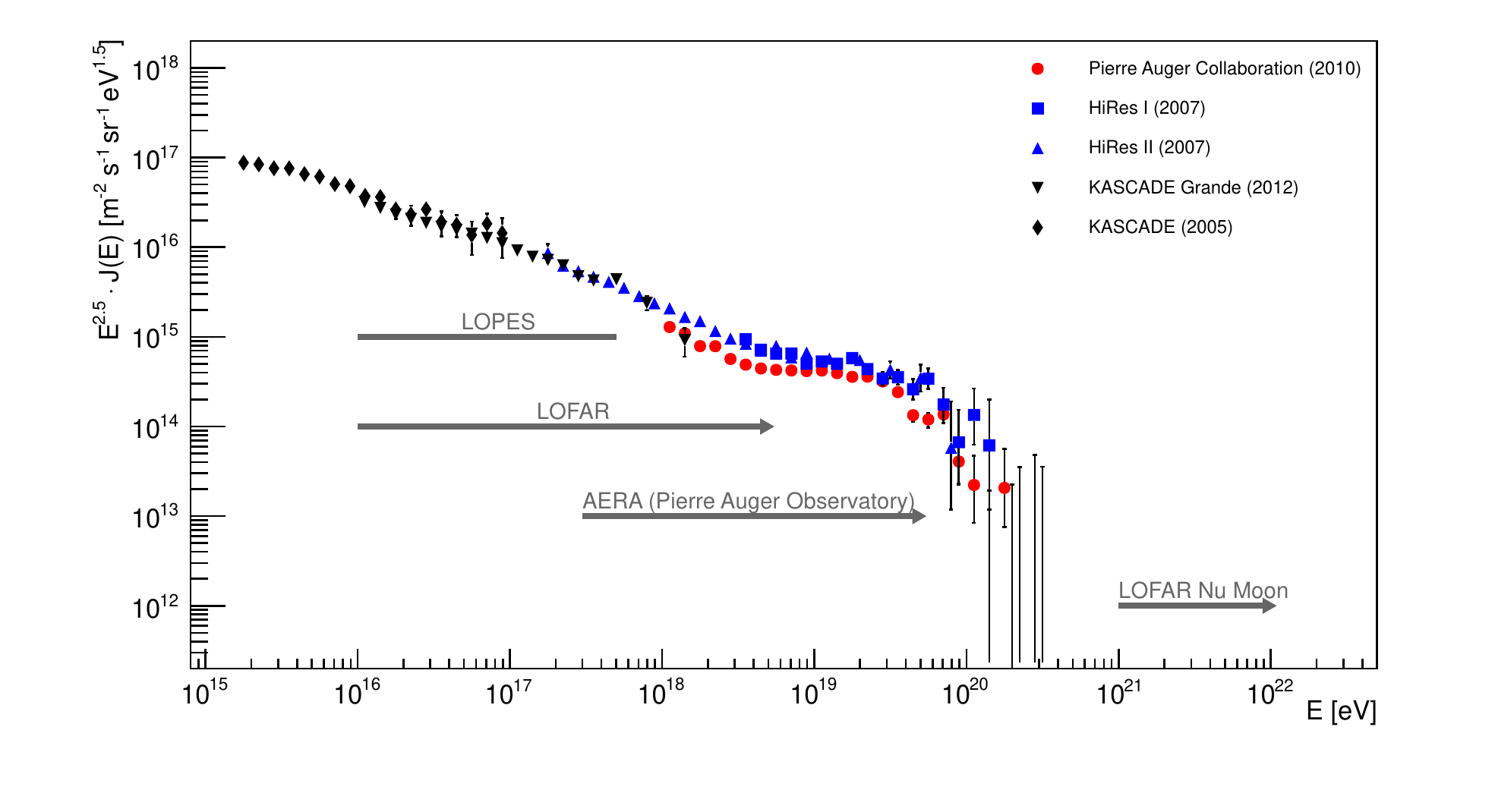}
\caption{\small High energy end of the spectrum of the CR flux as measured by a number of current experiments. The flux has been multiplied by a factor of $E^{2.5}$ to better show features in the spectrum, which are related to acceleration and propagation mechanisms. The gray bars indicate the energy range in which LOFAR will be sensitive to CRs. Furthermore, the energy ranges of other experiments detecting radio emission of CRs are shown. Among those are the Lofar Prototype Station \citep[LOPES;][]{Falcke05, LOPES2012} and the Auger Engineering Radio Array \citep[AERA;][]{Kelley2011}.}
\label{fig:spectrum}
\end{figure*}

\subsection{Astroparticle physics}
\label{sec:cr}

CR atomic nuclei and electrons have been detected with various methods on Earth, and indirectly in our Galaxy as well as other galaxies from their electromagnetic signature in gamma rays down to the radio domain, e.g. in supernova remnants (SNRs), radio pulsars, protostars, planetary magnetospheres, X-ray binaries, jets in radio galaxies and quasars, active galactic nuclei (AGN), and GRBs. 

Over a wide range of energies (E) the primary CR flux follows a simple power law $dN/dE \propto E^{-\gamma}$, as shown in Fig.\,\ref{fig:spectrum}. At 10$^{11}$ eV about one particle per second per square meter hits the Earth on average. This number changes to approximately one particle yr$^{-1}$ m$^{-2}$ at $5 \times 10^{15}$ eV, and above 10$^{19}$ eV only about one particle per century per square kilometer hits the Earth. These low fluxes require experiments with large effective areas in order to collect sufficient statistics. 

At an energy of $E \sim 5 \times 10^{15}$ eV the spectrum shows a turn-over where the power law index $\gamma$ changes from $\gamma \approx 2.7$ to $\gamma \approx 3.1$. This feature is called the knee of the CR spectrum. Up to the knee in the spectrum the composition of the primary CRs is dominated by protons, but at higher energies the composition still needs clarification \citep{Blumer2009, Kampert2012}. The question about the composition of these ultra-high energy CRs will be crucial for the understanding of acceleration and propagation mechanisms. At the highest energies above 10$^{19}$ eV there is a flattening of the spectrum, the so-called ankle which could be caused by the Greisen-Zatsepin-Kuz’min (GZK) effect \citep{Greisen1966, Zatsepin1966}. Ultra High Energy protons above $\sim 5 \times 10^{19}$ eV loose their energy quickly by producing pions in collisions with photons from the CMB. This effect accumulates protons that had been accelerated to higher energies at energies below the reaction threshold, and implies that any observed CR of this energy finds its origin in the near Universe ($<$\,50 Mpc).

Because of the smoothness of the spectrum, much effort has gone into identifying a universal acceleration process. It is believed that diffusive shock acceleration –- a first-order Fermi-type acceleration process –- is this universal mechanism. It operates in strong collisionless shocks such as occur in a multitude of explosive objects in the Universe and produces a differential power law spectrum in energy with power law index of -2, close to and somewhat flatter than is observed, for any shock as long as it is both strong and non-relativistic. Up to the knee diffusive shock acceleration in SNRs is believed to be the main acceleration process. Above the knee, possible candidate sources of high energy CRs are shocks in radio lobes of powerful radio galaxies, intergalactic shocks created during the epoch of galaxy formation, magnetars, so-called hyper-novae, and GRBs. So far no conclusive evidence has been found that clearly identifies the source of the highest energy CRs. 

The identification of the sources of CRs is not only hindered by the low statistics of events measured at Earth, but also by the lack of knowledge of the Galactic and intergalactic magnetic fields. CRs will propagate a considerable amount of time through the Galaxy and intergalactic space before finally reaching Earth. Magnetic fields of different strengths and degrees of turbulence will obscure their original direction. As this effect is energy dependent, there is hope that the most energetic particles will still indicate their sources and their paths will then provide information about the magnetic field structure. But also in reverse: an improved knowledge about the magnetic fields in the Universe will help to solve the open questions about the origin of the CRs. 

When a CR hits the nucleus of an atom in the terrestrial atmosphere it undergoes a nuclear reaction and produces several secondary particles. These secondary particles again react with atmospheric nuclei and produce more secondary particles. Together these particles form an extensive air shower. If the energy of the primary particle was high enough this air shower can be measured at ground level. The highest energies observed fall outside the domain which is currently being studied with Earth bound particle accelerators. Thus air showers not only are messengers from the distant Universe but also form a laboratory to study new particle physics \citep{ThePierreAugerCollaboration2012}.

LOFAR will observe CRs above 10$^{16}$ eV up to 10$^{19}$ eV from their bright radio flashes in the terrestrial atmosphere. These flashes are caused by the deflection of particles in the Earth magnetic field and charged processes within the development of the air shower. The theories explaining this phenomenon have developed rapidly during the last ten years, e.g. \citet{2005A&A...430..779H}, \citet{2011APh....34..438L}, \citet{2008APh....29...94S} or \citet{2012APh....37....5W}. They indicate that the radio emission will also be sensitive to the height of the shower development and thereby able to identify the particle type of the primary CR. However, to really confirm the predictions, data of higher quality and abundance is needed. The high numbers of antennas at LOFAR are essential to measure every shower in highest possible detail. 

In addition to conclusively explaining the exact mechanisms of these radio emissions, the observations with LOFAR will aim to answer a number of fundamental questions in astroparticle physics such as the composition and origin of these particles. The measurements of the radio emission of air showers will be complemented with LOFAR observations of the Moon to detect (or put upper limits to) Cherenkov flashes from CRs from $10^{21}$ to $10^{22}$ eV in the lunar regolith (LOFAR Nu Moon). For a more detailed description refer to \citet{2009PhRvL.103s1301S}, \citet{Buitink2010} and  \citet{Mevius2012}.
Together with LOFAR observations of supernova remnants and other explosive events in the Universe, the study of magnetic fields, and, ultimately, with radio observations done at the Pierre Auger Observatory in Argentina at (well- calibrated) energies above a few 10$^{18}$ eV one aims at a full picture of CR physics.

\begin{figure*}
\centering
\includegraphics[width=3.5in]{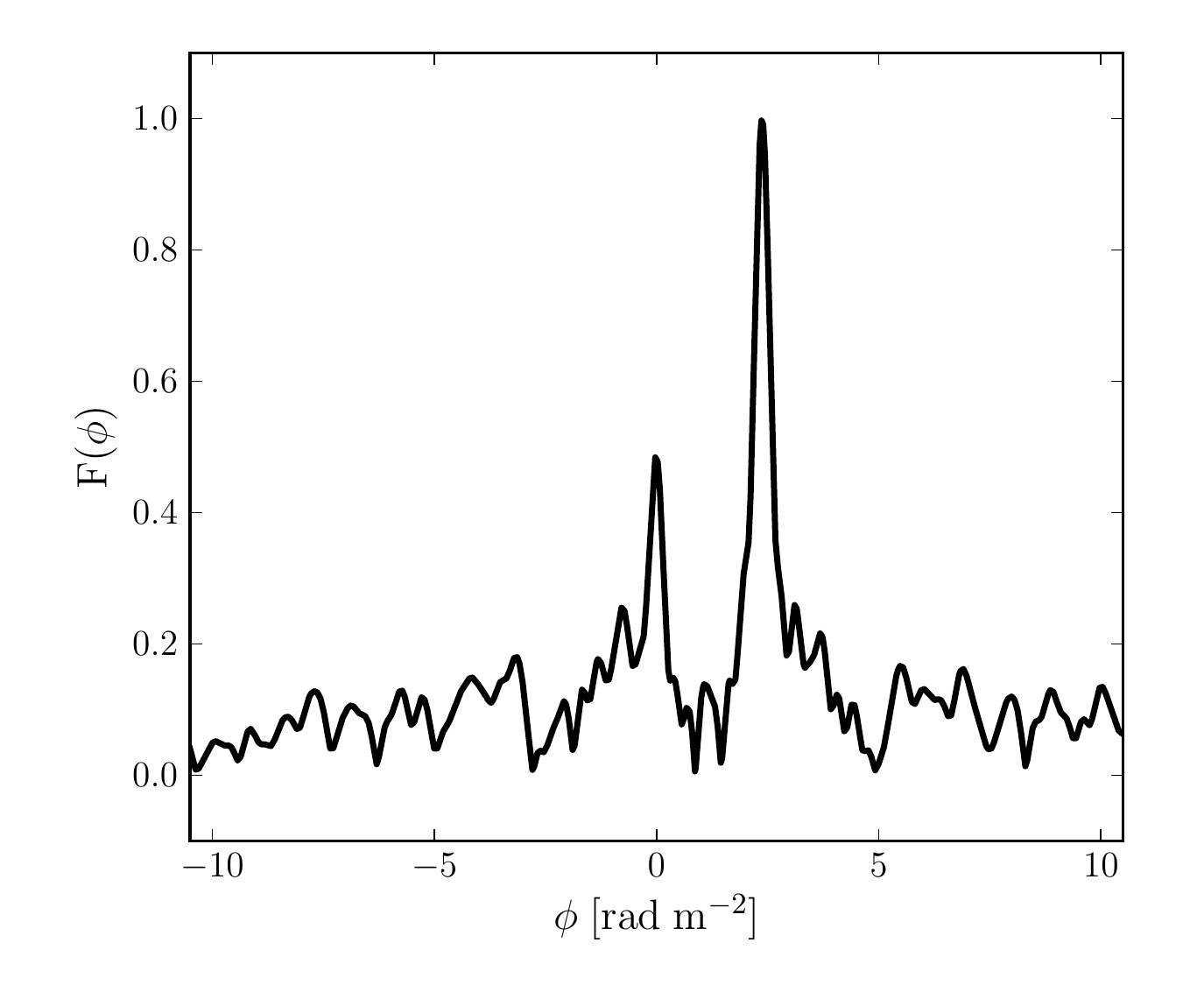}
\includegraphics[width=3.5in]{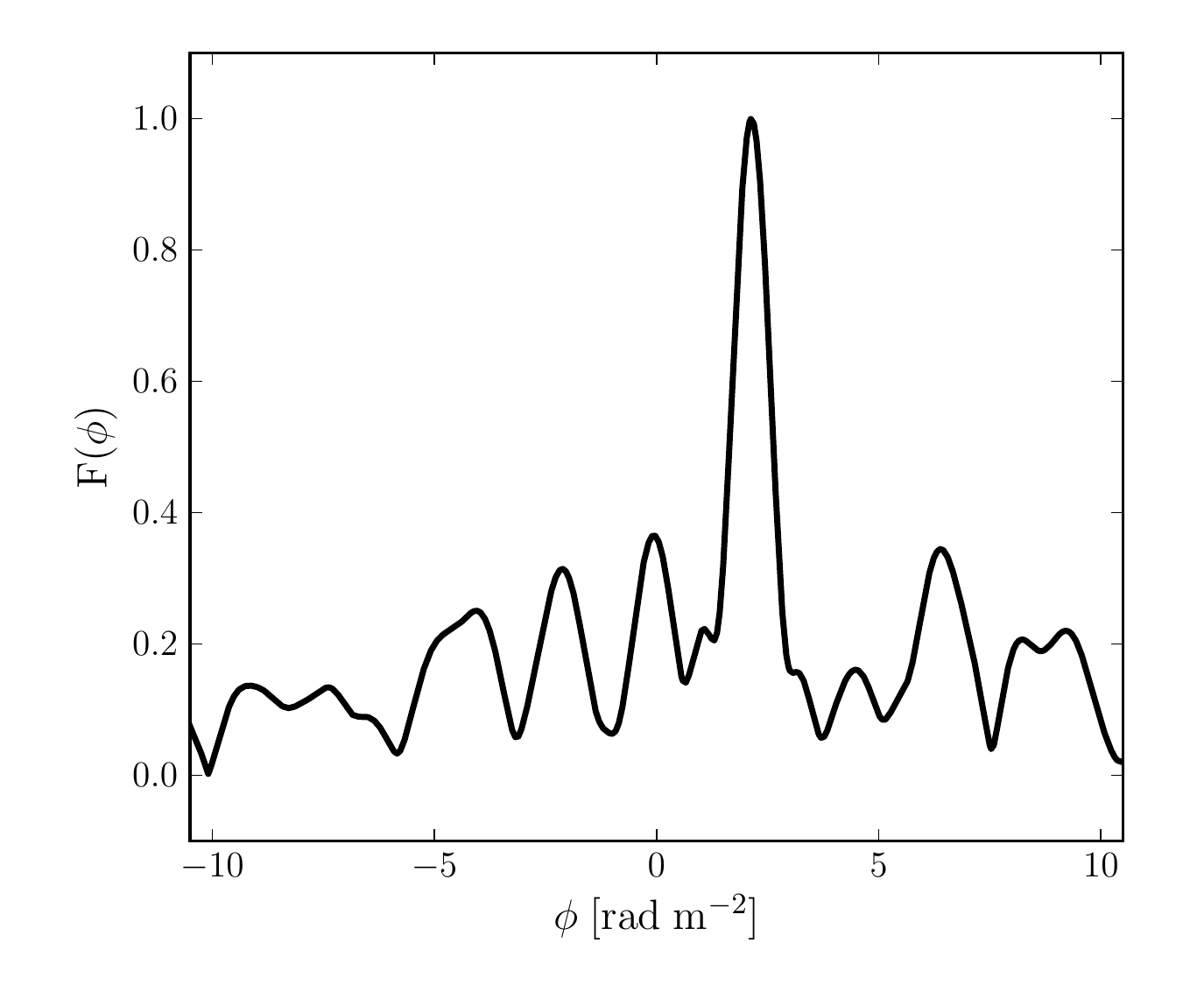}
\caption{\small Faraday dispersion functions (FDFs, or Faraday spectra) obtained from LOFAR observations of the polarised pulsar B0950+08 (normalized absolute value). {\bf Left:} 27-minute LBA tied-array beam-formed observation using coherent addition of the six LOFAR Superterp stations with a center frequency of 56 MHz, 10 MHz bandwidth, MJD 55901, and using data from obsID L36787. {\bf Right:} 10-minute HBA tied-array beam-formed observation using coherent addition of 20 LOFAR core stations with a center frequency of 150 MHz, 90 MHz bandwidth, MJD 56260, and using data from obsID L78234. The narrow FWHM of the functions allows the peaks associated with the pulsar ($2.373 \pm 0.011$ and $2.136 \pm 0.061$ rad m$^{-2}$, respectively) and instrumental response ($\sim 0$ rad m$^{-2}$) to be individually resolved, despite the very low absolute rotation measure (RM). These RMs were corrected for ionospheric Faraday rotation ($0.899 \pm 0.042$ and $0.665 \pm 0.059$ rad m$^{-2}$, respectively) using the {\tt ionFR} code which employs International GNSS Service vertical total electron content (VTEC) maps \citep{IGSVTEC2009} and data from the International Geomagnetic Reference Field \citep{IGRF2010} \citep[see][]{Sotomayor-Beltran13}. The resulting RM of the ISM toward B0950+08 was determined to be $1.47 \pm 0.04$ and $1.47 \pm 0.08$ rad m$^{-2}$, from LBA and HBA observations respectively. These results are significantly more precise and in good agreement with the value of $1.35 \pm 0.15$ rad m$^{-2}$ previously measured \citep{Taylor93}.}
\label{fig:rmsf}
\end{figure*}

\subsection{Magnetic fields in the universe}
\label{sec:magnetism}

Understanding the Universe is impossible without understanding magnetic fields. Magnetic fields are present in almost every place in the Universe but in spite of their importance the evolution, structure and origin of magnetic fields all remain open fundamental problems. Most of our knowledge of astrophysical magnetic fields has come from radio-frequency observations of synchrotron radiation from relativistic cosmic-ray leptons (mostly electrons). These observations trace the total field strength (from the synchrotron intensity), the orientation and degree of ordering of fields in the plane of the sky (from the polarized component of the radiation), and the component of ordered fields along the line of sight (via Faraday rotation).

LOFAR's exceptionally wide bandwidth at low frequencies is extremely useful for the study of magnetic fields, for several complementary reasons: (i) it provides excellent leverage on the spectral characteristics of the synchrotron radiation, which allows study of the synchrotron losses of the emitting electrons; (ii) low energy synchrotron-emitting electrons are detectable {\it only} at low frequencies, so LOFAR can uniquely trace magnetic fields far away from CR acceleration sites; and (iii) studies of Faraday rotation have the best precision when the range of measured wavelength is wide -- LOFAR will thus trace weak magnetic fields \citep{Beck10}. A very powerful tool for detection and characterization of polarized emission with LOFAR will be the Rotation Measure Synthesis \citep[RM Synthesis;][]{Brentjens2005} technique. This provides the Faraday dispersion function, or, in short, the Faraday spectrum (Fig.\,\ref{fig:rmsf}) that gives information about the structure of the magneto-ionic medium along the line of sight.

The magnetism key science project (MKSP) aims to investigate cosmic magnetic fields in a variety of astrophysical sources, including an initial target list of galaxies, followed by deep observations of galaxies and galaxy groups. These deep fields will also serve as targets to investigate magnetic fields in the Milky Way foreground. The structure of small-scale magnetic fields will be studied the lobes of giant radio galaxies. Polarized synchrotron emission and rotation measures from pulsars and polarized jets from young stars will be observed.

At high latitudes above the Milky Way plane, LOFAR will be uniquely sensitive to synchrotron emission of low-energy electrons in the Galactic halo, which will allow investigations of the propagation and evolution of matter and energy far from the Galactic disk. Weak magnetic fields or small density fluctuations of thermal electrons will become visible through Faraday rotation, leading to a better understanding of the turbulent ISM, and allowing a three-dimensional model of the gas and magnetic fields in the solar neighborhood to be constructed.

With present-day radio telescopes, GHz synchrotron emission from electrons in a $5~\mu$G magnetic field can be detected in external galaxies, or a $1~\mu$G field in clusters. The minimum detectable magnetic field strength varies with $\nu^{-\alpha/(3-\alpha)}$ (where $\alpha$ is the synchrotron spectral index, $\alpha\simeq-0.8$) so that all else being equal, observing at a 10$\times$ lower frequency permits the detection of $\simeq2\times$ weaker magnetic fields. The observable extent of radio emitters is limited by the propagation speed of CRs away from their sources and by the extent of the magnetic fields. At high radio frequencies (1--10\,GHz) the radio emission from disks of star-forming galaxies is restricted to about 1~kpc from the sources of CRs. Low-frequency radio emission traces low-energy CRs which suffer less from energy losses and hence can propagate further away from their sources into regions with weak magnetic fields. The lifetime of CRs due to synchrotron losses increases with decreasing frequency and decreasing total field strength. In a $5~\mu$G field electrons emitting in the LOFAR bands have a lifetime of $2-5\times10^{8}$~yr and can travel several tens of kpc in a magnetic field of about $3~\mu$G.

Many of the traditional depolarization effects expected from the technical limitations of low frequency radio observing are mitigated by the long baselines and high spectral resolution of the LOFAR instrument. As at higher frequencies, beam depolarization will limit polarization studies of sources with rapid image plane field reversals for any instrument of finite resolution. More important at these frequencies, however, is the internal depolarization of radio sources caused by field fluctuations along the line of sight. The effect of this internal Faraday dispersion has been discussed in the case of specific source morphologies by a number of authors \citep[e.g.,][]{Cioffi1980, Laing1981} as well as for analytic geometries and random fluctuations \citep[e.g.,][]{Tribble1991, Sokoloff1998}. Although these effects are expected to become increasingly important at longer wavelengths a directed use of Faraday Rotation Measure Synthesis \citep{Brentjens2005} will mitigate them in many circumstances. Indeed this technique is now becoming standard not only for recovering widespread polarized emission in a variety of environments but also for characterizing the line of sight medium itself \citep[e.g.,][]{deBruyn2005, Heald2009}.

LOFAR's sensitivity to regions of low density and weak field strengths will allow us to measure the magnetic structure in the halos of galaxy clusters, in the intergalactic medium of galaxy groups, in wider halos and in outer disks of spiral galaxies. It is here that star formation activity is low, and processes additional to dynamo action, such as gas outflows from the inner disk, the magneto-rotational instability, gravitational interaction and ram pressure by the intergalactic medium are imprinted on this magnetic structure. The low frequencies provided by LOFAR will be highly sensitive to such steep-spectrum shock-like features, resembling relics in clusters, and knowledge of their 3-D magnetic field structures from RM Synthesis will allow us vastly improved understanding of intergalactic gas dynamics.

Grids of RM measurements of polarized background sources are powerful tools to study magnetic field patterns in foreground galaxies and clusters of galaxies \citep{2008A&A...480...45S}. Greater leverage on Faraday RM values is expected at lower frequencies; thus, LOFAR will observe tenuous ionized gas and/or very weak magnetic fields. It is even possible that LOFAR will directly detect magnetic fields in the filamentary intergalactic medium of the cosmic web. Unlike evolved clusters of galaxies, where highly efficient turbulent amplification is expected to have lead to saturation of the fields, the filamentary structures of the cosmic web are anticipated to be far more sensitive to the original seed mechanism responsible for cosmic magnetism. Importantly, detection of this field, or placing stringent upper limits on it, will provide powerful observational constraints on the origin of cosmic magnetism.

LOFAR pulsar searches will benefit from both high sensitivity and an increasing pulsar brightness at low frequencies. This is expected to result in the discovery of a new population of dim, nearby and high-latitude pulsars too weak to be found at higher frequencies: roughly 1,000 pulsar discoveries are expected from LOFAR. Polarization observations of these pulsars will approximately double the current sample of Faraday rotation measures (RMs) (see Fig.\,\ref{fig:rmsf}). This will provide the strength and direction of the regular magnetic field in previously unexplored directions and locations in the Galaxy; e.g. very little is known about the magnetic field properties of the Milky Way beyond a few hundred parsecs from the Galactic plane. RMs of high-latitude pulsars and extragalactic sources are crucial for determining fundamental properties such as the scale height and geometry of the magnetic field in the thick disk and halo, as well as providing the exciting prospect of discovering magnetic fields in globular clusters.

\subsection{Solar physics and space weather}
\label{sec:solar}

The Sun is an active star which exerts a strong influence on the space environment around Earth. This {\it Space Weather} can strongly affect global communication technology on which we increasingly rely. The Sun is an intense and variable source of radio emission: The strong thermal radiation of the quiet Sun is interspersed with intense radio bursts associated with solar activity such as flares and coronal mass ejections (CMEs).  By combining the imaging and beam-forming observational modes, LOFAR can serve as a highly effective solar monitoring and imaging system. Thus, the study of the Sun by LOFAR is of great interest in solar physics and Space Weather.
 
The nonthermal radio radiation of the Sun is generated by energetic electrons produced by flares and/or CMEs. These energetic electrons excite high-frequency plasma waves (Langmuir and/or upper-hybrid waves) leading to the emission of radio waves by nonlinear plasma processes near the local electron plasma frequency and/or its harmonics \citep{Melrose1985}. Since the plasma frequency only depends on the electron number density, and due to the gravitational stratification of the corona, each frequency corresponds to a certain height level in the corona \citep{1999A&A...348..614M}. Thus, LOFAR enables the study of plasma processes associated with energetic electrons at different heights in the corona.
 
\begin{figure*}
\centering
\includegraphics[width=\textwidth]{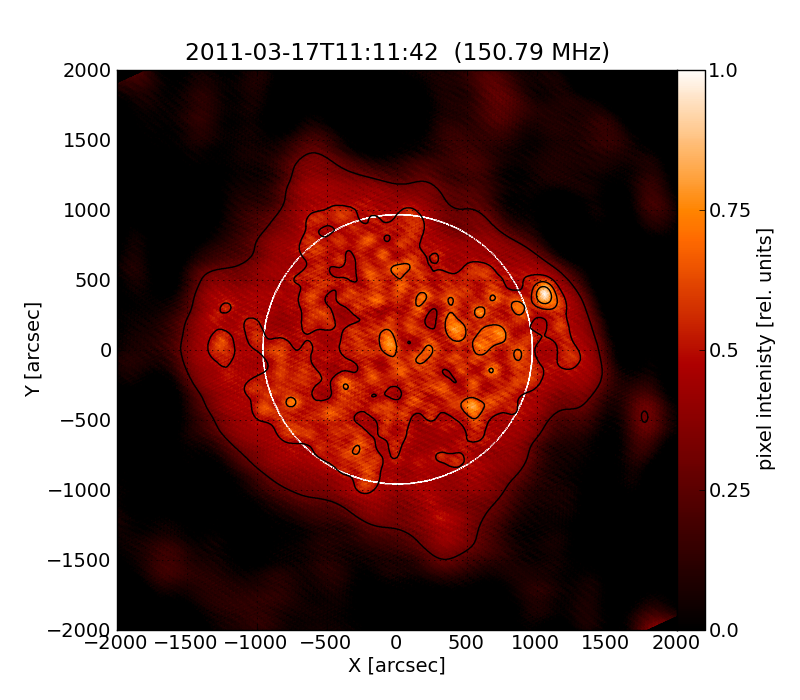}
\caption{\small LOFAR image of a Type I solar radio burst, observed at 150\,MHz on 17 March 2011. The white circle indicates the edge of the solar photosphere. Further study has revealed that this radio burst was located above an active region on the Sun (Mann et al., in prep.).}
\label{fig:solar}
\end{figure*}

In March 2011, LOFAR observed its first solar radio burst, a so-called type I burst \citep{McLean1985}, seen at 150 MHz on the West limb of the Sun (see Fig.\,\ref{fig:solar}). A detailed study of the radio morphology indicates that the burst is located above an active region. During the flare, as a signature of magnetic reconnection, a hot plasma jet is injected into the corona leading to the acceleration of electrons and subsequent radio emission \citep[][Mann et al., in prep.]{2007A&A...461.1127M}.
 
Solar type III radio bursts were observed by LOFAR in October 2011, in both radio images and dynamic spectra. They manifest as a rapid drift from high to low frequencies in the dynamic radio spectrum \citep{McLean1985, 2011pre7.conf..373B}. In this case, the type III burst source is located at the east limb above an active region (Mann et al., in prep.). Type III radio bursts are signatures of electron beams propagating along open magnetic field lines through the solar corona and sometimes into interplanetary space. They arise from electrons accelerated within a solar flare being injected into open magnetic field geometries.
 
In the corona, a shock wave is produced by a flare and/or driven by a CME. Signatures of such shock waves appear as type II radio bursts in solar dynamic radio spectra \citep{Mann95, Aurass97}. These shock waves are able to accelerate electrons up to supra-thermal velocities, resulting in type II radio bursts. Such bursts have also been detected in LOFAR observations.

\begin{figure*}
\centering
\includegraphics[width=\textwidth]{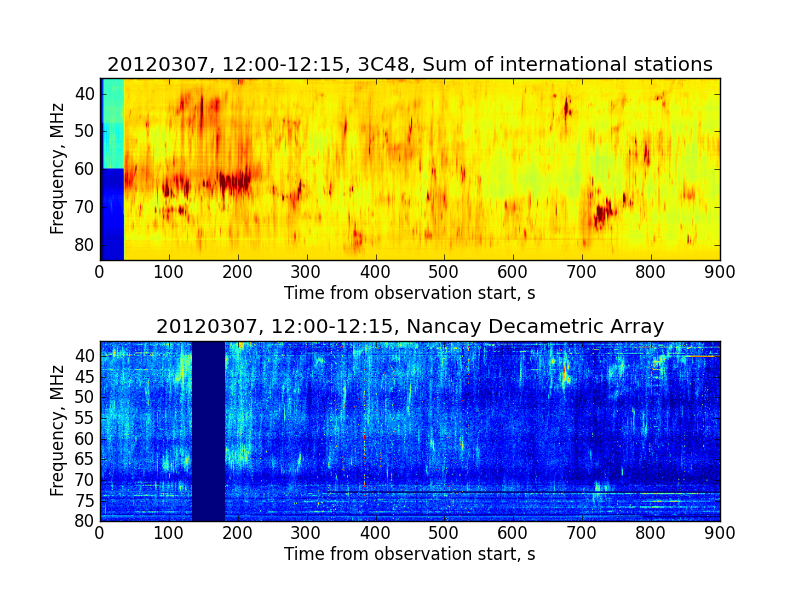}
\caption{\small 
Dynamic spectrum of LOFAR LBA data from an observation of 3C48 (J0137+331) at an elongation of 50 degrees from the Sun (top panel) taken on 07 March 2012.  These data have been averaged to a time resolution of ~1.24\,s and a frequency resolution of ~180\,kHz, matching the full-resolution data for the corresponding time period from the Nancay Decametric Array (lower panel). The Nancay Decametric Array is an array dedicated to solar observation (data courtesy A. Kerdraon, Meudon-Paris). The radio emission appears to be consistent with a Type II radio burst from solar flare activity.}
\label{fig:ips}
\end{figure*}

Only a few instruments currently observe the Sun at low radio frequencies. The radioheliograph at Nancay
\citep{Kerdraon97}, for example, is one of the few solar observatories currently operated at selected frequencies ranging from 150 to 432\,MHz and yields a typical image resolution of 2 -- 5 arcminutes at the lower end of its frequency band. In recent years, however, a number of new instruments well suited for low frequency solar observations have come online. Besides LOFAR itself, these new instruments include the recently commissioned LWA (10$-$88 MHz) \citep[see][and examples therein]{Kassim2010, Lazio2010, LWA2012b} and MWA (80$-$300 MHz) \citep[see][and examples therein]{Rightley2009, Oberoi2011, Bowman13, Tingay2013}, both of which provide powerful solar observing capabilities. Although not specficially intended for solar observations, the PAPER \citep[see][for descriptions of the PAPER instrument]{Parsons2010, Stefan2013, Pober2013} operates in the 100$-$200 MHz range and also provides similar potential capablities. In the Ukraine, the radio telescopes UTR2 \citep{2005KFNTS...5...57S} and URAN2 \citep{2005KFNTS...5...43B} also work at low frequencies. 

Like these other instruments, LOFAR expands the observable solar frequency range down to 10 MHz. Unlike these other facilities, however, it can also achieve angular resolutions of a few 10 arcseconds, scattering in the solar corona becoming the limiting factor for resolution rather than baseline length. The broad low-frequency coverage combined with high resolution imaging makes LOFAR a powerful tool for probing previously unexplored solar coronal structures.

The excellent uv-coverage available using the Superterp stations also enable direct snapshot imaging of radio emission from CMEs.  Further, it is possible to use a grid of many tied-array beams (as illustrated in Fig.\,\ref{fig:ta-beam}) to form ``maps'' of radio emission covering a broad area around the Sun. These maps will in practice have a lower spatial resolution than those obtained by direct imaging, but have the advantage of the high time and frequency resolution available using the beam-formed mode and enable the direct deduction via dynamic spectra of the types of radio burst formed within the CME.  
 
The solar wind, the expansion of the solar corona through interplanetary space, can be probed by observing the interplanetary scintillation (IPS) of compact radio sources \citep{Hewishetal:1964}. Observations of IPS provide the ability to systematically study the solar wind at nearly all heliographic latitudes over a wide range of distances from the Sun. Various analysis techniques are used to probe different aspects of solar wind structure:
\begin{itemize}
\item Cross-correlation of the signals from two antennas, taken at times of suitable geometrical alignment between the antennas, Sun and radio source, can be used to resolve multiple solar wind streams in the lines of sight between antennas and radio source, yielding solar wind speeds and flow direction \citep[e.g.][]{Breenetal:2006, Fallows13}.
\item Multiple IPS observations over several days or more can be combined to produce three-dimensional reconstructions of solar wind speed and density throughout the inner heliosphere \citep[e.g.][]{springerlink1, 2010ApJ...715L.104B}.
\item The combination of IPS and white light observations of the solar corona permit a far improved understanding of solar wind processes than would be possible with either technique alone \citep{springerlink2, Hardwick11}.
\item The high bandwidth capabilities of LOFAR enable the dynamic spectrum of IPS to be studied (see Fig.\,\ref{fig:ips}). The opportunity for such studies is available on few other instruments and may yield further information on solar wind micro-structure \citep{Fallows13}.
\end{itemize}

The low-frequency bands available with LOFAR are best suited to using observations of interplanetary scintillation to study the solar wind from the orbit of Mercury out to beyond Earth orbit. This region is of particular interest in space weather as this is where much of the evolution of solar wind and CME structure occurs - information on which is essential for the accurate timing of the impact of such structures on the space environment around Earth.  

Thus, with LOFAR, both the corona of the Sun and interplanetary space can be observed with unprecedented spatial and temporal resolution. This allows plasma processes in the corona and the solar wind to be studied in a manner that could not be achieved with other (e.g. optical) instruments. Both the imaging of the corona and the observation of IPS can contribute to the investigation of processes from initiation and launch of a CME to its subsequent development and propagation through interplanetary space, topics of great importance for understanding many aspects of Space Weather.

\section{Current and future developments}
\label{sec:future}

\subsection{Final construction}
\label{sec:rollout}

Construction of the LOFAR array has been underway for the past five years and began in 2006 with the placement of several test stations on the site of the array core near Exloo in the Netherlands. This core is located in an area rich in peat deposits that were extensively harvested between 1850 and 1950 leaving behind a landscape used primarily for starch production from potato farming. Starting in 2008, the core area was extensively reshaped and established as a nature reserve with dedicated locations for the LOFAR stations. 

Due to its agricultural history and the extensive landscaping required to establish the nature area, a large effort was required to stabilize the soil in the region.This work combined with the required $\pm 3$--6 cm tolerances on the flatness of the fields delayed the start of the large-scale civil engineering effort until the spring of 2009. Once begun however, progress has been rapid since with the deployment of 22 stations in 2009 and a further 11 stations in 2010. For the remaining 7 remote NL stations, additional effort was required to obtain the necessary planning permission and building permits. Nonetheless, construction for the majority of these remaining stations was completed in 2012. At the time of writing, only one final station is as yet unfinished and, subject to obtaining final building permits, the NL array should be fully complete in 2013.

The international LOFAR stations have been built in parallel with those in the Netherlands beginning with the construction of the LBA field of the first German station near Effelsberg in 2007. The Effelsberg station was augmented with HBA tiles in 2009, and additional stations in Germany, each under different ownership, were completed throughout 2009-2011 in Unterweilenbach, Tautenburg, Potsdam, and J\"{u}lich. A further station near Hamburg has recently been funded and construction is planned to start in early 2013. In 2010, a station was also built in Nan\c{c}ay, France, and in Chilbolton in the UK. Finally, a station near Onsala, Sweden was completed in 2011. Further expansion of the international array, in particular with a view to filling gaps in the $uv$-plane or extending the array, resulting in higher resolution, is currently under study.

\subsection{Functionality enhancements}
\label{sec:enhance}

One of the great strengths of the LOFAR system is its capacity for enhancement. It is of course common for astronomical facilities to increase their capabilities through continued software development. For LOFAR however, the system design is sufficiently flexible that scientific capacity can be added relatively straightforwardly at both the software and hardware levels. In the simplest case, this capacity increase can be achieved through the addition of more stations to the array resulting in improved $uv$ coverage, longer baselines, and increased sensitivity. Such extensions to the array would, however, also require the addition of significant additional compute capacity. Due to practical I/O limits set by the BG/P configuration, only 64 stations of the 72\footnote{Assuming all core HBA sub-stations are being correlated as independent stations.} total possible can be correlated any given time with the current configuration. Increasing this number would require substantial changes to the current LOFAR computing infrastructure (see \ref{sec:online}). We note that even adding smaller numbers of additional stations would require increasing the computing and storage capacity of the post-processing cluster in order to keep up with the increased data flow. 

Similarly, the capabilities of individual LOFAR stations can also be expanded. With minimal modifications, the data-stream from a given station can be replicated and processed independently of the standard LOFAR processing. A number of such ``stand-alone'' or single station enhancements are already in development. The first of these, called ARTEMIS, implements a real-time dedispersion search engine to detect pulsars using the data-streams from one or more LOFAR stations \citep[][and Karastergiou et al., in prep.]{Serylak12, Armour12}. A second, EU-funded project named AARTFAAC will expand upon LOFAR's ability to monitor radio transients by correlating the signals from all dipoles on the Superterp in real-time \citep{Prasad12}. Finally, a design for an expanded station concept has been proposed by the French LOFAR consortium. This design would add significant numbers of additional dipoles as well as computing capability to the current French station at Nan\c{c}ay resulting in a ``SuperStation'' optimized for beam-formed observations with high instantaneous sensitivity in the 10-80 MHz range \citep{Zarka12}. Although the design has yet to be finalized, the proposed SuperStation would provide a factor of $\sim20$ increase in effective area relative to a standard international LOFAR station. For comparison, the French SuperStation would deliver $\sim7$ times the effective area of the current LWA station \citep{LWA2012b}.

\section{Conclusions}
\label{sec:conclude}

In this paper, we have presented an overview and brief introduction to the LOFAR telescope. LOFAR represents a step-change in the evolution of radio astronomy technology. As one of the first of a new generation of radio instruments, LOFAR provides a number of unique capabilities for the astronomical community. These include among others remote configuration and operation, data processing that is both distributed and parallel, buffered retrospective all-sky imaging, dynamic real-time system response, and the ability to provide multiple simultaneous streams of data to a community whose scientific interests run the gamut from radio aurorae in the magnetospheres of distant planets to the origins of the Universe itself. Due to the tremendous data rates generated, LOFAR will also be one of the first radio observatories to feature automated processing pipelines to deliver fully calibrated scientific products to the community. Many of the technological solutions developed for LOFAR, in particular the calibration of phased-arrays as well as large-scale data transport and processing, will be highly relevant for future radio telescope projects such as the SKA.

\begin{acknowledgements}
The authors would like to thank the referee for the careful reading and many constructive comments that helped improve the paper. The LOFAR facilities in the Netherlands and other countries, under different ownership, are operated through the International LOFAR Telescope foundation (ILT) as an international observatory open to the global astronomical community under a joint scientific policy. In the Netherlands, LOFAR is funded through the BSIK program for interdisciplinary research and improvement of the knowledge infrastructure. Additional funding is provided through the European Regional Development Fund (EFRO) and the innovation program EZ/KOMPAS of the Collaboration of the Northern Provinces (SNN). ASTRON is part of the Netherlands Organization for Scientific Research (NWO). C. Ferrari and G. Macario acknowledge financial support by the {\it ``Agence Nationale de la Recherche''} through grant ANR-09-JCJC-0001-01.
\end{acknowledgements}

\bibliographystyle{aa}
\bibliography{refs}

\begin{appendix}
\onecolumn
\section{LOFAR station field center positions}
\label{sec:positions}

\longtab{1}{
\begin{longtable}{lrrr}
\caption{\label{tab:fields} LOFAR station field center positions} \\
\hline\hline
Station      &        ETRS-X     &     ETRS-Y     &       ETRS-Z      \\
             &          (m)      &       (m)      &         (m)       \\ 
\hline\hline
\endfirsthead
\caption{continued.} \\
\hline\hline
Station      &        ETRS-X     &     ETRS-Y     &       ETRS-Z      \\
             &          (m)      &       (m)      &         (m)       \\ 
\hline\hline
\endhead
\hline\hline
\endfoot
CS001LBA   &   3826923.942  &   460915.117  &   5064643.229  \\
CS001HBA   &   3826938.206  &   460938.202  &   5064630.436  \\
CS001HBA0  &   3826896.631  &   460979.131  &   5064657.943  \\
CS001HBA1  &   3826979.780  &   460897.273  &   5064602.929  \\
CS002LBA   &   3826577.462  &   461022.624  &   5064892.526  \\
CS002HBA0  &   3826601.357  &   460953.078  &   5064880.876  \\
CS002HBA1  &   3826565.990  &   460957.786  &   5064906.998  \\
CS003LBA   &   3826517.144  &   460929.742  &   5064946.197  \\
CS003HBA0  &   3826471.744  &   460999.814  &   5064973.941  \\
CS003HBA1  &   3826518.208  &   461034.934  &   5064935.890  \\
CS004LBA   &   3826654.593  &   460939.252  &   5064842.166  \\
CS004HBA0  &   3826586.022  &   460865.520  &   5064900.301  \\
CS004HBA1  &   3826579.882  &   460917.156  &   5064900.242  \\
CS005LBA   &   3826669.146  &   461069.226  &   5064819.494  \\
CS005HBA0  &   3826701.556  &   460988.926  &   5064802.425  \\
CS005HBA1  &   3826631.590  &   461021.491  &   5064851.999  \\
CS006LBA   &   3826597.126  &   461144.854  &   5064866.718  \\
CS006HBA0  &   3826654.179  &   461136.116  &   5064824.683  \\
CS006HBA1  &   3826612.895  &   461079.974  &   5064860.746  \\
CS007LBA   &   3826533.757  &   461098.642  &   5064918.461  \\
CS007HBA0  &   3826479.111  &   461083.396  &   5064960.857  \\
CS007HBA1  &   3826538.417  &   461169.407  &   5064908.567  \\
CS011LBA   &   3826667.465  &   461285.525  &   5064801.332  \\
CS011HBA   &   3826643.587  &   461290.469  &   5064818.809  \\
CS011HBA0  &   3826637.817  &   461227.021  &   5064828.874  \\
CS011HBA1  &   3826649.357  &   461353.917  &   5064808.743  \\
CS013LBA   &   3826346.661  &   460791.787  &   5065086.876  \\
CS013HBA   &   3826360.925  &   460814.872  &   5065074.083  \\
CS013HBA0  &   3826319.350  &   460855.801  &   5065101.590  \\
CS013HBA1  &   3826402.499  &   460773.943  &   5065046.576  \\
CS017LBA   &   3826462.450  &   461501.626  &   5064935.567  \\
CS017HBA   &   3826452.835  &   461529.655  &   5064940.251  \\
CS017HBA0  &   3826405.491  &   461507.136  &   5064977.823  \\
CS017HBA1  &   3826500.179  &   461552.174  &   5064902.678  \\
CS021LBA   &   3826406.939  &   460538.280  &   5065064.610  \\
CS021HBA   &   3826416.554  &   460510.252  &   5065059.927  \\
CS021HBA0  &   3826463.898  &   460532.770  &   5065022.354  \\
CS021HBA1  &   3826369.209  &   460487.733  &   5065097.499  \\
CS024LBA   &   3827161.630  &   461409.084  &   5064420.786  \\
CS024HBA   &   3827171.245  &   461381.055  &   5064416.102  \\
CS024HBA0  &   3827218.589  &   461403.574  &   5064378.530  \\
CS024HBA1  &   3827123.900  &   461358.537  &   5064453.675  \\
CS026LBA   &   3826391.312  &   461869.528  &   5064955.653  \\
CS026HBA   &   3826377.049  &   461846.443  &   5064968.446  \\
CS026HBA0  &   3826418.623  &   461805.513  &   5064940.939  \\
CS026HBA1  &   3826335.474  &   461887.372  &   5064995.953  \\
CS028LBA   &   3825600.841  &   461260.269  &   5065604.065  \\
CS028HBA   &   3825615.105  &   461283.354  &   5065591.272  \\
CS028HBA0  &   3825573.530  &   461324.283  &   5065618.779  \\
CS028HBA1  &   3825656.679  &   461242.425  &   5065563.765  \\
CS030LBA   &   3826014.662  &   460387.065  &   5065372.068  \\
CS030HBA   &   3826000.399  &   460363.979  &   5065384.861  \\
CS030HBA0  &   3826041.973  &   460323.050  &   5065357.354  \\
CS030HBA1  &   3825958.824  &   460404.909  &   5065412.368  \\
CS031LBA   &   3826440.392  &   460273.509  &   5065063.334  \\
CS031HBA   &   3826430.777  &   460301.538  &   5065068.018  \\
CS031HBA0  &   3826383.433  &   460279.019  &   5065105.590  \\
CS031HBA1  &   3826478.121  &   460324.057  &   5065030.445  \\
CS032LBA   &   3826891.969  &   460387.586  &   5064715.032  \\
CS032HBA   &   3826906.233  &   460410.671  &   5064702.239  \\
CS032HBA0  &   3826864.658  &   460451.600  &   5064729.746  \\
CS032HBA1  &   3826947.807  &   460369.742  &   5064674.732  \\
CS101LBA   &   3825843.362  &   461704.125  &   5065381.213  \\
CS101HBA   &   3825852.977  &   461676.097  &   5065376.530  \\
CS101HBA0  &   3825900.321  &   461698.615  &   5065338.957  \\
CS101HBA1  &   3825805.632  &   461653.578  &   5065414.102  \\
CS103LBA   &   3826304.675  &   462822.765  &   5064934.074  \\
CS103HBA   &   3826290.412  &   462799.679  &   5064946.867  \\
CS103HBA0  &   3826331.986  &   462758.750  &   5064919.360  \\
CS103HBA1  &   3826248.837  &   462840.609  &   5064974.374  \\
CS201LBA   &   3826709.325  &   461913.423  &   5064713.578  \\
CS201HBA   &   3826685.447  &   461918.367  &   5064731.055  \\
CS201HBA0  &   3826679.677  &   461854.919  &   5064741.120  \\
CS201HBA1  &   3826691.217  &   461981.815  &   5064720.989  \\
CS301LBA   &   3827413.261  &   460992.019  &   5064269.684  \\
CS301HBA   &   3827437.139  &   460987.076  &   5064252.208  \\
CS301HBA0  &   3827442.908  &   461050.523  &   5064242.143  \\
CS301HBA1  &   3827431.369  &   460923.628  &   5064262.273  \\
CS302LBA   &   3827946.312  &   459792.315  &   5063989.756  \\
CS302HBA   &   3827932.048  &   459769.230  &   5064002.547  \\
CS302HBA0  &   3827973.622  &   459728.300  &   5063975.040  \\
CS302HBA1  &   3827890.473  &   459810.159  &   5064030.053  \\
CS401LBA   &   3826766.502  &   460100.064  &   5064836.210  \\
CS401HBA   &   3826790.378  &   460095.120  &   5064818.736  \\
CS401HBA0  &   3826796.148  &   460158.570  &   5064808.669  \\
CS401HBA1  &   3826784.607  &   460031.669  &   5064828.802  \\
CS501LBA   &   3825626.175  &   460641.786  &   5065640.512  \\
CS501HBA   &   3825616.560  &   460669.815  &   5065645.196  \\
CS501HBA0  &   3825569.216  &   460647.296  &   5065682.768  \\
CS501HBA1  &   3825663.904  &   460692.334  &   5065607.623  \\
\hline
RS106LBA   &   3829261.821  &   469161.961  &   5062137.050  \\
RS106HBA   &   3829205.994  &   469142.209  &   5062180.742  \\
RS205LBA   &   3831438.959  &   463435.116  &   5061025.206  \\
RS205HBA   &   3831480.066  &   463487.205  &   5060989.643  \\
RS208LBA   &   3847810.446  &   466929.381  &   5048356.961  \\
RS208HBA   &   3847753.705  &   466962.484  &   5048396.983  \\
RS210LBA   &   3877847.841  &   467456.599  &   5025437.344  \\
RS210HBA   &   3877827.956  &   467536.277  &   5025445.321  \\
RS305LBA   &   3828721.154  &   454781.087  &   5063850.822  \\
RS305HBA   &   3828733.107  &   454692.080  &   5063850.055  \\
RS306LBA   &   3829792.203  &   452829.524  &   5063221.330  \\
RS306HBA   &   3829771.644  &   452761.378  &   5063242.921  \\
RS307LBA   &   3837941.343  &   449560.431  &   5057381.027  \\
RS307HBA   &   3837964.914  &   449626.936  &   5057357.324  \\
RS310LBA   &   3845433.443  &   413580.563  &   5054755.909  \\
RS310HBA   &   3845376.681  &   413616.239  &   5054796.080  \\
RS406LBA   &   3818468.029  &   451974.278  &   5071790.337  \\
RS406HBA   &   3818425.334  &   452019.946  &   5071817.384  \\
RS407LBA   &   3811596.257  &   453444.359  &   5076770.170  \\
RS407HBA   &   3811649.851  &   453459.572  &   5076728.693  \\
RS409LBA   &   3824756.246  &   426178.523  &   5069289.608  \\
RS409HBA   &   3824813.014  &   426130.006  &   5069251.494  \\
RS503LBA   &   3824090.848  &   459437.959  &   5066897.930  \\
RS503HBA   &   3824138.962  &   459476.649  &   5066858.318  \\
RS508LBA   &   3797202.513  &   463087.188  &   5086604.779  \\
RS508HBA   &   3797136.881  &   463114.126  &   5086651.028  \\
RS509LBA   &   3783579.528  &   450178.562  &   5097830.578  \\
RS509HBA   &   3783537.922  &   450129.744  &   5097865.889  \\
\hline
DE601LBA   &   4034038.635  &   487026.223  &   4900280.057  \\
DE601HBA   &   4034101.901  &   487012.401  &   4900230.210  \\
DE602LBA   &   4152561.068  &   828868.725  &   4754356.878  \\
DE602HBA   &   4152568.416  &   828788.802  &   4754361.926  \\
DE603LBA   &   3940285.328  &   816802.001  &   4932392.757  \\
DE603HBA   &   3940296.126  &   816722.532  &   4932394.152  \\
DE604LBA   &   3796327.609  &   877591.315  &   5032757.252  \\
DE604HBA   &   3796380.254  &   877613.809  &   5032712.272  \\
DE605LBA   &   4005681.742  &   450968.282  &   4926457.670  \\
DE605HBA   &   4005681.407  &   450968.304  &   4926457.940  \\
FR606LBA   &   4323980.155  &   165608.408  &   4670302.803  \\
FR606HBA   &   4324017.054  &   165545.160  &   4670271.072  \\
SE607LBA   &   3370287.366  &   712053.586  &   5349991.228  \\
SE607HBA   &   3370272.092  &   712125.596  &   5349990.934  \\
UK608LBA   &   4008438.796  &  -100310.064  &   4943735.554  \\
UK608HBA   &   4008462.280  &  -100376.948  &   4943716.600  \\
\hline
\end{longtable}
\tablefoot{For the stations in the Netherlansds, the nomenclature CS and RS are used to refer to "core stations" and "remote stations", respectively. See Sect.\,\ref{sec:layouts} for a description of the distinction between the two types. International LOFAR stations use a nomenclature based on the host country.}
}

\section{LOFAR performance metrics}
\label{sec:metrics}

\begin{table*}[t]                                                                                                
\begin{minipage}[t]{\textwidth}
\caption{LOFAR primary beams}
\label{tab:beams}                                                                                           
\centering
\renewcommand{\footnoterule}{}  
\begin{tabular}{lrrrrrrrrrrrrr}
\hline \hline                                                                                                
Freq. & $\lambda$ & $D$ & $A_\mathrm{eff}$ & $FWHM$ & $FOV$ & $D$ & $A_\mathrm{eff}$ & $FWHM$ & $FOV$ & $D$ & $A_\mathrm{eff}$ & $FWHM$ & $FOV$ \\
(MHz)     & (m)       & (m) & (m$^2$) & (deg)  & (deg$^2$) & (m) & (m$^2$) & (deg) & (deg$^2$) & (m) & (m$^2$) & (deg) & (deg$^2$) \\
\hline \hline \\
  &      & \multicolumn{4}{c}{NL Inner} & \multicolumn{4}{c}{NL Outer}  & \multicolumn{4}{c}{EU} \\
  &      & \multicolumn{4}{c}{\hrulefill}  & \multicolumn{4}{c}{\hrulefill} & \multicolumn{4}{c}{\hrulefill} \\
15 & 20.0 &  32.25 & 1284.0 & 39.08 & 1199.83 & 81.34 & 4488.0 & 15.49 & 188.62 & 65.00 & 3974.0 & 19.39 & 295.36 \\
30 & 10.0 &  32.25 & 848.9 & 19.55 &  299.96 & 81.34 & 1559.0 &  7.75 &  47.15 & 65.00 & 2516.0 & 9.70 & 73.84 \\
45 & 6.67 &  32.25 & 590.2 & 13.02 &  133.31 & 81.34 & 708.3 &  5.16 &  20.96 & 65.00 & 1378.0 &  6.46 &  32.82 \\
60 & 5.00 &  32.25 & 368.5 & 9.77 &  74.99 & 81.34 & 399.9 &  3.88 &  11.78 & 65.00 & 800.0 &  4.85 &  18.46 \\
75 & 4.00 &  32.25 & 243.6 &  7.82 &  47.99 & 81.34 & 256.0 &  3.10 &  7.55 & 65.00 & 512.0 &  3.88 &  11.81 \\
\\
  &      & \multicolumn{4}{c}{NL core}  & \multicolumn{4}{c}{NL Remote} & \multicolumn{4}{c}{EU} \\
  &      & \multicolumn{4}{c}{\hrulefill}  & \multicolumn{4}{c}{\hrulefill} & \multicolumn{4}{c}{\hrulefill} \\
120 & 2.50 &  30.75 & 600.0 & 4.75 & 17.73 & 41.05 & 1200.0 & 3.56 & 9.95 & 56.50 & 2400.0 & 2.59 & 5.25 \\ 
150 & 2.00 &  30.75 & 512.0 & 3.80 & 11.35 & 41.05 & 1024.0 & 2.85 & 6.37 & 56.50 & 2048.0 & 2.07 & 3.36 \\
180 & 1.67 &  30.75 & 355.6 & 3.17 & 7.88 & 41.05 & 711.1 & 2.37 &  4.42 & 56.50 & 1422.0 & 1.73 & 2.33 \\
200 & 1.50 &  30.75 & 288.0 & 2.85 & 6.38 & 41.05 & 576.0 & 2.13 &  3.58 & 56.50 & 1152.0 & 1.55 & 1.89 \\
210 & 1.43 &  30.75 & 261.2 & 2.71 &  5.79 & 41.05 & 522.5 & 2.03 &  3.25 & 56.50 & 1045.0 & 1.48 & 1.72 \\
240 & 1.25 &  30.75 & 200.0 & 2.38 &  4.43 & 41.05 & 400.0 & 1.78 &  2.49 & 56.50 & 800.0 & 1.29 & 1.31 \\
\hline \hline
\end{tabular}
\tablefoot{The full-width half-maximum (FWHM) in radians of a LOFAR Station beam is determined by $FWHM = \alpha \lambda / D$ where $\lambda$ denotes the wavelength and $D$ denotes the station diameter. The value of $\alpha$ will depend on the final tapering of the station. For these values, we have used a value of $\alpha = 1.1$ for LBA, and $\alpha = 1.02$ for HBA, as described in Sect.\,\ref{sec:beam}.}
\end{minipage}
\end{table*}

\begin{table*}[t]                                                                                                
\begin{minipage}[t]{\textwidth}
\caption{LOFAR angular resolution}
\label{tab:resolution}                                                                                           
\centering
\renewcommand{\footnoterule}{}  
\begin{tabular}{lrrrrr}                                                                                       
\hline \hline                                                                                                 
       &  & \multicolumn{4}{c}{Resolution} \\
       &  & \multicolumn{4}{c}{\hrulefill} \\
Freq.  & $\lambda$     & $L$ = 320 m & $L$ = 2 km  & $L$ = 100 km  & $L$ = 1000 km  \\
(MHz) & (m)  & (arcsec)    & (arcsec)    & (arcsec)      & (arcsec) \\
\hline \hline
 15 & 20.0 & ~~~~~10310.00 & ~~~~~1650.00 & ~~~~~33.00 & ~~~~~3.30 \\						
 30 & 10.0 & 5157.00 & 825.00 & 16.50 & 1.65 \\						
 45 & 6.67 & 3438.00 & 550.00 & 11.00 & 1.10 \\						
 60 & 5.00 & 2578.00 & 412.50 &  8.25 & 0.83 \\						
 75 & 4.00 & 2063.00 & 330.00 &  6.60 & 0.66 \\						
120 & 2.50 & 1289.00 & 206.30 &  4.13 & 0.41 \\						
150 & 2.00 & 1031.00 & 165.00 &  3.30 & 0.33 \\						
180 & 1.67 &  859.40 & 137.50 &  2.75 & 0.28 \\						
200 & 1.50 &  773.50 & 123.80 &  2.48 & 0.25 \\						
210 & 1.43 &  736.70 & 117.90 &  2.36 & 0.24 \\						
240 & 1.25 &  644.60 & 103.10 &  2.06 & 0.21 \\						
\hline \hline
\end{tabular}
\tablefoot{The resolution of the LOFAR array is given by $\alpha \lambda / L$, where $L$ denotes the longest baseline. The value of $\alpha$ depends on the array configuration and the weighting scheme used during imaging, i.e. natural, uniform, or robust. The values computed here assume a value of $\alpha = 0.8$ corresponding to a uniform weighting scheme.}
\end{minipage}
\end{table*}

\begin{table*}[t]                                                                                                
\begin{minipage}[t]{\textwidth}
\caption{LOFAR sensitivities}
\label{tab:sensitivity}                                                                                           
\centering
\renewcommand{\footnoterule}{}  
\begin{tabular}{lrrrrr}                                                                                       
\hline \hline                                                                                                 
       &  & \multicolumn{4}{c}{Sensitivity} \\
       &  & \multicolumn{4}{c}{\hrulefill} \\
Freq.  & $\lambda$     & ~~~~~~Superterp & ~~~~~~NL Core  & ~~~~~~Full NL  & ~~~~~~Full EU  \\
(MHz) & (m)  & (mJy)    & (mJy)    & (mJy)      & (mJy) \\
\hline \hline
 15 & 20.0  &     ... &    ... &   ... &   ...  \\
 30 & 10.0  &      36 &    9.0 &   5.7 &   3.8  \\
 45 & 6.67  &      29 &    7.4 &   4.7 &   3.1  \\
 60 & 5.00  &      25 &    6.2 &   3.9 &   2.6  \\
 75 & 4.00  &      44 &   10.8 &   6.8 &   4.5  \\
120 & 2.50  &     1.5 &   0.38 &  0.30 &  0.20  \\
150 & 2.00  &     1.3 &   0.31 &  0.24 &  0.16  \\
180 & 1.67  &     1.5 &   0.38 &  0.30 &  0.20  \\
200 & 1.50  &   (2.5) & (0.62) & (0.48) &  (0.32)  \\
210 & 1.43  &   (2.5) & (0.62) & (0.48) &  (0.32)  \\
240 & 1.25  &   (5.6) &  (1.4) &  (1.1) &  (0.73)  \\
\hline \hline
\end{tabular}
\tablefoot{
The quoted sensitivities are for image noise calculated assuming 8 hours of integration and an effective bandwidth of 3.66 MHz (20 subbands). The sensitivities are based on the zenith SEFD's derived from 3C295 in the Galactic halo as presented in Fig.\,\ref{fig:sensitivity}. These values assume a factor of 1.3 loss in sensitivity due to time-variable station projection losses for a declination of 30 degrees, as well as a factor 1.5 to take into account losses for ``robust'' weighting of the visibilities, as compared to natural weighting. Values for 15 MHz have not yet been determined awaiting a good station calibration. Similarly values at 200, 210, and 240 MHz should be viewed as preliminary and are expected to improve as the station calibration is improved. The procedure for determining these values along with associated caveats are discussed in more detail in Sect.\,\ref{sec:sensitivity}.}
\end{minipage}
\end{table*}

\end{appendix}

\end{document}